\documentclass[aps,prd,floatfix,showpacs,superscriptaddress,nofootinbib,10pt]{revtex4-2}
\usepackage{amsmath}
\usepackage{amssymb}
\usepackage{graphicx}
\usepackage{braket}
\usepackage{bbm}
\usepackage{xcolor}

\usepackage{subfigure}
\usepackage{tikz}
\usetikzlibrary{shapes}
\usepackage{bbm}
\usepackage{siunitx,pslatex}
\newcommand\TR{\tilde{r}}
\newcommand\TG{\tilde{g}}

\begin{document}

\title{Interplay between the Lyapunov exponents and phase transitions of charged AdS black holes}
\author{Bhaskar Shukla}\email{bhasker\_shukla@nitrkl.ac.in}\affiliation{Department of Physics and Astronomy, National Institute of Technology Rourkela, Rourkela - 769008, India}
\author{Pranaya Pratik Das}\email{pranayapratik\_das@nitrkl.ac.in}\affiliation{Department of Physics and Astronomy, National Institute of Technology Rourkela, Rourkela - 769008, India}
\author{David Dudal}\email{david.dudal@kuleuven.be}\affiliation{KU Leuven Campus Kortrijk -- Kulak, Department of Physics Etienne Sabbelaan 53 bus 7657, 8500 Kortrijk, Belgium}\affiliation{
	Ghent University, Department of Physics and Astronomy, Krijgslaan 281-S9, 9000 Gent, Belgium}
\author{Subhash Mahapatra}\email{mahapatrasub@nitrkl.ac.in}\affiliation{Department of Physics and Astronomy, National Institute of Technology Rourkela, Rourkela - 769008, India}

\begin{abstract}
We study the relationship between the standard or extended thermodynamic phase structure of various AdS black holes and the Lyapunov exponents associated with the null and timelike geodesics. We consider dyonic, Bardeen, Gauss-Bonnet, and Lorentz-symmetry breaking massive gravity black holes and calculate the Lyapunov exponents of massless and massive particles in unstable circular geodesics close to the black hole. We find that the thermal profile of the Lyapunov exponents exhibits distinct behaviour in the small and large black hole phases and can encompass certain aspects of the van der Waals type small/large black hole phase transition. We further analyse the properties of Lyapunov exponents as an order parameter and find that its critical exponent is 1/2, near the critical point for all black holes considered here.
\end{abstract}

\maketitle

\section{Introduction}
\label{sec:intro}
Chaos theory is a branch of physics and mathematics that explores the behaviour of dynamic systems highly sensitive to initial conditions, leading to ostensibly random and unpredictable outcomes. The theory originated from the pioneering work of mathematician and meteorologist Edward Lorenz~\cite{Lorenz:1963yb, Lorenzelli2014Apr}, who discovered that small changes in the initial conditions could lead to vastly different outcomes in non-integrable systems, giving rise to the famous term ``butterfly effect''. It has applications in many scientific fields and is also extensively used in the nascent field of holography~\cite{Shenker:2013pqa, Grozdanov:2017ajz, Blake:2016wvh, Lucas:2016yfl, Kudler-Flam:2020yml, Jeong:2017rxg, Ling:2017jik, Ling:2016ibq, Dong:2022ucb, Alishahiha:2016cjk, Maldacena:2015waa, Kan:2021blg, Gwak:2022xje}. For a summary of recent developments in holographic chaos, let us refer to the review paper~\cite{Jahnke:2018off}. Central to chaos theory is the concept of the Lyapunov exponent~\cite{Lyapunov1992Mar}, which measures the rate of divergence or convergence of nearby trajectories in a dynamical system~\cite{sandri1996}. Positive Lyapunov exponents indicate divergence, suggesting sensitive dependence on initial conditions, which is synonymous with chaos, while negative exponents imply stability and convergence. Lyapunov exponents provide a quantitative measure of predictability in a system; higher values signify greater unpredictability and chaotic dynamics. This mathematical tool has applications across various scientific disciplines, including physics, biology, economics, and meteorology, helping researchers understand and model the inherent complexity of natural phenomena~\cite{Colangelo:2021kmn, Shukla:2023pbp, ferriere1995lyapunov, dechert1992lyapunov, vannitsem2017predictability}.

There has been a significant interest in investigating phase transitions using chaos diagnosis tools in recent years. This investigation extends to both classical and quantum domains and has attracted great interest in many areas of physics, such as condensed matter and quantum information physics. The prime examples include the interplay of chaos and the quantum phase transition in the Dicke model \cite{Emary:2003zza}, Sachdev-Ye-Kitaev model (SYK) and its variants \cite{Sorokhaibam:2019qho, Davis:2022iqi}, long-range models of coupled oscillators \cite{Miritello:2008zd}, finite Fermi and quantum dot systems \cite{Heiss:1991zza,emary2003chaos}, etc. Two diagnostic tools for quantum chaos, Out-of-Time Order Correlators (OTOCs) and Krylov complexity have also gathered prominence as subjects of current scholarly interest~\cite{daug2019detection,sun2020out,wang2019probing,huh2021diagnosing,anegawa2024krylov}.

Lately, this correlation of chaos and phase transitions has also left its footprints in general relativity and black hole physics. The bonafide table-top laboratories that investigate such correlation between chaos and phase transition are the black holes in AdS (Anti-de Sitter) spaces. In particular, black holes in AdS spaces are not only usually thermodynamically stable but also undergo interesting phase transitions, such as Hawking/Page or small/large black hole phase transitions, and exhibit a rich phase structure \cite{hawking1983thermodynamics,chamblin1999holography}. The latter transition, in particular, is of great interest as it generally involves a first-order transition line terminating at a second-order critical point, with critical exponents identified with the mean-field type, thereby emulating the standard van der Waals type phase transition of the liquid-gas system in black hole context. These interesting phase transitions are not typically associated and observed with their asymptotically flat counterparts. After the initial finding of \cite{chamblin1999holography} on charged RN-AdS black hole, such small/large phase transitions have been observed in various AdS gravity models in different contexts. The examples are higher curvature Gauss-Bonnet black holes \cite{Cai:2013qga, Dolan:2014vba}, dyonic black holes \cite{lu2013ads, Hartnoll:2007ai, rasheed1995rotating, tseytlin1996extreme, cheng1993dyonic, priyadarshinee2021analytic, chaturvedi2017thermodynamic, dutta2013dyonic}, nonlinear electrodynamics based regular Bardeen black holes \cite{Balart:2014cga}, $R$-charged gauge supergravity inspired STU black holes \cite{Behrndt:1998jd, Duff:1999gh, Cvetic:1999xp, Sahay:2010yq}, hairy lower-dimensional black holes \cite{Priyadarshinee:2023cmi}, quasi-topological gravity \cite{Hennigar:2015esa}, $f(R)$ gravity \cite{Chen:2013ce}, etc. The phase structure of AdS black holes becomes even more interesting in the context of extended phase space thermodynamics, where the cosmological constant is interpreted as thermodynamic pressure and treated as another thermodynamic variable \cite{Kubiznak:2012wp, Dolan:2014jva, Kastor:2009wy, Dolan:2011xt, Dolan:2012jh}. See \cite{Kubiznak:2016qmn, Altamirano:2014tva} for reviews on this interesting topic.

Free energies are the standard approach to analysing the black hole phase transitions. Lately, researchers have actively sought to explore phase transitions of black holes through indirect methods, see for instance~\cite{Liu:2014gvf, Shen:2007xk, Rao:2007zzb, He:2010zb, Mahapatra:2016dae, Priyadarshinee:2023exb, Belhaj:2020nqy, Cai:2021uov, Zhang:2019glo, Wei:2017mwc, Chabab:2019kfs, Han:2018ooi, Johnson:2013dka, Dey:2015ytd, Dudal:2018ztm, Mahapatra:2019uql, Jain:2022hxl, Jain:2020rbb} etc. In such indirect methods, the physical quantities exhibit both multi-valuedness and discontinuity at the phase transition point, analogous to the thermodynamic variables in ordinary thermodynamics. Lyapunov exponents are another step in searching for indirect black hole phase transition probes. Indeed, the fact that recent studies have actively demonstrated a close relationship between the imaginary part of quasinormal modes (QNMs) and the Lyapunov exponents of unstable null geodesics~\cite{Cardoso:2008bp, Guo:2021enm} and that QNMs do exhibit noticeable changes near the phase transition indicates that pronounced modifications may occur in the structure of Lyapunov exponents near the black hole phase transition.

\begin{figure}[htb!]
	\centering
	\includegraphics[width=0.45\linewidth]{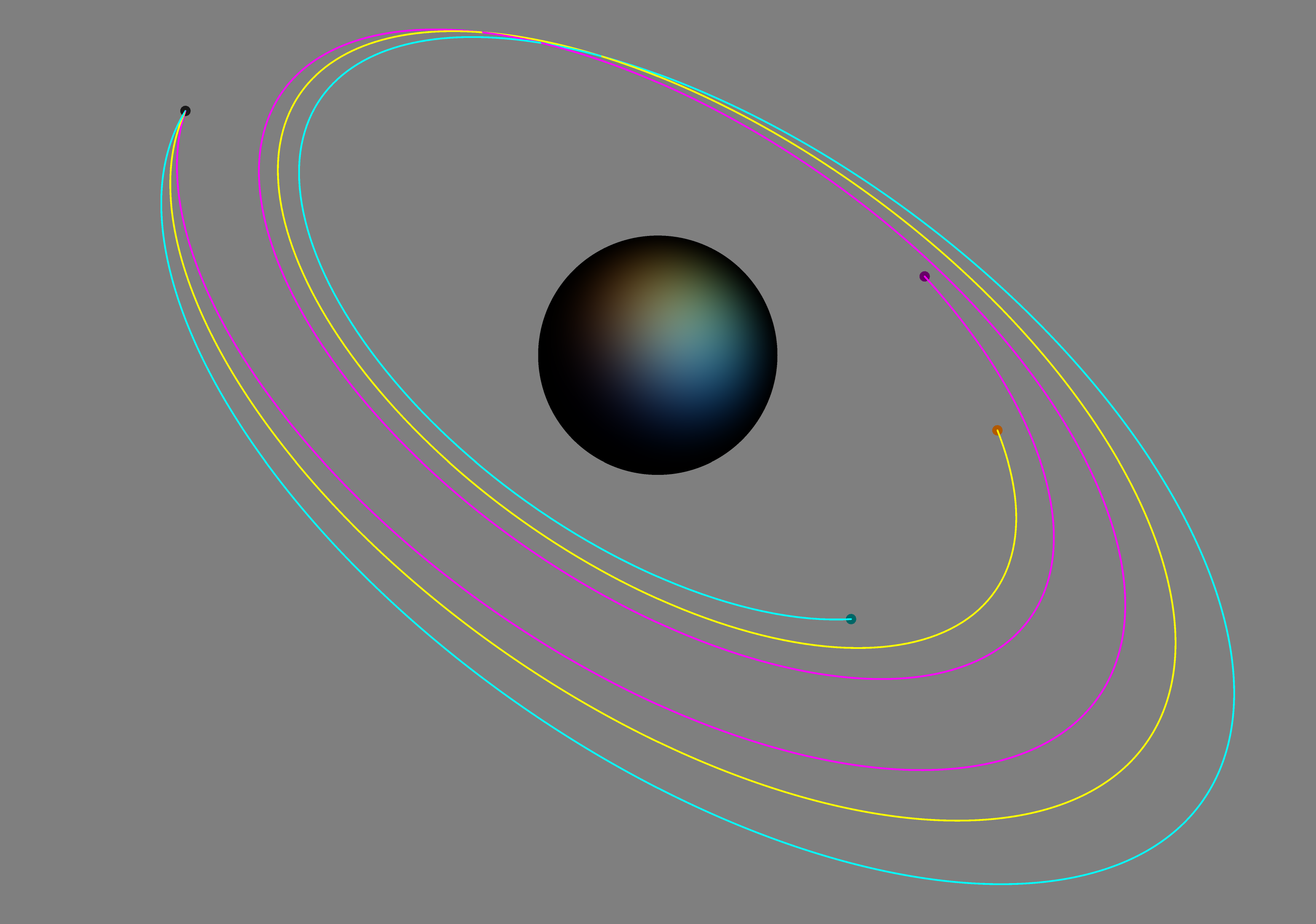}
	\caption{\label{fig:circular_geodesic}This shows a time-slice of the unstable circular geodesics of the massive particles designated by \textcolor{magenta}{$\bullet$}, \textcolor{yellow}{$\bullet$} and \textcolor{cyan}{$\bullet$} colours (with slightly perturbed specific initial positions and velocities) around a Schwarzschild black hole (with mass $M=0.5$ in natural units). The black sphere in the centre represents the black hole.}
\end{figure}

Intriguing investigations have recently started on the correlation of black hole phase transition and Lyapunov exponent \cite{guo2022probing, yang2023lyapunov, lyu2023probing, kumara2024lyapunov, Du:2024uhd, Gogoi:2024akv}.  It has been observed that the structure of black hole phase transitions is imprinted on the Lyapunov exponents of massive and massless probe particles moving around the black hole backgrounds in unstable circular geodesics. For example, we have shown a timeslice of unstable circular geodesics of massive particles around a Schwarzschild black hole in Fig.~\ref{fig:circular_geodesic}. In particular, the Lyapunov exponent exhibits multi-valuedness as a function of temperature, a scenario usually associated with the phase transition, and its thermal profile simulates the thermal profile of black hole entropy. Interestingly, the Lyapunov exponent also becomes single-valued at some critical model-dependent parameter value, thereby indirectly encoding the information about the second-order critical point.

This is fascinating, considering that in the context of gauge/gravity duality, the small/
large black hole phase transition in AdS spaces corresponds to some large $N$ phase transition, $N$ being the number of colours, such as confined/deconfined phase transition, in the dual boundary field theory at strong couplings.\footnote{The small black hole phase does not correspond to the confined phase; however, it approximately shares many of its features. For more discussion on this, see \cite{Dudal:2017max}.} Therefore, the behaviour of the Lyapunov exponent provides another way to look at the structure of confined/deconfined phases at strong couplings. This, in turn, may also be useful for studying chaotic properties of strongly coupled QCD systems, an issue not directly accessible via standard computations like lattice QCD.

In the present work, we further test this interplay of black hole phase transition and Lyapunov exponents in various AdS black holes. Specifically, we concentrate on dyonic black holes, Bardeen black holes, Gauss-Bonnet black holes, and massive gravity black holes. These choices of black holes are motivated by various physical reasons. For instance, the phase structure of dyonic black holes, which contain additional magnetic charge, is much richer than RN-AdS black holes \cite{lu2013ads, Hartnoll:2007ai, rasheed1995rotating, tseytlin1996extreme, cheng1993dyonic, priyadarshinee2021analytic, chaturvedi2017thermodynamic, dutta2013dyonic}. Hence, it would be interesting to see how the magnetic charge modifies the thermal profile of the Lyapunov exponent. Similarly, the everywhere regular Bardeen black holes are sourced by non-linear electrodynamics and do not contain any curvature singularity \cite{Ayon-Beato:2000mjt}. Therefore, it would be interesting to see how the Lyapunov exponent behaves near the phase transition involving regular black holes.
On the other hand, the Gauss-Bonnet black holes are the simplest prototype of higher-order curvature gravity theories \cite{Boulware:1985wk, Deser:2002jk}. Such higher curvature theories are expected to play an essential role in quantum gravity and arise naturally from the low-energy expansion of string theory \cite{Gross:1986mw}. The Gauss-Bonnet black holes also exhibit a van der Waals-type phase transition, and it is undoubtedly interesting to investigate whether the Lyapunov exponent can probe such a phase transition in higher curvature gravity theories or not \cite{Cai:2013qga, Dolan:2014vba,lyu2023probing}. Similarly, massive gravity theory corresponds to a non-trivial modification of general relativity in which the graviton acquires a mass. This modification changes general relativity by weakening it at large scales. Still, it leads to the exact prediction as general relativity at small scales \cite{Fierz:1939ix, deRham:2010ik, deRham:2010kj, Dvali:2000hr, Bergshoeff:2009hq, deRham:2014zqa, Hinterbichler:2011tt}. The thermodynamics of AdS massive gravity further reveals novel phase transitions in this system \cite{Cai:2014znn, Xu:2015rfa, Hendi:2017fxp, Zou:2016sab, Hendi:2016vux, Fernando:2016sps}.

In all the above-mentioned black holes, a critical value of a parameter (usually the black hole charge) appears below, and these black holes exhibit the small/large black hole phase transition. However, above this critical value, no such phase transition occurs. This can be true with and without extended thermodynamics and in different ensembles. We compute the Lyapunov exponent associated with both massive and massless particles propagating in these black hole backgrounds and find that, in all cases, the thermal profile of the Lyapunov exponent captures the essence of the phase transition. In particular, the thermal profile of the Lyapunov exponent is multi-valued for smaller parameter values, whereas it is single-valued for larger parameter values.
Interestingly, this structure of the Lyapunov exponent also allows us to pinpoint the critical value of the parameter, thereby indirectly providing information about the second-order critical point. Following \cite{lyu2023probing}, we further define the Lyapunov exponent and compute the corresponding critical exponent at the phase transition point. This critical exponent has a value of one-half in all black hole backgrounds considered here. This is true for both standard and extended-phase space thermodynamics.

The paper is organised as follows. In Sec.~\ref{sec:effective_potential}, we discuss the methodology and main recipe for computing the Lyapunov exponents associated with massless and massive particles. We then discuss the black hole phase transition and its interplay with the Lyapunov exponent (both for massless and massive particles) for various charged black holes in Sec.~\ref{sec:phase_structure}. The use of discontinuity in the Lyapunov exponent as an order parameter is illustrated in Sec.~\ref{sec:order-parameter}. Finally, we conclude our paper by highlighting the principal results in Sec.~\ref{sec:conclusions}.


\section{Methodology}
\label{sec:effective_potential}
In this section, we briefly establish the calculation relating the principal Lyapunov exponent $\lambda$ of massless and massive particles to the effective potential of unstable orbits. We will mainly follow the algorithm suggested in \cite{Cardoso:2008bp,guo2022probing}, and more details can be found there.  For this purpose, we start with the following spherically symmetric static AdS black hole background
\begin{equation}
\label{eq:general-metric}
	ds^{2} = -f(r)dt^{2}+\frac{dr^{2}}{f(r)}+r^{2}[d\theta^{2} +\sin^{2}\theta(d\phi^{2}+\sin^{2}\phi d\psi^{2})] \,,
\end{equation}
where $f(r)$ is the blackening function. Here, the metric is written in five dimensions. However, we will mainly concentrate on unstable geodesics lying on the equatorial plane with $\theta=\phi=\pi/2$, in which case the expression of the particle Lagrangian, and hence the associated principal Lyapunov exponent, will become the same in four and five dimensions. Also, we have taken the coefficients of $g_{tt}$ and $g_{rr}$ to be reciprocal, as in the case of dyonic, Bardeen, Gauss-Bonnet, and massive gravity black holes.

We follow the procedure used by Chandrasekhar~\cite{chandrasekhar1991selected} to calculate the Lagrangian of the particle's geodesic motion for the above spacetime (\ref{eq:general-metric}),
\begin{equation}
\label{eq:lagrangian}
	2\mathcal{L} = -f(r)\dot{t}^{2}+\frac{\dot{r}^{2}}{f(r)}+r^{2}\dot{\theta}^{2}+r^{2}\sin^{2}\theta\dot{\phi}^{2}+r^{2}\sin^{2}\theta \sin^{2}\phi\dot{\psi}^{2} \,,
\end{equation}
where a dot represents a derivative with respect to the proper time ($\tau$). The canonical momenta can be derived from the Lagrangian as
\begin{eqnarray}
\label{eq:canonical-momenta}
	p_{t}&=&\frac{\partial\mathcal{L}}{\partial\dot{t}}=-f(r)\dot{t}=-E \,,\\
	p_{r}&=&\frac{\partial\mathcal{L}}{\partial\dot{r}}=\frac{\dot{r}}{f(r)} \,,\\
	p_{\theta}&=&\frac{\partial\mathcal{L}}{\partial\dot{\theta}}=r^{2}\dot{\theta} \,,\\
	p_{\phi}&=&\frac{\partial\mathcal{L}}{\partial\dot{\phi}}=r^{2}\sin^{2}\theta\dot{\phi} \,,\\
	p_{\psi}&=&\frac{\partial\mathcal{L}}{\partial\dot{\psi}}=r^{2}\sin^{2}\theta \sin^{2}\phi\dot{\psi}=L \,.
\end{eqnarray}
where $E$ and $L$ are the particle's conserved energy and angular momentum, respectively. Using the above relations, the Hamiltonian $\mathcal{H}$ of the particle can be written as
\begin{equation}
\label{eq:hamiltonian1}
	\mathcal{H}=p_{t}\dot{t}+p_{r}\dot{r}+p_{\theta}\dot{\theta}+p_{\phi}\dot{\phi}+p_{\psi}\dot{\psi}-\mathcal{L} \,,
\end{equation}
which simplifies to
\begin{equation}
\label{eq:hamiltonian}
	2\mathcal{H}=-E\dot{t}+\frac{\dot{r}^{2}}{f(r)}+r^{2}\dot{\theta}^{2}+r^{2}\sin^{2}\theta\dot{\phi}^{2}+L\dot{\psi}
\end{equation}
The Hamiltonian is also equal to the norm of the tangent vector to the particle's worldline. This can be verified as
\begin{equation}
\label{eq:norm}
	\begin{split}
		\eta_{\mu}\eta^{\mu}&=g_{\mu\nu}\frac{dx^{\mu}}{d\tau}\frac{dx^{\nu}}{d\tau}\\ &
		=g_{tt}\dot{t}^{2}+g_{rr}\dot{r}^{2}+g_{\theta\theta}\dot{\theta}^{2}+g_{\phi\phi}\dot{\phi}^{2}+g_{\psi\psi}\dot{\psi}^{2}\\&
		=-f(r)\dot{t}^{2}+\frac{\dot{r}^{2}}{f(r)}+r^{2}\dot{\theta}^{2}+r^{2}\sin^{2}\theta\dot{\phi}^{2}+r^{2}\sin^{2}\theta \sin^{2}\phi\dot{\psi}^{2}\\&
		=-E\dot{t}+\frac{\dot{r}^{2}}{f(r)}+r^{2}\dot{\theta}^{2}+r^{2}\sin^{2}\theta\dot{\phi}^{2}+L\dot{\psi}\\&
		=\epsilon
	\end{split}
\end{equation}
Note that $\epsilon = -1$, resp.~$0$ corresponds to massive, resp.~massless particles moving along timelike, resp.~null geodesics. Now, restricting our attention to the equatorial plane $\theta=\phi=\pi/2$, we have
\begin{equation}
\label{eq:epsilon}
	\epsilon=-E\dot{t}+\frac{\dot{r}^{2}}{f(r)}+L\dot{\psi} \,,
\end{equation}
which can be rewritten in the form of a radial equation of motion as
\begin{equation}
\label{eq:radial-motion}
	\dot{r}^{2}+V_{\text{eff}}(r)=E^{2} \,,
\end{equation}
where we introduce the effective potential
\begin{equation}
\label{eq:Veffective}
	V_{\text{eff}}(r)=f(r)\left[\frac{L^{2}}{r^{2}}-\epsilon\right] \,,
\end{equation}
To determine the radius ($r_0$) of an unstable circular geodesic, we use the conditions
\begin{equation}
\label{eq:Veffective-primes}
\begin{aligned}
	V^{'}_{\text{eff}}(r_0) = 0 \,,
	\qquad
	V^{''}_{\text{eff}}(r_0) < 0 \,.
\end{aligned}
\end{equation}
Now, using Eq.~(\ref{eq:Veffective}) we can rewrite the Hamiltonian~(\ref{eq:hamiltonian1}) as:
\begin{equation}
\label{eq:hamiltonian-simplified}
\mathcal{H} = \frac{V_{\text{eff}}(r)-E^{2}}{2f(r)}+\frac{f(r)p^{2}_{r}}{2}+\frac{\epsilon}{2} \,,
\end{equation}
which gives us the following equations of motion,
\begin{equation}
\label{eq:equations-of-motion}
\begin{aligned}
    \dot{r} &= f(r) p_{r} \,, \\
	\dot{p_{r}} &= -\frac{V^{'}_{\text{eff}}(r)}{2f(r)}-\frac{f'(r)p_{r}^{2}}{2}+\frac{V_{\text{eff}}(r)-E^{2}}{2f^{2}(r)}f'(r) \,.
\end{aligned}
\end{equation}
Now, we linearise the above equations around the unstable circular orbit at $r=r_0$, which gives us,
\begin{equation}
\label{eq:linearized-equations}
\begin{pmatrix}
\frac{d(\delta r)}{d\tau} \\
\frac{d(\delta p_r)}{d\tau}
\end{pmatrix}
=
K
\begin{pmatrix}
\delta r \\
\delta p_r
\end{pmatrix}
\,,
\end{equation}
where $K$ is the linear stability matrix~\cite{Cornish:2003ig} and is given by,
\begin{equation}
\label{eq:Kmatrix}
K = \begin{pmatrix}
    0 & K_1 \\ K_2 & 0
\end{pmatrix}
\,,
\end{equation}
The values of $K_1$ and $K_2$ can be found using Eqs.~(\ref{eq:radial-motion}) and~(\ref{eq:Veffective-primes}) and are given as,
\begin{equation}
\label{eq:Kvalues}
\begin{aligned}
    K_1 &= f(r_0)\dot{t}^{-1} \,, \\
    K_2 &= -\frac{V^{''}_{\text{eff}}(r_0)}{2f(r_0)}\dot{t}^{-1} \,.
\end{aligned}
\end{equation}
The relation which gives the principal Lyapunov exponent for the circular orbits is then derived as \cite{Cardoso:2008bp}
\begin{equation}
\label{eq:lyapunov-relation}
\lambda = \pm \sqrt{K_1 K_2} \,.
\end{equation}
Using the values of $K_1$ and $K_2$ from Eq.~(\ref{eq:Kvalues}) in Eq.~(\ref{eq:lyapunov-relation}) and dropping the $\pm$ sign, we get
\begin{equation}
\label{eq:lyapunov-general-formula}
\lambda = \sqrt{-\frac{V^{''}_{\text{eff}}(r_0)}{2\dot{t}^{2}}} \,.
\end{equation}

Now, we use Eqs.~(\ref{eq:radial-motion}) and~(\ref{eq:Veffective-primes}) to find the expression of $\dot{t}$ in both massless and massive cases. For the massless particles ($\epsilon=0$), we find  that
\begin{equation}
\label{eq:tdot-massless}
    \dot{t}=\frac{L}{r_0\sqrt{f(r_0)}} \,,
\end{equation}
which gives us the Lyapunov exponent of the massless particle in an unstable circular orbit~(\ref{eq:lyapunov-general-formula}) as
\begin{equation}
\label{eq:lyapunov-massless}
	\lambda=\sqrt{-\frac{r_{0}^{2}f(r_{0})}{2 L^{2}}V''_{\text{eff}}(r_{0})} \,.
\end{equation}
For the massive particles ($\epsilon=-1$), we have
\begin{equation}
\label{eq:tdot-massive}
	\dot{t}=\frac{2}{2 f(r_0) - r_0 f'(r_0)} \,,
\end{equation}
from which we can write the formula for the Lyapunov exponent of the massive particles in an unstable circular orbit~(\ref{eq:lyapunov-general-formula}) as
\begin{equation}
\label{eq:lyapunov-massive}
	\lambda=\frac{1}{2}\sqrt{\left(r_0 f'(r_0) - 2 f(r_0)\right)V''_{\text{eff}}(r_0)} \,.
\end{equation}

\section{Thermodynamics and Phase Structure of charged black holes based on Lyapunov exponents}\label{sec:phase_structure}
In this section, we will review the thermodynamic properties of four important classes of black holes, namely dyonic, Bardeen, Gauss-Bonnet, and massive gravity in AdS spaces, and test the conjectured relationship between phase transitions and Lyapunov exponents.

\subsection{Dyonic Black Holes}
Because of electromagnetic duality in four dimensions, it is feasible to obtain black hole solutions with electric and magnetic charges. Such black holes are called dyonic black holes and can be considered as the simplest generalisations of RN-AdS black holes \cite{lu2013ads, rasheed1995rotating, tseytlin1996extreme, cheng1993dyonic, priyadarshinee2021analytic}. The magnetic charge not only enriches the thermodynamic phase diagram of RN-AdS black holes but also acquaints the dual boundary theory with a background magnetic field, thereby making these solutions relevant for the holographic study of, for instance, the Hall effect \cite{Hartnoll:2007ai}, ferromagnetism \cite{dutta2013dyonic} etc.

Dyonic black holes are solutions of the Einstein-Maxwell gravity system. The details of the solutions can be found in \cite{dutta2013dyonic}. Its line element reads as
\begin{equation}\label{eq:dyonic_metric}
	ds^2 = -f(r)dt^2 + \frac{1}{f(r)}dr^2 + r^2(d\theta^2 + \sin^2\theta d\phi^2)\,,
\end{equation}
where the blackening function $f(r)$ is given by
\begin{equation}\label{eq:dyonic_metric_function}
	f(r) = 1 + \frac{r^2}{l^2} - \frac{2M}{r} + \frac{(q_e^2 + q_m^2)}{r^2}\,,
\end{equation}
here $q_e$, $q_m$, $l$, and $M$ are the electric charge, magnetic charge, AdS radius and the mass of the black hole, respectively. From the gauge field ($A$) solution
\begin{equation}\label{gaugesolutiondyonic}
A = q_e\bigg(\frac{1}{r_h} - \frac{1}{r}\bigg)dt + q_m \cos{\theta}d\phi ,
\end{equation}
one can obtain the relation between the electric potential ($\phi_e$) and charge $q_e$
\begin{equation}\label{eq:dyonic_electric_potential}
	\phi_e = \frac{q_e}{r_h},
\end{equation}
where $r_h$ is the radius of the black hole event horizon specified from the condition $f(r_h)=0$. The entropy ($S$) of the dyonic black hole is
\begin{equation}\label{eq:dyonic_entropy}
	S = \frac{\pi r_h^2}{G_4} \,,
\end{equation}
and its temperature is given by
\begin{equation}\label{eq:dyonic_temperature}
T = \frac{f'(r_h)}{4 \pi} =  \frac{1}{4 \pi  r_h^3} \left(-\phi _e^2 r_h^2+\frac{3 r_h^4}{l^2}+r_h^2-q_m^2 \right)\,,
\end{equation}
where $G_4$ is the four-dimensional Newton's constant.  We set $G_4=1$ from here on.

We are mainly interested in the dyonic black hole thermodynamics in the grand-canonical ensemble, where we keep the potential $\phi_e$ fixed. The corresponding Gibbs free energy can be obtained from the standard holographic renormalisation procedure and is given  by
\begin{eqnarray}
\label{eq:dyonic_free_energy_formula}
G &=& M - TS - \phi_e q_e \nonumber \\
&=& \frac{1}{4 G_4} \left(-\phi _e^2 r_h - \frac{r_h^3}{l^2} + \frac{3q_m^2}{r_h} + r_h \right)\,.
\end{eqnarray}
Before we explicitly discuss the phase structure of the dyonic black hole in the grand-canonical ensemble, it is important to mention that here, we are treating the magnetic field as a constant parameter and not a thermodynamic variable. This is a consistent thermodynamic setup as it leads to standard thermodynamic relations. Moreover, the fact that the pressure is equal to the negative of Gibbs free energy, i.e., $P=-G$, also signifies the correctness of this thermodynamic setup.

Below, we discuss the standard thermodynamics of the dyonic black hole in the grand-canonical ensemble. One can perform a similar analysis in the extended phase space as well. Since the thermodynamic results of the extended phase space are quite identical to the above-discussed results, we, therefore, concentrate only on standard thermodynamics without losing any generality.

\begin{figure}[htb!]
	\centering
	\subfigure[Temperature vs horizon radius]{\label{fig:dyonic_Tvsrh_collage}
		\includegraphics[width=0.45\linewidth]{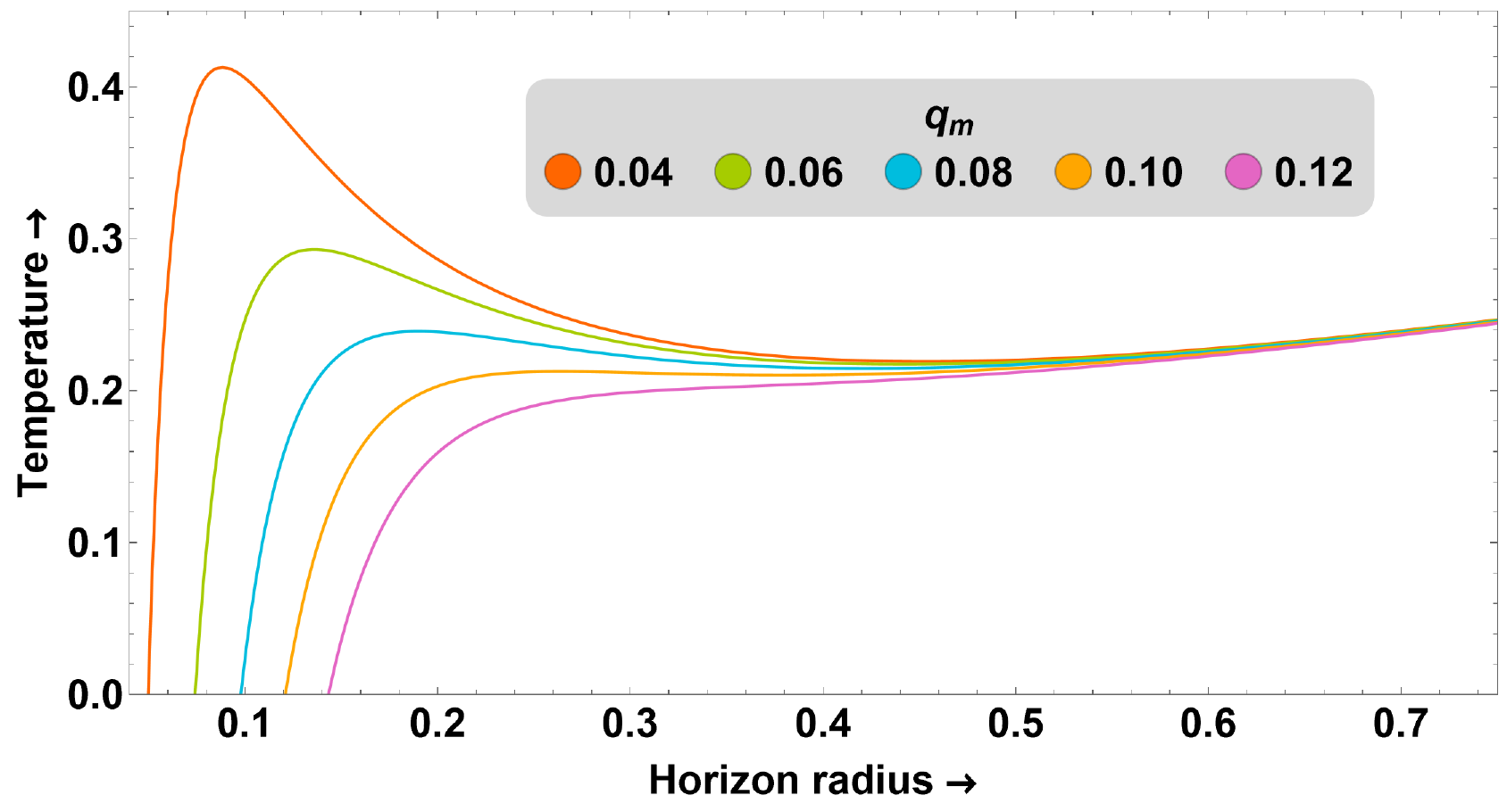}}
	\hfill
	\subfigure[Gibbs free energy vs temperature]{\label{fig:dyonic_GvsT_collage}
		\includegraphics[width=0.45\linewidth]{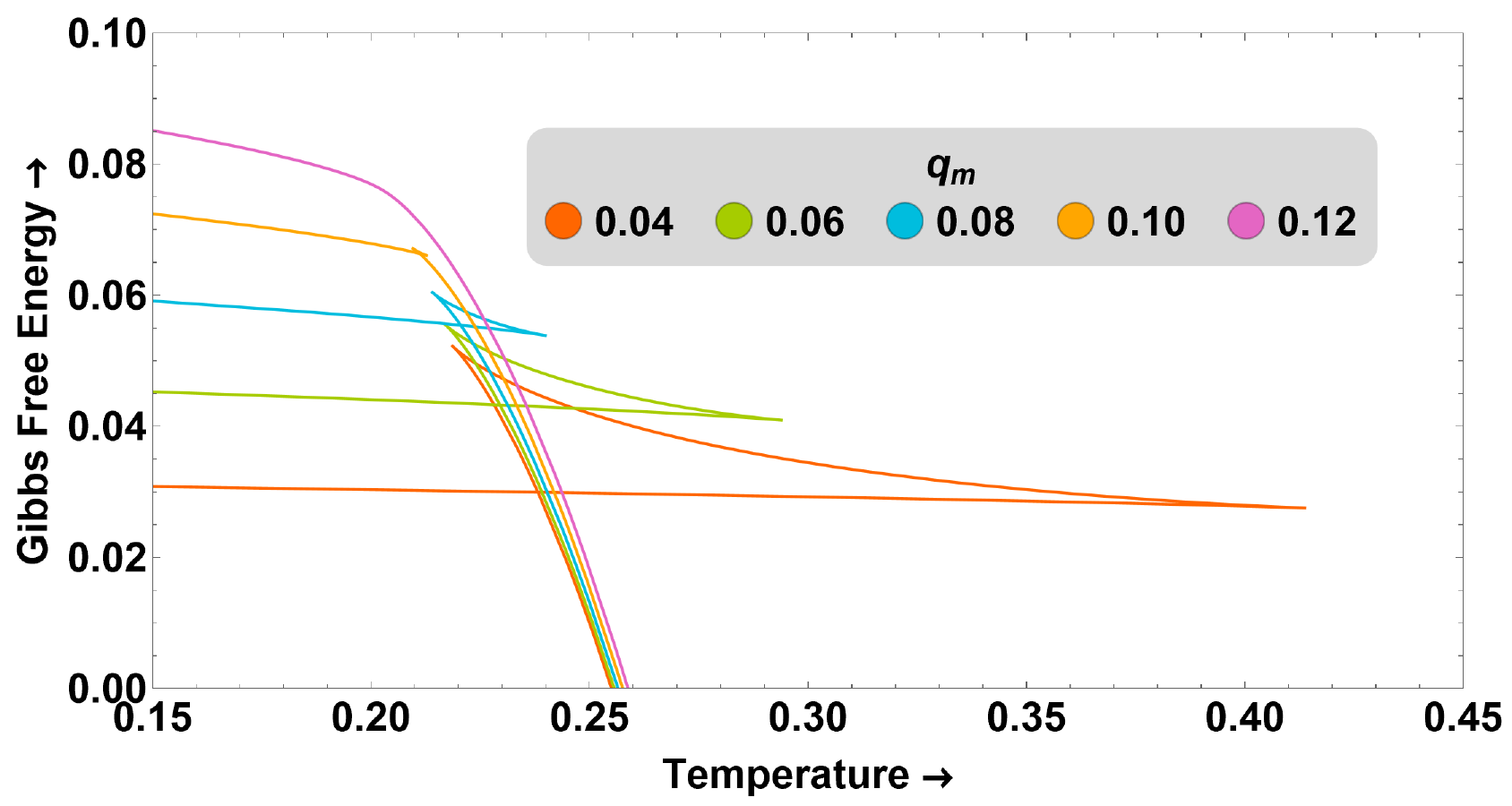}}
    \vfill
    \subfigure[$q_m = 0.05 < q_{mc}$]{\label{fig:dyonic_GvsT_1}
		\includegraphics[width=0.45\linewidth]{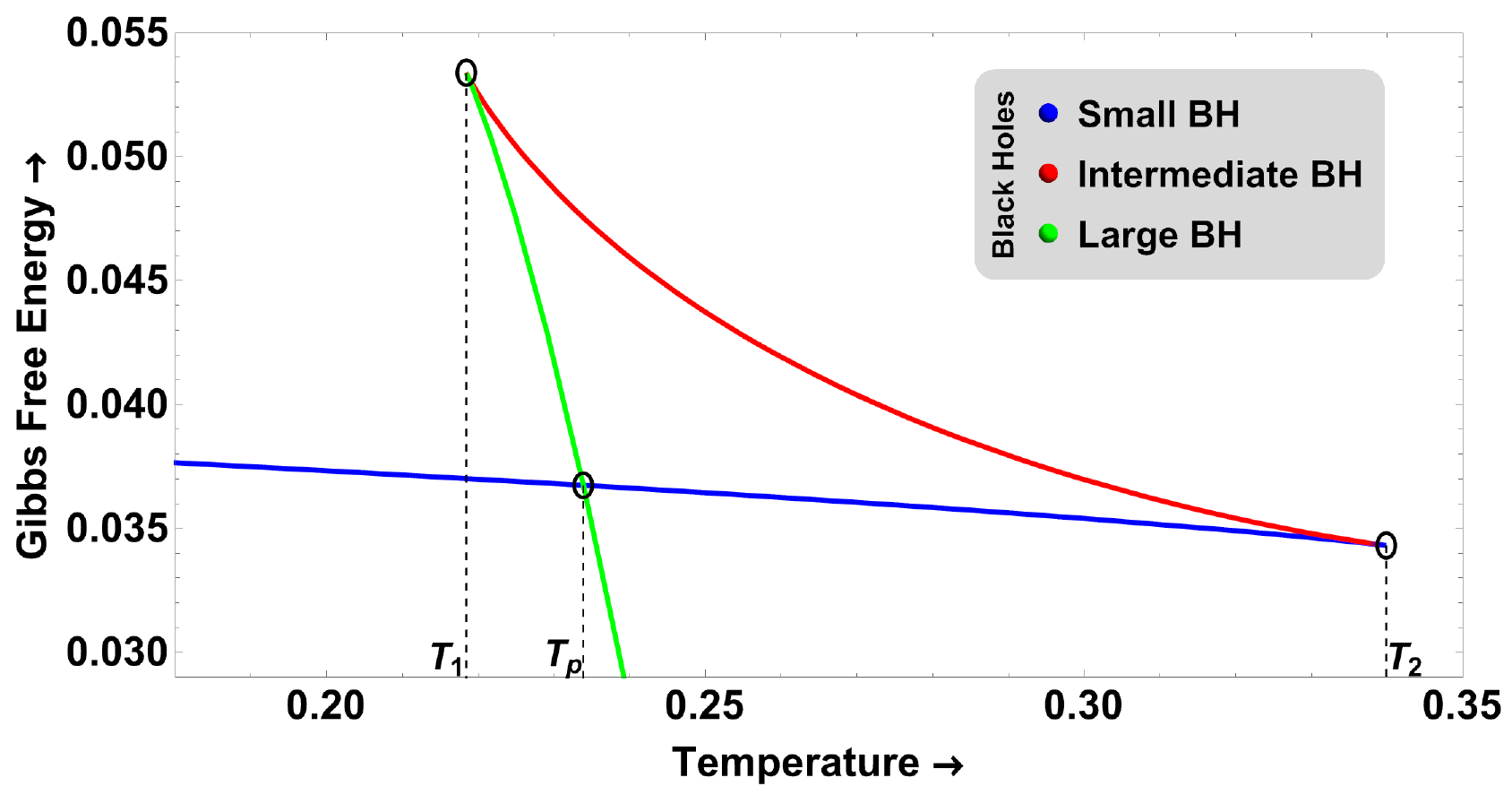}}
	\hfill
	\subfigure[$q_m = 0.12 > q_{mc}$]{\label{fig:dyonic_GvsT_2}
		\includegraphics[width=0.45\linewidth]{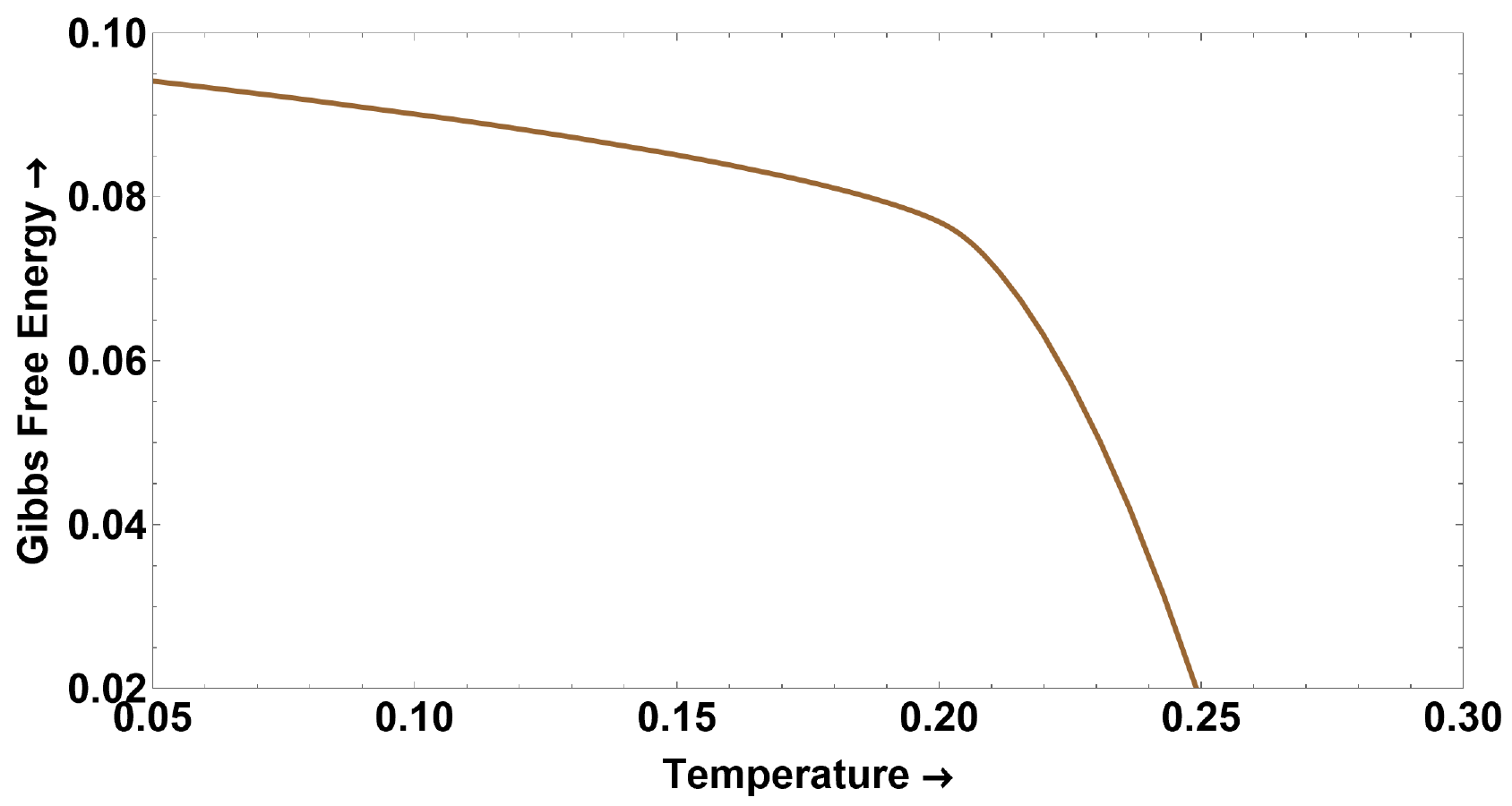}}
	\caption{\label{fig:dyonic_GvsTandTvsrh}The thermodynamic phase structure of the dyonic black hole. Here, $\phi_e = 0.6$ is used.}
\end{figure}

The thermodynamic phase structure of the dyonic black hole in the grand-canonical ensemble is shown in Fig.~\ref{fig:dyonic_GvsTandTvsrh}.\footnote{Here we have set $l=1$ and $G_4=1$.}  For $\lvert\phi_e\rvert < 1$, there appear three black hole branches -- large, intermediate, and small -- for small values of $q_m$. These three branches are explicitly shown in Fig.~\ref{fig:dyonic_GvsT_1}. The large black hole (present when $T>T_2$) and small black hole (appear when $T<T_1$) phases are thermodynamically stable, whereas the intermediate phase (present when $T_1 \leq T \leq T_2$) is thermodynamically unstable. The behaviour of the corresponding Gibbs free energy as a function of temperature is shown in Fig.~\ref{fig:dyonic_GvsT_collage}. It exhibits the standard swallow-tail-like structure, a bonafide behaviour generally associated with the first-order phase transition. The free energy is single-valued when $T<T_1$ or $T>T_2$ is multi-valued in the temperature range $T_1 \leq T \leq T_2$.  The Gibbs free energy of small and large black hole phases also exchange dominance as the temperature varies. In particular, the small black hole phase is thermodynamically favoured at low temperatures, whereas the large black hole phase is favoured at high temperatures. This suggests a phase transition between the large and small black hole phases. This phase transition happened at $T=T_p$. The free energy of the intermediate phase is always higher than the small and large black hole phases and is, therefore, always disfavoured.

By increasing $q_m$ but keeping $\phi_e$ fixed, there appears a critical value $q_{mc}$ above which the three black hole branches merge to form a single black hole branch that remains stable at all temperatures. This implies that the small/large black hole phase transition continues to exist for $q_m<q_{mc}$, whereas no such phase transition appears above $q_{mc}$. Therefore, the $q_{mc}$ defines a second-order critical point at which the first-order phase transition line between the small and large black hole phases stops. This behaviour is analogous to the van der Waals-type phase transition generally seen in liquid/gas systems. Also, note that the small/large transition temperature $T_p$ decreases with $q_m$, as can be explicitly seen from Fig.~\ref{fig:dyonic_GvsTandTvsrh}.

Moreover, this $q_{mc}$ is a $\phi_e$ dependent quantity. The magnitude of the critical point can be determined from the condition of the inflection point
\begin{equation}\label{eq:dyonic_critical_relation}
\begin{aligned}
\frac{\partial T}{\partial r_h} = 0\,,
\qquad
\frac{\partial^2 T}{\partial r_h^2} = 0\,.
\end{aligned}
\end{equation}
This gives us $\phi_e$ dependent critical values
\begin{equation}\label{eq:dyonic_critical_values}
\begin{aligned}
q_{mc}= \frac{l}{6}\left(1-\phi_e^2\right)\,,
\qquad
r_{hc}= \frac{l \sqrt{1-\phi _e^2}}{\sqrt{6}}\,.
\end{aligned}
\end{equation}
These critical relations are true only for $\lvert\phi_e\rvert<1$. Therefore, the small/large phase transition exists only for small $\lvert\phi_e\rvert<1$, whereas no such phase transition exists for large $\lvert\phi_e\rvert>1$. Let us also mention the special case where $\phi_e=0$ and $q_m=0$ correspond to the AdS-Schwarzschild black hole. In this case, only the Hawking/Page phase transition between a stable black hole and thermal-AdS occurs, and no small/large phase transition appears.

\begin{figure}[htb!]
	\centering
	\includegraphics[width=0.4\linewidth]{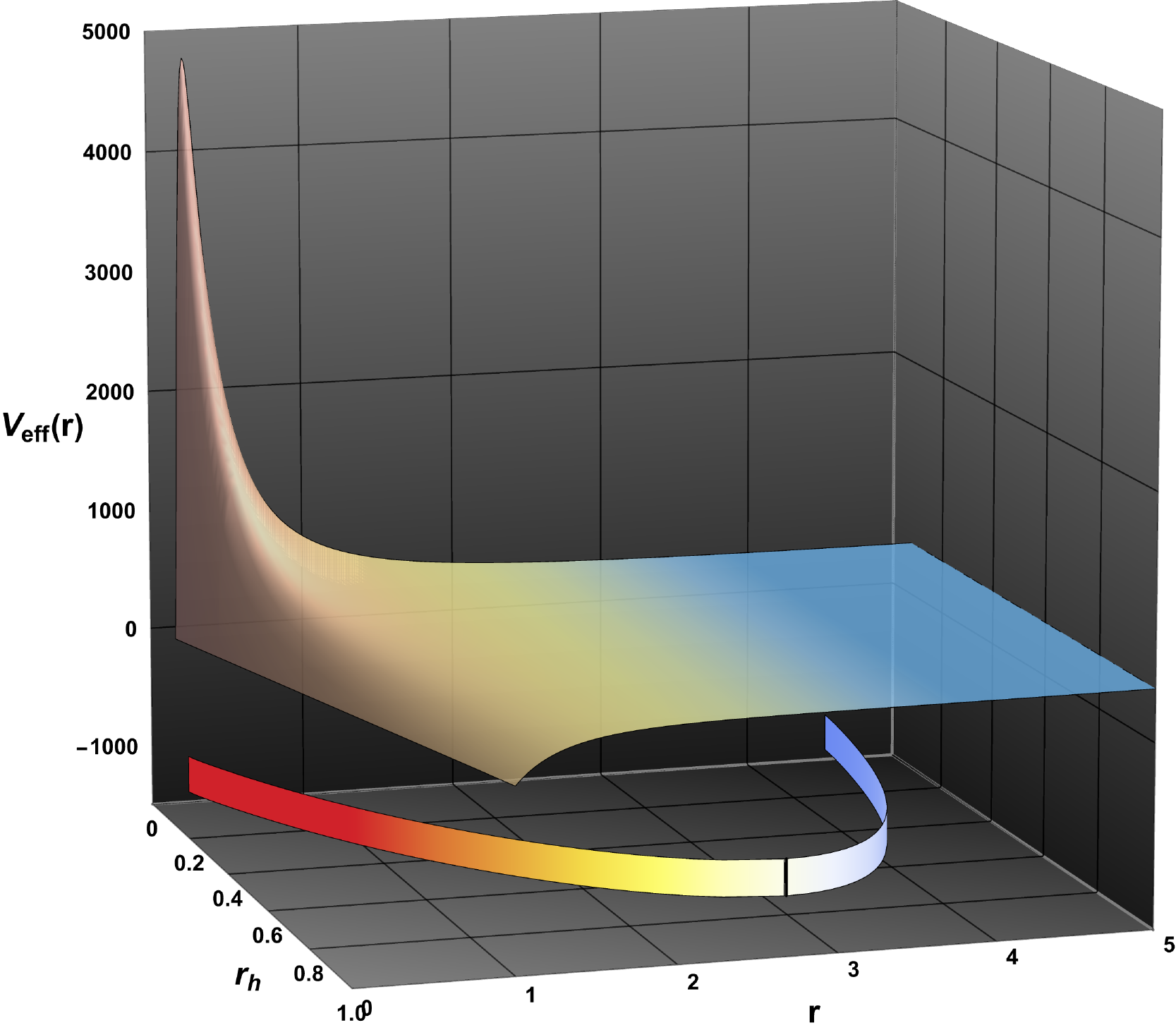}
	\caption{\label{fig:dyonic_veffective3d}The 3D plot of the effective potential $V_\text{eff}(r)$ as a function of horizon radius $r_h$ and  orbit radius $r$ of the massive particle when $L = 20$, $\phi_e = 0.6$, and $q_m = 0.06$. The red and blue curves projected below correspond to the unstable and stable equilibria of the circular geodesic, respectively.}
\end{figure}

Before explicitly analysing the Lyapunov exponent associated with the massless and massive particles moving in the dyonic black hole background, let us briefly discuss their effective potential. The expression for the effective potential~(\ref{eq:Veffective}) using Eq.~(\ref{eq:dyonic_metric_function}) can be simplified as,
\begin{equation}\label{eq:dyonic_Veffective}
   V_{\text{eff}}(r) = \frac{\left(r-r_h\right) \left(L^2-r^2 \epsilon \right) \left(r_h \left(r \left(r_h \left(r_h+r\right)+l^2+r^2\right)-l^2 \phi _e^2 r_h\right)-l^2 q_m^2\right)}{l^2 r^4 r_h}\,,
\end{equation}
where $L$ and $r$ are the angular momentum of the particle and the radius of the particle's orbit, respectively. The three-dimensional plot of the effective potential $V_{\text{eff}}(r)$ as a function of $r_h$ and $r$ is shown in Fig.~\ref{fig:dyonic_veffective3d}. The 3D plot remains the same for both the massless and massive particles. Here we have set $L=20$, $\phi_e=0.6$ and $q_m=0.06$ for a unit AdS radius ($l=1$).

The curve projected onto the $V_{\text{eff}}(r)=-1500$ plane represents the extrema of the effective potential for the massive particle ($\epsilon=-1$). The reddish yellow part represents the maxima or the radii of the unstable timelike circular geodesics given by $V^{'}_{\text{eff}}(r_0)=0$ and $V^{''}_{\text{eff}}(r_0)<0$, while the blueish-white part represents the minima or the radii of the stable time like circular geodesic given by $V^{'}_{\text{eff}}(r_0)=0$ and $V^{''}_{\text{eff}}(r_0)>0$. As we increase the $r_h$ slowly, the maxima and the minima shift. At a particular $r_h=0.827$, they become equal, represented by the black line on the curve, after which there are no extrema for the massive particle. We are only interested in the unstable circular geodesics, which are important in calculating the Lyapunov exponents.

\subsubsection{Massless particles}
\begin{figure}[htb!]
	\centering
    \textbf{Massless particles}\par\medskip
	\subfigure[$q_m = 0.05 < q_{mc}$]{\label{fig:dyonic_Lmassless_vsT_1}
		\includegraphics[width=0.45\linewidth]{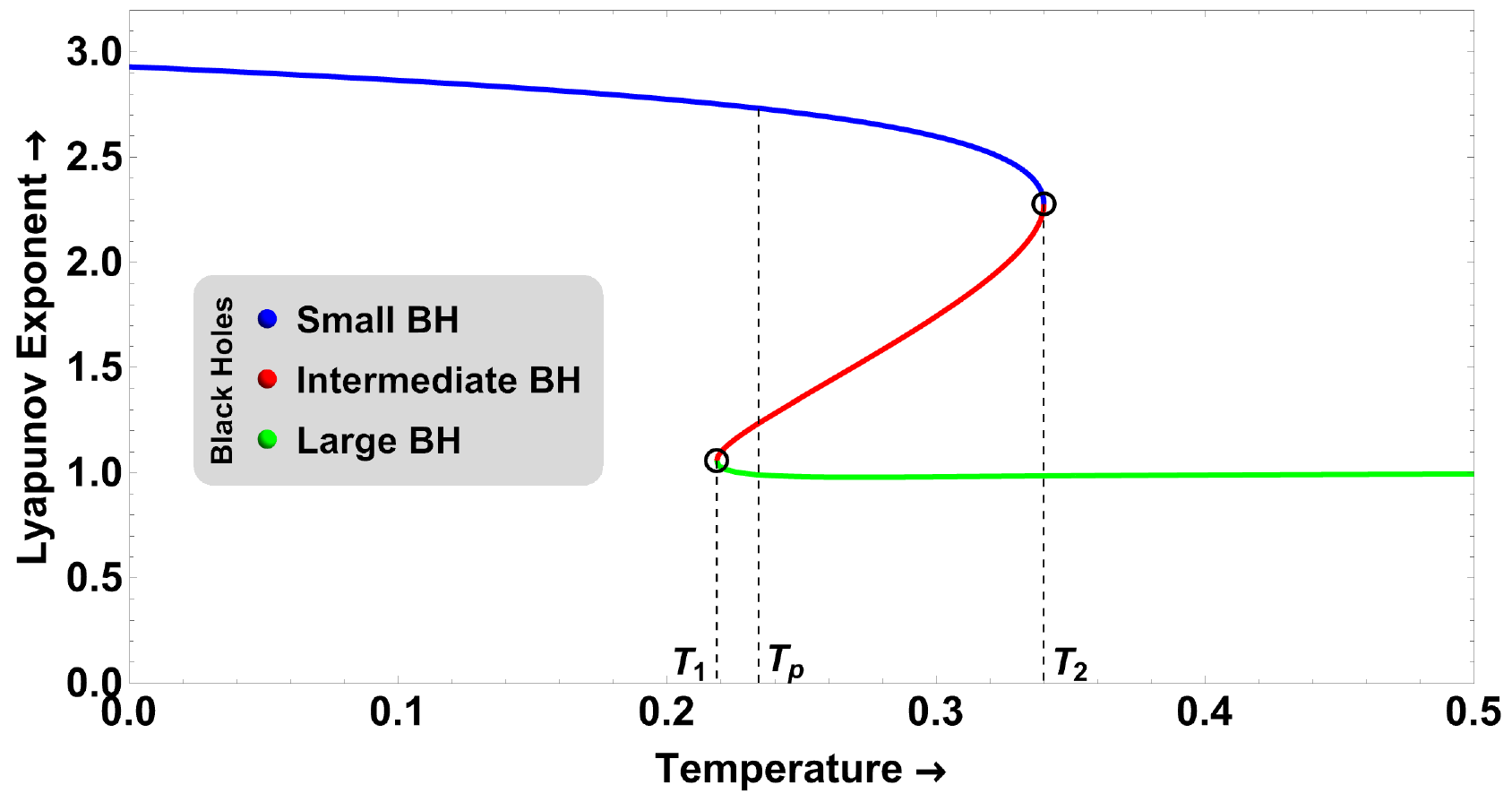}}
	\hfill
	\subfigure[$q_m = 0.12 > q_{mc}$]{\label{fig:dyonic_Lmassless_vsT_2}
		\includegraphics[width=0.45\linewidth]{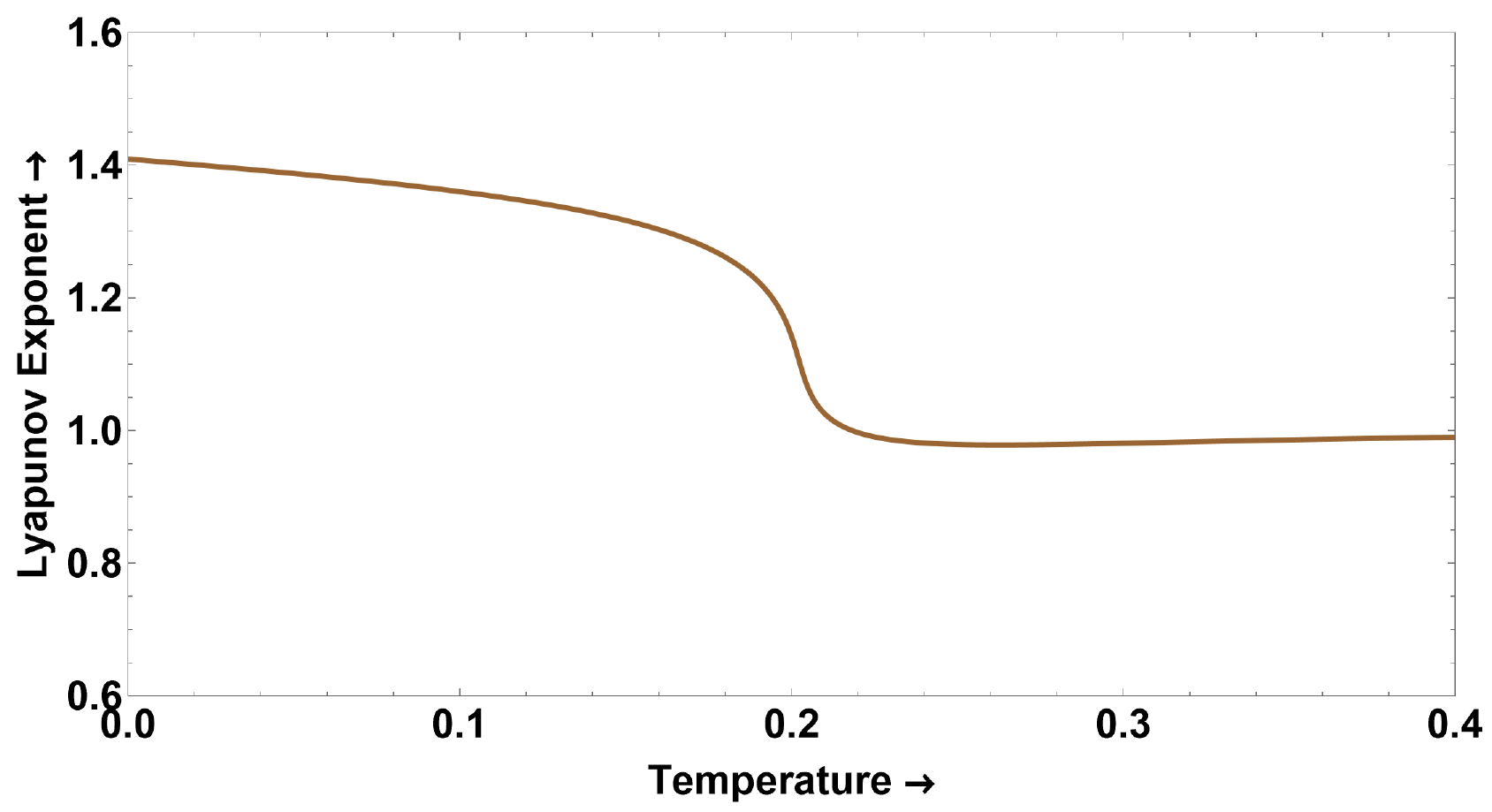}}
	\vfill
	\subfigure[Lyapunov exponent vs temperature]{\label{fig:dyonic_Lmassless_vsT_collage}
		\includegraphics[width=0.45\linewidth]{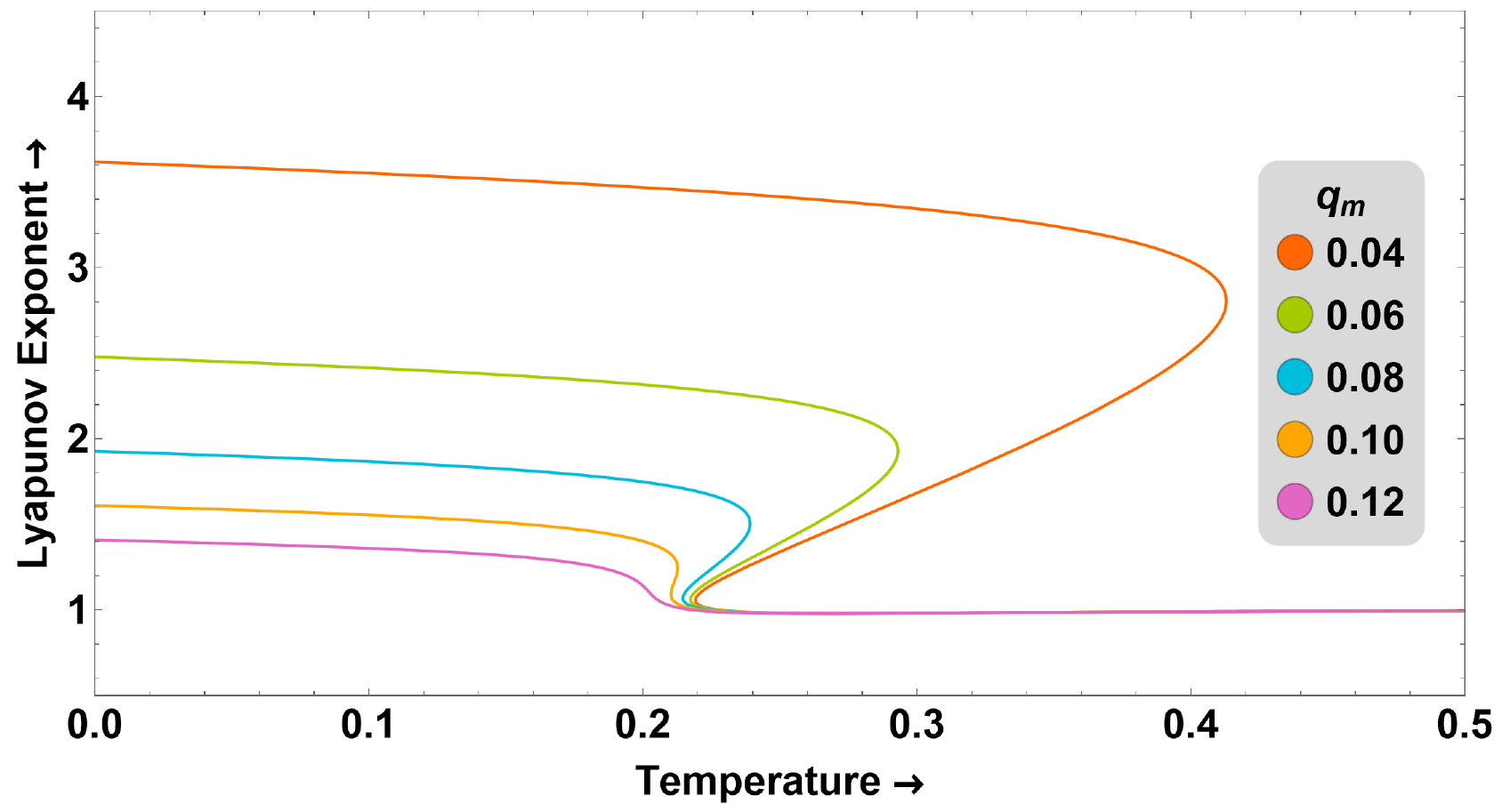}}
    \hfill
	\subfigure[Lyapunov exponent vs horizon radius]{\label{fig:dyonic_Lmassless_vsrh_collage}
		\includegraphics[width=0.45\linewidth]{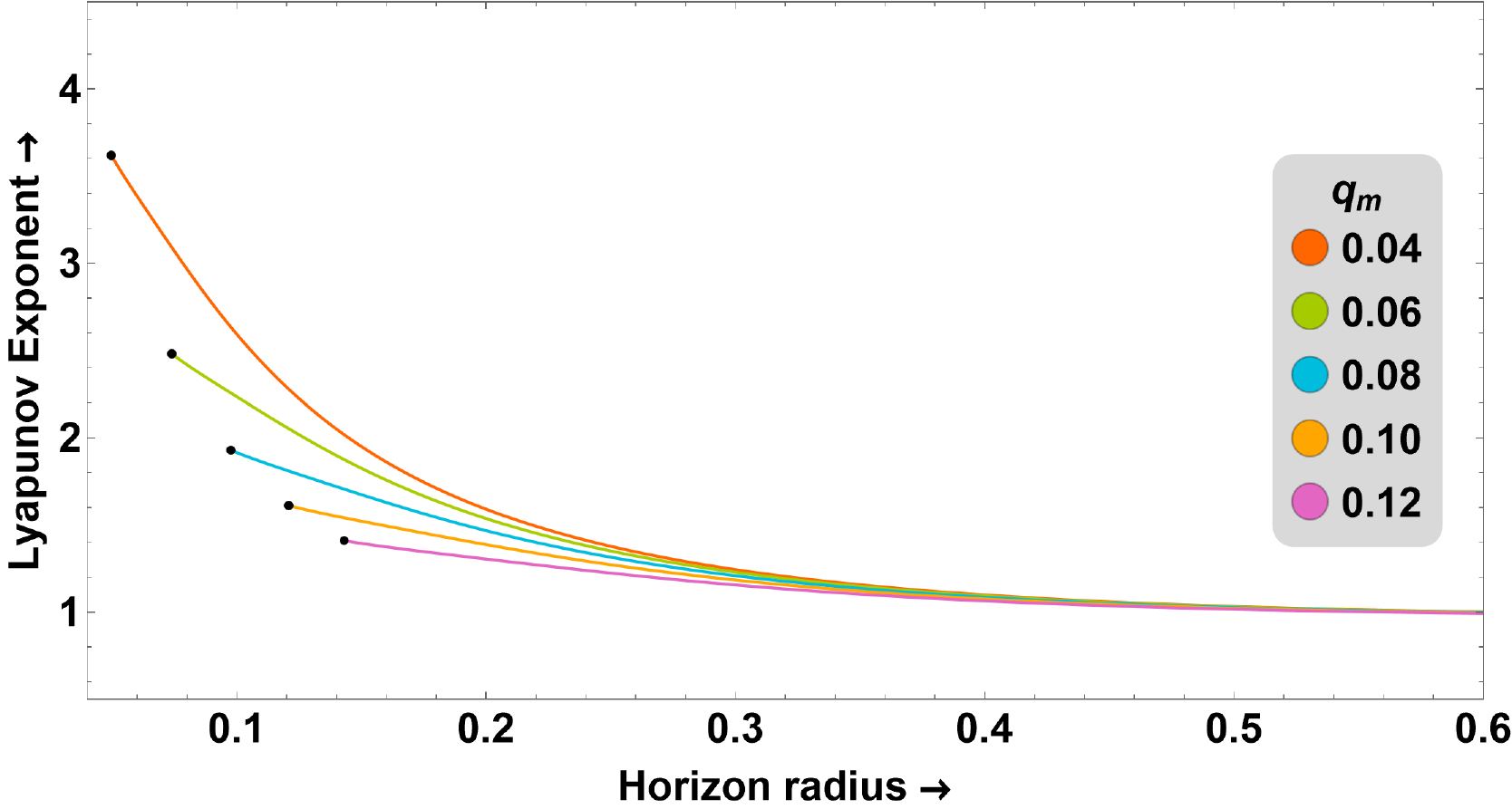}}
	\caption{\label{fig:dyonic_Lmassless}Lyapunov exponent of the massless particle $\lambda$ as a function of temperature $T$ and horizon radius $r_h$ for the dyonic black hole. Here, $\phi_e = 0.6$ is used.}
\end{figure}
We now compute the Lyapunov exponent's thermal profile associated with massless or massive particles moving in the dyonic black hole background. For the massless particle, this can be done using Eqs.~(\ref{eq:Veffective}) and (\ref{eq:lyapunov-massless}). The Lyapunov exponent as a function of dyonic black holes' temperature and horizon radius is shown in Fig.~\ref{fig:dyonic_Lmassless}. Here, the same colour coding and parameter values as in Fig.~\ref{fig:dyonic_GvsTandTvsrh} are used. When the magnetic charge $q_m$ is less than the critical value $q_{mc}$, the Lyapunov exponent exhibits multi-valuedness in some temperature range. As we increase $T$ from $0$ to $T_1$, $\lambda$ decreases gradually, as shown by the blue curve in Fig.~\ref{fig:dyonic_Lmassless_vsT_1}. At $T_1$, the Lyapunov exponent becomes multi-valued. In particular, in the temperature range $T_1\leq T \leq T_2$, there are two profiles: one for which the Lyapunov exponent decreases with temperature and one for which it increases with temperature. The former two profiles appear in the small and large black hole phases, whereas the latter profile appears in the intermediate phase. The Lyapunov exponents in the small, intermediate, and large black hole phases are represented by blue, red, and green lines, respectively, in Fig.~\ref{fig:dyonic_Lmassless_vsT_1}. The multi-valued nature of $\lambda$ corresponds to the swallow tail in the $G-T$ diagram. For convenience, we have also shown the small/large black hole transition temperature $T_p$ in Fig.~\ref{fig:dyonic_Lmassless}. The Lyapunov exponent becomes single-valued for $T>T_2$ when $q<q_{mc}$. This behaviour is again similar to the free energy behaviour. For large temperatures, the Lyapunov exponent approaches a constant value.

For $q>q_{mc}$, when there is no small/large black hole phase transition, the Lyapunov exponent remains single-valued for all temperatures. In particular, it gradually decreases and then attains a constant value as we increase the temperature. This is shown in Fig.~\ref{fig:dyonic_Lmassless_vsT_2}. The overall behaviour of the Lyapunov exponent as a function of temperature for different magnetic charge values $q_m$ is shown in Fig.~\ref{fig:dyonic_Lmassless_vsT_collage}. Here, we can see that, just like for the free energy, the thermal profile of the Lyapunov exponent becomes multi-valued for only those values of $q_m$ which are less than $q_{mc}$. Importantly, this behaviour of the Lyapunov exponent, specifically the transition from multi-valuedness to single-valuedness, correctly pinpoints the second-order critical point $q_{mc}$, i.e., the critical point is reflected in the thermal behaviour of the Lyapunov exponent. This suggests that the information about the thermodynamic phase structure and transition of dyonic black holes, to some extent, are encoded in the Lyapunov exponent. For completion, in Fig.~\ref{fig:dyonic_Lmassless_vsrh_collage}, we have shown the plot of the Lyapunov exponent as a function of horizon radius. Here, the black dot points correspond to the extremal horizon radius. We observe that the maximum value of $\lambda$ decreases as $q_m$ increases at the extremal $r_h$. This is true for all $\phi_e$. Moreover, for large $r_h$, the $\lambda$ values for different $q_m$ attain a constant value, which is unity for the unit AdS radius $l$ considered here. This saturation value varies as we vary the AdS-radius, but it is always unity for a dimensionless model, i.e., by scaling various variables by $l$. This is an interesting result, which is also true in other gravity models in the following sections, implying some universality in $\lambda$.

\begin{figure}[htb!]
	\centering
	\includegraphics[width=0.8\linewidth]{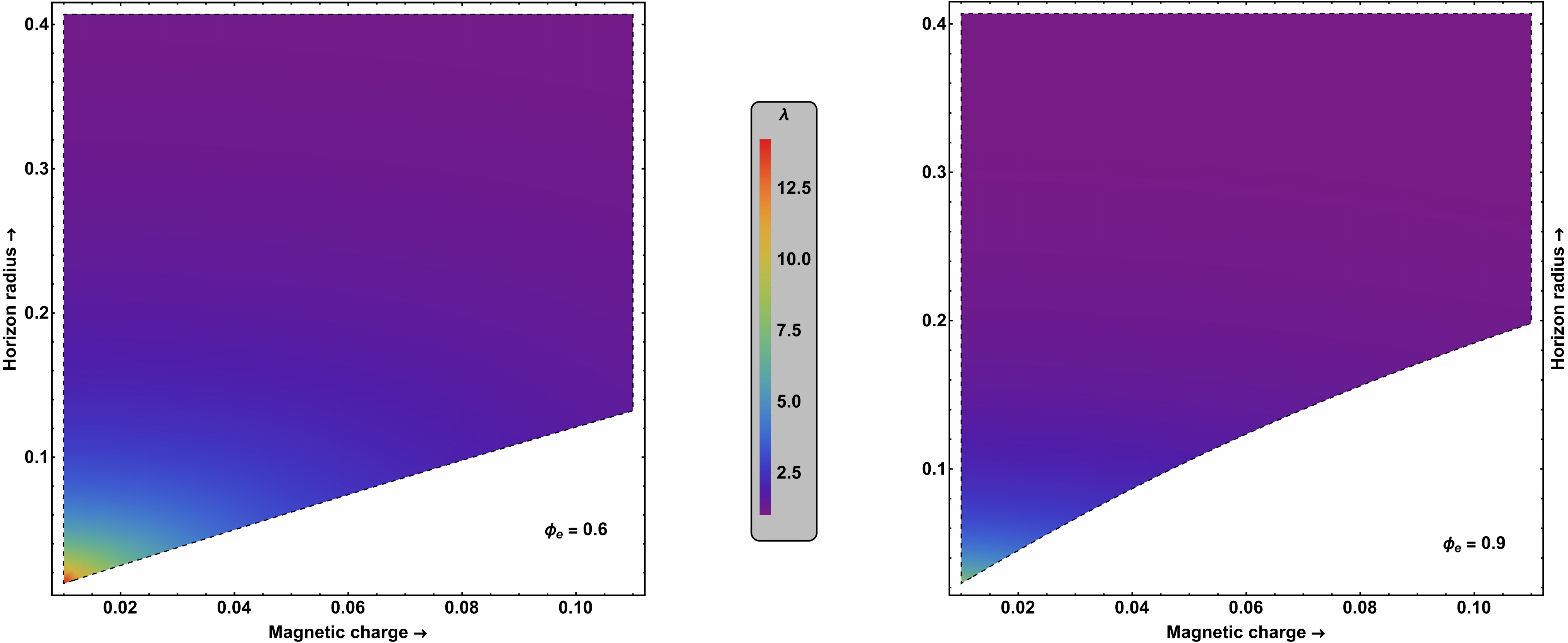}
	\caption{\label{fig:dyonic_Lmassless_contour}Density plot of $\lambda$ for the massless particle as a function of the magnetic charge $q_m$ and horizon radius $r_h$ for the dyonic black hole. Here fixed $\phi_e = 0.6$ (left) and $\phi_e = 0.9$ (right) are used.}
\end{figure}

To gain a comprehensive understanding of how the magnetic charge impacts the Lyapunov exponent for massless particles, we have also shown the density plot of $\lambda$ in Fig.~\ref{fig:dyonic_Lmassless_contour} for two different $\phi_e$ values. On the left, we have fixed the electrical potential $\phi_e=0.6$, so the result is similar to that we obtained in Fig.~\ref{fig:dyonic_Lmassless_vsrh_collage} but with some extra details like the continuous decrease of $\lambda$ as we increase $r_h$ or $q_m$. We get some white space in our density plots because the extremal horizon radius increases as we increase the magnetic charge, so the allowed range of $r_h$ varies for different $q_m$. One more striking observation is that the maximum value that $\lambda$ can reach decreases as we increase the electrical potential. This can be verified by the above density plots for two different potentials. 

\subsubsection{Massive particles}
\begin{figure}[htb!]
	\centering
    \textbf{Massive particles}\par\medskip
	\subfigure[$q_m = 0.05 < q_{mc}$]{\label{fig:dyonic_Lmassive_vsT_1}
		\includegraphics[width=0.45\linewidth]{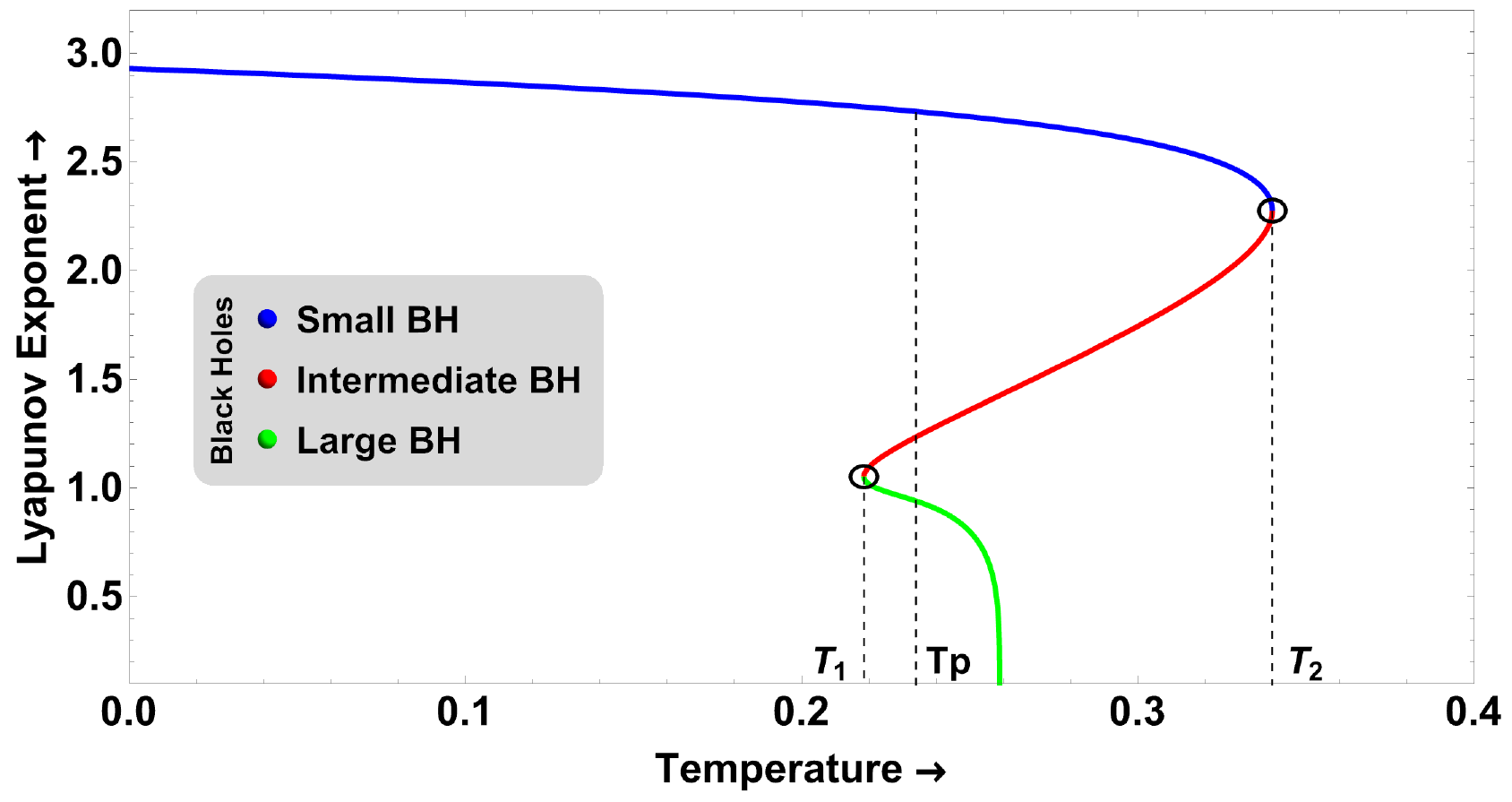}}
	\hfill
	\subfigure[$q_m = 0.12 > q_{mc}$]{\label{fig:dyonic_Lmassive_vsT_2}
		\includegraphics[width=0.45\linewidth]{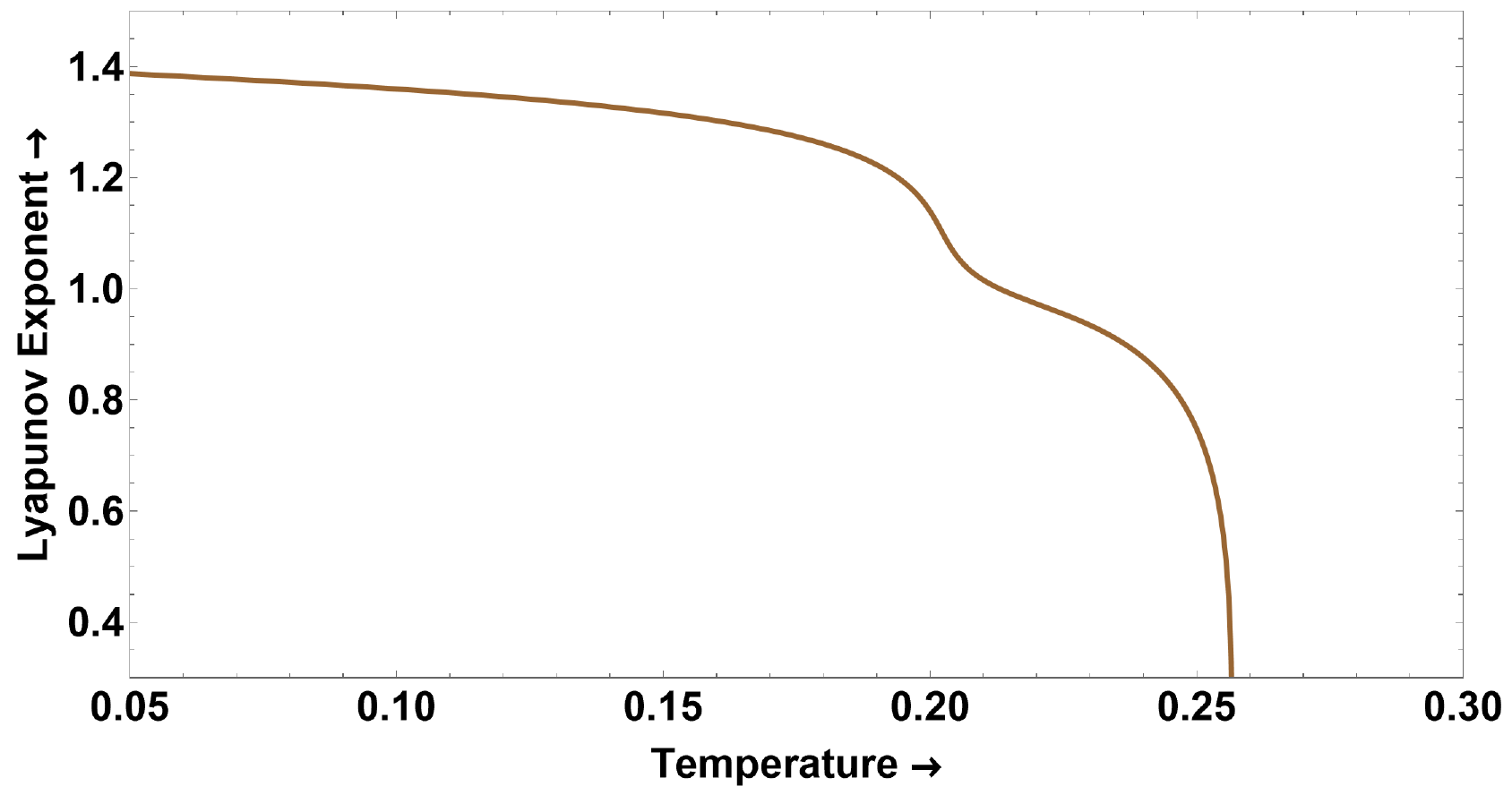}}
	\vfill
	\subfigure[Lyapunov exponent vs temperature]{\label{fig:dyonic_Lmassive_vsT_collage}
		\includegraphics[width=0.45\linewidth]{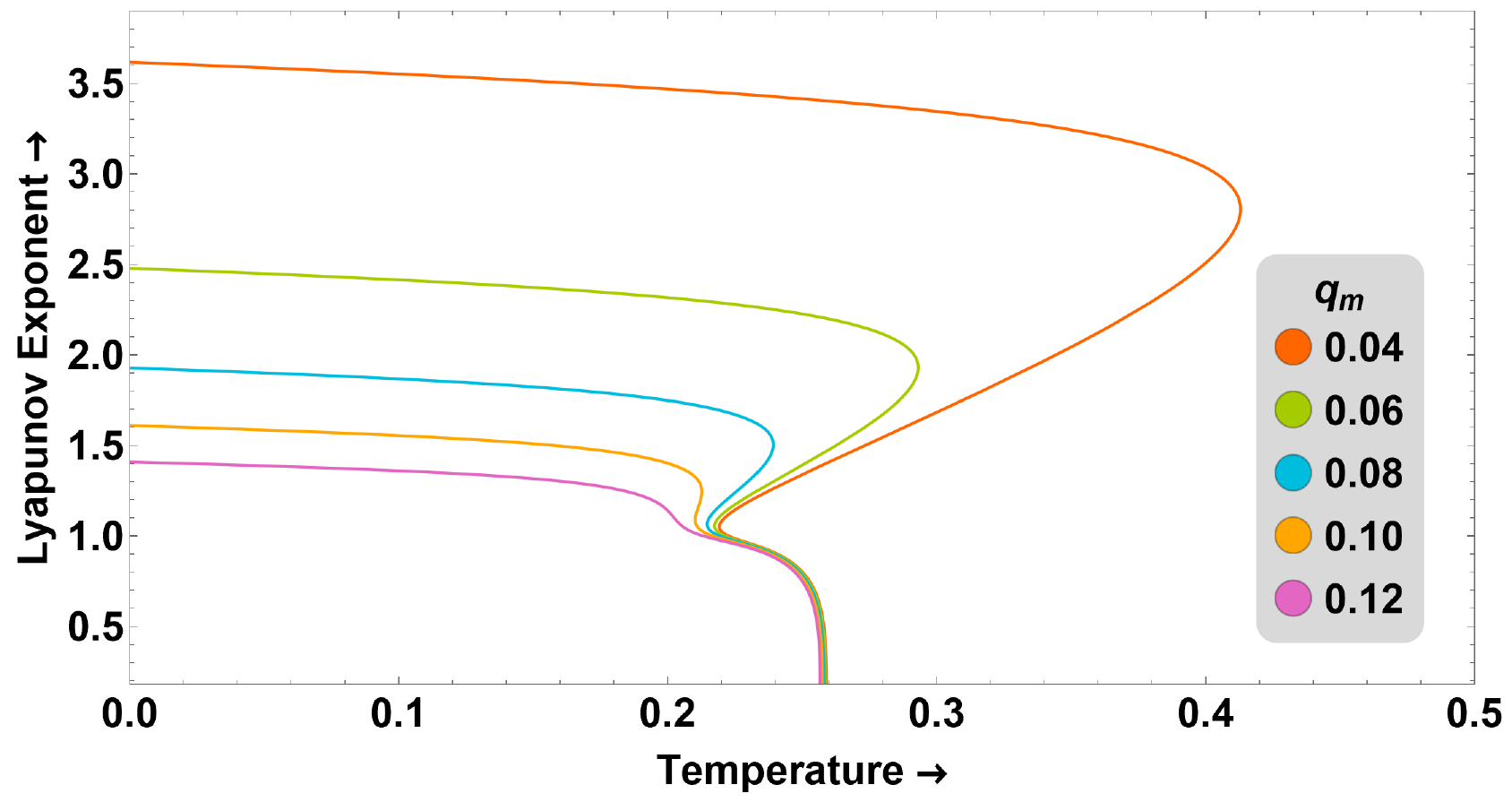}}
    \hfill
	\subfigure[Lyapunov exponent vs horizon radius]{\label{fig:dyonic_Lmassive_vsrh_collage}
		\includegraphics[width=0.45\linewidth]{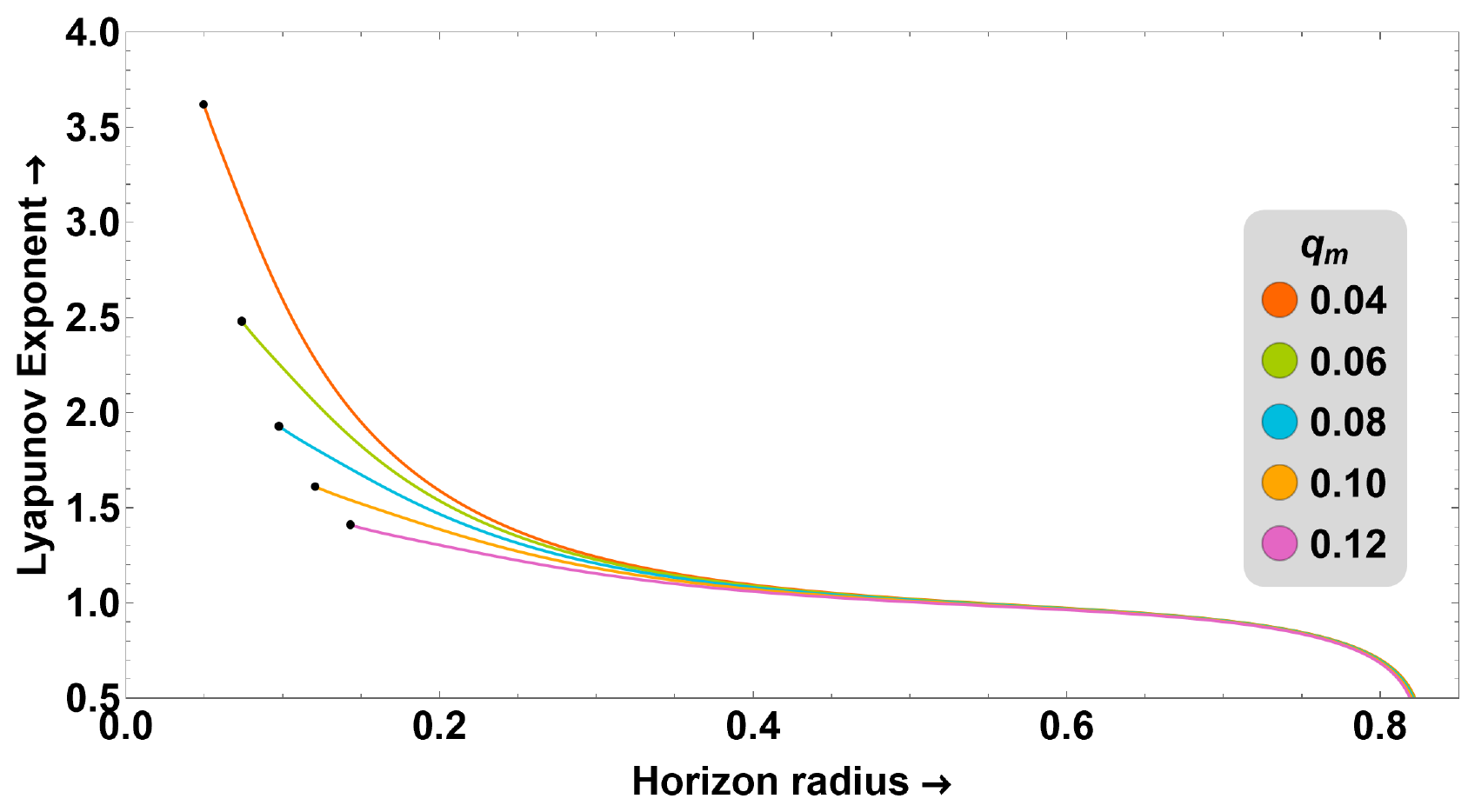}}
	\caption{\label{fig:dyonic_Lmassive}Lyapunov exponent of the massive particle $\lambda$ as a function of temperature $T$ and horizon radius $r_h$ for the dyonic black hole. Here, $\phi_e = 0.6$ is used.}
\end{figure}

We can similarly compute the thermal behaviour of the Lyapunov exponent associated with the massive particle. For this, we use Eqs.~(\ref{eq:Veffective}) and (\ref{eq:lyapunov-massive}). Our numerical results for the Lyapunov exponent are shown in Fig.~\ref{fig:dyonic_Lmassive}. Here, we have taken the angular momentum $L=20$ for illustration purposes; however, our main results of the Lyapunov exponent are quite robust for different values of $L$. We again observe that when the magnetic charge $q_m$ is less than the critical value $q_{mc}$, $\lambda$ is a multi-valued function of $T$ with three distinct thermal profiles, i.e., in the small and large stable dyonic black hole background the Lyapunov exponent decreases with temperature.
In contrast, the intermediate unstable black hole background increases with temperature. On the contrary, when $q_m$ exceeds the critical value, $\lambda$ is single-valued and decreases monotonically with temperature. Therefore, we again see that the transition from multi-valued to single-valued behaviour of the Lyapunov exponent precisely captures the details of the critical point $q_{mc}$.

\begin{figure}[htb!]
	\centering
	\includegraphics[width=0.8\linewidth]{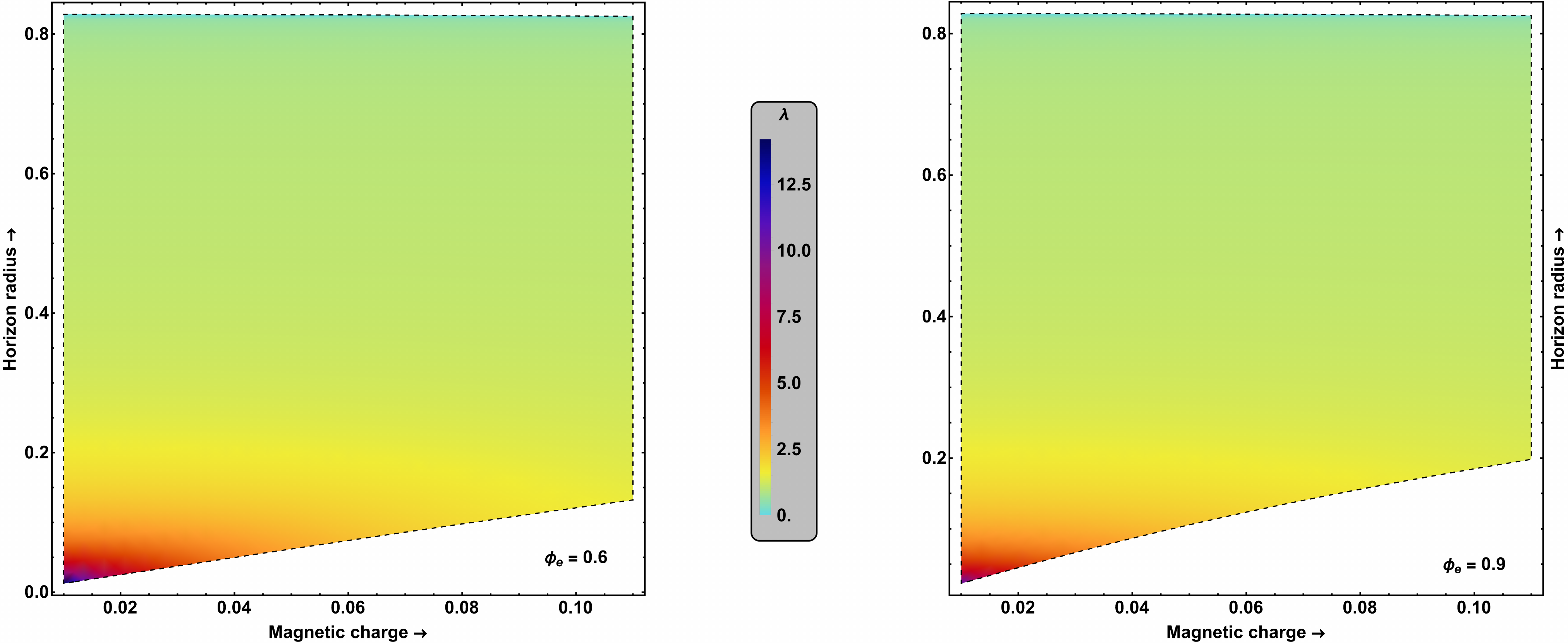}
	\caption{\label{fig:dyonic_Lmassive_contour}Density plot of $\lambda$ for the massive particle as a function of the magnetic charge $q_m$ and horizon radius $r_h$ for the dyonic black hole. Here fixed $\phi_e = 0.6$ (left) and $\phi_e = 0.9$ (right)  are used.}
\end{figure}

The above results of the Lyapunov exponent associated with the massive particle are similar to those associated with the massless particle. For instance, just like in the massless particle, it decreases with $q_m$ in the small black hole phase, whereas the effects of $q_m$ are minimal in the large black hole phase.  However, there are also a few subtle differences. In particular, the Lyapunov exponent of the massive particle approaches zero in the large black hole phase at some temperature, i.e., at some horizon radius, the Lyapunov exponent tends to zero for all $q_m$. This can be explicitly seen in Figs.~\ref{fig:dyonic_Lmassive_vsT_collage} and \ref{fig:dyonic_Lmassive_vsrh_collage}.
Moreover, this horizon radius is almost the same for different $q_m$. It implies that the behaviour of the massive particle becomes non-chaotic in the large black hole phase. The vanishing Lyapunov exponent is due to the disappearance of unstable equilibrium at those horizon radii. This behaviour should be contrasted with the Lyapunov exponent of the massless particle where $\lambda$ approaches a nonzero constant value at large horizon radii for all $q_m$.

The density plot of the Lyapunov exponent for massive particles is shown in Fig.~\ref{fig:dyonic_Lmassive_contour} for two different $\phi_e$ values. Once again, these plots add more detail to Fig.~\ref{fig:dyonic_Lmassive_vsrh_collage}, suggesting that $\lambda$ decreases as we increase $r_h$ or $q_m$. The green shade near $r_h=0.8$ means that $\lambda$ eventually drops to zero as we increase the horizon radius for a fixed $q_m$. This implies that for the time like circular geodesic, we do not have any extrema after a fixed value of $r_h$, which can also be seen in Fig.~\ref{fig:dyonic_veffective3d} and is explained in the previous section. Moreover, the allowed range of $r_h$ for which $\lambda$ is nonzero decreases with $q_m$ and $\phi_e$. Also, the maximum value of $\lambda$ for $\phi_e=0.6$ is greater than that of $\phi_e=0.9$,  which means that the increase of the potential $\phi_e$ decreases the range of values that $\lambda$ can have.

\subsection{Bardeen Black Holes}
Bardeen black holes are the first examples of a regular black hole
satisfying the weak energy conditions \cite{bardeen1968non}. They have a distinct ``regular'' centre, meaning they do not contain a singularity at their core, unlike the more well-known Schwarzschild and Kerr black holes. This is because matter cannot collapse to a point as a zone of repulsive gravity surrounds Bardeen black holes. The Bardeen black holes are usually obtained from the Einstein equation with nonlinear electrodynamics sources \cite{Ayon-Beato:2000mjt}.\footnote{Recently, it has been suggested that an infinite tower of higher-curvature corrections can also give rise to regular black hole solutions~\cite{Bueno:2024dgm}.} The various ways to construct such regular black hole solutions can be found in \cite{Hayward:2005gi, Bronnikov:2000vy, Berej:2006cc, Dymnikova:2004zc, Ayon-Beato:1998hmi, Bronnikov:2005gm}. Although no astrophysical Bardeen black holes have been discovered to date, theoretical physicists still find them an important subject of study as they provide a unique viewpoint on the nature of black holes and the underlying laws of physics \cite{tzikas2019bardeen, singh2020thermodynamics, guo2020joule}.

Here, we focus on the Bardeen black hole obtained in \cite{Ayon-Beato:2000mjt} from a particular nonlinear electromagnetic source. The line element of this Bardeen-AdS black hole is given by
\begin{equation}\label{eq:bardeen_metric}
	ds^2 = -f(r)dt^2 + \frac{1}{f(r)}dr^2 + r^2(d\theta^2 + \sin^2\theta d\psi^2),
\end{equation}
where the blackening function $f(r)$ is
\begin{equation}\label{eq:bardeen_metric_function}
	f(r) = 1 - \frac{2 M r^2}{(g^2 + r^2)^\frac{3}{2}} + \frac{r^2}{l^2}\,,
\end{equation}
with $g$ as the magnetic charge, $M$ as the mass of the black hole and $l$ as the AdS radius. The condition of the event horizon $f(r_h) = 0$, gives us the mass expression
\begin{equation}\label{eq:bardeen_mass}
	M = \bigg(1 + \frac{r_h^2}{l^2}\biggr) \frac{(r_h^2 + g^2)^\frac{3}{2}}{2 r_h^2} \,.
\end{equation}
The Hawking temperature of the Bardeen black hole is given by
\begin{equation}\label{eq:bardeen_temperature}
	T = \frac{f'(r_h)}{4\pi} = \frac{3 r_h}{4 \pi (r_h^2 + g^2)} + \frac{3 r_h^3}{4 \pi (r_h^2 + g^2) l^2} - \frac{1}{2\pi r_h}\,,
\end{equation}
and its entropy is
\begin{equation}\label{eq:bardeen_entropy}
	S = \frac{A}{4 G_4} = \frac{\pi r_h^2}{G_4}\,,
\end{equation}
with $A$ as the horizon area. The differential form of the first law takes the form
\begin{equation}\label{1stlawbardeen}
dM= \mathcal{T} dS +\Phi_g dg\,,
\end{equation}
where $\mathcal{T}=T/(1-\Pi)$, $\Phi_g=\phi_g/(1-\Pi)$, $\Pi=1-r_h^3(g^2+r_h^2)^{-3/2}$, and $\Phi_g$ is the potential associated with the charge $g$. Notice that this first-law expression differs from the usual first-law differential expression. The difference mainly arises due to the nonlinear nature of the gauge field sources \cite{Zhang:2016ilt, Guo:2021wcf}.\footnote{The standard version of the first law can also be satisfied in the Bardeen black hole. This requires logarithmic corrections to the black hole entropy \cite{Tzikas:2018cvs,li2019thermodynamics}. Unfortunately, this looks unphysical, as the nonlinear gauge field sources do not give any additional contribution to the Wald entropy. Therefore, one also expects the black hole entropy to be given by the horizon area in the Bardeen black hole.}

Here, we are mainly interested in black hole thermodynamics in a constant charge ensemble, as the system exhibits familiar van der Waals-type phase transitions in this ensemble. The relevant Helmholtz free energy is given by
\begin{equation}\label{eq:bardeen_free_energy_formula}
F = M - \mathcal{T} S \,,
\end{equation}
which simplifies to
\begin{equation}\label{eq:bardeen_free_energy}
F = \frac{\sqrt{g^2+r_h^2} \left(2 g^2 \left(r_h^2+2 l^2\right)+r_h^2
   \left(L^2-r_h^2\right)\right)}{4 L^2 r_h^2}\,.
\end{equation}
The above thermodynamic quantities can be further written in a dimensionless form as
\begin{equation}\label{eq:bardeen_dimensionless}
	\tilde{r_h} = \frac{r_h}{l},~~\tilde{g} = \frac{g}{l},~~\tilde{\mathcal{T}} = \mathcal{T} l,~~ \tilde{M} = \frac{M}{l},~~ \tilde{F} = \frac{F}{l}\,.
\end{equation}

\begin{figure}[htbp!]
	\centering
	\subfigure[Temperature ($\tilde{\mathcal{T}}$) vs horizon radius ($\tilde{r_h}$)]{\label{fig:bardeen_Tvsrh_collage}	\includegraphics[width=0.45\linewidth]{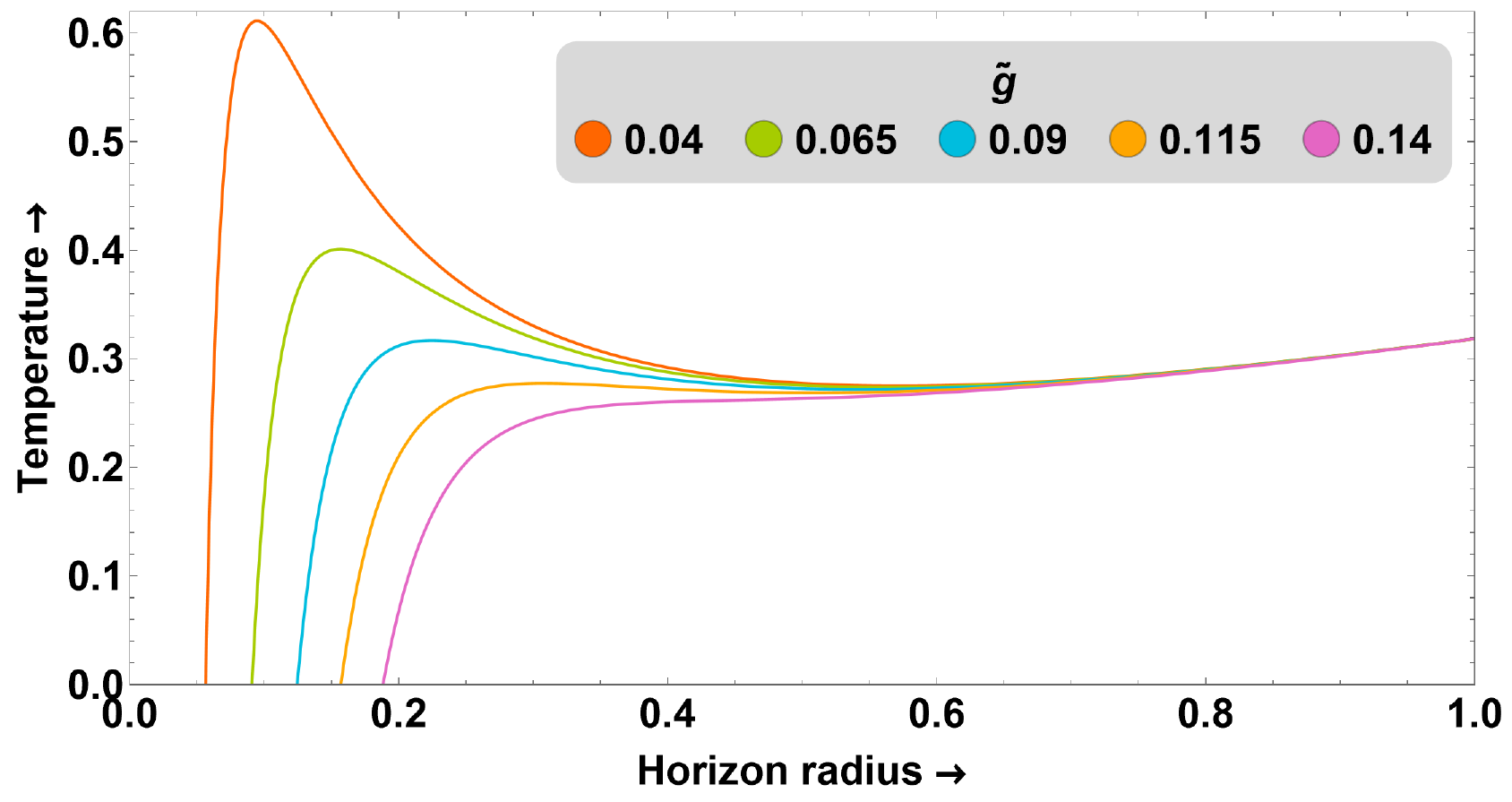}}
	\hfill
	\subfigure[Free energy ($\tilde{F}$) vs temperature ($\tilde{\mathcal{T}}$)]{\label{fig:bardeen_GvsT_collage}
		\includegraphics[width=0.45\linewidth]{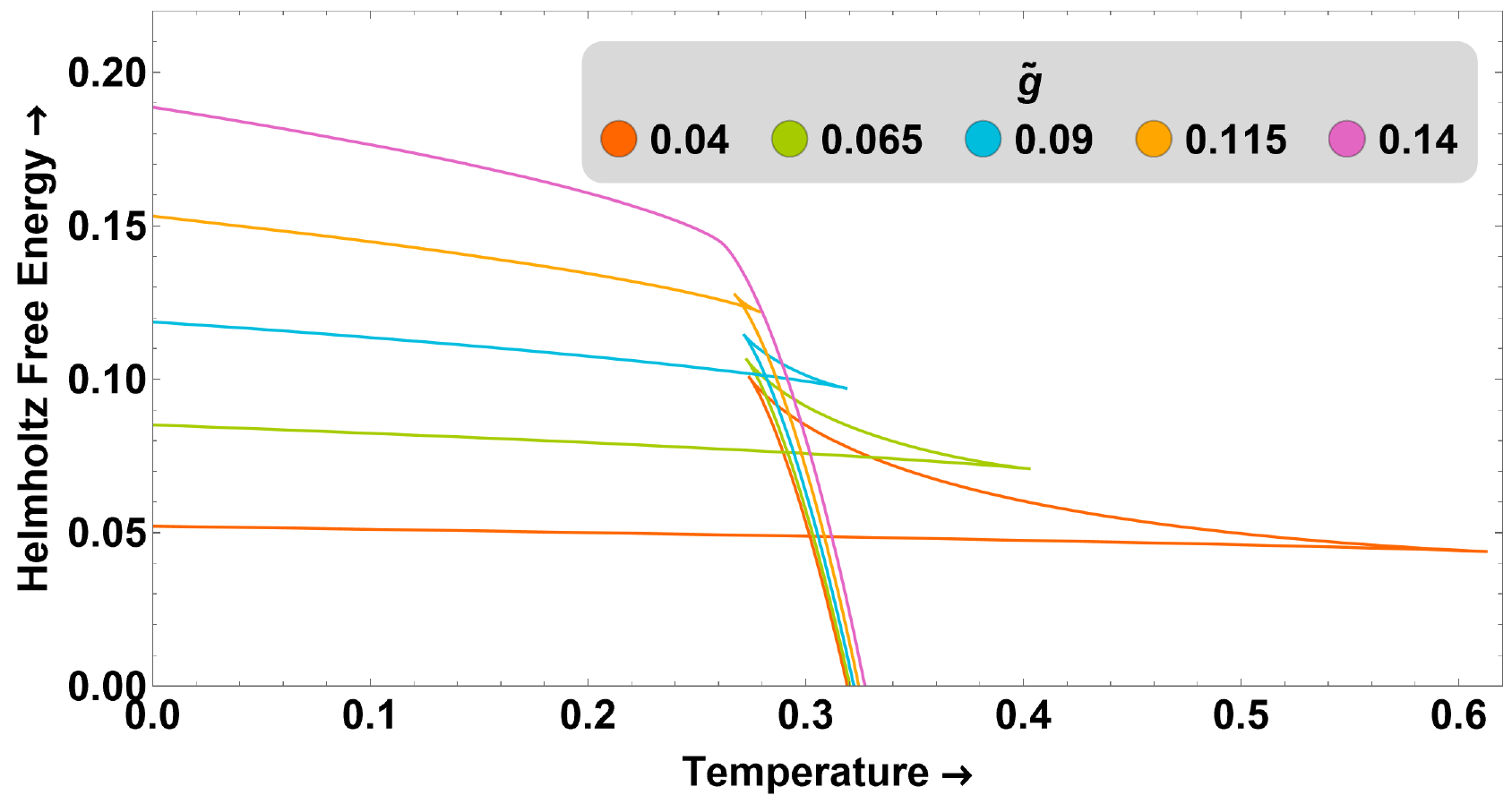}}
    \vfill
    \subfigure[$\tilde{g} = 0.04 < \tilde{g_c}$]{\label{fig:bardeen_GvsT_1}
		\includegraphics[width=0.45\linewidth]{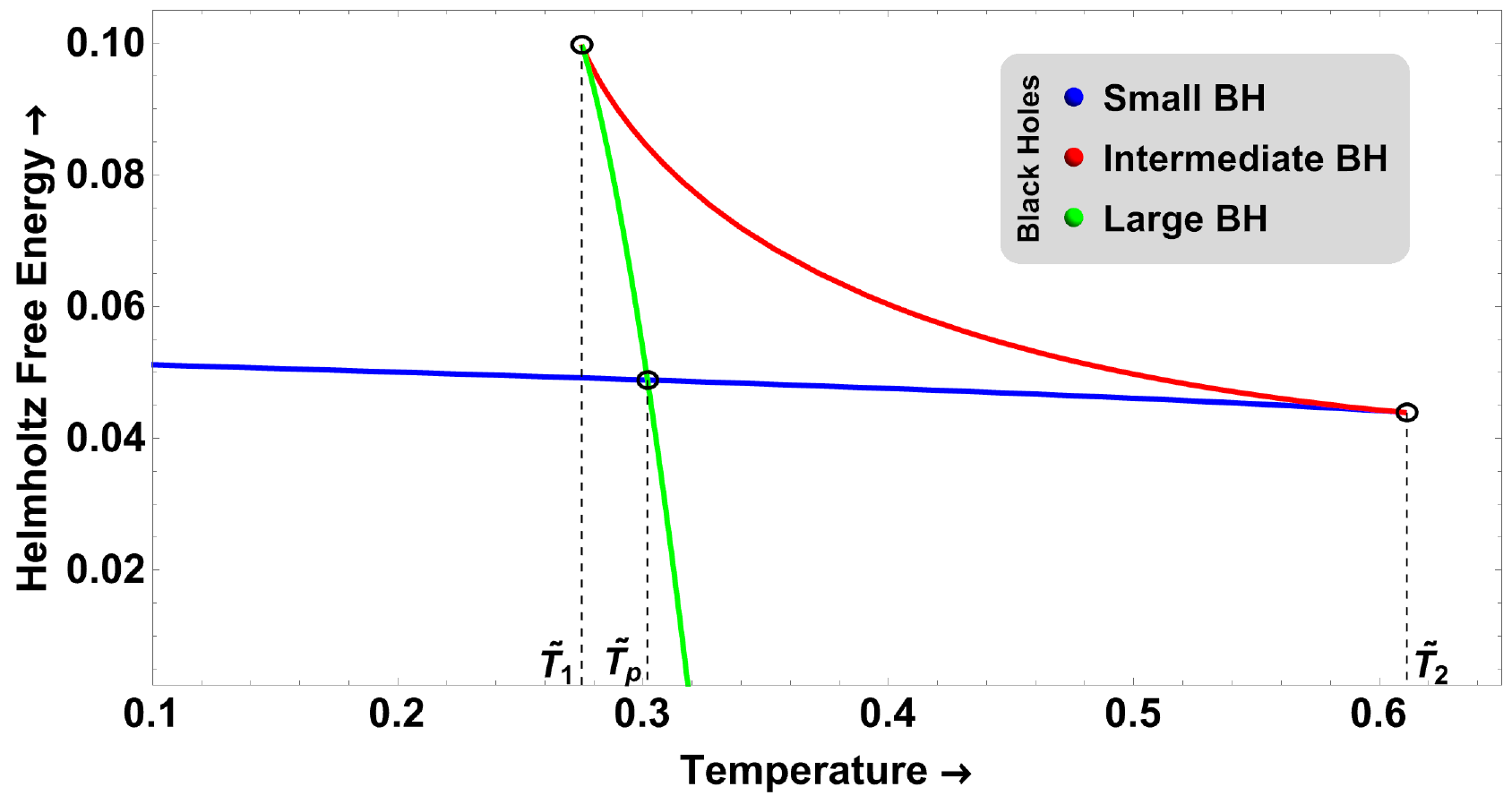}}
	\hfill
	\subfigure[$\tilde{g} = 0.15 > \tilde{g_c}$]{\label{fig:bardeen_GvsT_2}
		\includegraphics[width=0.45\linewidth]{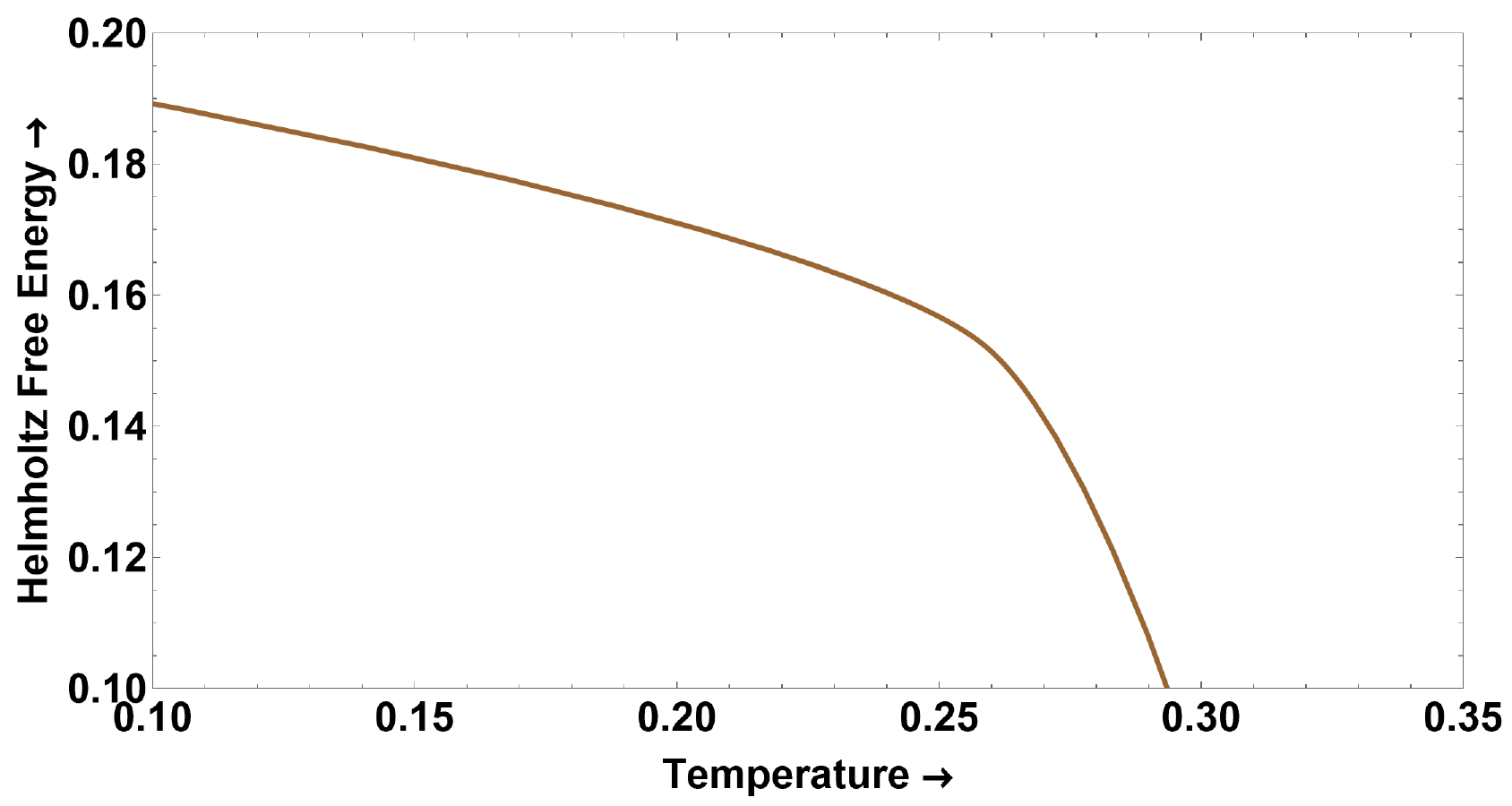}}
	\caption{\label{fig:bardeen_GvsTandTvsrh} The thermodynamic phase structure of the Bardeen black hole.}
\end{figure}

The thermodynamic phase structure of the Bardeen black hole in the fixed charge ensemble is shown in Fig.~\ref{fig:bardeen_GvsTandTvsrh}. For $\tilde{g}=0$, the Bardeen black hole becomes identical to the Schwarzschild-AdS black hole. Accordingly, it also exhibits the Hawking/Page phase transition for $\tilde{g}=0$. However, the essence of the Bardeen black hole appears only for finite $\tilde{g}$; that is, finite $\tilde{g}$ makes the whole spacetime regular and the thermodynamic phase structure much richer. For instance, for small but finite $\tilde{g}$, three coexisting black hole solutions exist for some temperature range. These solutions correspond to large, intermediate, and small black hole solutions. The free energy of these solutions further exhibits the swallow-tail-like structure, which exchanges dominance as the temperature varies. In particular, the large black hole phase has the lowest free energy at high temperatures, whereas the small black hole phase has the lowest free energy at low temperatures. The intermediate black hole, on the other hand, always has a free energy higher than that of the large or small black hole phases. Accordingly, a first-order phase transition exists between the small and large black hole phases as the temperature varies. This phase transition occurs at temperature $\tilde{\mathcal{T}}_p$, shown by a dashed vertical line in Fig.~\ref{fig:bardeen_GvsT_1}.

By further increasing the value of $\tilde{g}$, there appears a critical value $\tilde{g}_c$ at which the three black hole phases merge to form a single black hole phase, i.e., the size of the swallow-tail decreases with $\tilde{g}$ and completely vanishes at $\tilde{g}_c$. Therefore, the first-order small/large transition line stops at $\tilde{g}_c$. Analogously to the dyonic black hole case, this $\tilde{g}_c$ defines a second-order critical point. This implies that the thermodynamics of the Bardeen-AdS black hole is heavily affected by the magnetic charge $\tilde{g}$. Also, this behaviour is again completely analogous to the van der Waals-type phase transition of the liquid/gas system.  The magnitude of the critical point $\tilde{g}_c$ can be further determined from the condition of the inflexion point
\begin{equation}\label{eq:bardeen_critical_relation}
	\frac{\partial\tilde{\mathcal{T}}}{\partial\tilde{r_h}} = 0,~~\frac{\partial^2\tilde{\mathcal{T}}}{\partial\tilde{r_h}^2} = 0\,,
\end{equation}
which gives us the critical values of $\tilde{r_h}$ and $\tilde{g}$ as
\begin{eqnarray}\label{eq:bardeen_critical_points}
\tilde{g}_c & = & \frac{1}{9} \sqrt{\frac{1}{2} \left(5 \sqrt{10}-13\right)} \simeq  0.132\,, \nonumber \\
\tilde{r_{hc}} & = & \frac{1}{3} \sqrt{\frac{1}{3} \left(8-\sqrt{10}\right)} \simeq 0.423\,.
\end{eqnarray}

\begin{figure}[htb!]
	\centering
	\includegraphics[width=0.4\linewidth]{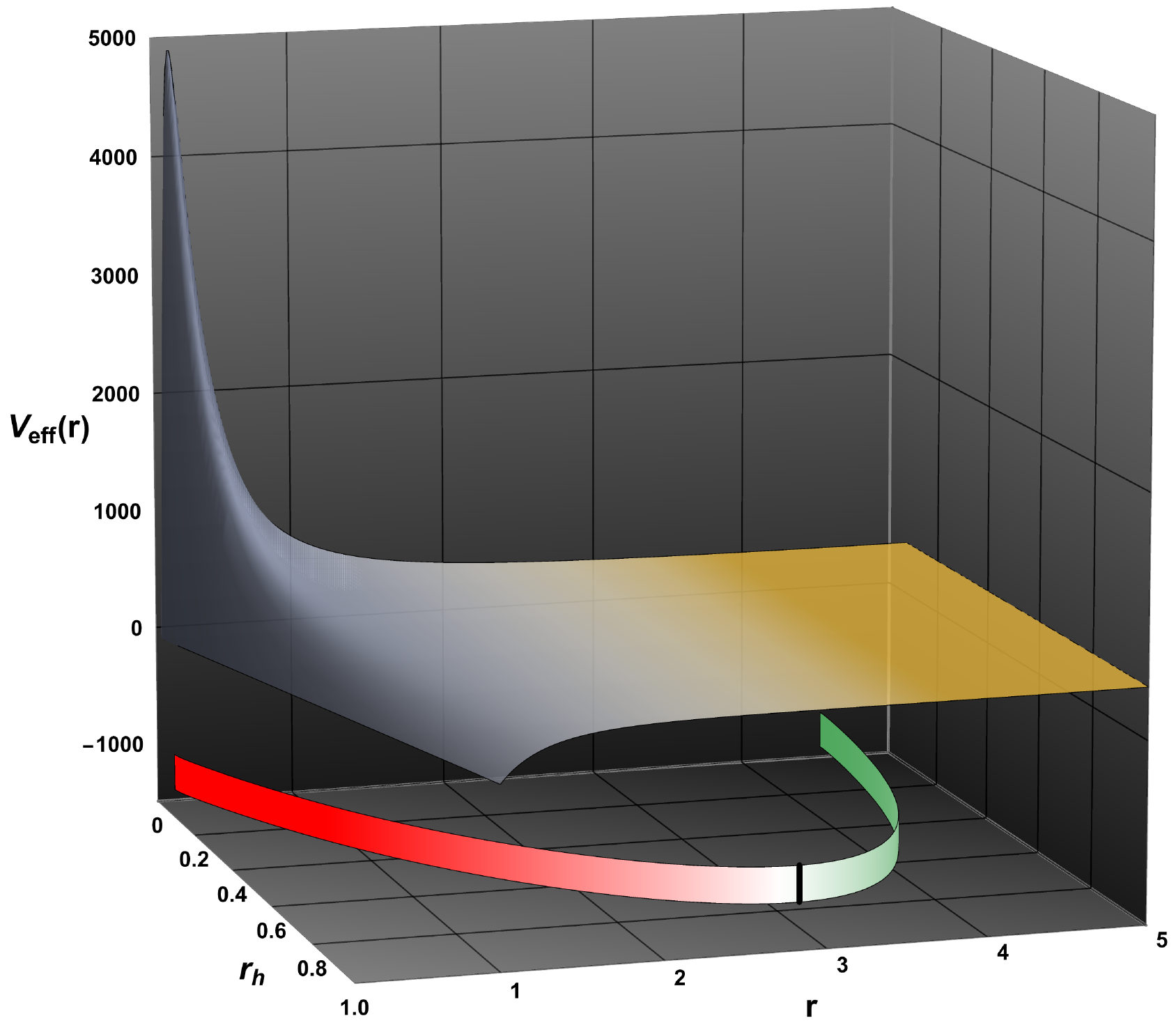}
	\caption{\label{fig:bardeen_veffective3d}The 3D plot of the effective potential $V_\text{eff}(r)$ as a function of horizon radius $\tilde{r_h}$ and orbit radius $r$ of the massive particle. Here, $L = 20$ and $\tilde{g} = 0.05$ are used. The red and green curves projected below correspond to the unstable and stable equilibria of the timelike circular geodesics, respectively.}
\end{figure}

Let us now discuss the behaviour of the effective potential in the Bardeen black hole background. This is given by
\begin{equation}\label{eq:bardeen_Veffective}
V_{\text{eff}}(r) = \left(-\frac{\tilde{r}^2 \left(\tilde{r_h}{}^2+1\right) \left(\tilde{g}^2+\tilde{r_h}{}^2\right){}^{3/2}}{\left(\tilde{g}^2+\tilde{r}^2\right)^{3/2}
   \tilde{r_h}{}^2}+\tilde{r}^2+1\right) \left(\frac{L^2}{\tilde{r}^2}-\epsilon \right)\,,
\end{equation}
where $L$ and $r$ are the angular momentum of the particle and the radius of the particle’s
orbit respectively. We can see that $V_{\text{eff}}(r)$ is a function of $r$ and $r_h$, and its 3D plot is shown in Fig.~\ref{fig:bardeen_veffective3d} where we have fixed $\tilde{g}=0.05$ and $L=20$. The curve with the extrema is projected onto the $V_{\text{eff}}(r) = -1500$ plane, where the red part indicates the unstable maxima and the green part indicates the stable minima. Beyond a certain value of $r_h = 0.884$, there are no extrema  (marked by the black line on the curve). We are mainly interested in the red part of the curve, and the analysis is done in the next subsections.

\subsubsection{Massless particles}
\begin{figure}[htb!]
	\centering
	\textbf{Massless particles}\par\medskip
	\subfigure[$\tilde{g} = 0.04 < \tilde{g_c}$]{\label{fig:bardeen_Lmassless_vsT_1}
		\includegraphics[width=0.45\linewidth]{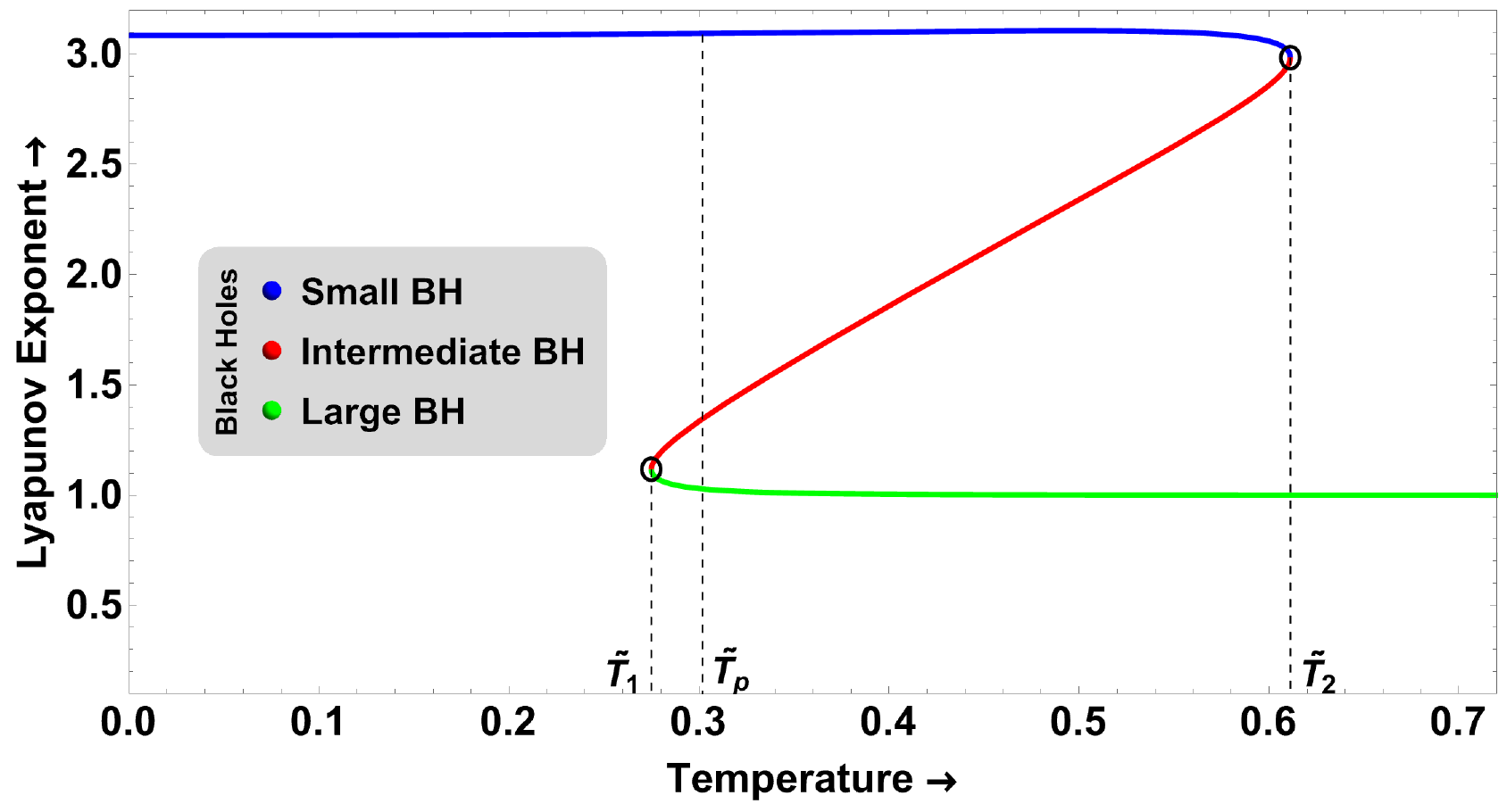}}
	\hfill
	\subfigure[$\tilde{g} = 0.15 > \tilde{g_c}$]{\label{fig:bardeen_Lmassless_vsT_2}
		\includegraphics[width=0.45\linewidth]{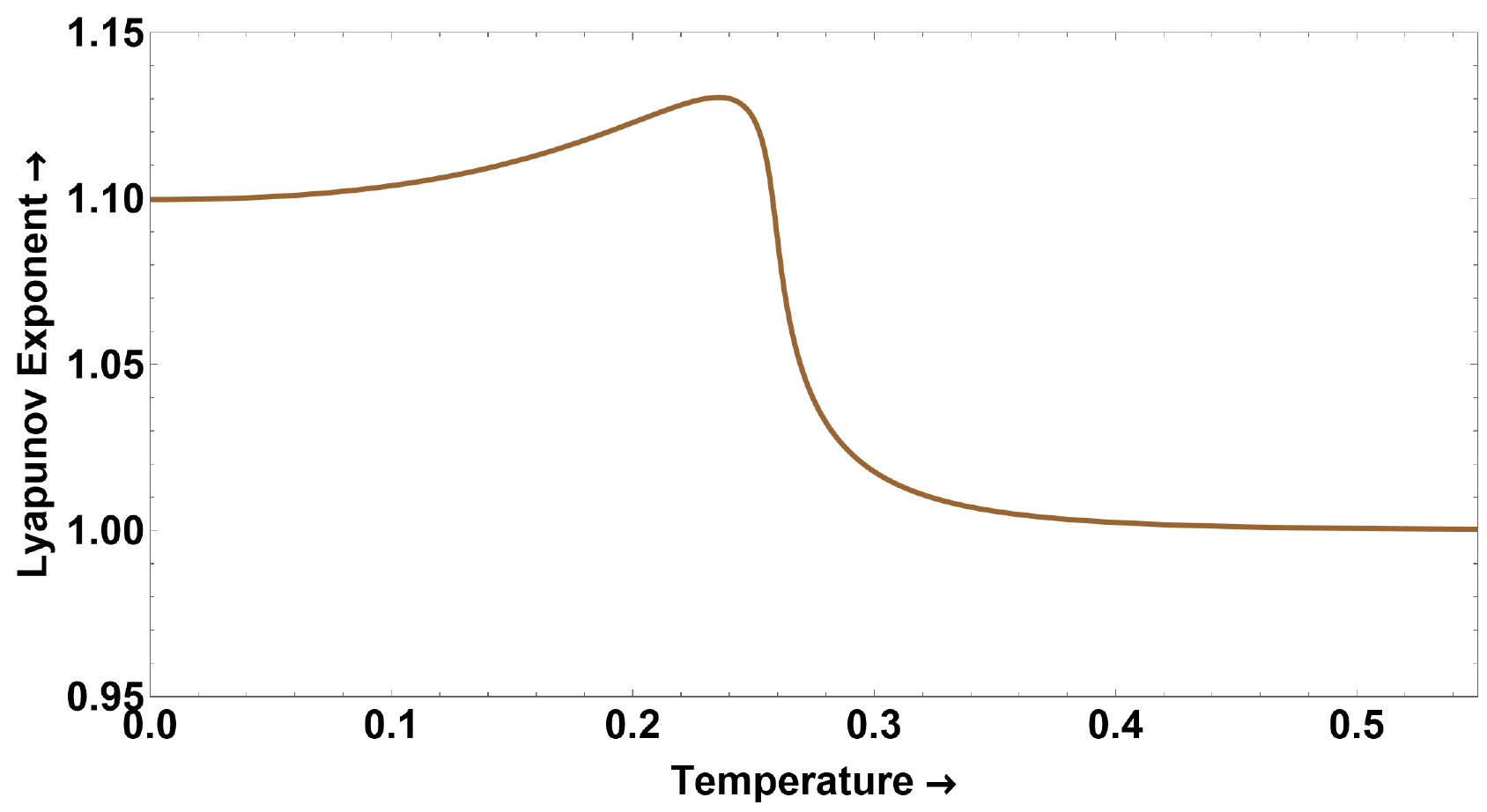}}
	\vfill
	\subfigure[Lyapunov exponent vs temperature]{\label{fig:bardeen_Lmassless_vsT_collage}
		\includegraphics[width=0.45\linewidth]{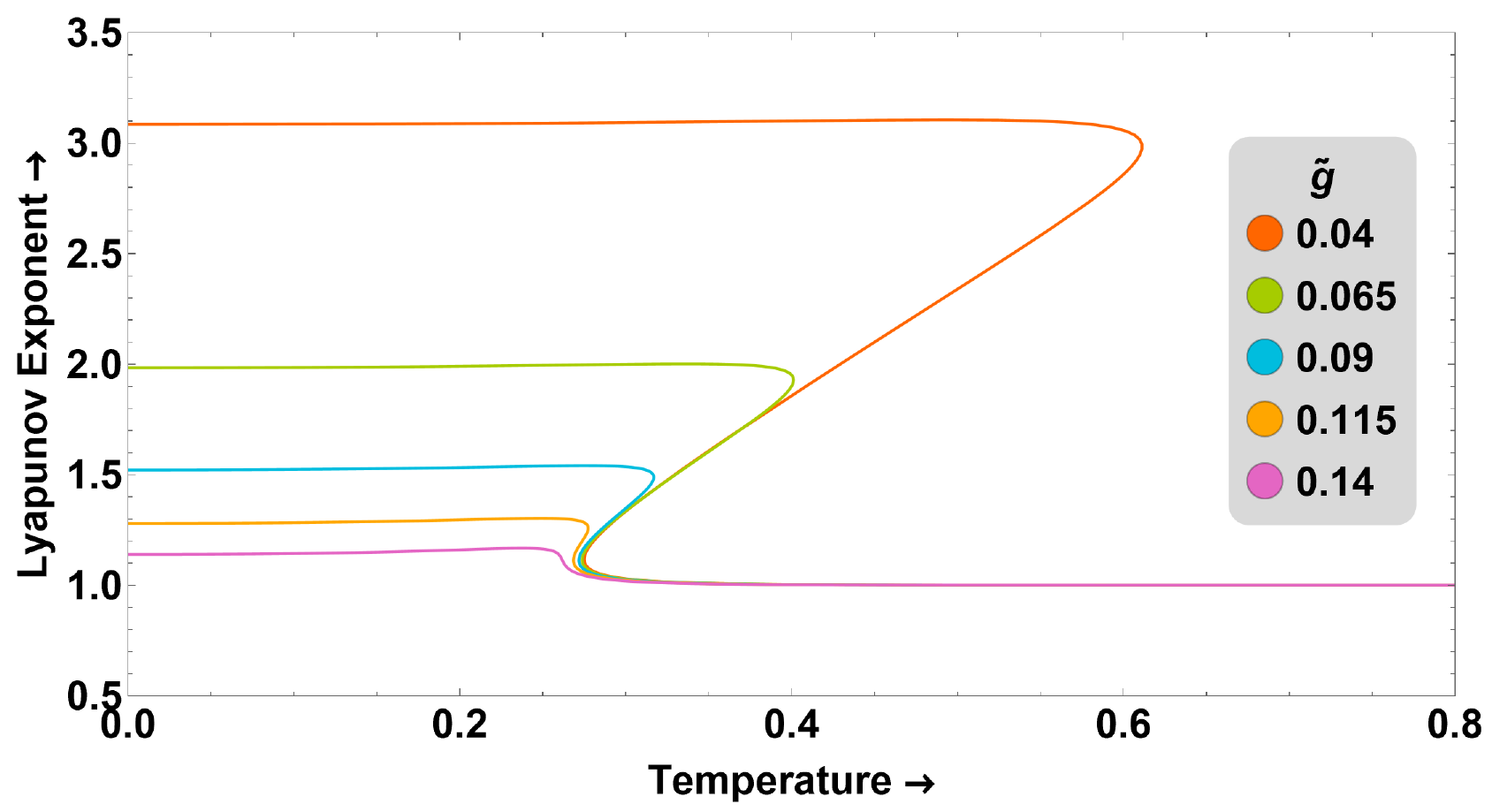}}
	\hfill
	\subfigure[Lyapunov exponent vs horizon radius]{\label{fig:bardeen_Lmassless_vsrh_collage}
		\includegraphics[width=0.45\linewidth]{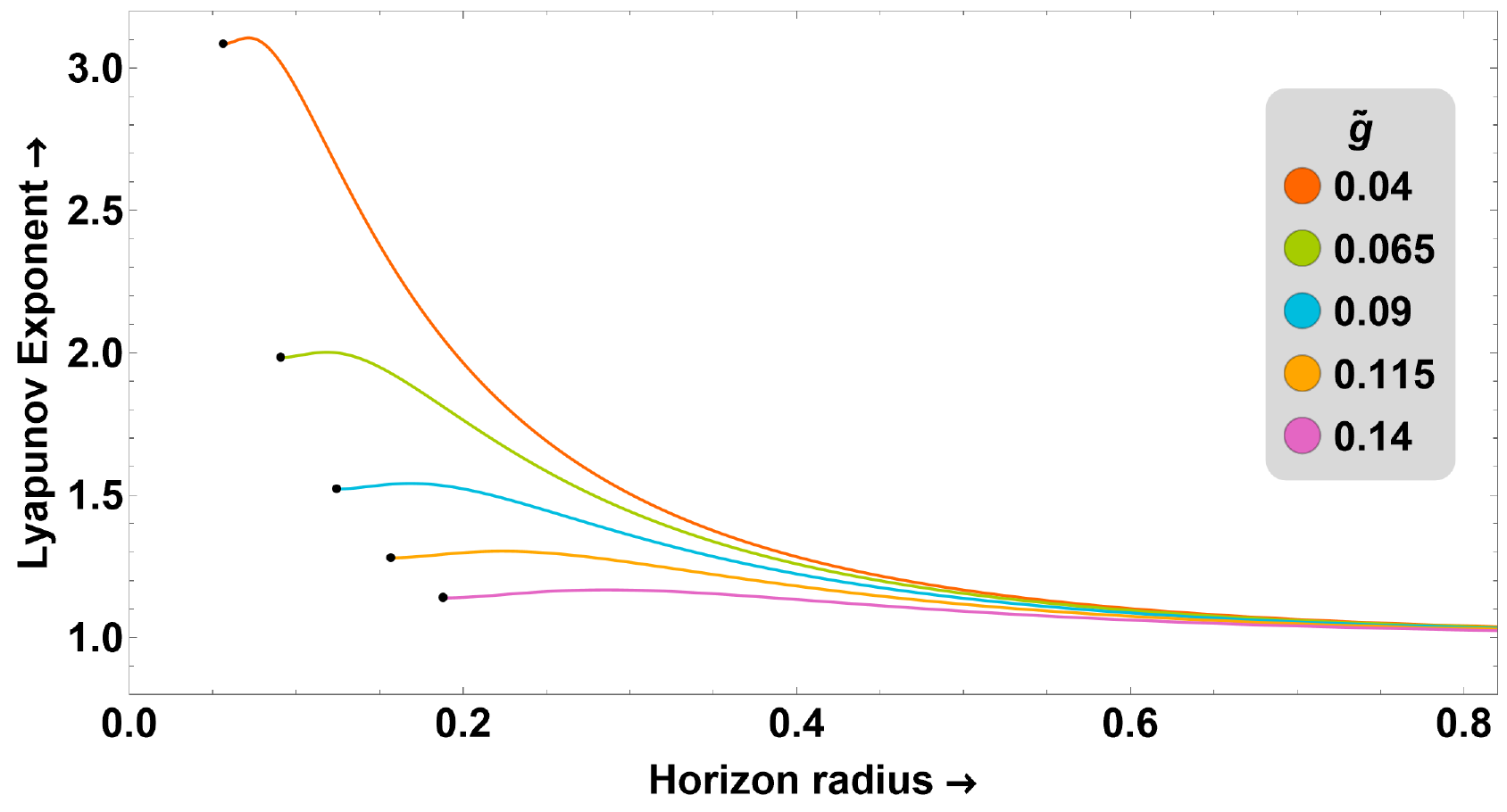}}
	\caption{\label{fig:bardeen_Lmassless}Lyapunov exponent $\lambda$ of the massless particle as a function of temperature $\tilde{\mathcal{T}}$ and horizon radius $r_h$ for the Bardeen black hole.}
\end{figure}

In this subsection, we probe the thermodynamics of Bardeen AdS black holes with Lyapunov exponents. The Lyapunov exponent for the massless particles can be found from Eqs.~(\ref{eq:lyapunov-massless}) and (\ref{eq:bardeen_Veffective}), and its expression is given by
\begin{equation}
\parbox{0.8\textwidth}{\hangafter=1\hangindent=2em\raggedright$\displaystyle
   \lambda = \sqrt{\frac{3}{2}}
   \Bigl(\frac{1}{\TR (m^2)^{7/4} \TR_h}\Bigr)
   \Bigl\{-
   \Bigl[-\bigl(\TR^2 k (n^2/m^2)^{3/2}\bigr)\big/ \TR_h^2+\TR^2+1\Bigr]
   \allowbreak
   \Bigl[2 \TG^6 m \TR_h^2
   +\TG^4 (\TR^4 k n+6 \TR^2 m \TR_h^2)
   +\TG^2 \TR^4 \bigl(\TR_h^4 n+\TR_h^2 (-4 \TR^2 n+n+6 m)
   -4 \TR^2 n\bigr)
   +2 \TR^6\TR_h^2 (-2 \TR_h^2 n-2 n+m)
   \Bigr]
   \Bigr\}^{1/2}
   $}\,
\end{equation}
where $k=\TR_h^2+1$, $m=\sqrt{\TG^2+\TR^2}$ and $n=\sqrt{\TG^2+\TR_h^2}$, which suggests that the Lyapunov exponent depends non-trivially on parameters $\tilde{g}$ and $\tilde{r}_h$ of the Bardeen black hole. Our numerical results for the Lyapunov exponent are shown in Fig.~\ref{fig:bardeen_Lmassless}.

\begin{figure}[htb!]
	\centering
	\includegraphics[width=0.4\linewidth]{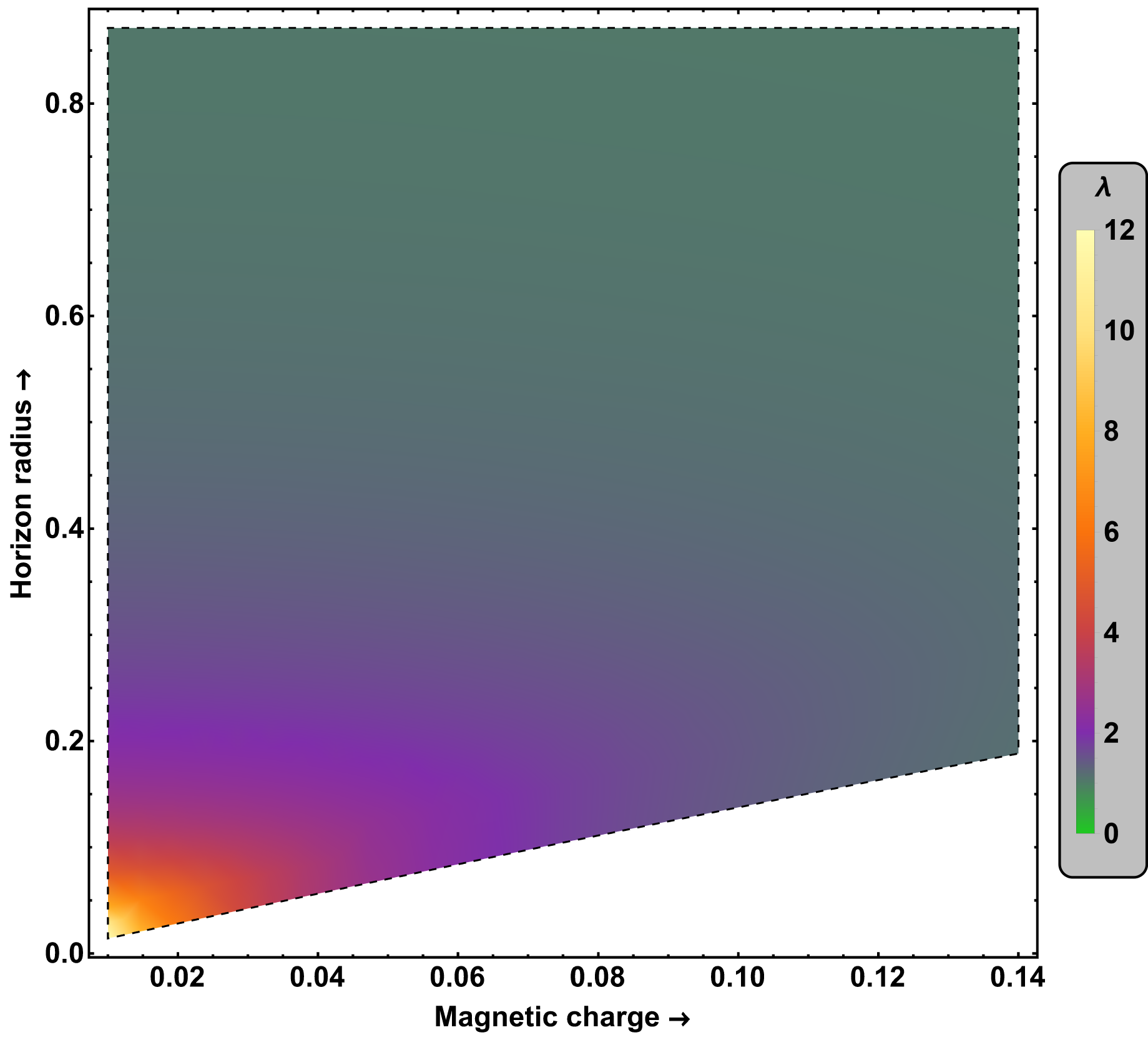}
	\caption{\label{fig:bardeen_Lmassless_contour} Density plot of $\lambda$ for massless particles as a function of the magnetic charge $\tilde{g}$ and horizon radius $\tilde{r_h}$ for the Bardeen black hole.}
\end{figure}

We again observe a distinct behaviour of $\lambda(\tilde{\mathcal{T}})$ depending on the magnitude of $\tilde{g}$. In particular, when $\tilde{g}$ is less than the critical value $\tilde{g_c}$, $\lambda$ is single-valued for $\tilde{\mathcal{T}} \in (0, \tilde{\mathcal{T}}_1)$, becomes multi-valued for $\tilde{\mathcal{T}} \in (\tilde{\mathcal{T}}_1, \tilde{\mathcal{T}}_2)$, and again becomes single-valued for $\tilde{\mathcal{T}} > \tilde{\mathcal{T}}_2$. The small/large transition temperature $\tilde{\mathcal{T}}_p$ is between $\tilde{\mathcal{T}}_1$ and $\tilde{\mathcal{T}}_2$ and is indicated by the dashed black line. Therefore, the Lyapunov exponent is single-valued in the thermodynamically favoured small and large black hole phases and exhibits multi-valuedness at the transition temperature. In contrast, when $\tilde{g}$ exceeds the critical value, $\lambda$ is single-valued, showing that only one black hole solution exists without any phase transition. In this case, the Lyapunov exponent first increases and gradually decreases to finally approach a constant value. In Fig.~\ref{fig:bardeen_Lmassless_vsT_collage}, we have shown the effects of the control parameter $\tilde{g}$ on the Lyapunov exponent. As we continue to increase the value of $\tilde{g}$ from $0.04$ to $0.14$, we observe that the multi-valued nature of the Lyapunov exponent gradually decreases and disappears completely beyond the critical value, which shows the typical nature of a first-order phase transition. These observations give weight to the idea that the Lyapunov exponents can also be used to study the black hole phase transitions.

The Lyapunov exponent also displays distinct behaviour in the large and small black hole phases, i.e., at low and high temperatures. In particular, the magnetic charge leaves imprints on the Lyapunov exponent more in the small black hole phase than in the large black hole phase. It decreases substantially with $\tilde{g}$ in the small black hole phase, whereas it remains almost the same with $\tilde{g}$ in the large black hole phase. Importantly, like in the dyonic black hole case, the Lyapunov exponent becomes constant, i.e., independent of $\tilde{g}$, in the large temperature limit. These results are explicitly shown in Fig.~\ref{fig:bardeen_Lmassless_vsT_collage}. For completion, in Fig.~\ref{fig:bardeen_Lmassless_vsrh_collage}, we have further demonstrated the behaviour of $\lambda$ as a function of horizon radius $\tilde{r}_h$. Here, the black dot points
correspond to the extremal horizon radius. The overall structure of the massless Lyapunov exponent in the parameter space of $\tilde{g}$ and $\tilde{r}_h$ is shown in the density plot in Fig.~\ref{fig:bardeen_Lmassless_contour}. The density plot shows that the Lyapunov exponent decreases substantially with $\tilde{g}$ for small horizon radii. In contrast, it remains almost the same with $\tilde{g}$ for large horizon radii. Interestingly, like in the case of the dyonic black hole, the Lyapunov exponent again saturates to a unit value in the large temperature or $r_h$ limit.

\subsubsection{Massive particles}
\begin{figure}[htb!]
	\centering
    \textbf{Massive particles}\par\medskip
	\subfigure[$\tilde{g} = 0.04 < \tilde{g_c}$]{\label{fig:bardeen_Lmassive_vsT_1}
		\includegraphics[width=0.45\linewidth]{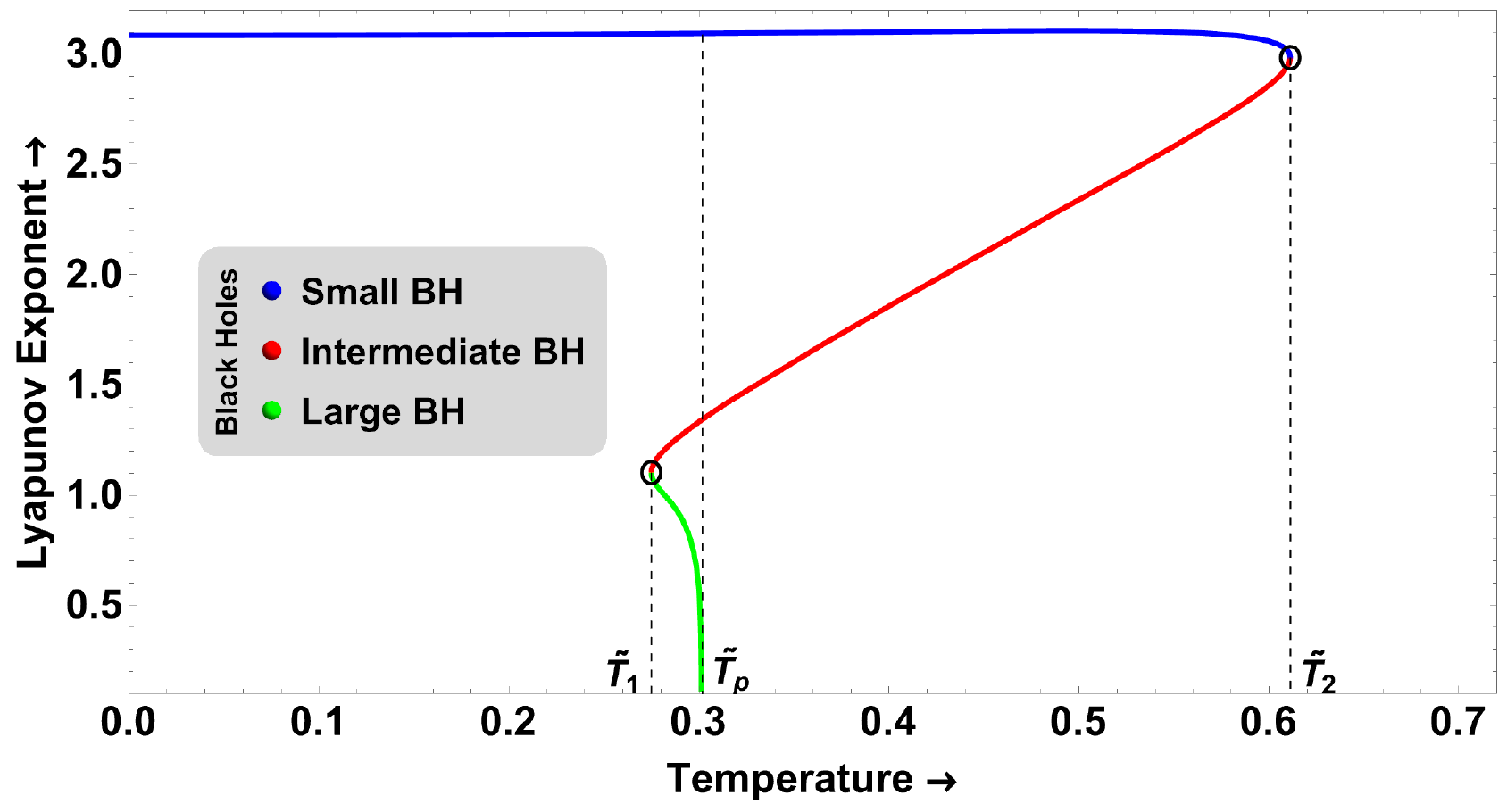}}
	\hfill
	\subfigure[$\tilde{g} = 0.15 > \tilde{g_c}$]{\label{fig:bardeen_Lmassive_vsT_2}
		\includegraphics[width=0.45\linewidth]{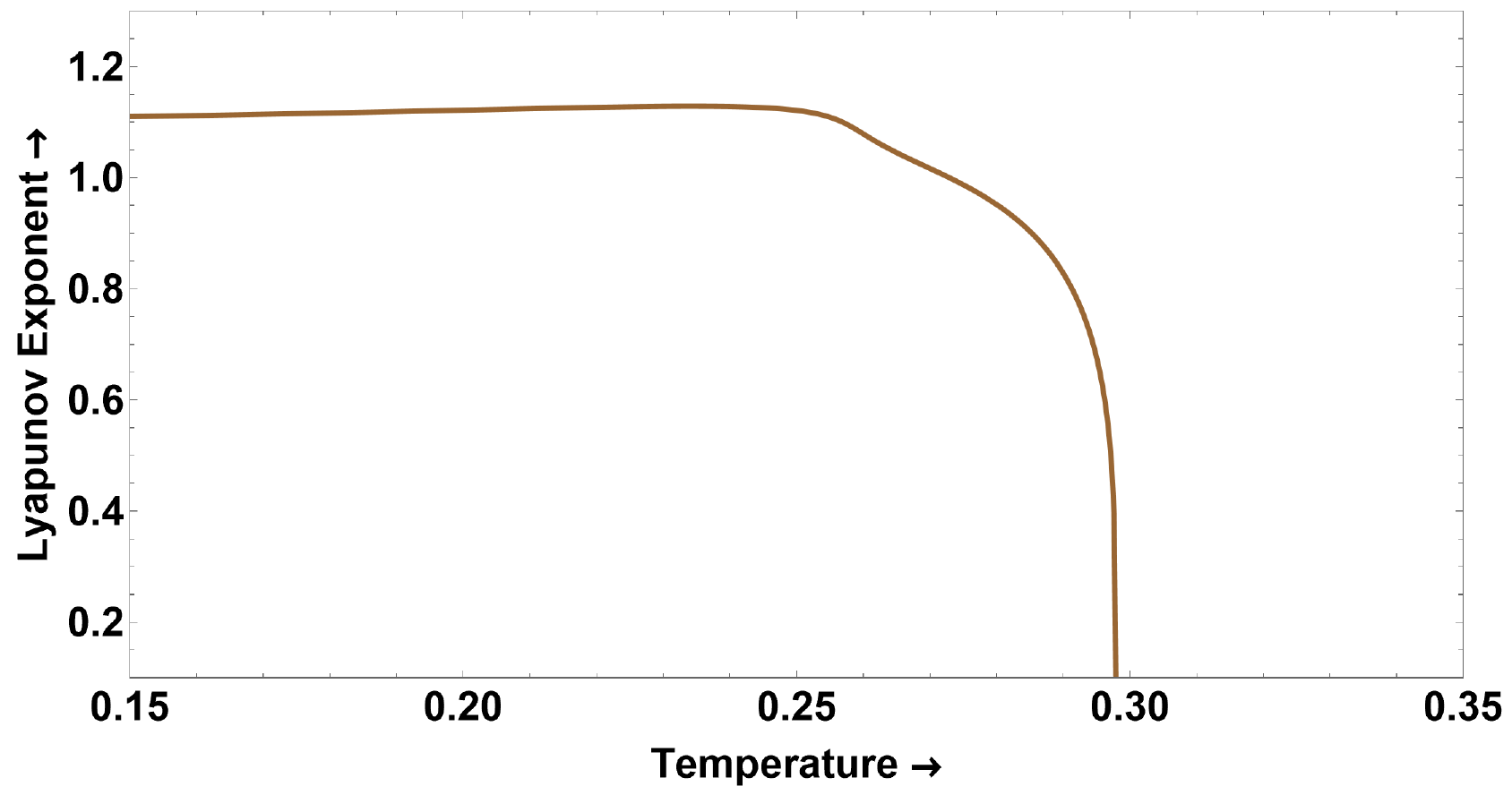}}
	\vfill
	\subfigure[Lyapunov exponent vs temperature]{\label{fig:bardeen_Lmassive_vsT_collage}
		\includegraphics[width=0.45\linewidth]{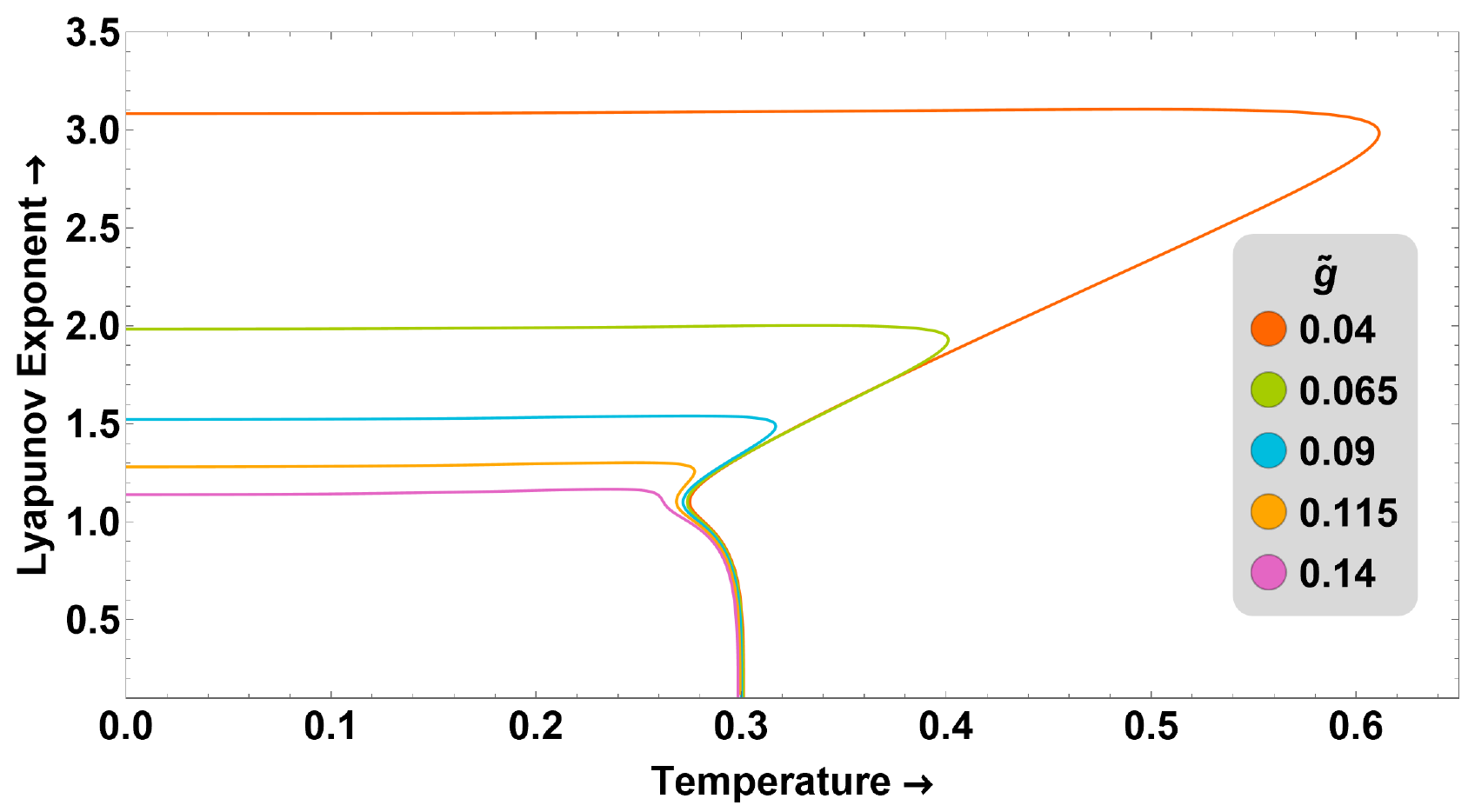}}
    \hfill
	\subfigure[Lyapunov exponent vs horizon radius]{\label{fig:bardeen_Lmassive_vsrh_collage}
		\includegraphics[width=0.45\linewidth]{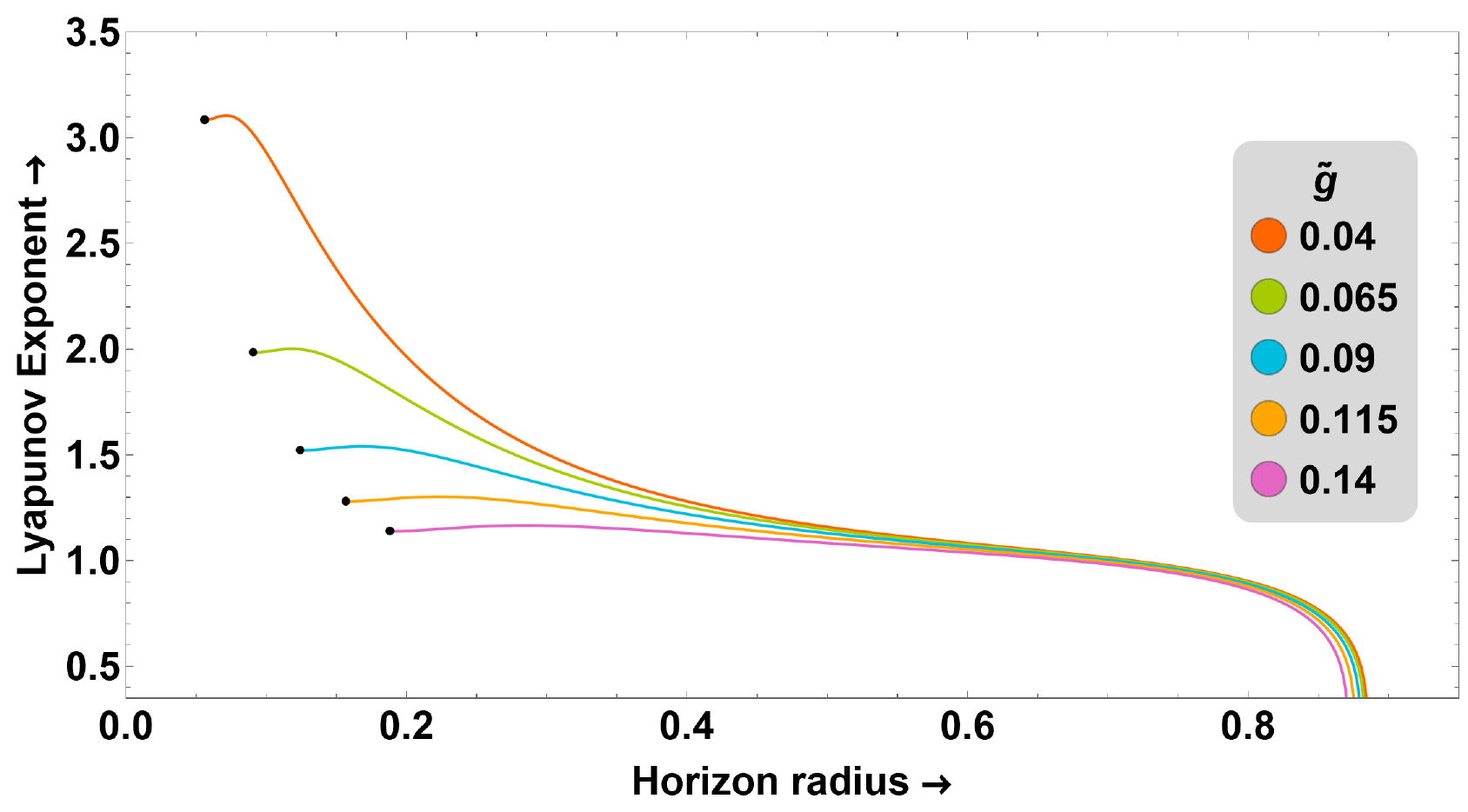}}
	\caption{\label{fig:bardeen_Lmassive}Lyapunov exponent $\lambda$ of the massless particle  as a function of temperature $\tilde{\mathcal{T}}$ and Horizon radius $r_h$ for the Bardeen black hole.}
\end{figure}
\begin{figure}[htb!]
	\centering
	\includegraphics[width=0.4\linewidth]{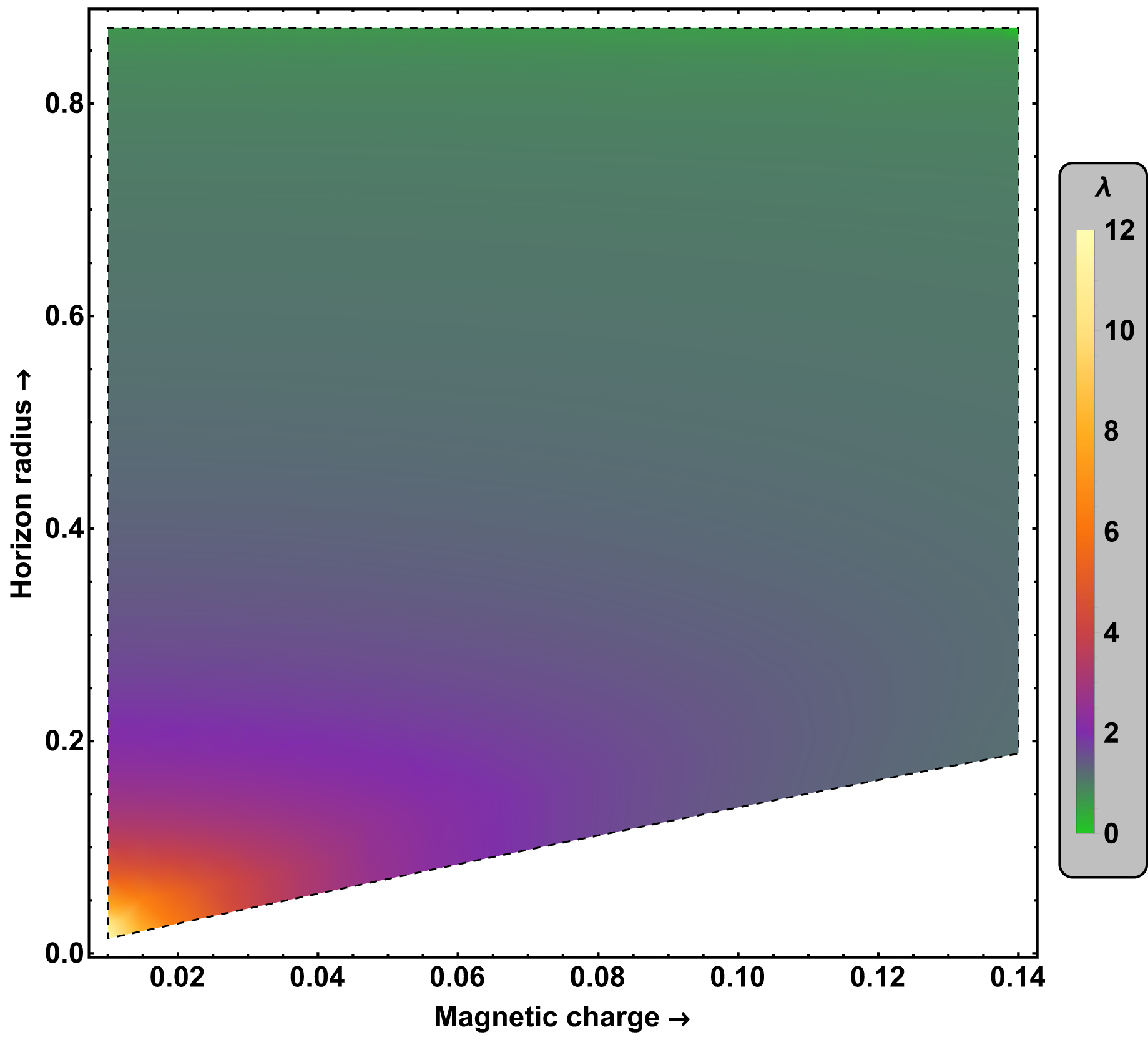}
	\caption{\label{fig:bardeen_Lmassive_contour} Density plot of $\lambda$ for massive particles as a function of the magnetic charge $\tilde{g}$ and horizon radius $\tilde{r_h}$ for the Bardeen black hole.}
\end{figure}
We can similarly analyse the thermal structure of the Lyapunov exponent of the massive particle in the Bardeen black hole background. The form of the corresponding effective potential also suggests that the massive Lyapunov exponent will again depend non-trivially on $\tilde{g}$ and $\tilde{r}_h$. Our numerical results are shown in Fig.~\ref{fig:bardeen_Lmassive}. We again observe that when the magnetic charge value $\tilde{g}$ is less than the critical value $\tilde{g_c}$, $\lambda$ is a multi-valued function of $\tilde{\mathcal{T}}$ in some temperature range, suggesting multiple black hole branches. On the contrary, when $\tilde{g}$ exceeds the critical value, $\lambda$ is single-valued, showing that only one black hole solution exists. In Fig.~\ref{fig:bardeen_Lmassive_vsT_collage}, we have demonstrated the effect of $\tilde{g}$ on the Lyapunov exponent. As we continue to increase the value of $\tilde{g}$ from $0.04$ to $0.14$, the multi-valued nature of the Lyapunov exponent gradually changes, and it disappears entirely beyond the critical value $\tilde{g}_c$. These observations are again similar to our previous results and suggest encoding of the phase transition critical points in the massive Lyapunov exponent.

Interestingly, as in the dyonic case, the Lyapunov exponent of the massive particle also goes to zero at a specific horizon radius. This behaviour is true irrespective of whether $\tilde{g}$ is less than or greater than $\tilde{g}$ and is explicitly shown in Fig.~\ref{fig:bardeen_Lmassive_vsrh_collage}. These parameter values always correspond to the large black hole phase, leading to the absence of chaos in certain large black holes. The Lyapunov exponent is always positive and finite for the small black hole phase. The Lyapunov exponent decreases with magnetic charge in the small black hole phase. The overall structure of the massive Lyapunov exponent in the parameter space of $\tilde{g}$ and $\tilde{r}_h$ is shown in the density plot in Fig.~\ref{fig:bardeen_Lmassive_contour}. Like for the massless particle, the magnetic charge again substantially affects the massive Lyapunov exponent when $\tilde{r}_h$ is small, whereas its effects are negligible for large $\tilde{r}_h$. Also, $\lambda$ drops to zero as we increase $\tilde{r}_h$ at a fixed $\tilde{g}$ in the case of massive particles for the reasons previously explained.

\subsection{Gauss-Bonnet Black Holes}
Gauss-Bonnet black holes are distinct black holes that can manifest in spacetime with more than four dimensions, owing to the presence of curvature terms that go beyond the standard Einstein-Hilbert action. The Gauss-Bonnet invariant, a fourth-order polynomial derived from the Riemann curvature tensor, gives rise to these curvature terms. The Gauss-Bonnet black holes are the simplest prototype of higher-order curvature gravity theories that arise naturally in quantum gravity theories such as string theory and are expected to play an important role in quantum gravity  \cite{Boulware:1985wk, Deser:2002jk, Gross:1986mw}. In addition to the typical features of black holes, such as a singularity and an event horizon, Gauss-Bonnet black holes exhibit attractive geometric qualities. For example, the Gauss-Bonnet theorem, which links the total curvature of a closed surface to its topology, has been extended to Gauss-Bonnet black holes. Specifically, the topology of the event horizon determines the Gauss-Bonnet coupling constant, which measures the strength of the curvature terms. These black holes have been the subject of extensive research in recent years, both for their mathematical characteristics and potential relevance to astrophysics \cite{Cai:2001dz, Cai:2013qga, Pani:2009wy, Clunan:2004tb, Wei:2014hba, Zhou:2020vzf}.

The Gauss-Bonnet black holes exist only in spacetime dimensions greater than four, i.e., the Gauss-Bonnet term becomes topological (or to a total derivative term) in four dimensions. It does not contribute to the field equations. However, in dimensions greater than or equal to five, the Gauss-Bonnet term does contribute to the Einstein equation and leads to interesting solutions and properties. For instance, the Gauss-Bonnet black hole solutions not only come with different horizon topologies, such as the planar, spherical, or hyperbolic topologies, but their thermodynamic phase structure also becomes much more interesting. In particular, the spherical charged Gauss-Bonnet black hole exhibits the Hawking/Page and small/large black hole phase transitions, and the associated critical points depend on the dimensions as well as the Gauss-Bonnet coupling constant \cite{Cai:2001dz, Cai:2013qga}. Similarly, the extended phase space thermodynamic of the Gauss-Bonnet black hole exhibits $P-V$ criticality and small/large black hole phase transitions in five dimensions that, depending on the magnitude of the Gauss-Bonnet coupling constant and charge, may change in higher dimensions \cite{Cai:2013qga}.

It is instructive to analyse the interplay of phase transitions and Lyapunov exponent in higher derivative Gauss-Bonnet black holes. This interplay was discussed in \cite{lyu2023probing} for the standard thermodynamics. Here, we concentrate on this interplay in the extended-phase space thermodynamics. We will also comment on this interplay in the standard thermodynamics, thereby complementing the work of \cite{lyu2023probing} and making nontrivial new observations not discussed there. For the extended phase thermodynamics of Gauss-Bonnet black holes, we follow \cite{Cai:2013qga, Wei:2014hba}. We will work in five dimensions for simplicity as the discussion can be easily generalised for higher dimensions.

The line element of five-dimensional charged Gauss-Bonnet-AdS black holes with spherical horizon topology is
\begin{equation}\label{eq:gb_metric}
	ds^{2} = -f(r)dt^{2}+\frac{dr^{2}}{f(r)}+r^{2}[d\theta^{2} +\sin^{2}\theta(d\phi^{2}+\sin^{2}\phi d\psi^{2})],
\end{equation}
where $f(r)$ is given by
\begin{equation}\label{gb_metric_function}
	f(r) = 1 + \frac{r^2}{4\alpha}\bigg(1-\sqrt{1+\frac{64\alpha M}{3\pi r^4}-\frac{8\alpha Q^2}{3 r^6}-\frac{32\pi \alpha P}{3}}\bigg),
\end{equation}
with $M$, $Q$, $\alpha$, and $P$ as the black hole mass, charge, Gauss-Bonnet coupling constant, and pressure, respectively. In the extended phase space, the pressure $P$ is equated with the cosmological constant $P = -\frac{\Lambda}{8\pi}=\frac{3}{8\pi l^2}$. The requirement of a well-defined vacuum solution, corresponding to $M=Q=0$, puts a constraint on $\alpha$ and $P$
\begin{equation}\label{gbconstraint}
0\leq \frac{32\pi \alpha P}{3} \leq 1 \,.
\end{equation}
The horizon radius $r_h$ is obtained by the largest real root of the equation $f(r_h)=0$, and the black hole mass can be expressed in terms of $r_h$ as
\begin{equation}\label{gbmass}
M = \frac{3\pi r_h^2}{8}\left(1+\frac{2\alpha}{r_h^2}+\frac{4\pi P r_h^2}{3}   \right) + \frac{\pi Q^2}{8r_h^2}\,.
\end{equation}
It is worth pointing out that the black hole mass $M$ acts as
the enthalpy $H$ rather than the internal energy of the gravitational system in the extended phase thermodynamics \cite{Kastor:2009wy}. Similarly, the expressions for temperature ($T$), entropy ($S$), volume ($V$), and electric potential ($\Phi$) are given by \footnote{We have set the five-dimensional Newton's constant to one in these expressions.}
\begin{eqnarray}\label{eq:gb_S_V_phi}
T &=& \frac{16\pi P r_h^6 + 6 r_h^4 - 2Q^2 }{12 \pi r_h^3(4\alpha+r_h^2)}, \\
	S &=& 6\alpha \pi^2 r_h + \frac{1}{2}\pi^2 r_h^3, \\
	V &=& \frac{\pi^2 r_h^4}{2}, \\
	\Phi &=& \frac{\pi Q}{4 r_h^2}.
\end{eqnarray}
These thermodynamic variables satisfy the differential first law
\begin{eqnarray}\label{bg1stlaw}
dH=TdS + \Phi dQ + VdP\,,
\end{eqnarray}
and the Gibbs free energy is defined as
\begin{eqnarray}\label{gb_free_energy_formula}
& & G = H - TS\,, \\ \nonumber
& & = \frac{\pi  \left(r_h^6 (3-144 \pi  \alpha  P)-4 \pi
   P r_h^8+72 \alpha ^2 r_h^2-18 \alpha
   r_h^4\right)}{24 r_h^2 \left(4 \alpha
   +r_h^2\right)}+\frac{\pi  Q^2 \left(36 \alpha +5
   r_h^2\right)}{24 r_h^2 \left(4 \alpha
   +r_h^2\right)}
\end{eqnarray}
From the above equations, one can find the equation of state
\begin{eqnarray}\label{gbEOS}
P =\frac{3}{4 r_h} \left( 1+\frac{4\alpha}{r_h^2} \right)T -\frac{3}{8\pi r_h^2}+\frac{Q^2}{8 \pi r_h^6} \,.
\end{eqnarray}
The above equation of state can be compared with the van der Waals equation by expanding the latter with the inverse specific volume $v$
\begin{eqnarray}\label{gbvdw}
P =\frac{T}{v-b} - \frac{a}{v^2} \approx \frac{T}{v} + \frac{b T - a}{v^2} +\mathcal{O}(v^{-3}) \,,
\end{eqnarray}
which also allows us to identify the specific volume $v$ with the horizon radius $r_h$. In the $P-V$ diagram, the critical point, as usual, is determined from the condition of the inflexion point
\begin{eqnarray}\label{inflectionpoint-equation}
\frac{\partial P}{\partial r_h}\bigg\rvert_{r_h=r_{hc},T=T_c} =0,~~ \frac{\partial^2 P}{\partial r_h^2}\bigg\rvert_{r_h=r_{hc},T=T_c}=0 \,.
\end{eqnarray}
These equations give us the critical temperature and pressure
\begin{eqnarray}\label{inflectionpoint}
T_c &=& \frac{r_{hc}^4-Q^2}{\pi  r_{hc}^3 \left(12 \alpha +r_{hc}^2\right)}\,, \\ \nonumber
P_c &=& -\frac{5 Q^2 r_{hc}^2+12 \alpha  r_{hc}^4-3
   r_{hc}^6+12 \alpha  Q^2}{8 \pi  r_{hc}^6 \left(12
   \alpha +r_{hc}^2\right)}\,,
\end{eqnarray}
and lead to the following equation
\begin{eqnarray}\label{gbcritical}
r_{hc}^6 - 12 r_{hc}^4 \alpha - 5 r_{hc}^2 Q^2 - 36\alpha Q^2 =0 \,,
\end{eqnarray}
from which the critical horizon radius $r_{hc}$ can be obtained. The above equations suggest the critical points are $\alpha$ and $Q$ dependent quantities. Although the above equation can be solved analytically for $r_{hc}$, it is too long to reproduce here and is not very informative.

\begin{figure}[htb!]
	\centering
	\subfigure[Temperature vs horizon radius]{\label{fig:GB_Tvsrh_collage}
		\includegraphics[width=0.45\linewidth]{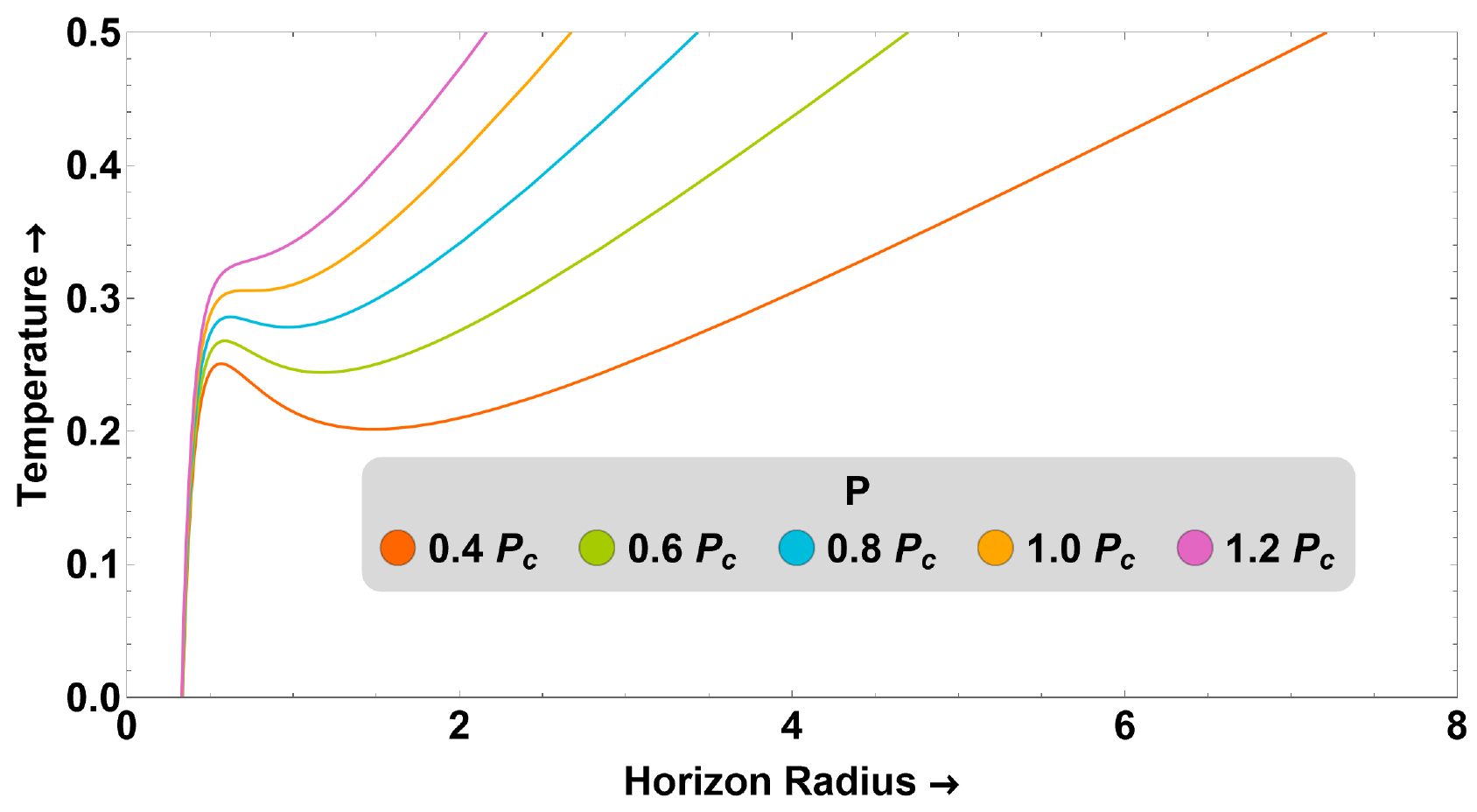}}
	\hfill
	\subfigure[Gibbs free energy vs temperature]{\label{fig:GB_GvsT_collage}
		\includegraphics[width=0.45\linewidth]{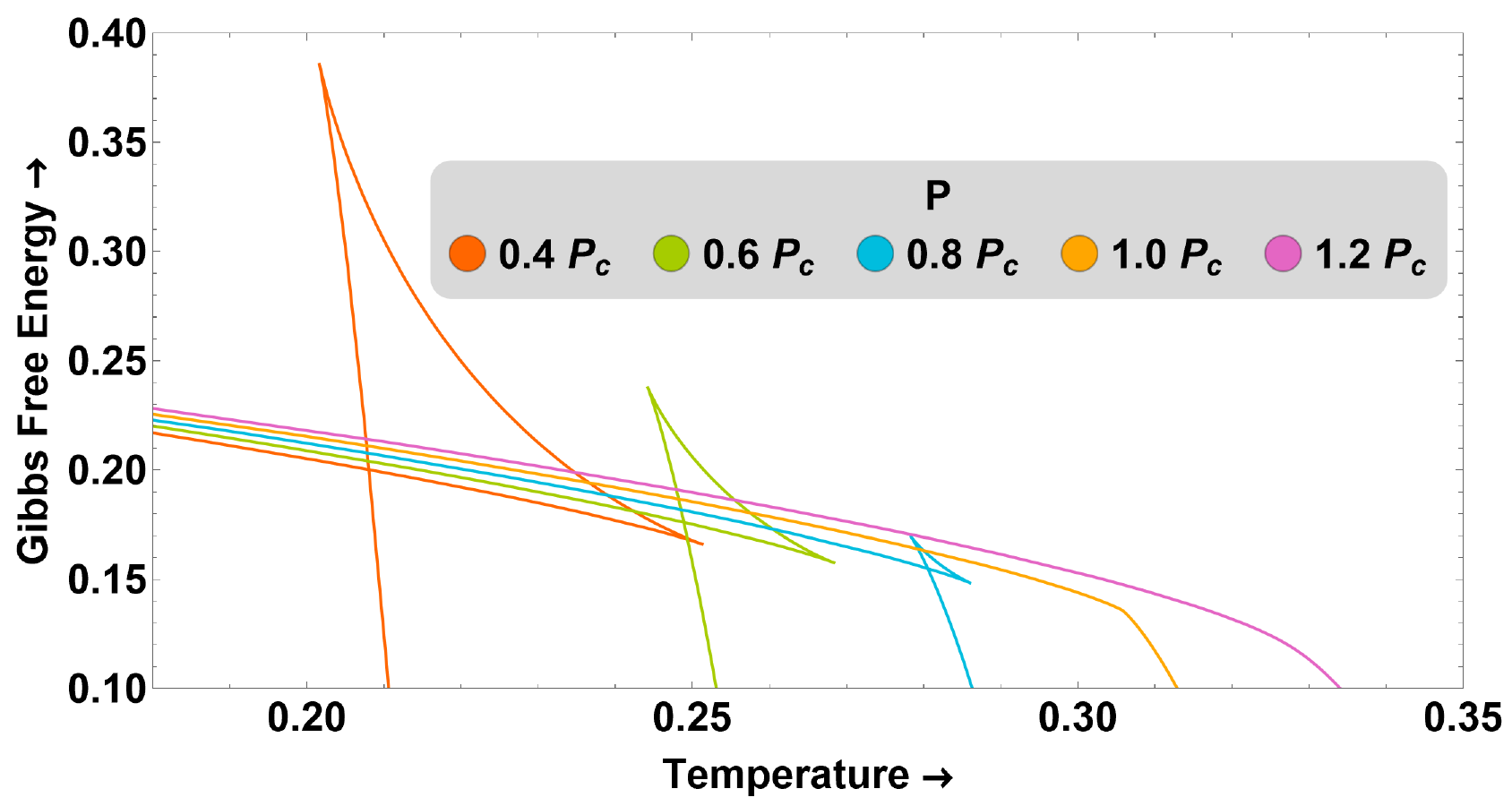}}
    \vfill
    \subfigure[$P = 0.5 P_c$]{\label{fig:GB_GvsT_1}
		\includegraphics[width=0.45\linewidth]{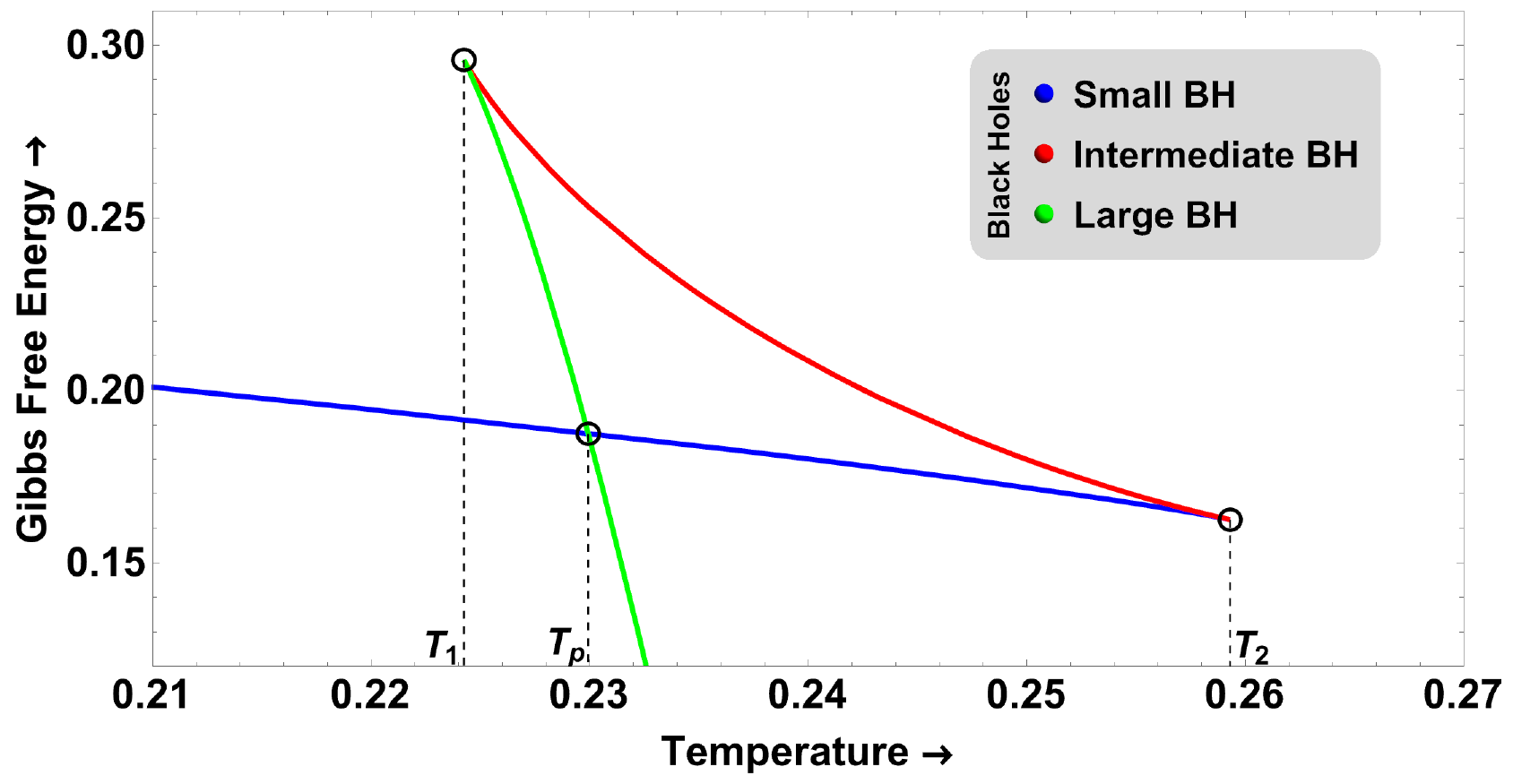}}
	\hfill
	\subfigure[$P = 1.1 P_c$]{\label{fig:GB_GvsT_2}
		\includegraphics[width=0.45\linewidth]{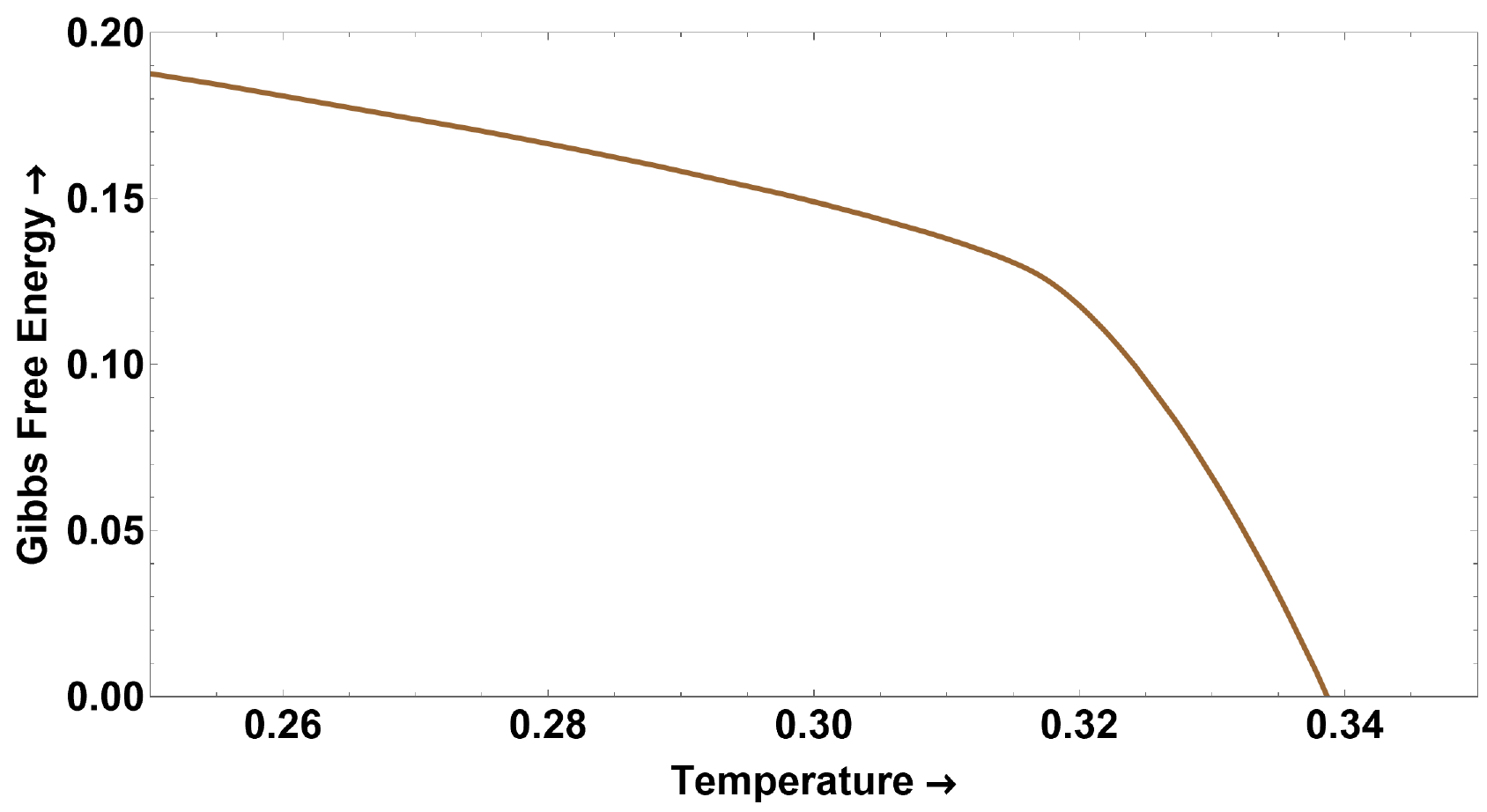}}
	\caption{\label{fig:GB_GvsTandTvsrh}Thermodynamic phase structure of the charged Gauss-Bonnet black hole. Here, $\alpha=0.01$ and $Q=0.2$ are used.}
\end{figure}

In the following discussion, we take $\alpha=0.01$ and $Q=0.2$ without loss of generality. The critical points for these values of $\alpha$ and $Q$ are $r_{hc}\simeq0.735$, $P_c\simeq0.124$, and $T_c\simeq0.0161$. The condition (\ref{gbconstraint}) also constrains the $P$ value to be $P<2.984$. The thermodynamic phase structure is shown in Fig.~\ref{fig:GB_GvsTandTvsrh}, where the orange line denotes the critical isotherm. Three black hole phases exist in some temperature range for a fixed pressure lower than the critical one. Depending upon the relative magnitude of $r_h$, these phases correspond to large, intermediate, and small phases. The small and large black hole phases have a positive specific heat and compression coefficient, thus corresponding to stable phases. Between them is an intermediate unstable phase with a negative specific heat and compression coefficient. Therefore, the isothermal line allows two physical horizon radii for appropriate temperature values. This leads to a phase transition between the small and large horizon radii phases.

\begin{figure}[htb!]
	\centering
	\subfigure[Temperature vs horizon radius]{\label{fig:GB_Tvsrh_collageforQ0}
		\includegraphics[width=0.45\linewidth]{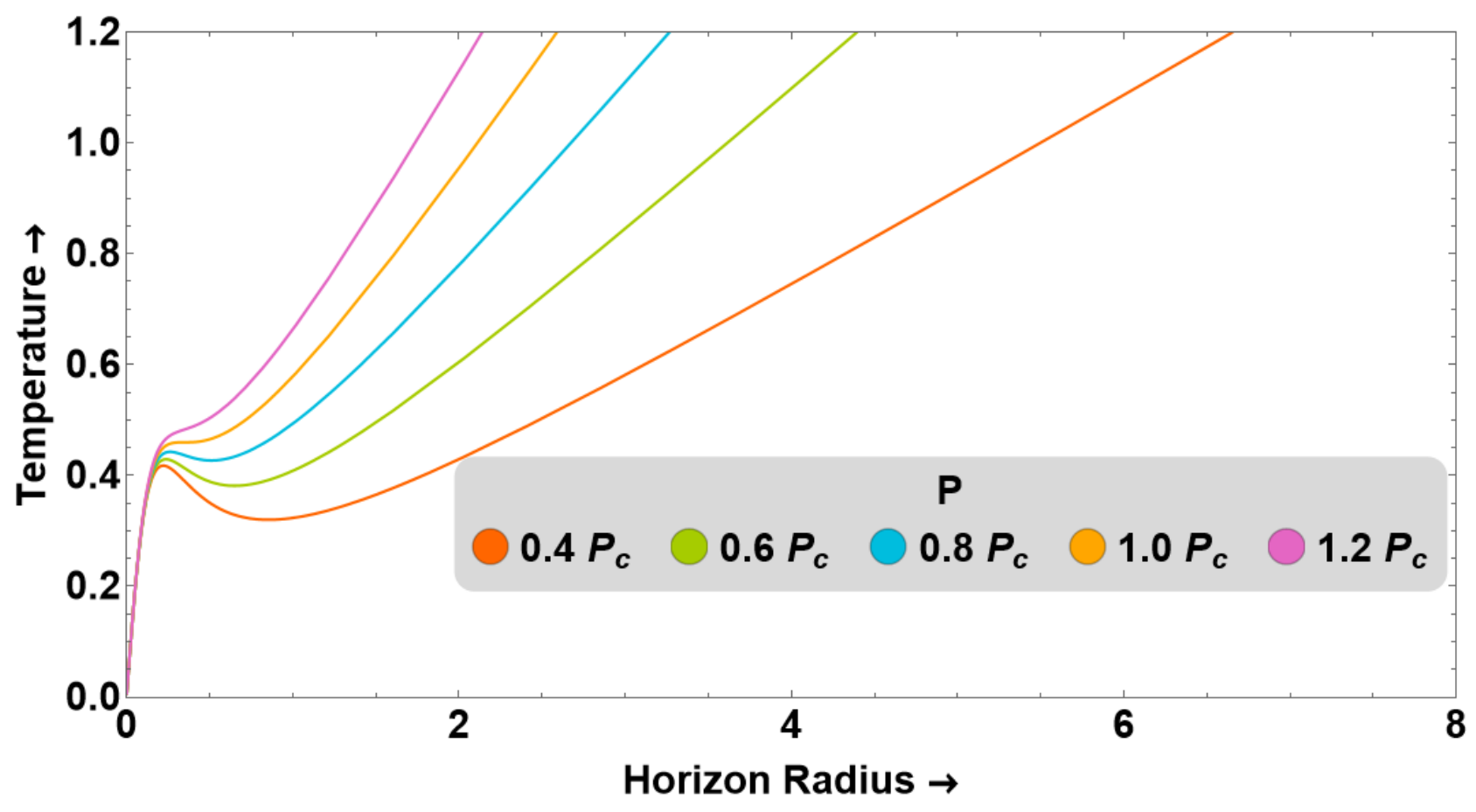}}
	\hfill
	\subfigure[Gibbs free energy vs temperature]{\label{fig:GB_GvsT_collageforQ0}
		\includegraphics[width=0.45\linewidth]{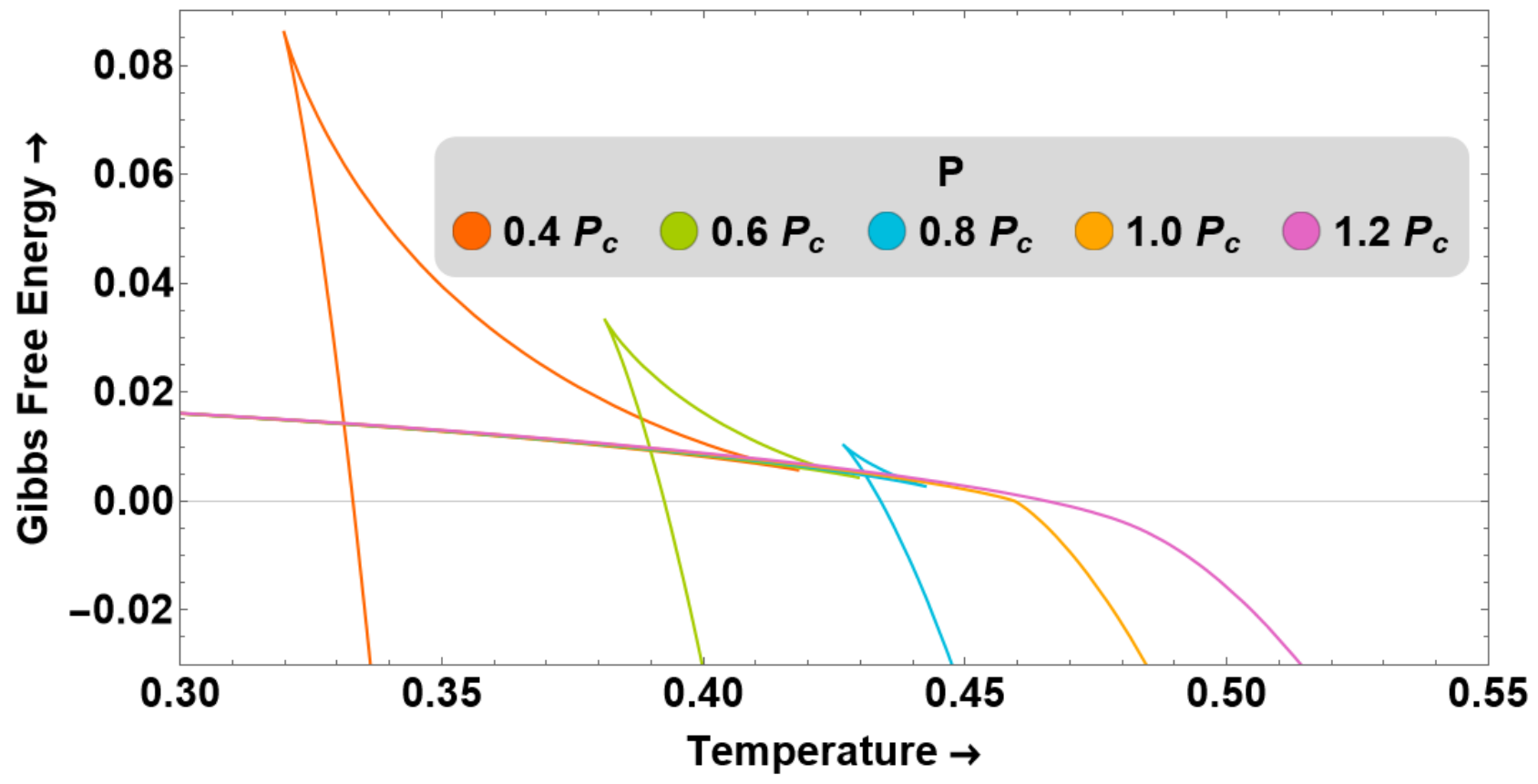}}
	\caption{\label{fig:GB_GvsTandTvsrhforQ0}Temperature as a function of horizon radius (left) and Gibbs free energy $G$ as a function of the temperature $T$ (right) for Gauss-Bonnet black holes when $Q=0$ and $\alpha=0.01$.}
\end{figure}

The corresponding Gibbs free energy behaviour is shown in Fig.~\ref{fig:GB_GvsT_collage}. As expected, it exhibits the swallow-tail-like structure for $P<P_c$ and exchange dominance as the temperature varies. In particular, the small/large black hole phase has the lowest free energy at small/large temperatures, indicating the existence of small/large black hole phase transition when $P<P_c$. When $P=P_c$, the two physical horizon radii coincide, resulting in coexistence. Meanwhile, for $P>P_c$, only one black hole phase appears, which always has a positive specific heat and compression coefficient. This small/large black hole phase transition phenomenon is reminiscent of the van der Waals system's liquid/gas phase transition. Such phase transition is first order for $P<P_c$, while it becomes second order at $P_c$, i.e., a first-order phase transition line exists between small and large black hole phases, which terminates at a second-order point.

The above phase structure of the Gauss-Bonnet black hole is quite similar to the dyonic or Bardeen black holes discussed earlier. However, there is one major difference. In the Gauss-Bonnet case, the small/large black hole phase transition appears even without charge, i.e., when $Q=0$. In dyonic or Bardeen black holes, the small/large phase transition occurs only when the charge is switched on, and in the absence of it, instead, the Hawking/Page phase transition between the black hole and thermal-AdS phases appears. For $Q=0$, the thermodynamic phase structure of the Gauss-Bonnet black hole is shown in Fig.~\ref{fig:GB_GvsTandTvsrhforQ0}, where one can explicitly see the existence of the first-order small/large black hole phase transition and the second-order critical point. For $Q=0$, the magnitude of pressure, temperature, and horizon radii at the critical point is $P_c=1/(96\pi)$, $T_c=1/(4\pi\sqrt{3})$, and $r_{hc}=2\sqrt{3}$.

\begin{figure}[htb!]
	\centering
	\includegraphics[width=0.4\linewidth]{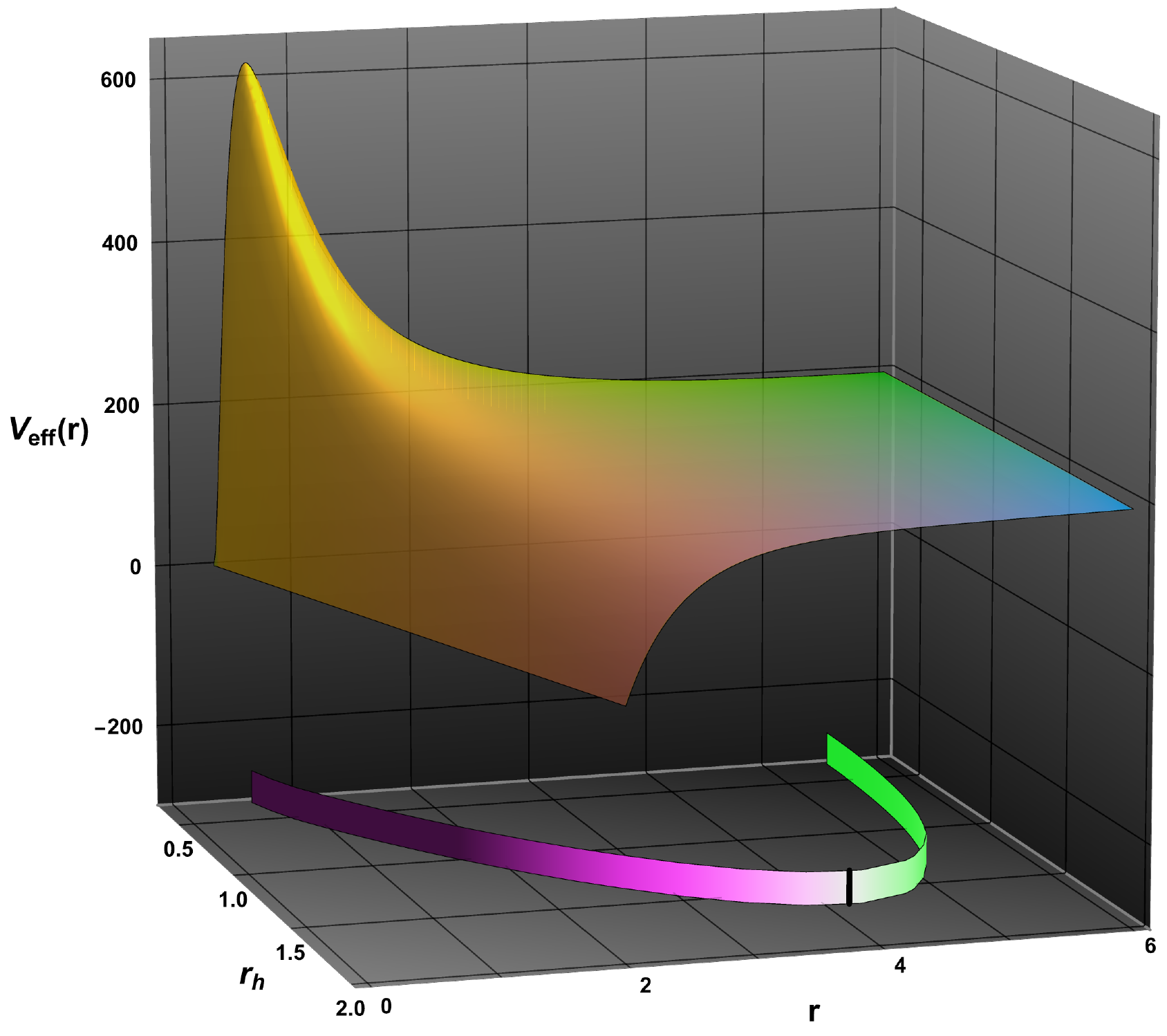}
	\caption{\label{fig:GB_veffective3d}The 3D plot of the effective potential $V_\text{eff}(r)$ as a function of horizon radius $r_h$ and orbit radius $r$ for the massive particle when $L = 20$, $P = 0.1$, $\alpha=0.01$ and $Q = 0.2$. The pink and green curves projected in the lower part correspond to the unstable and stable equilibria of the timelike circular geodesics.}
\end{figure}

Now, we will discuss the effective potential for massless and massive particles in the background of the Gauss-Bonnet black hole. The potential is given by,
\begin{equation}\label{eq:GB_Veffective}
    V_{\text{eff}}(r) = -\frac{1}{12 \alpha  r^2}\left(L^2-r^2 \epsilon \right) \left(r^2 \left(\sqrt{3} \left(\sqrt{\begin{aligned}3+\frac{8 \alpha
   \left(4 \pi  P r_h^6+3 r_h^4+Q^2\right)}{r^4 r_h^2}\\
   -\frac{8 \alpha  \left(4 \pi  P
   r^6+Q^2-6 \alpha  r^2\right)}{r^6}\end{aligned}}\right)-3\right)-12 \alpha \right)
\end{equation}
where, once again, $L$ is the particle's angular momentum, and $r$ is the radius of the particle's orbit. The 3D Plot of $V_{\text{eff}}(r)$ as a function of $r$ and $r_h$ is shown in Fig.~\ref{fig:GB_veffective3d} for a fixed $L = 20$, $P = 0.1$, $\alpha$ = 0.01 and $Q = 0.2$.

Again, the extrema are projected below the 3D plot onto the plane $ V_{\text{eff}}(r)=-300$. Here, the part with the pink gradient represents the maxima, the unstable stationary points, and the green gradient represents the minima, the stable stationary points of the circular geodesic. We are mainly concerned with the pink region, which appears in the calculation of the Lyapunov exponents.

\subsubsection{Massless particles}
\begin{figure}[htb!]
	\centering
    \textbf{Massless particles}\par\medskip
	\subfigure[$P = 0.5 P_c$]{\label{fig:GB_Lmassless_vsT_1}
		\includegraphics[width=0.45\linewidth]{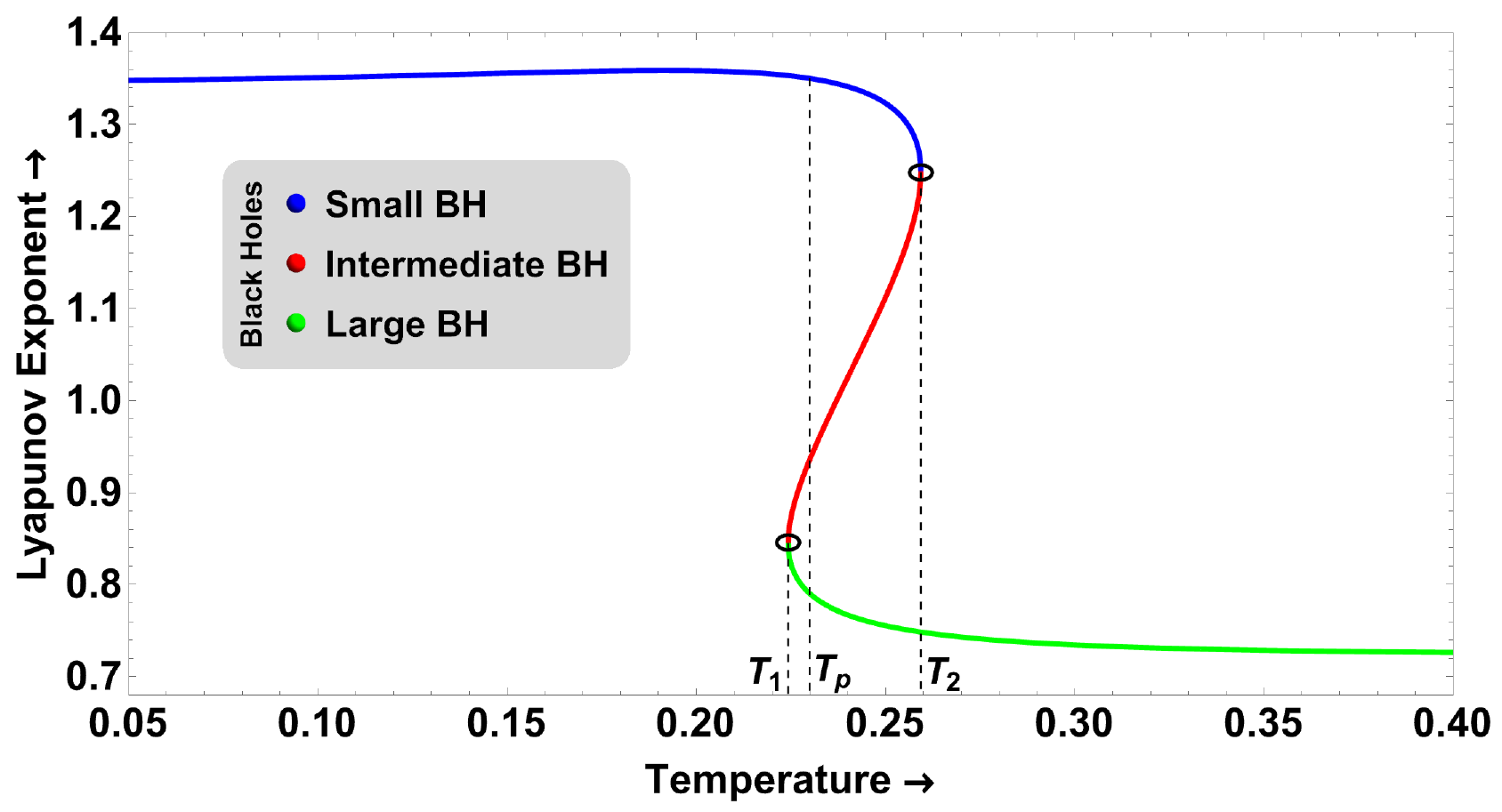}}
	\hfill
	\subfigure[$P = 1.1 P_c$]{\label{fig:GB_Lmassless_vsT_2}
		\includegraphics[width=0.45\linewidth]{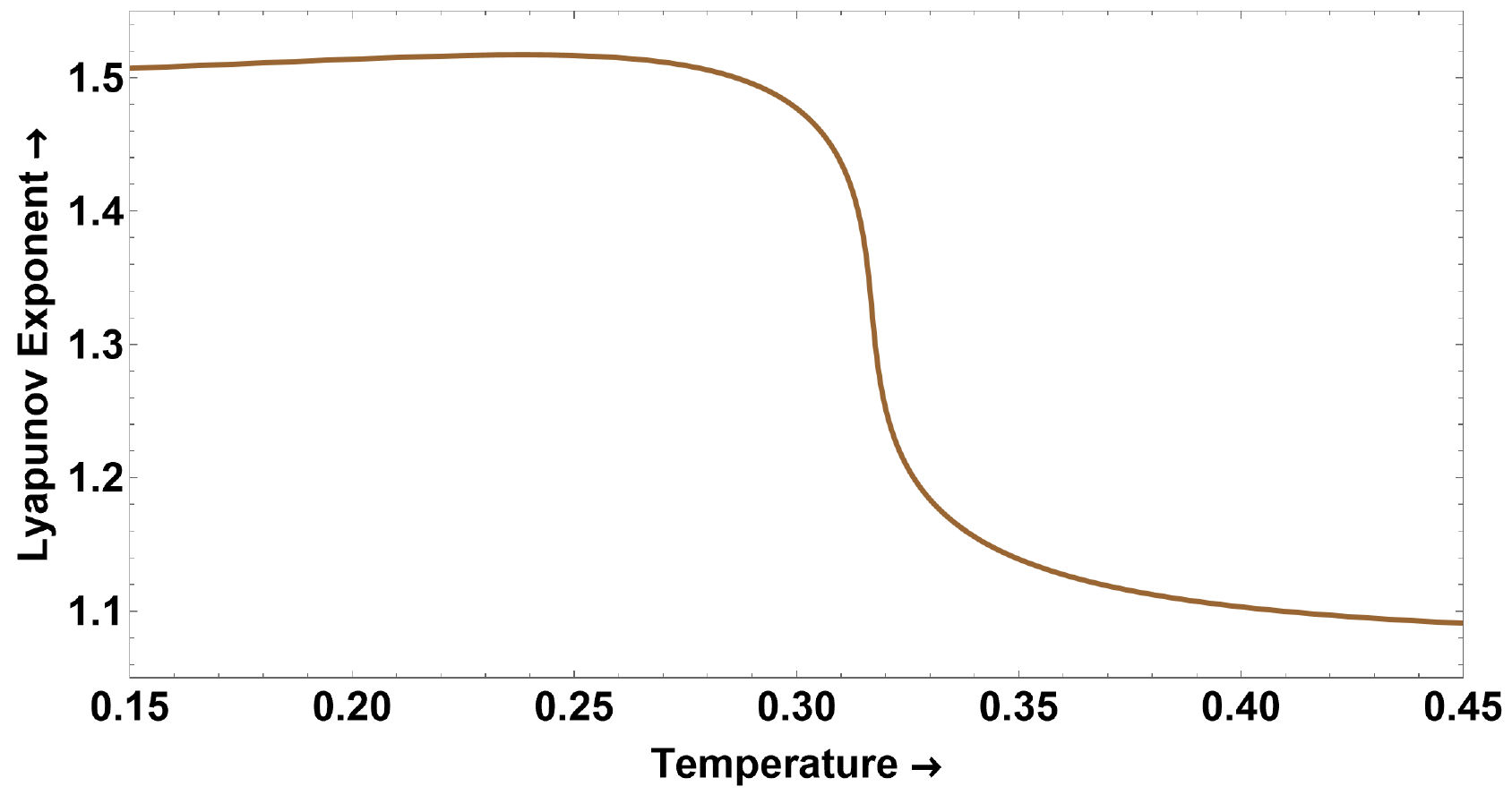}}
	\vfill
	\subfigure[Lyapunov exponent vs temperature]{\label{fig:GB_Lmassless_vsT_collage}
		\includegraphics[width=0.45\linewidth]{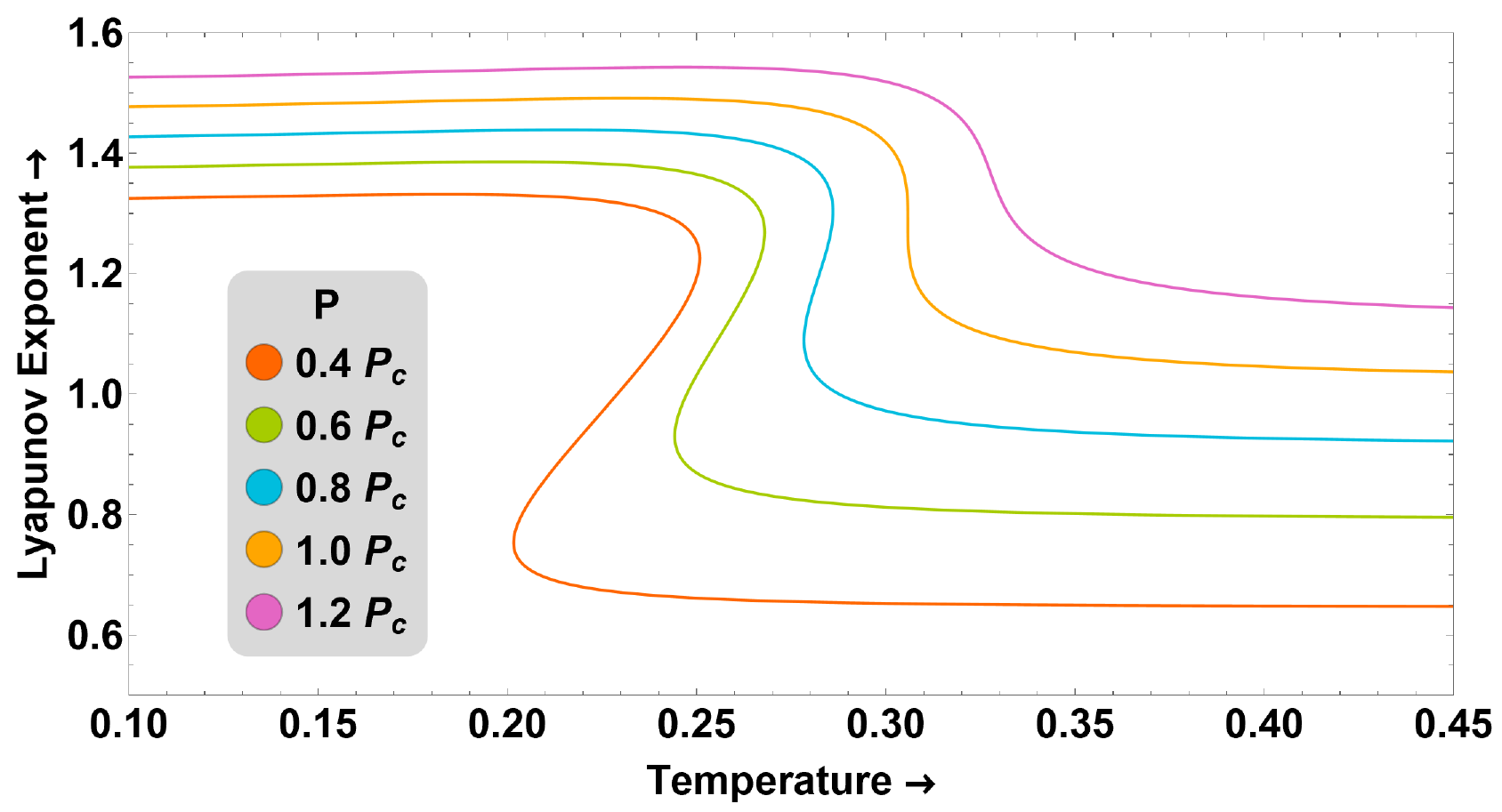}}
    \hfill
	\subfigure[Lyapunov exponent vs horizon radius]{\label{fig:GB_Lmassless_vsrh_collage}
		\includegraphics[width=0.45\linewidth]{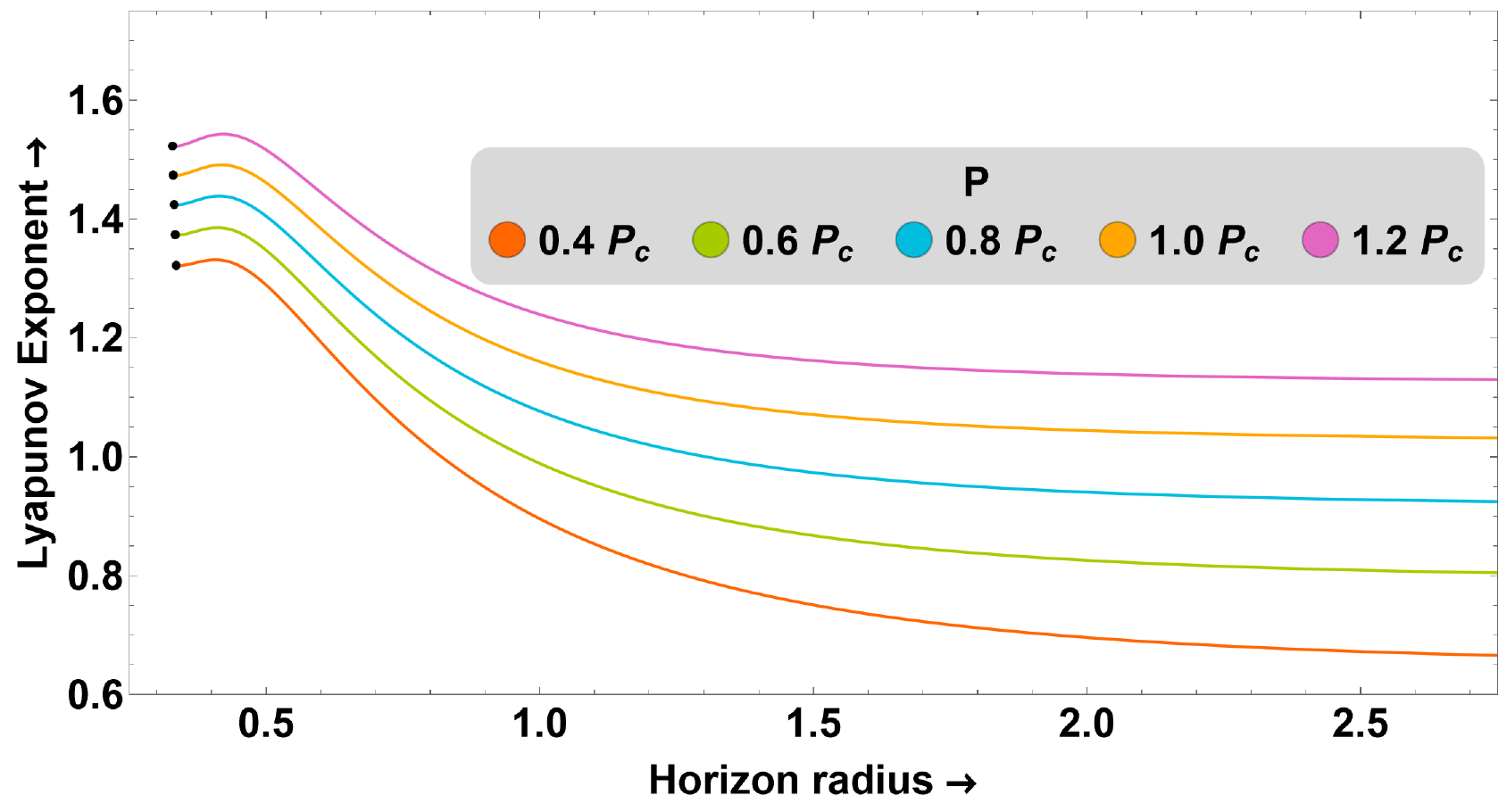}}
	\caption{\label{fig:GB_Lmassless}Lyapunov exponent of the massless particle $\lambda$ as a function of temperature $T$ and horizon radius $r_h$ for the Gauss-Bonnet black hole. Here, $\alpha = 0.01$ and $Q=0.2$ are used.}
\end{figure}
Using Eq.~(\ref{eq:GB_Veffective}) for $\epsilon=0$ and (\ref{eq:lyapunov-massless}), we can probe the thermodynamics of Gauss-Bonnet black holes with Lyapunov exponents for massless particles. It is shown in Fig.~\ref{fig:GB_Lmassless} for a fixed $\alpha=0.01$ and $Q=0.2$. We work in the extended phase space where the cosmological constant is treated like the pressure $P$. When $P$ is less than the critical pressure $P_c$, as calculated in Eq.~(\ref{inflectionpoint}), the Lyapunov exponent again shows a multi-valued nature. Specifically, $\lambda$ is single-valued when $T<T_1$ (small black hole phase) or $T>T_2$ (large black hole phase) but is multi-valued when $T_1<T<T_2$, with the small/large black hole phase transition occurring at $T_p$. As we increase the pressure beyond the critical value, this multi-valuedness disappears, and we have a single thermal profile of $\lambda$. This is explicitly shown in Fig.~\ref{fig:GB_Lmassless_vsT_2}. This means that even in the extended phase space, the Lyapunov exponent can be a useful tool in studying black hole phase transition, and the second-order critical point can be precisely obtained from its thermal structure. For completion, the plot of $r_h$ vs $\lambda$ is also shown in Fig.~\ref{fig:GB_Lmassless_vsrh_collage}. Here, we can observe that $\lambda$ first slightly increases and gradually decreases until it attains different saturation values for different $P$.

Unlike the dyonic and Bardeen black holes studied in the previous subsections, since we are slowly increasing the pressure and taking it beyond the critical value, the Lyapunov exponent does not attain the same fixed constant value at large $T$. This is illustrated in Fig.~\ref{fig:GB_Lmassless_vsT_collage}. The reason is that changing the pressure (or cosmological constant) ultimately changes the AdS radius. If we take $l=1$ and work in the standard phase space, then $\lambda$ saturates to a constant value at large $T$ and $r_h$ even in the Gauss-Bonnet black hole background as well.

\begin{figure}[htb!]
	\centering
	\includegraphics[width=0.8\linewidth]{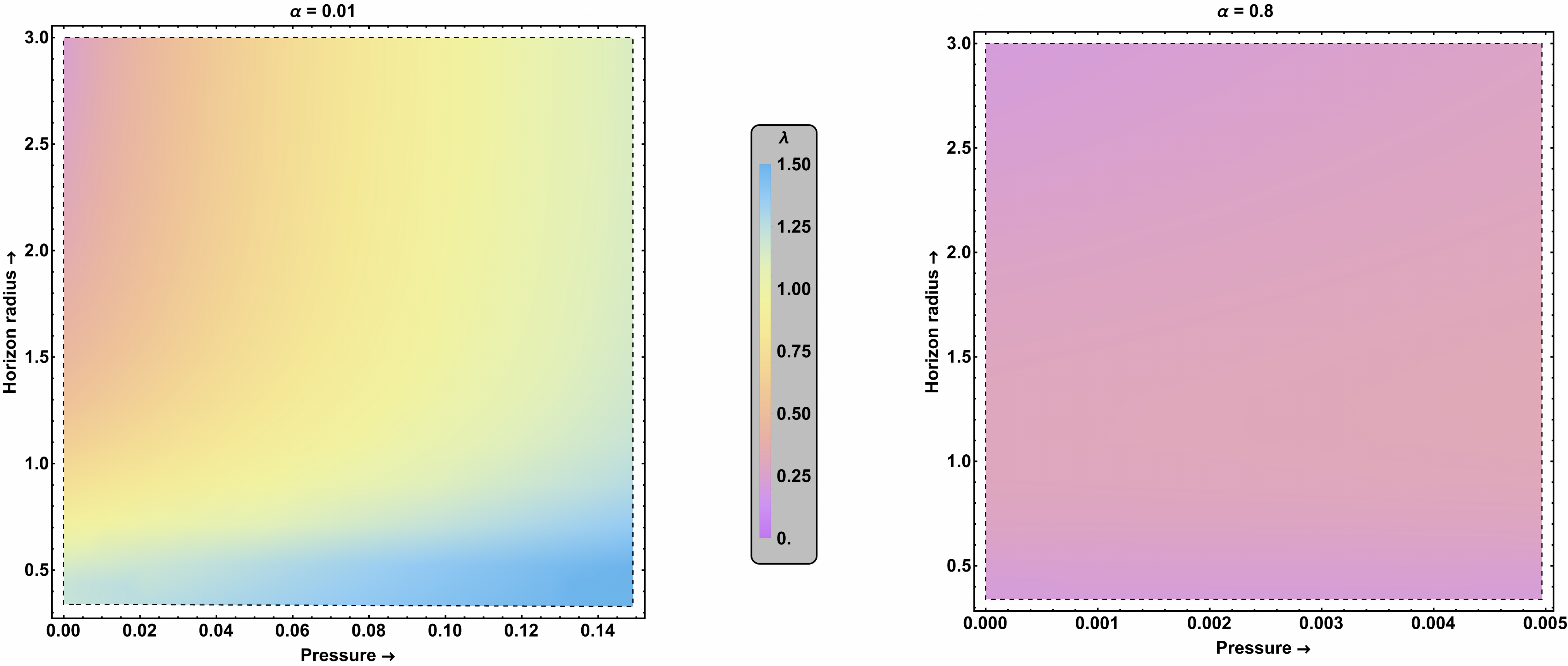}
	\caption{\label{fig:GB_Lmassless_contour} Density plot of $\lambda$ for the massless particle as a function of the pressure $P$ and horizon radius $r_h$ for the Gauss-Bonnet black hole. Here, fixed $\alpha = 0.01$ (left) and $\alpha = 0.8$ (right)  are used.}
\end{figure}

The density plot of $\lambda$ is shown in Fig.~\ref{fig:GB_Lmassless_contour} for two different $\alpha$ values. On the left, we have used $\alpha=0.01$ to observe that the maximum value that $\lambda$ attains lies in the sky blue region, $\lambda\simeq 1.50$. In this case, at a fixed horizon radius, as we move from left to right in the density plot, i.e., increasing the pressure, $\lambda$ slightly increases. On the contrary, at a fixed pressure, as we move from bottom to top in the density plot, i.e., increasing the horizon radius, $\lambda$ gradually decreases and finally attains a constant value. Coming to the right, for $\alpha=0.8$, we observe similar behaviour as we go left to right or bottom to top for a fixed horizon radius/pressure. The only difference is in the range of values that $\lambda$ can have. Here, the maximum value of $\lambda$ lies in the light pink region, where $\lambda\simeq0.32$. Thus, increasing the Gauss-Bonnet coupling parameter $\alpha$ makes particle motion less chaotic.

\subsubsection{Massive particles}

\begin{figure}[htb!]
	\centering
	\textbf{Massive particles}\par\medskip
	\subfigure[$Q = 0.4 < Q_c$]{\label{fig:GB_Lmassive_vsT_1}
		\includegraphics[width=0.45\linewidth]{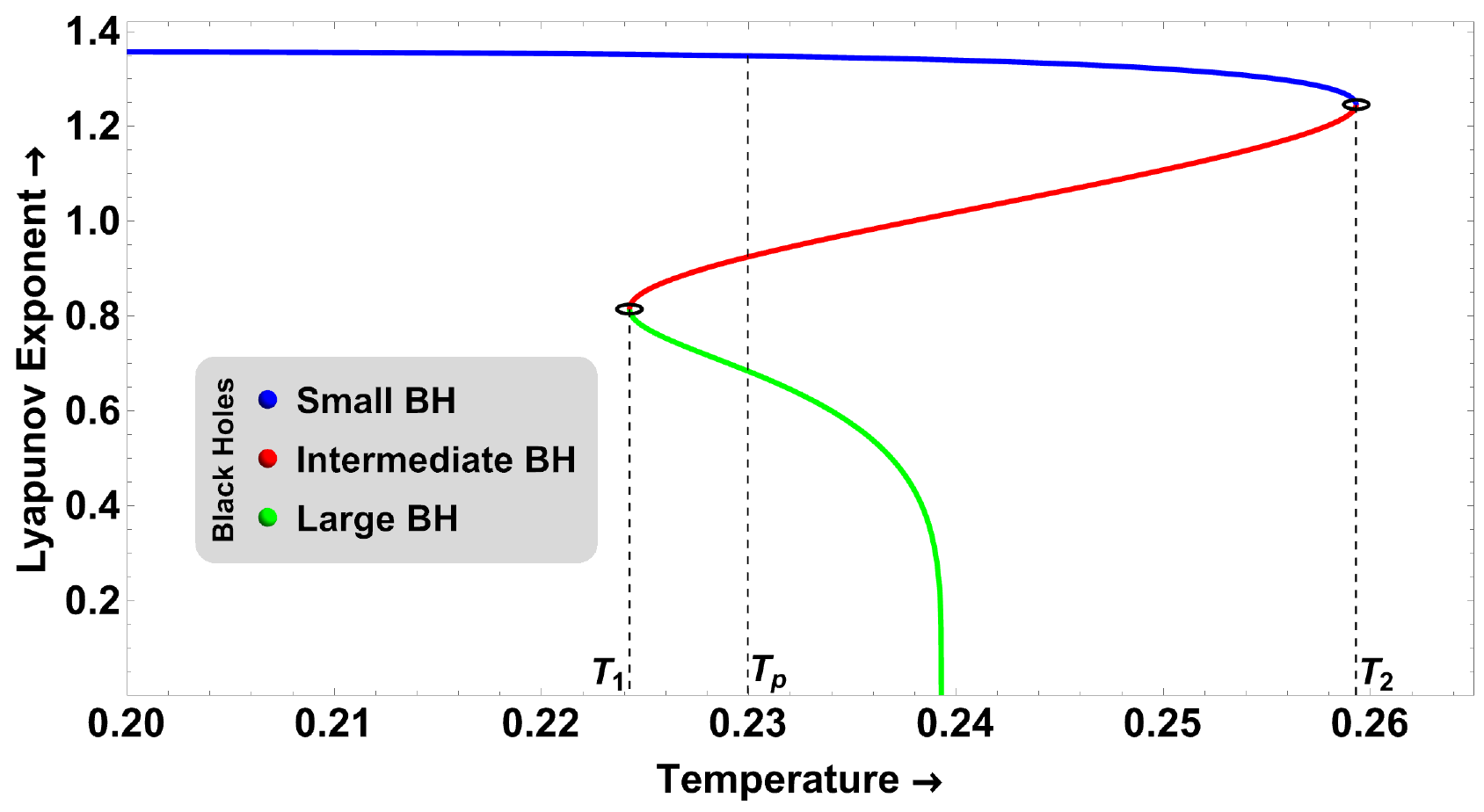}}
	\hfill
	\subfigure[$Q = 0.85 > Q_c$]{\label{fig:GB_Lmassive_vsT_2}
		\includegraphics[width=0.45\linewidth]{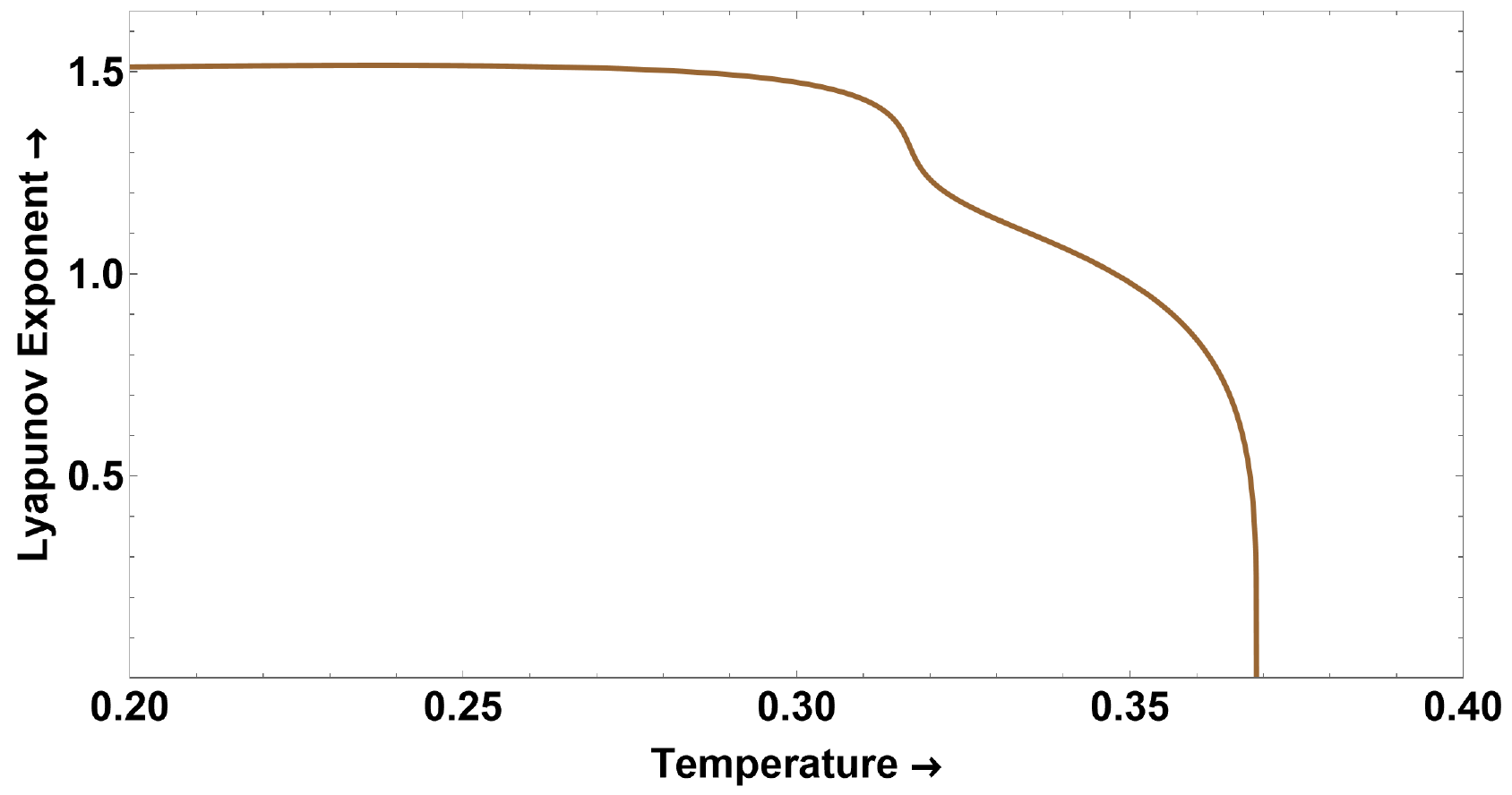}}
	\vfill
	\subfigure[Lyapunov exponent vs temperature]{\label{fig:GB_Lmassive_vsT_collage}
		\includegraphics[width=0.45\linewidth]{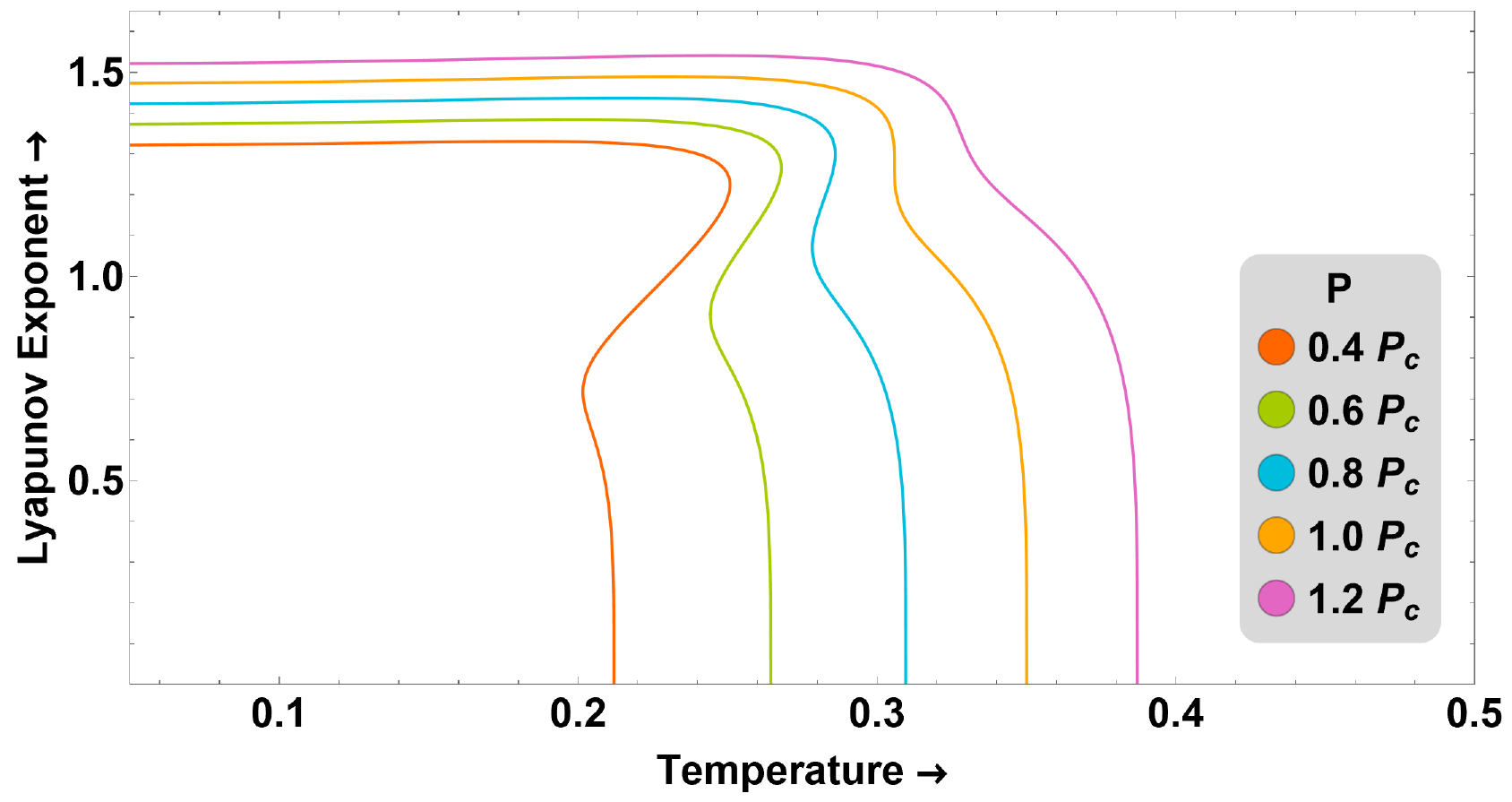}}
	\hfill
	\subfigure[Lyapunov exponent vs horizon radius]{\label{fig:GB_Lmassive_vsrh_collage}
		\includegraphics[width=0.45\linewidth]{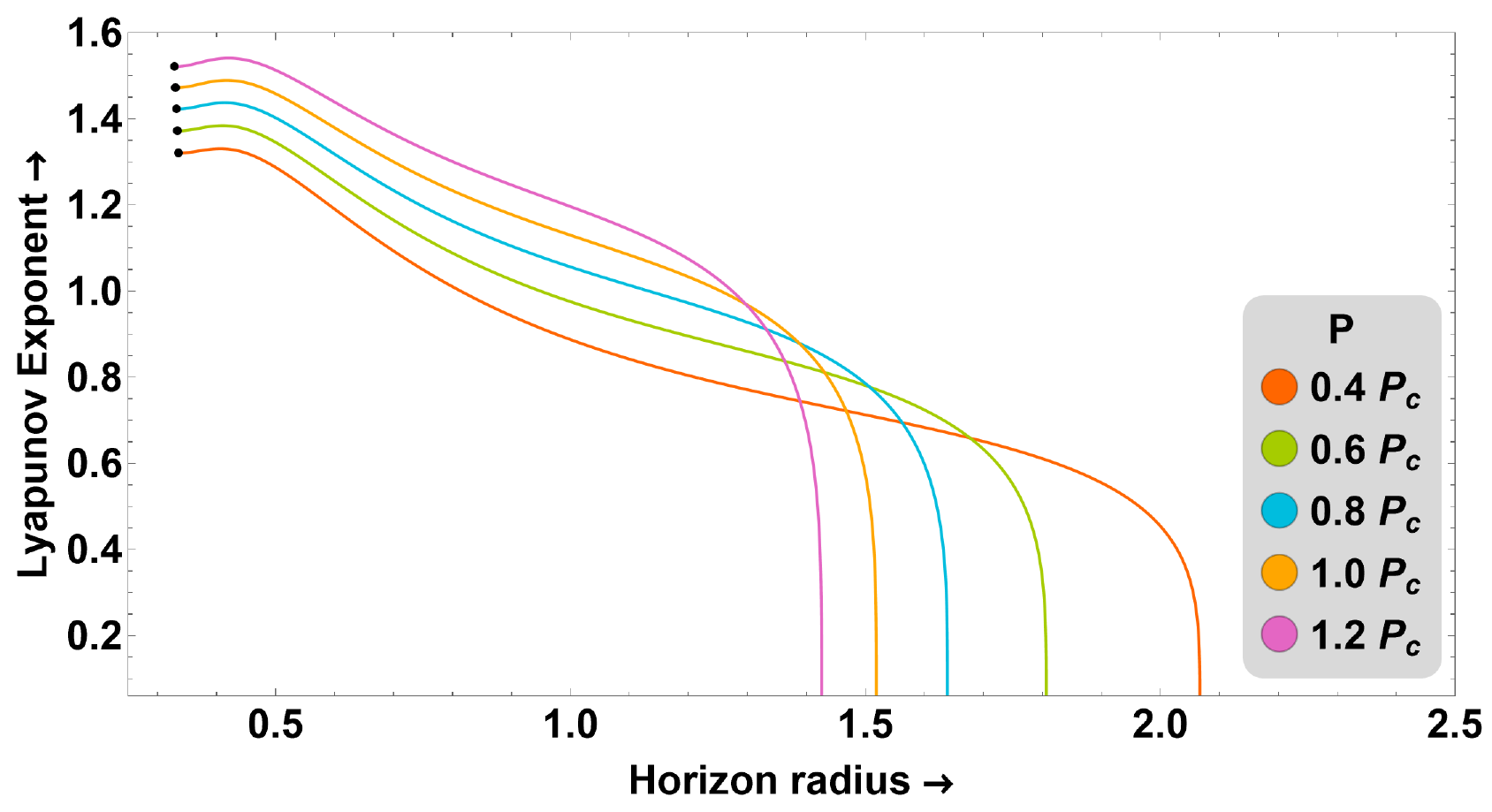}}
	\caption{\label{fig:GB_Lmassive}Lyapunov exponent of the massive particle $\lambda$ as a function of temperature $T$ and horizon radius $r_h$ for the Gauss-Bonnet black hole.}
\end{figure}

Using Eq.~(\ref{eq:GB_Veffective}) for $\epsilon=-1$ and (\ref{eq:lyapunov-massive}), we can similarly analyse the phase structure of the Gauss-Bonnet black hole via the Lyapunov exponents of massive particles. This is shown in Fig.~\ref{fig:GB_Lmassive}. The behaviour of the Lyapunov exponent, in this case, is quite similar to the massless particle, with one major difference being that here $\lambda$ drops to zero in the large black hole phase at some temperature (or horizon radius) for all $P$ values. This is illustrated
in Figs.~\ref{fig:GB_Lmassive_vsT_collage} and \ref{fig:GB_Lmassive_vsrh_collage}. The reason for this drop is the disappearance of the unstable equilibrium at those horizon radii, an example of which can be seen from the 3D plot of the effective potential in Fig.~\ref{fig:GB_veffective3d} where $P=0.1$ is used. There is no extremum on the right-hand side of the marked black line, and hence, the Lyapunov exponent approaches zero as $r_h$ comes close to the black line in the 3D plot.

\begin{figure}[htb!]
	\centering
	\includegraphics[width=0.8\linewidth]{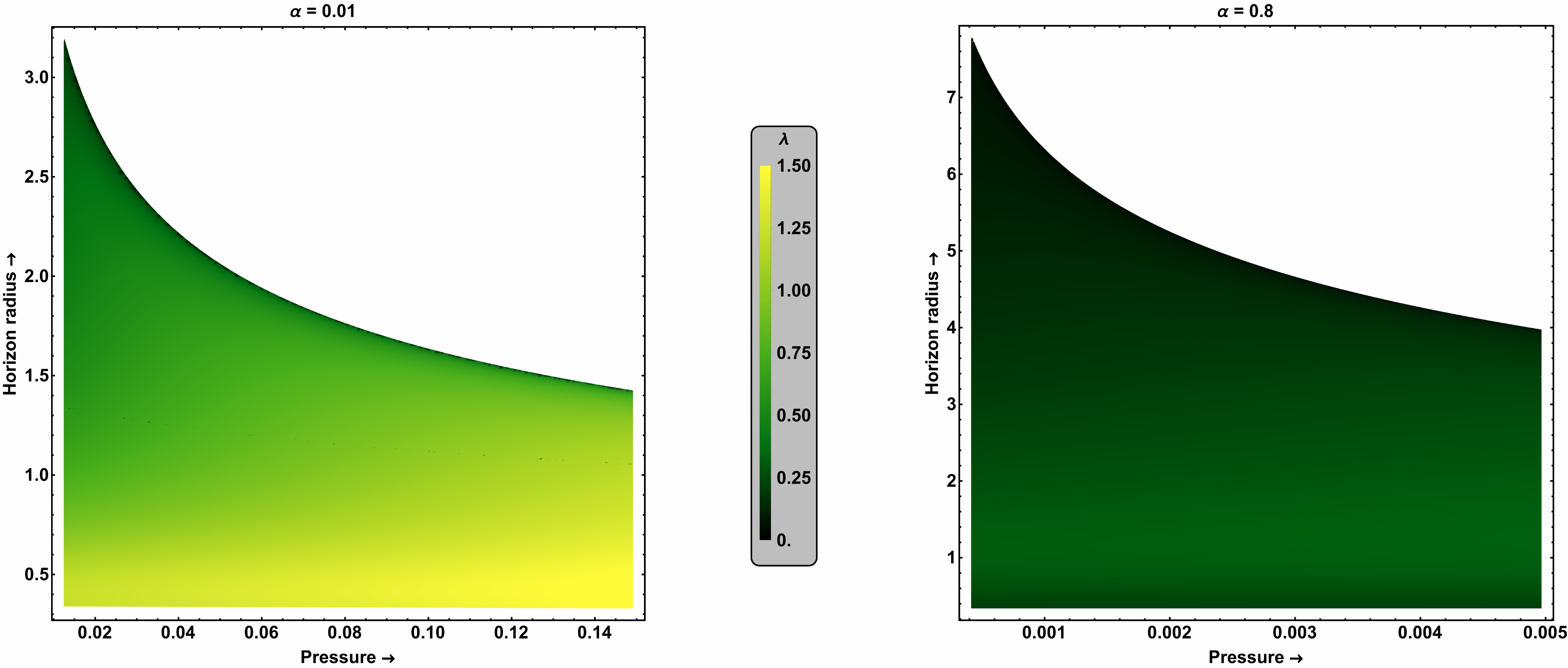}
	\caption{\label{fig:GB_Lmassive_contour} Density plot of $\lambda$ for the massive particle as a function of the pressure $P$ and horizon radius $r_h$ for the Gauss-Bonnet black hole. Here, fixed $\alpha = 0.01$ (left) and $\alpha = 0.8$ (right)  are used.}
\end{figure}

The density plot of $\lambda$, shown in Fig.~\ref{fig:GB_Lmassive_contour}, further confirms the above results. The top boundary for $\alpha=0.01$ and $\alpha=0.8$ cases is dark green, corresponding to $\lambda=0$. Another important observation from the density plot is that the horizon radius at which $\lambda$ drops to zero decreases as we increase the pressure for both $\alpha$ values. The maximum allowed value of $r_h$ at a fixed pressure is greater for $\alpha=0.8$ than $\alpha=0.01$. This implies that by increasing the value of $\alpha$, the allowed $r_h$ range for which $\lambda$ is nonzero increases. Similarly, like for the massless case, the magnitude of $\lambda$ decreases as we increase $\alpha$, implying less chaos in the massive particle motion with the Gauss-Bonnet coupling.


\subsection{AdS Black Holes in Massive Gravity}
The study of AdS black holes with massive gravity is an exciting and dynamic field of theoretical research \cite{Fierz:1939ix, deRham:2010ik, deRham:2010kj, Dvali:2000hr, Bergshoeff:2009hq, deRham:2014zqa, Hinterbichler:2011tt}. Giving the graviton a mass considerably changes general relativity by weakening it at large scales, but nonetheless, it leads to the same predictions as general relativity at small scales. In principle, the massive gravity systems could explain the acceleration of the Universe without introducing the dark energy component, thereby making them highly popular in gravitation studies. By adding an effective mass to the graviton, the large graviton modifies the gravitational potential and affects how the black hole solutions behave. Compared to their general relativity counterparts, these black holes have unique characteristics such as altered horizon structures, thermodynamics, and dynamics. The study of massive gravity in AdS spaces has also revealed new information on thermodynamic phase transitions, the nature of gravitational waves, and the holographic dualities between massive gravity and their boundary theories. For more information and the current status of the massive gravity theories, we refer the readers to review papers \cite{deRham:2014zqa, Hinterbichler:2011tt, deRham:2016nuf}. From the perspective of this paper, the AdS massive gravity also exhibits interesting phase transitions. Therefore, it is worthwhile to explore further the interplay between phase transitions and the Lyapunov exponent in these gravity systems.

Of the many avatars of massive gravity theories, we will examine a theory with Lorentz symmetry breaking. See \cite{Dubovsky:2004sg, Rubakov:2008nh}, for a review of Lorentz violating massive gravity theory. The metric of such massive gravity black holes was obtained in \cite{Bebronne:2009mz, Comelli:2010bj}, and their extended phase space thermodynamics were discussed in \cite{Fernando:2016sps}.

The metric of AdS black holes in massive gravity is given by
\begin{equation}\label{eq:mg_metric}
	ds^{2} = -f(r)dt^{2}+\frac{dr^{2}}{f(r)}+r^{2}(d\theta^{2} +\sin^{2}\theta d\phi^{2}),
\end{equation}
where
\begin{equation}\label{eq:mg_metric_function}
	f(r) = 1-\frac{2M}{r}-\gamma\frac{Q^2}{r^\omega}-\frac{\Lambda r^2}{3}\,,	
\end{equation}
The above solution approaches the usual AdS-Schwarzschild black hole for large distances when $\omega > 1$, whereas the Arnowitt-Deser-Misner mass of such solutions becomes divergent when $\omega < 1$. Hence, here we will consider $\omega > 1$.  The parameter $\gamma$ can take two values $1$ or $-1$ \cite{Fernando:2016sps}. In our analysis, we choose $\gamma = -1$, for which the geometry matches the RN-AdS charged black hole with two horizons.

The expressions of the Hawking temperature, mass, and entropy of the black hole are
\begin{eqnarray}\label{eq:mg_temperature}
& & T = \frac{1}{4 \pi}\biggl(\frac{2M}{r_h^2}+\frac{\gamma Q^2 \omega}{r_h^{\omega+1}}-\frac{2 \Lambda r_h}{3}\biggr)\,, \\ \nonumber
& & M = \frac{r_h}{2}-\frac{\gamma Q^2}{2 r_h^{\omega -1}}-\frac{r_h^3 \Lambda}{6}\,, \\ \nonumber
& & S = \pi r_h^2\,.
\end{eqnarray}
The potential corresponding to the charge $Q$ is given by
\begin{equation}\label{eq:mg_phi}
	\Phi = - \frac{\gamma Q}{r_h^{\omega -1}}\,.
\end{equation}
Again, we would like to probe the black hole thermodynamics in the extended phase space. The pressure is related to the cosmological constant as
\begin{equation}\label{eq:mg_pressure}
	P = - \frac{\Lambda}{8 \pi}=\frac{3}{8 \pi l^2}\,,
\end{equation}
with the corresponding conjugate volume $V$ given by
\begin{equation}\label{eq:mg_volume}
V = \frac{4 \pi r_h^3}{3}\,.
\end{equation}
These thermodynamic variables satisfy the difference form
\begin{equation}\label{eq:1stlawmassive}
dM=TdS + \Phi dQ + VdP \,,
\end{equation}
with the usual condition that $M$ should be identified with the enthalpy, $H$ of the system. Accordingly, the Gibbs free energy is defined as
\begin{equation}\label{eq:mg_free_energy}
G = H-TS = \frac{r}{4}-\frac{\gamma Q^2}{4 r^{\omega -1}}(1+\omega)-\frac{2 \pi P r^3}{3}\,.
\end{equation}
For the above equations, we can further find the equation of state
\begin{equation}\label{eq:mg_EOS}
P = \frac{T}{2r_h}+\frac{\gamma Q^2 (1-\omega)}{8 \pi r_h^{\omega+2}}-\frac{1}{8 \pi r_h^2}\,.
\end{equation}
Comparing this with Eq.~(\ref{gbvdw}) allows us to identify the specific volume $v=2 r_h$.  The condition of the inflexion point [Eq.~(\ref{inflectionpoint-equation})] gives following the critical points in the $P-V$ diagram
\begin{eqnarray}\label{eq:mg_critcal}
& & T_c = \frac{\omega}{\pi\left(1+\omega\right)}\biggl(Q^2 (\omega^2-1)(\omega+2)2^{\omega-1}  \biggr)^{-1/\omega}\,, \\ \nonumber
& & P_c = \frac{\omega}{2\pi\left(2+\omega\right)}\biggl(Q^2 (\omega^2-1)(\omega+2)2^{\omega-1}  \biggr)^{-2/\omega}\,, \\ \nonumber
& & r_{hc} = \frac{1}{2}\biggl(Q^2 (\omega^2-1)(\omega+2)2^{\omega-1}  \biggr)^{1/\omega}\,,
\end{eqnarray}
 where we have used $\gamma=-1$. One can easily check that no inflexion point exists when $\gamma=1$. Therefore, at the critical point, we have
\begin{equation}\label{eq:mg_EOScritical}
\frac{P_c v_c}{T_c} = \frac{\omega+1}{2(\omega+2)}\,.
\end{equation}
As expected, the right-hand side of the above ratio simplifies to $3/8$ for $\omega=2$, which has the same value obtained for the van der Waals gas-liquid system. Notice that the critical points depend non-trivially on $\omega$ and $Q$.

\begin{figure}[htb!]
	\centering
	\subfigure[$Q = 0.02 < Q_c$]{\label{fig:MG_GvsT_1}
		\includegraphics[width=0.45\linewidth]{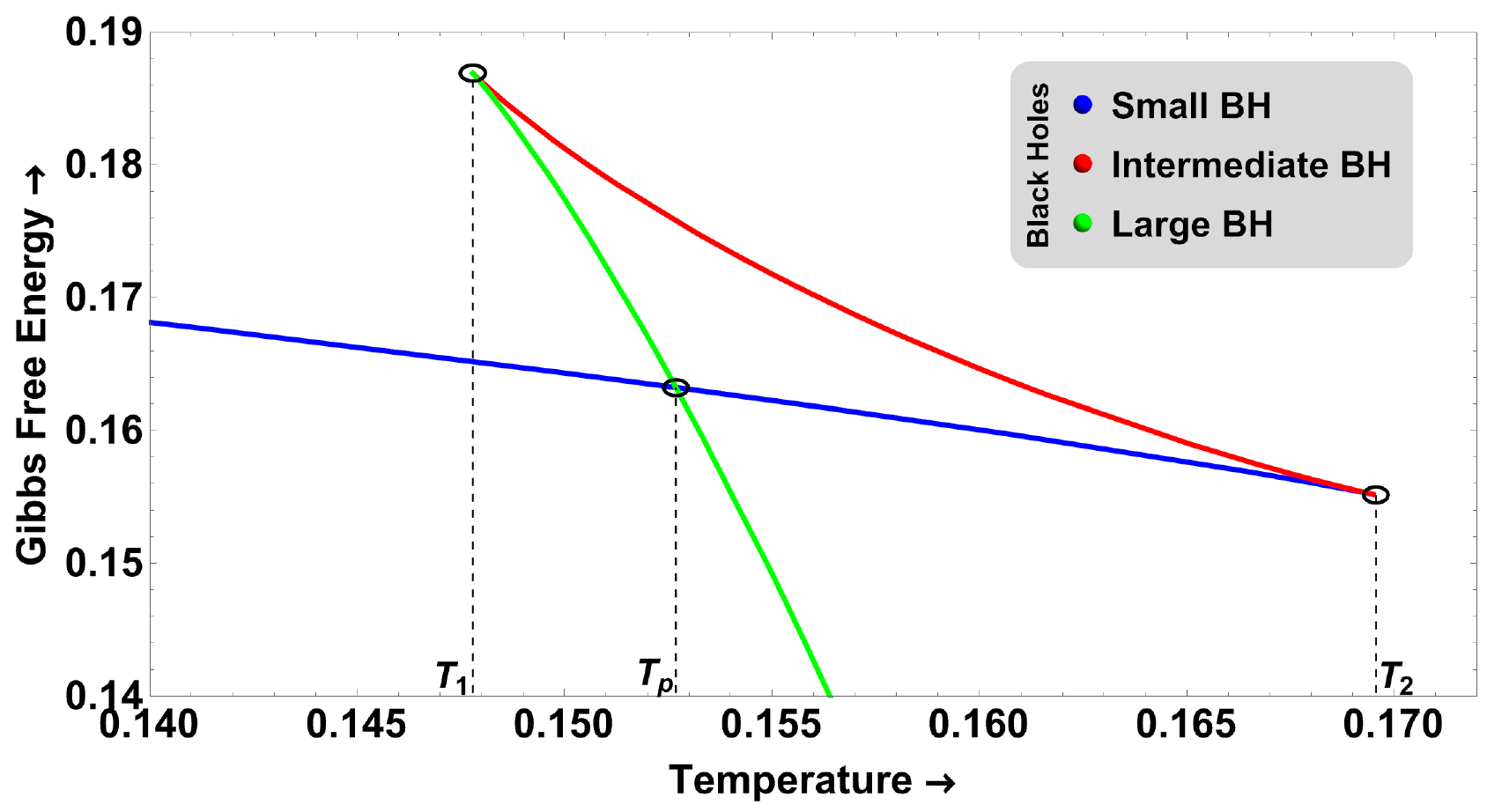}}
	\hfill
	\subfigure[$Q = 0.09 > Q_c$]{\label{fig:MG_GvsT_2}
		\includegraphics[width=0.45\linewidth]{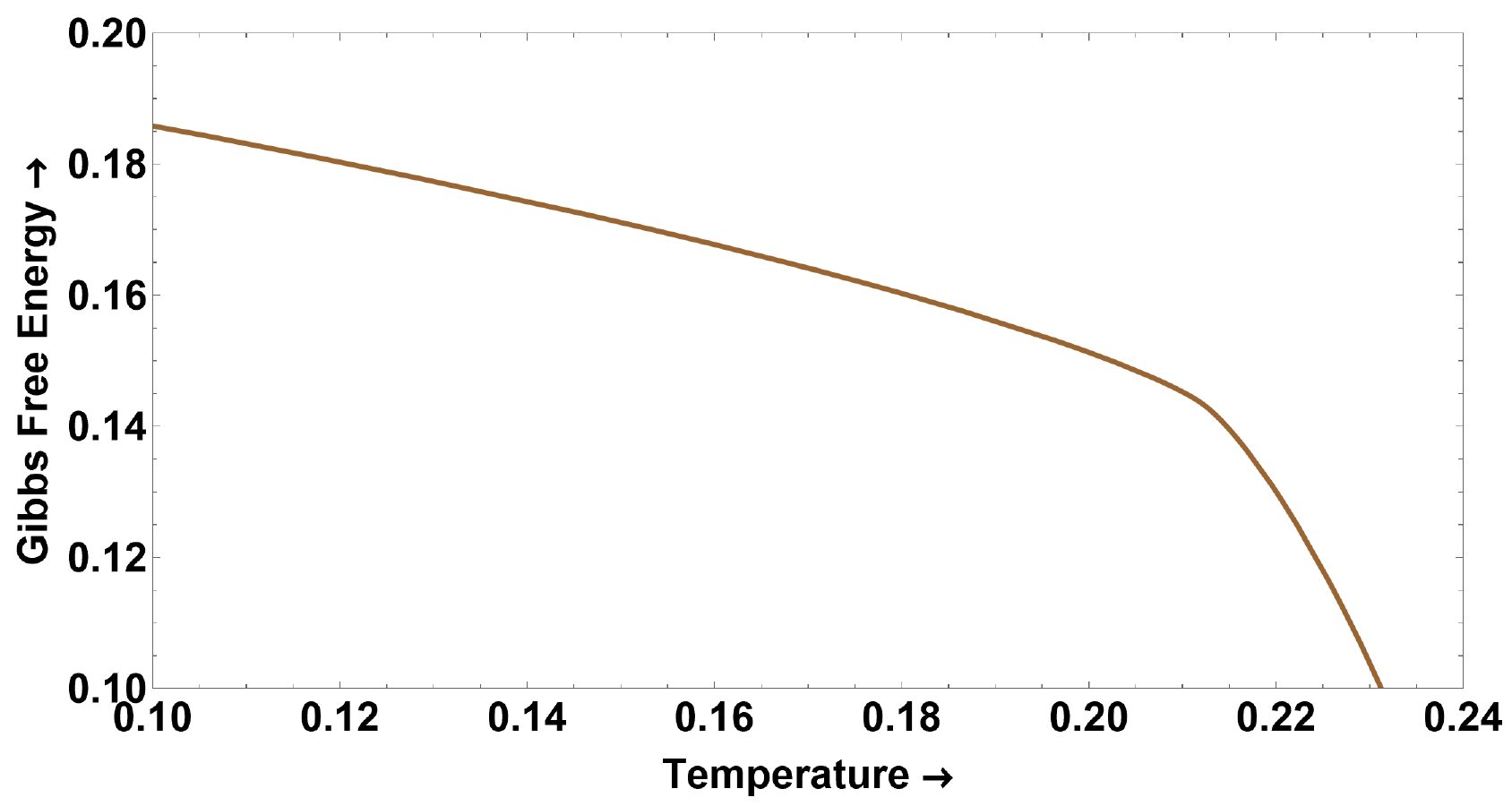}}
    \vfill
    \subfigure[Temperature vs horizon radius]{\label{fig:MG_Tvsrh_collage}
		\includegraphics[width=0.45\linewidth]{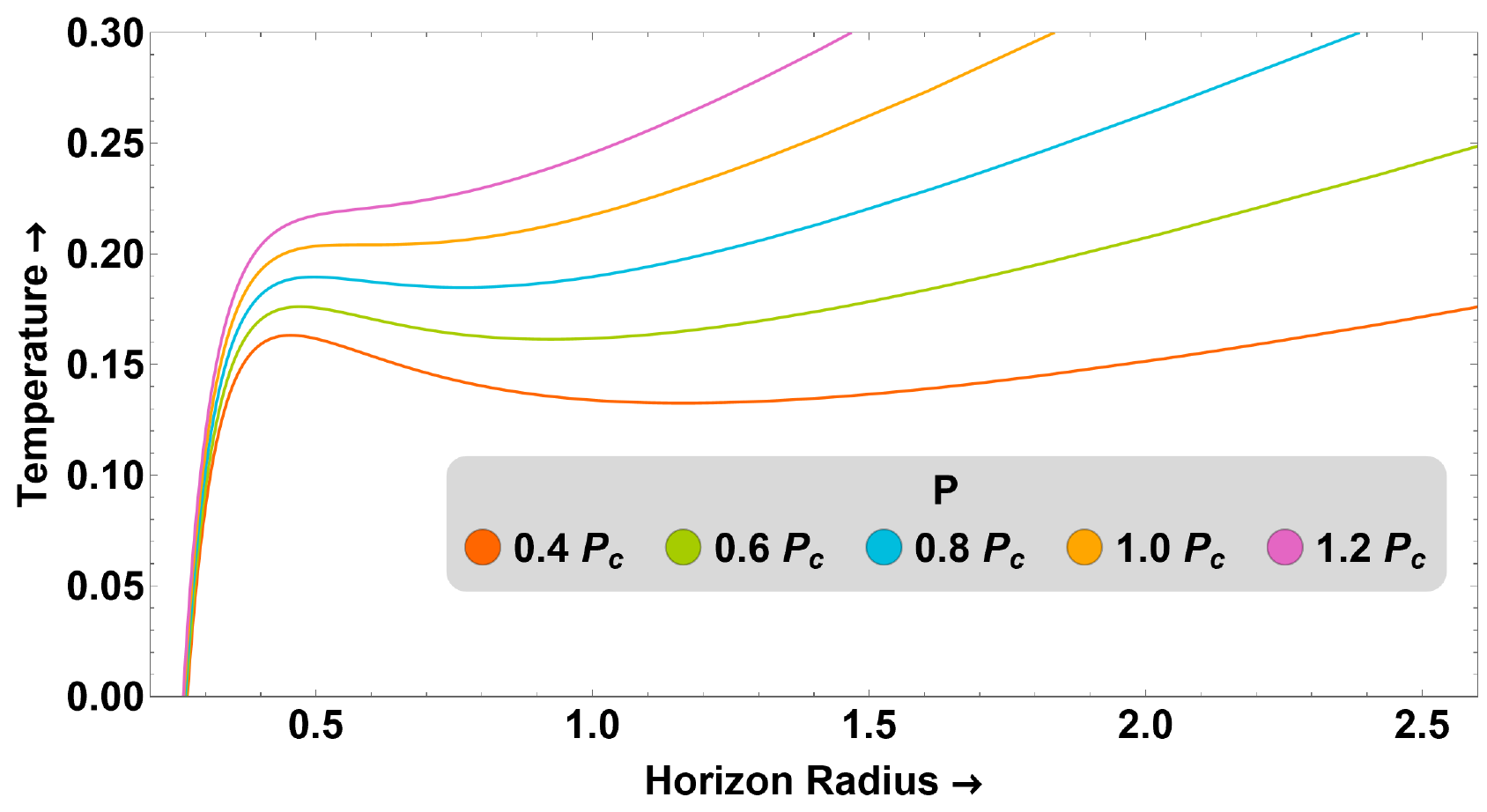}}
	\hfill
	\subfigure[Gibbs Free energy vs temperature]{\label{fig:MG_GvsT_collage}
		\includegraphics[width=0.45\linewidth]{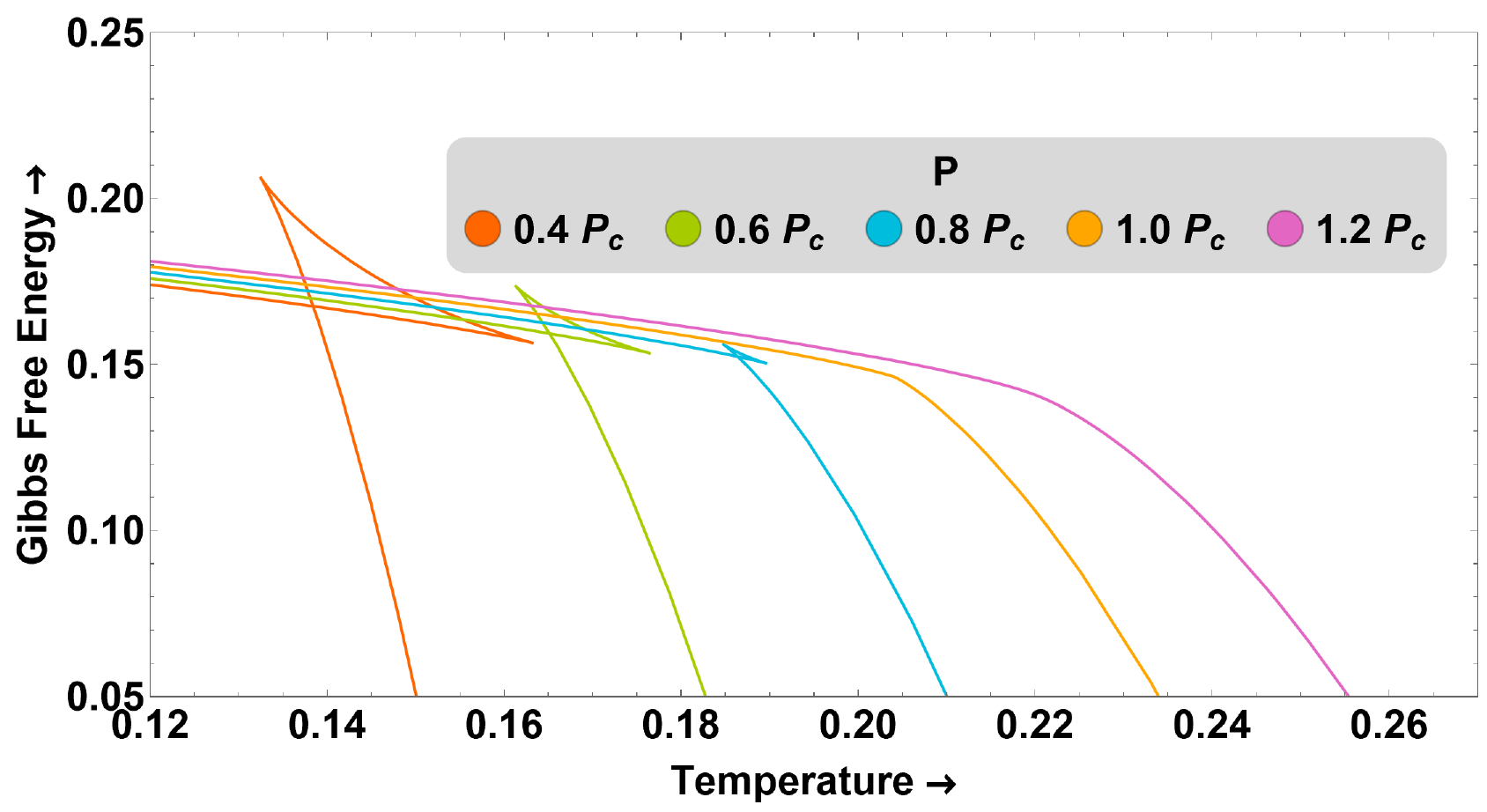}}
	\caption{\label{fig:MG_GvsTandTvsrh} Thermodynamic phase structure of the massive gravity black hole. Here, $\omega=3$ and $Q=0.1$ are used.}
\end{figure}

The thermodynamic phase structure of the Lorentz symmetry-breaking massive black hole is shown in Fig.~\ref{fig:MG_GvsTandTvsrh}. Here, we have set $\omega=3$ and $Q=0.1$. For these values, the critical points are $r_{hc}\simeq0.585$, $P_c\simeq0.0698$, and $T_c\simeq0.204$. The orange line denotes the critical isotherm. Again, three black hole phases exist when $P<P_c$ in some temperature range. The small and large black hole phases have a positive specific heat and compression coefficient, thus corresponding to stable phases. In contrast, the intermediate phase connecting them has an unstable negative specific heat and compression coefficient. Similarly, the Gibbs free energy exhibits the swallow-tail-like structure for $P<P_c$ and exchange dominance as the temperature is varied, i.e., the small/large black hole phase has the lowest free energy at small/large temperatures, indicating the existence of a first-order small/large black hole phase transition when $P<P_c$. When $P=P_c$, the small and large physical horizon radii coincide, resulting in coexistence.
Meanwhile, for $P>P_c$, only one stable black hole phase appears, always with a positive specific heat and compression coefficient. Thus, this thermodynamic behaviour is reminiscent of the van der Waals system's liquid/gas phase transition, with $P_c$ acting as a second-order critical point. Although the critical points are explicit $Q$ dependent, the van der Waals behaviour of a massive gravity black hole is true for all $Q$ as long as it is non-zero. For $Q=0$, the massive gravity geometry reduces the AdS-Schwarzschild black hole, and in this case, we only get the Hawking/Page phase transition between the AdS-Schwarzschild black hole and thermal-AdS.

\begin{figure}[htb!]
	\centering
	\includegraphics[width=0.4\linewidth]{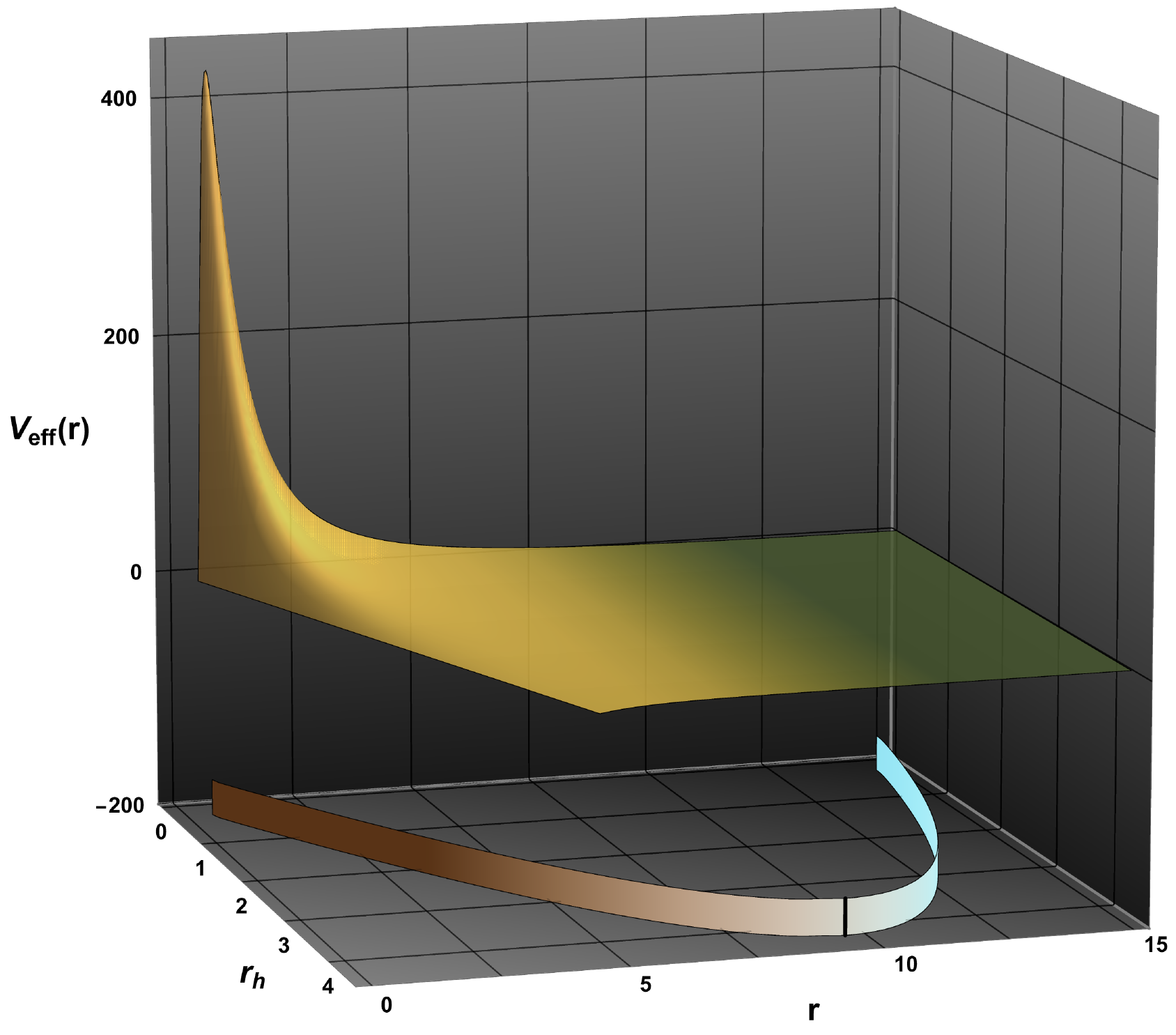}
	\caption{\label{fig:MG_veffective3d}The 3D plot of the effective potential $V_\text{eff}(r)$ as a function of horizon radius $\tilde{r_h}$ and orbit radius $r$ for the massive particle when $L = 20$ , $P = 0.001$, $\gamma = -1$, $Q = 0.1$, and $\omega=3$. The brown and cyan curves projected in the lower part correspond to the unstable and stable equilibria of the timelike circular geodesics.}
\end{figure}

The effective potential of particles [Eq.~(\ref{eq:Veffective})] is given by
\begin{equation}\label{eq:MG_Veffective}
    V_{\text{eff}}(r) = \frac{1}{3} \left(\frac{L^2}{r^2}-\epsilon \right) \left(\frac{r_h \left(3 \gamma Q^2 r_h^{-\omega }-8 \pi  P r_h^2-3\right)}{r}-3 \gamma Q^2 r^{-\omega }+8 \pi  P r^2+3\right)
\end{equation}
which shows that it is a nontrivial function of $L, Q, P, \gamma, \omega, r$ and $r_h$. For the massive particle, the behaviour of $V_{\text{eff}}$ is shown in Fig.~\ref{fig:MG_veffective3d} for a fixed $L=20$, $P=0.001$, $\gamma=-1$, $Q=0.1$, and $\omega=3$. Similar behaviour appears for the massless particle as well.

The orbit radius of the massive particle is shown in the lower part of Fig.~\ref{fig:MG_veffective3d}, which is the projection of stationary points of the effective potential ($V_{\text{eff}}'(r)=0$) onto the $V_{\text{eff}}(r)=-200$ plane. The unstable/stable orbit parts are shown in brown/cyan colours. As we slowly increase $r_h$, the brown and cyan parts move closer to each other and finally meet at the black-marked line, beyond which there is no stationary point for the effective potential. Again, we mainly focus on the brown region and the unstable equilibria, for which we will calculate the Lyapunov exponents in the next section.

\subsubsection{Massless particles}
\begin{figure}[htb!]
	\centering
    \textbf{Massless particles}\par\medskip
	\subfigure[$P = 0.5 P_c$]{\label{fig:MG_Lmassless_vsT_1}
		\includegraphics[width=0.45\linewidth]{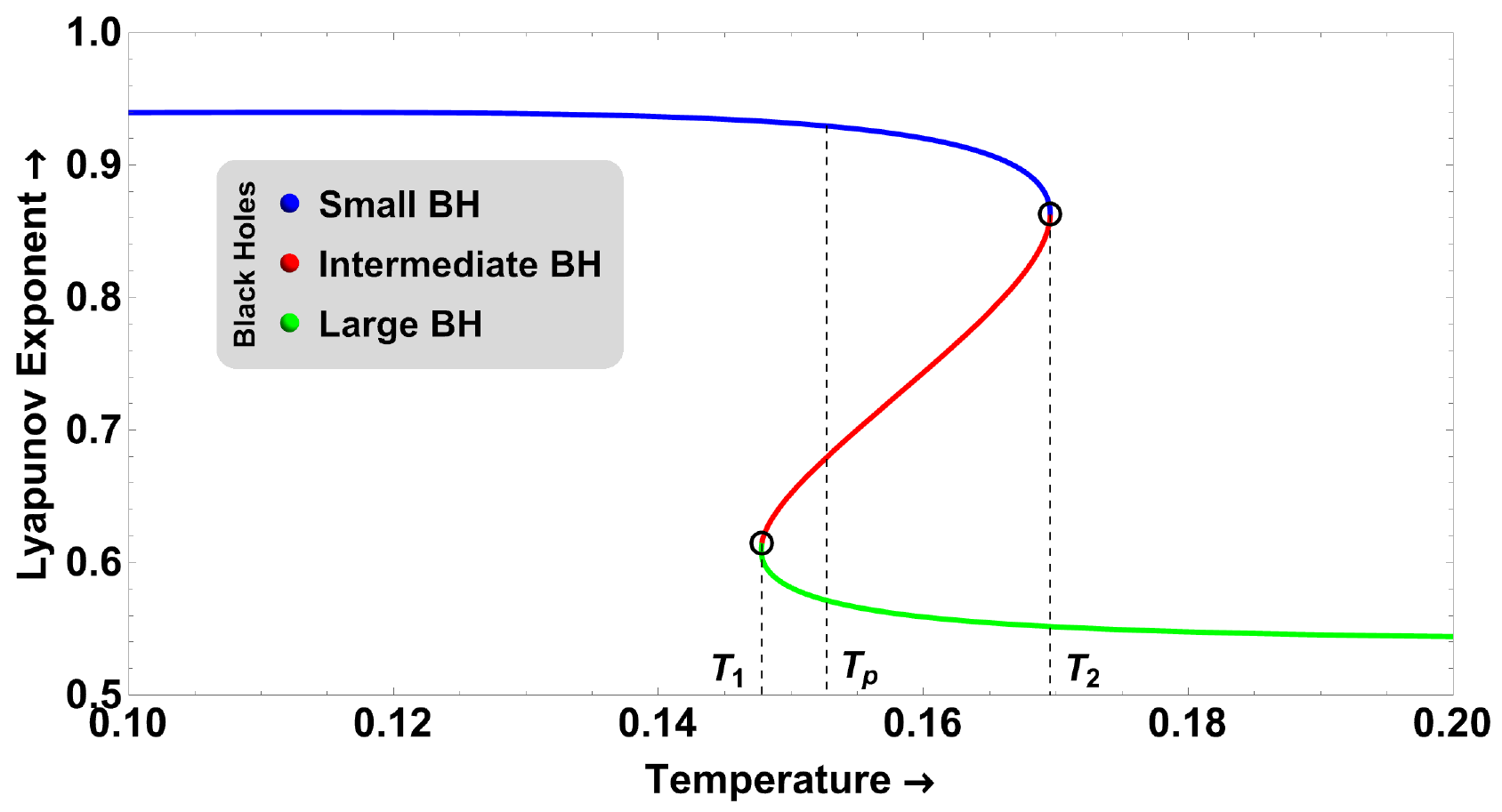}}
	\hfill
	\subfigure[$P = 1.1 P_c$]{\label{fig:MG_Lmassless_vsT_2}
		\includegraphics[width=0.45\linewidth]{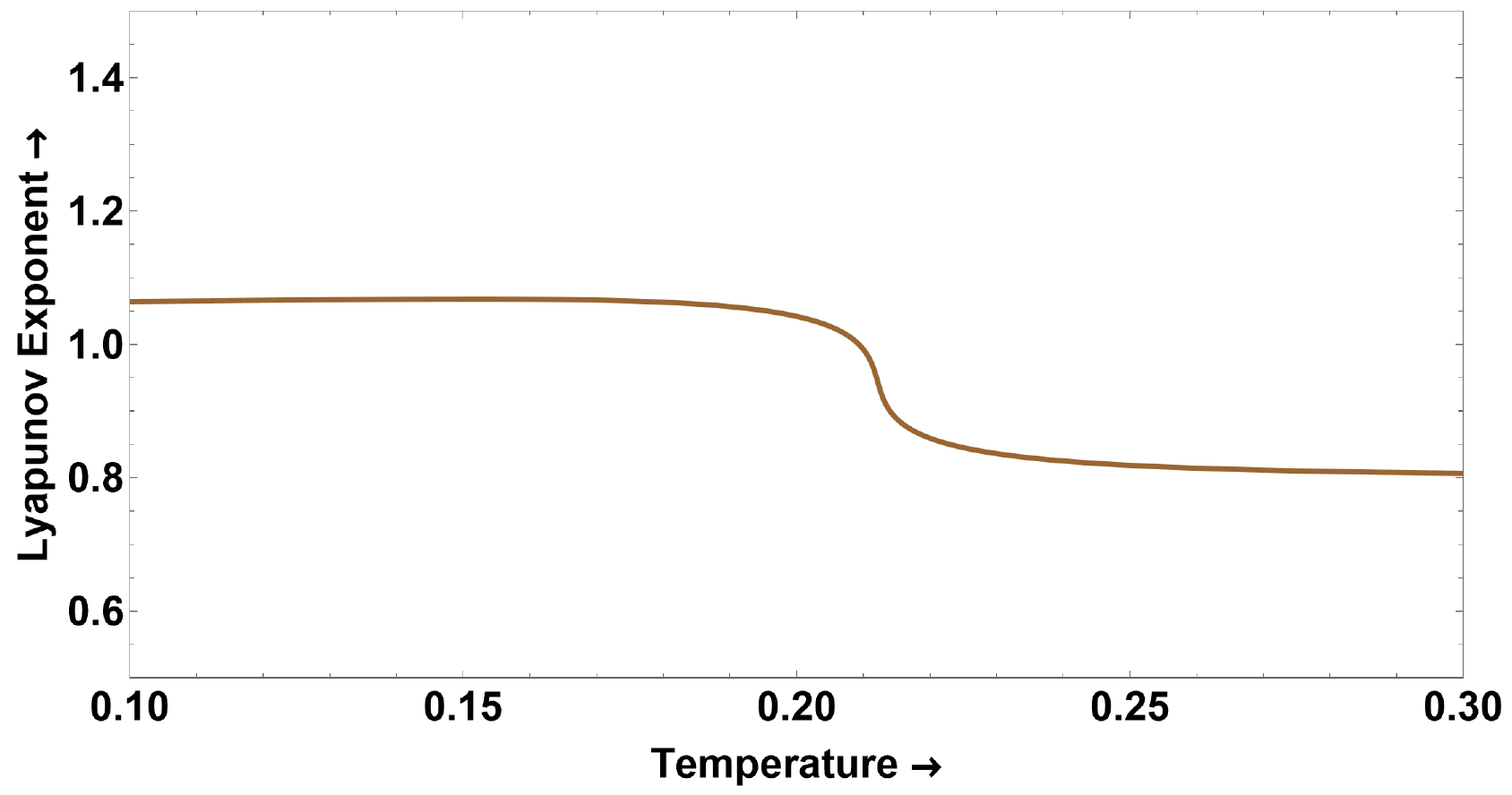}}
	\vfill
	\subfigure[Lyapunov exponent vs temperature]{\label{fig:MG_Lmassless_vsT_collage}
		\includegraphics[width=0.45\linewidth]{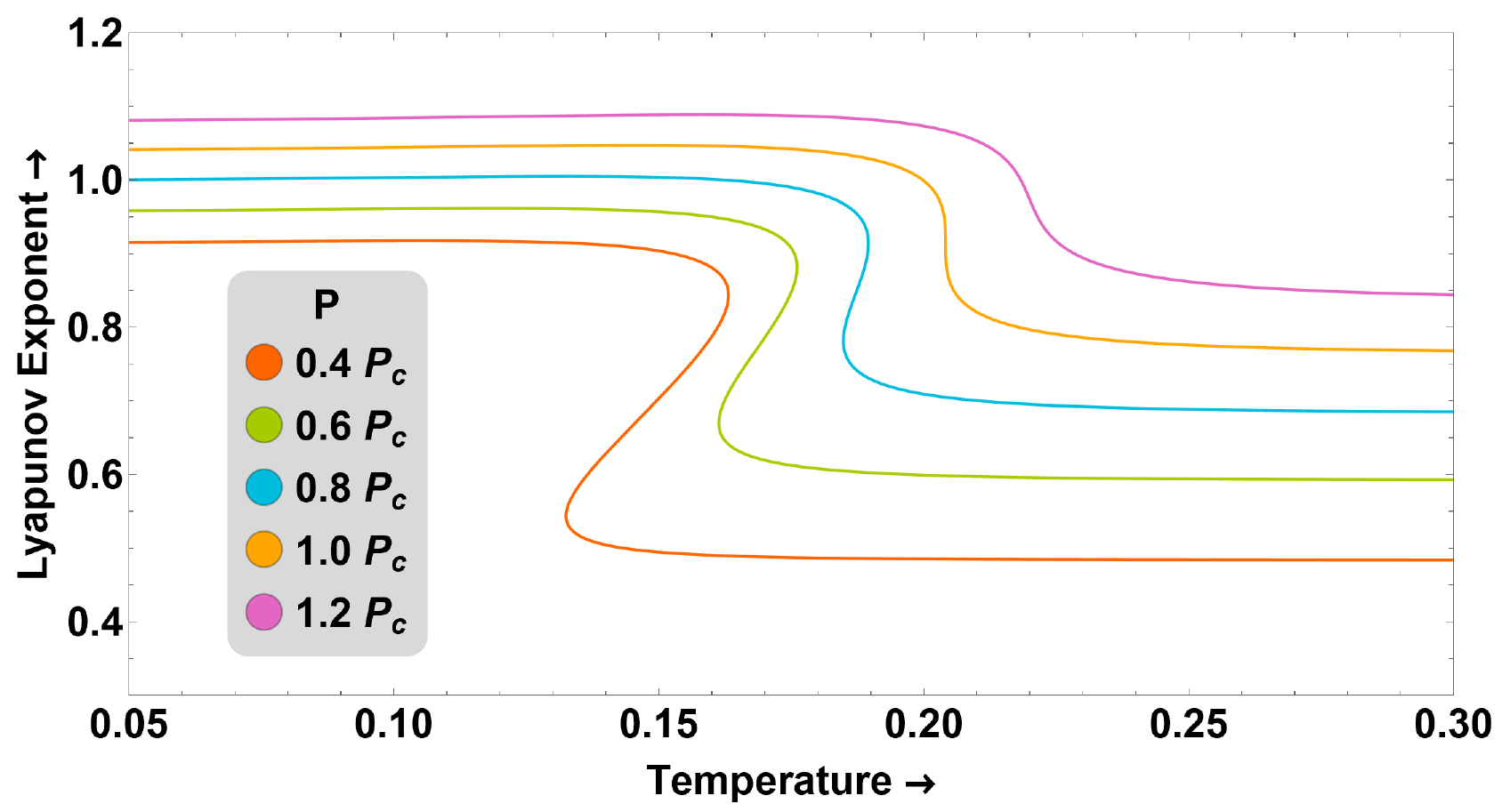}}
    \hfill
	\subfigure[Lyapunov exponent vs horizon radius]{\label{fig:MG_Lmassless_vsrh_collage}
		\includegraphics[width=0.45\linewidth]{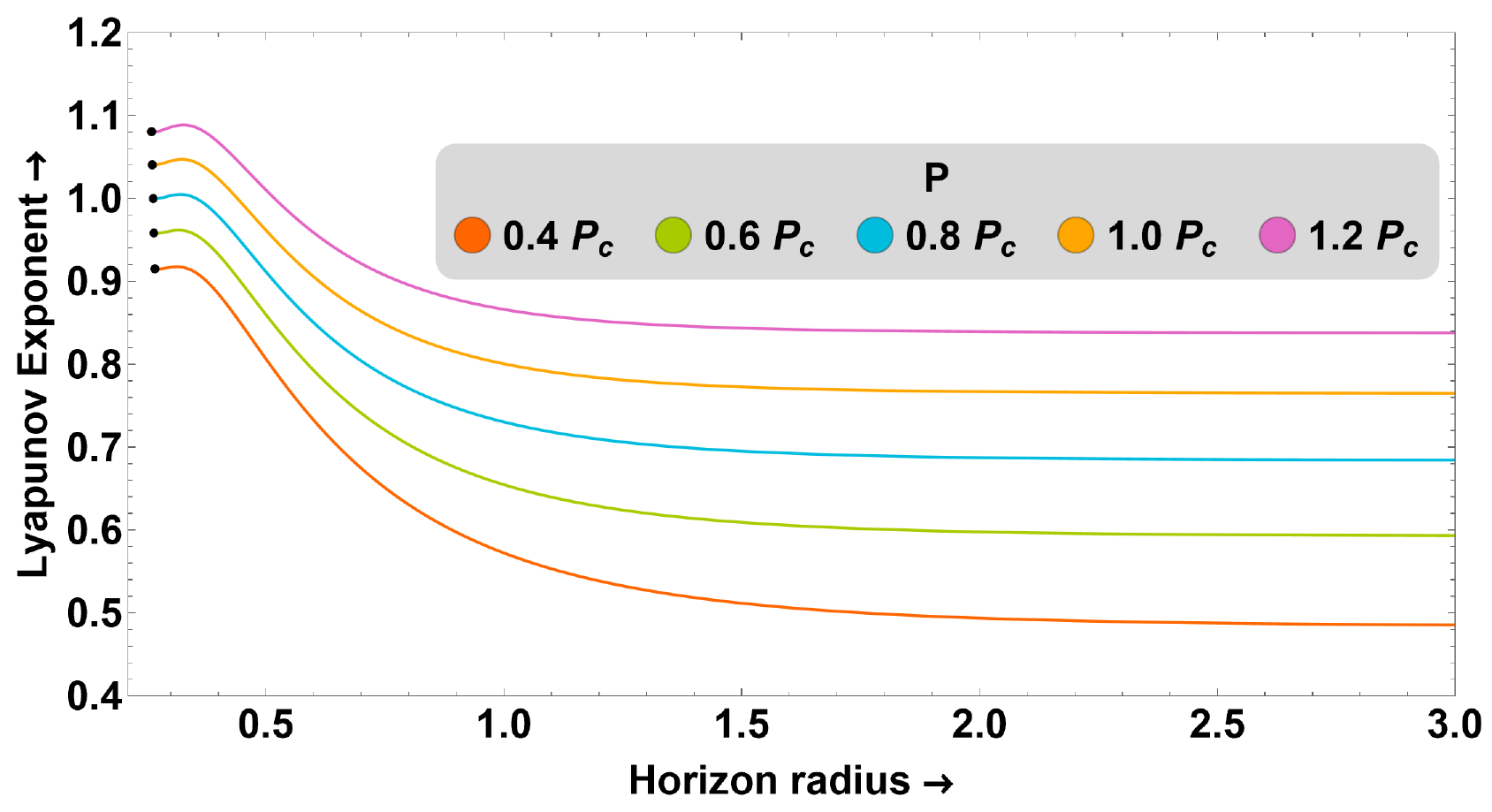}}
	\caption{\label{fig:MG_Lmassless}Lyapunov exponent $\lambda$ of the massless particle as a function of temperature $T$ and Horizon radius $r_h$ for the AdS black hole in massive gravity. Here, $k=-1, \omega=3$, and $Q=0.1$ are used.}
\end{figure}
The thermal profile of the Lyapunov exponent associated with the massless particle is shown in Fig.~\ref{fig:MG_Lmassless} for a fixed $\gamma=-1, \omega=3$, and $Q=0.1$ in the extended phase space. Once again, we observe that below the critical value $P_c$, the Lyapunov exponent is multi-valued between $T_1$ and $T_2$, where once again, $T_p$ stands for the phase transition temperature. Whereas, above $P_c$, the multi-valued nature of $\lambda$ disappears, and $\lambda$ becomes single-valued. This is shown in Figs.~\ref{fig:MG_Lmassless_vsT_1} and \ref{fig:MG_Lmassless_vsT_2}. As we increase $P$, the maximum value that $\lambda$ can achieve increases, which comes with the additional price of the disappearance of the multi-valuedness in $\lambda$. For a fixed $P$, $\lambda$ stays constant in the small black hole branch and gradually decreases as we increase the temperature and then finally attain a constant value in the large black hole branch, similar to what we observed in the Gauss-Bonnet case. For the sake of completion and to further scrutinise the role of horizon radius in the analysis of the Lyapunov exponent, we have also added the $r_h$ vs $\lambda$ plot in Fig.~\ref{fig:MG_Lmassless_vsrh_collage}. For a fixed $P$, $\lambda$ starts from the extremal $r_h$ (represented by black dots). As we slowly increase $r_h$ beyond the extremal value, $\lambda$ first slowly increases and then gradually decreases till it attains a constant saturation value. This is true for all $P$ values, with the difference being that the constant saturation value differs for different $P$. This difference arises because we are working in an extended phase space. Therefore, different $P$ signifies different cosmological constant values and, hence, different AdS radii.

\begin{figure}[htb!]
	\centering
	\includegraphics[width=\linewidth]{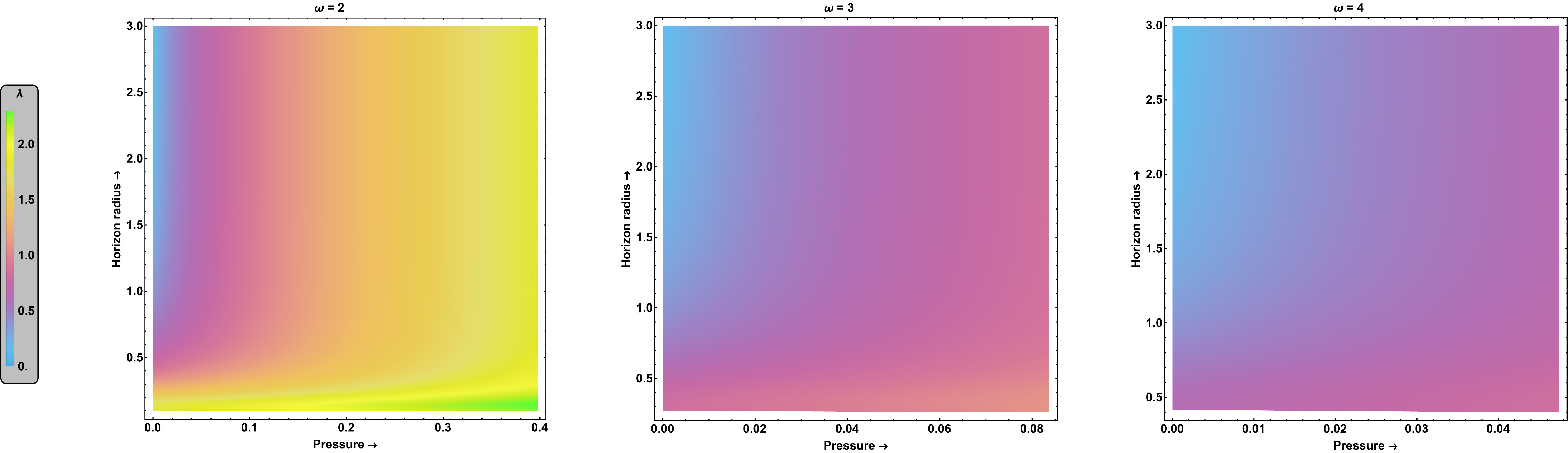}
	\caption{\label{fig:MG_Lmassless_contour} Density plot of $\lambda$ for the massless particle as a function of the pressure $P$ and horizon radius $r_h$ for the AdS black hole in massive gravity. Here $\omega = 2$ (left), $\omega = 3$ (centre) and $\omega = 4$ (right) with fixed $Q = 0.1$ and $\gamma = -1$ are used.}
\end{figure}

The density plot of $\lambda$ is shown in Fig.~\ref{fig:MG_Lmassless_contour}, where a comparison is made between different values of $\omega$, keeping $Q=0.1$ and $\gamma=-1$ fixed. The left density plot is for $\omega=2$, which shows that increasing the pressure $P$ while keeping the horizon radius $r_h$ constant increases the Lyapunov exponent $\lambda$. In contrast, by increasing $r_h$ while keeping $P$ fixed, the Lyapunov exponent increases slowly and then decreases until it attains a constant saturation value. This is also true for $\omega=3$ (middle) and $\omega=4$ (right). The effect of increasing $\omega$ can be seen in the decreased available range for $\lambda$; the higher the $\omega$, the lower the values that $\lambda$ can achieve.

\subsubsection{Massive particles}
\begin{figure}[htb!]
	\centering
    \textbf{Massive particles}\par\medskip
	\subfigure[$P = 0.5 P_c$]{\label{fig:MG_Lmassive_vsT_1}
		\includegraphics[width=0.45\linewidth]{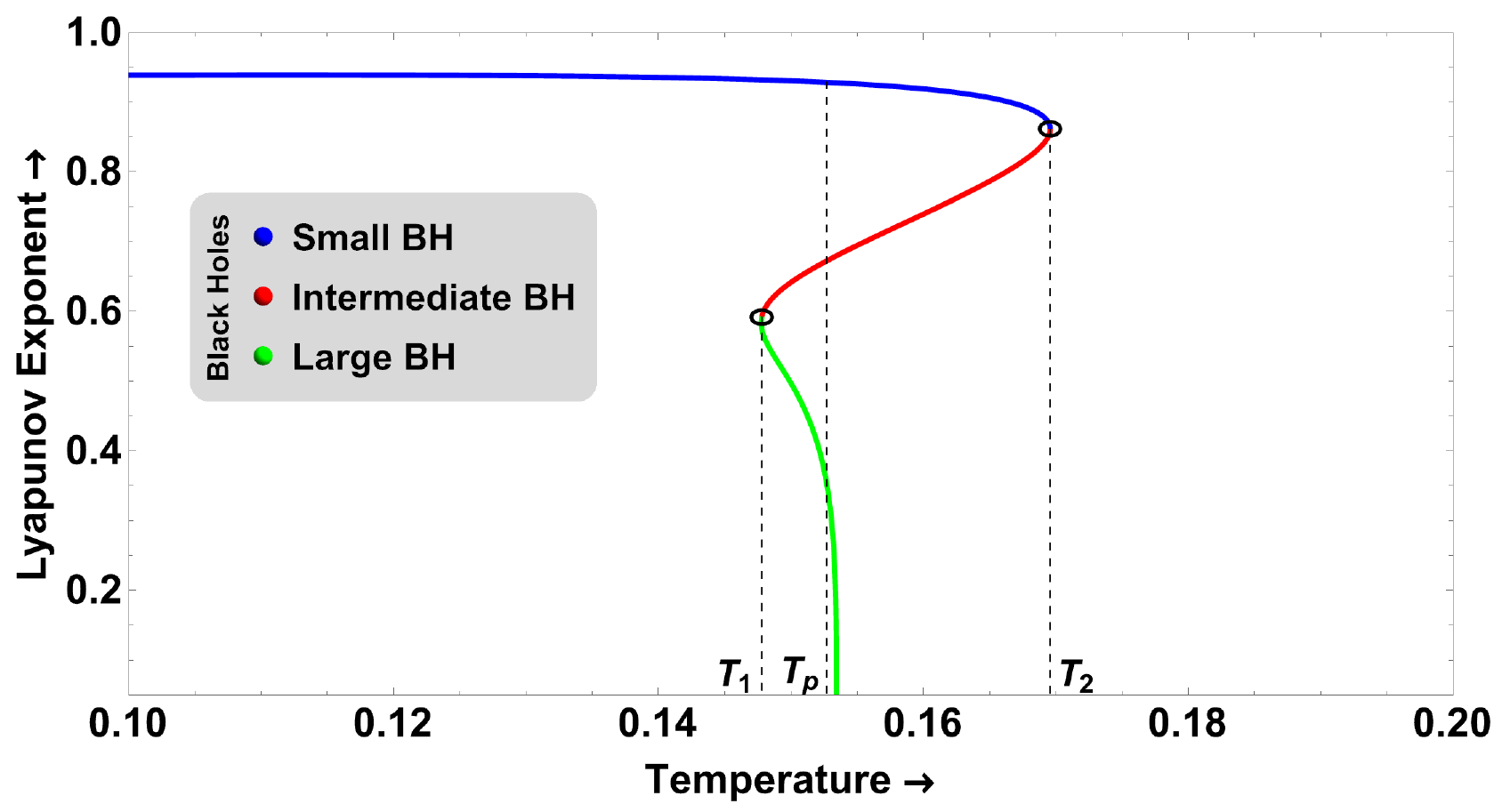}}
	\hfill
	\subfigure[$P = 1.1 P_c$]{\label{fig:MG_Lmassive_vsT_2}
		\includegraphics[width=0.45\linewidth]{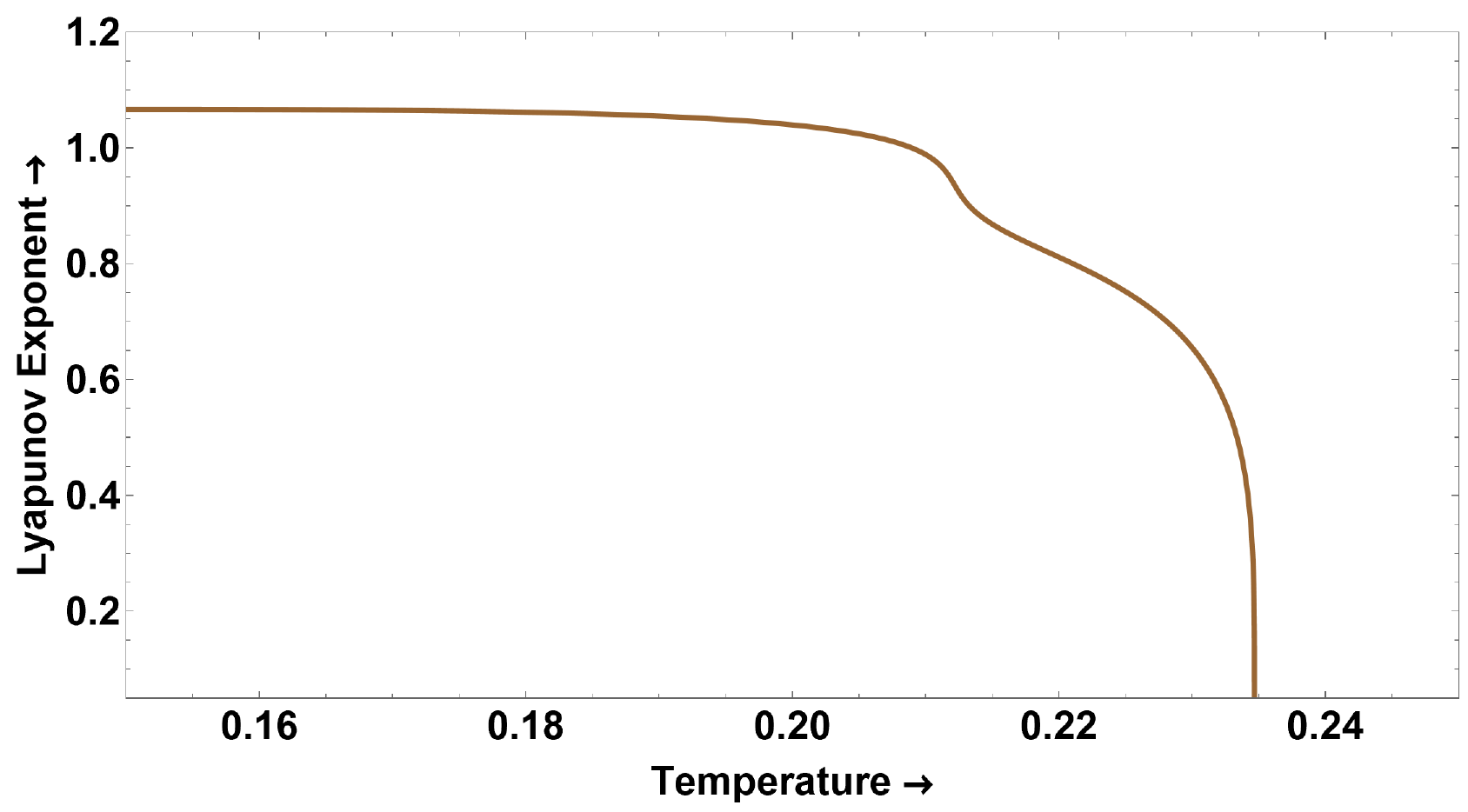}}
	\vfill
	\subfigure[Lyapunov exponent vs temperature]{\label{fig:MG_Lmassive_vsT_collage}
		\includegraphics[width=0.45\linewidth]{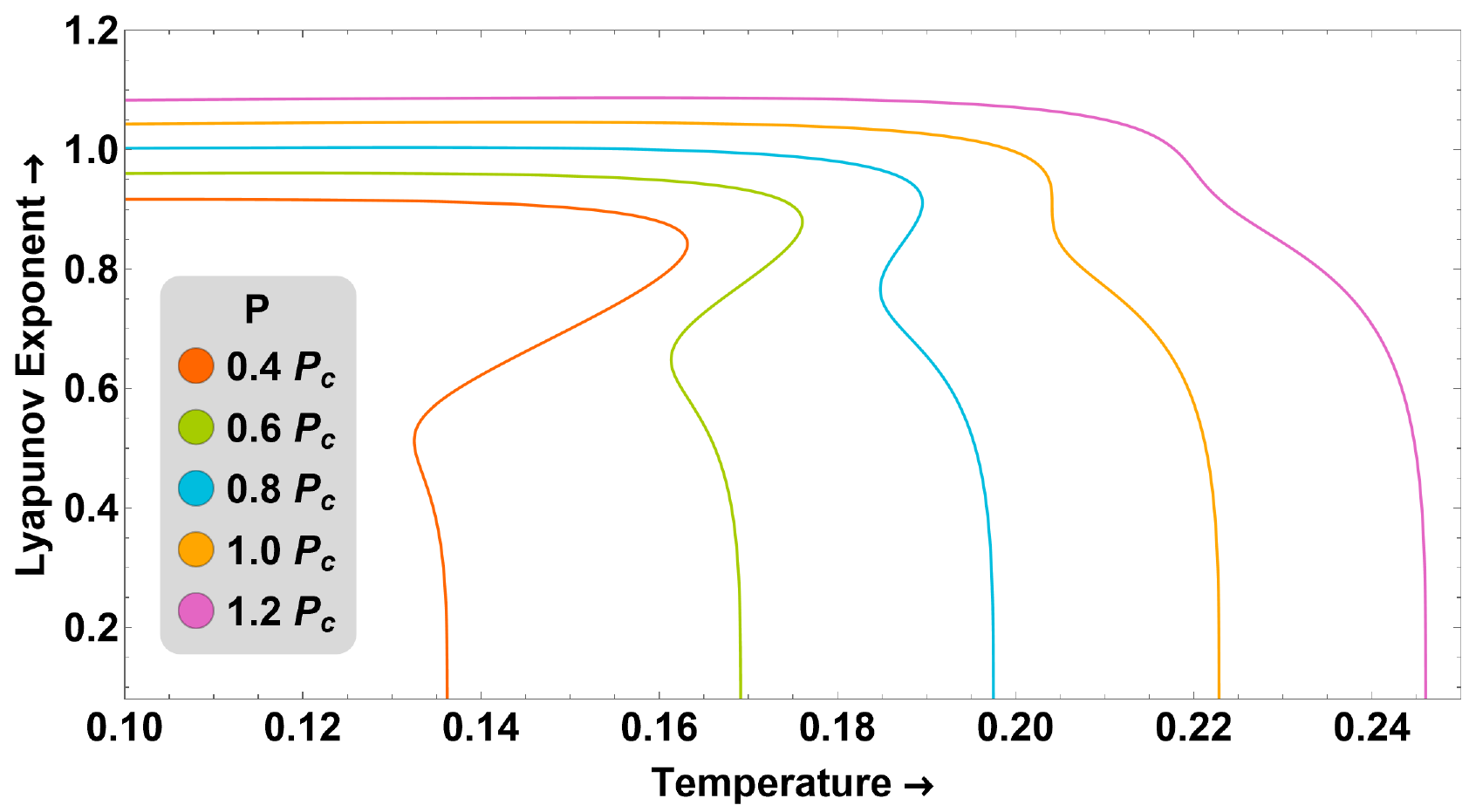}}
    \hfill
	\subfigure[Lyapunov exponent vs horizon radius]{\label{fig:MG_Lmassive_vsrh_collage}
		\includegraphics[width=0.45\linewidth]{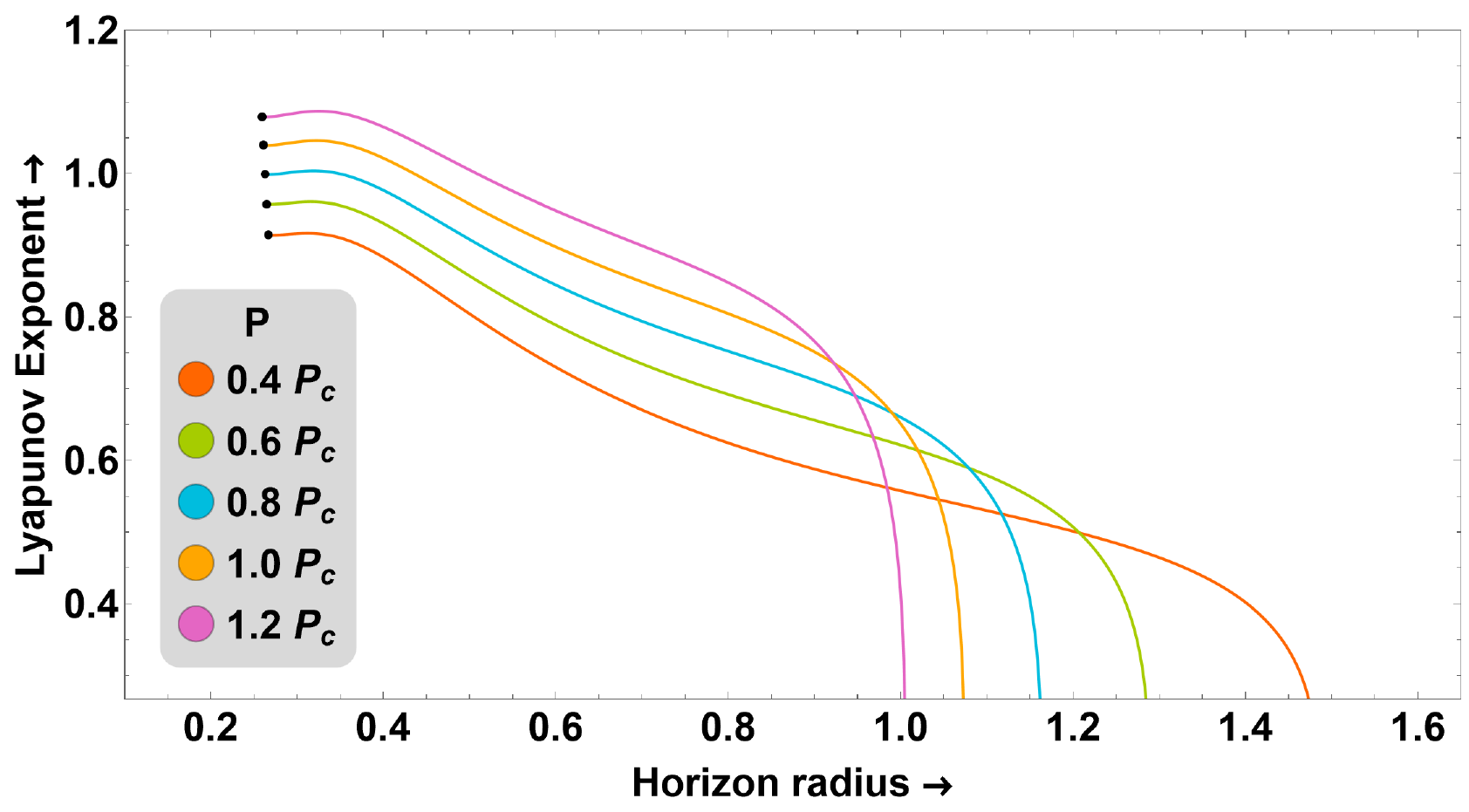}}
	\caption{\label{fig:MG_Lmassive}Lyapunov exponent of the massive particle $\lambda$ as a function of temperature $T$ and horizon radius $r_h$ for the AdS black hole in massive gravity. Here, $\gamma=-1, \omega=3$, and $Q=0.1$ are used.}
\end{figure}

Using Eqs.~(\ref{eq:MG_Veffective}) and (\ref{eq:lyapunov-massive}) for the massive particle, we can investigate the thermal behaviour of the Lyapunov exponent in AdS massive gravity background. The results are shown in Fig.~\ref{fig:MG_Lmassive}. Again, the results are similar to the Lyapunov exponent behaviour of the massive particle discussed in the earlier gravity backgrounds. We again observe that when the pressure $P$ is less than the critical value $P_c$, $\lambda$ is a multi-valued function of $T$ with three black hole branches. In contrast, when the value exceeds the critical value, $\lambda$ is single-valued with only one black hole branch. Once again, the Lyapunov exponent approaches zero for the massive particle as we gradually increase $T$ or $r_h$. This is again since after a certain $r_h$; there are no stationary points, also visible in Fig.~\ref{fig:MG_veffective3d}, indicated by the black line. This maximal $r_h$, beyond which there are no stationary points, varies as we increase the pressure.

\begin{figure}[htb!]
	\centering
	\includegraphics[width=\linewidth]{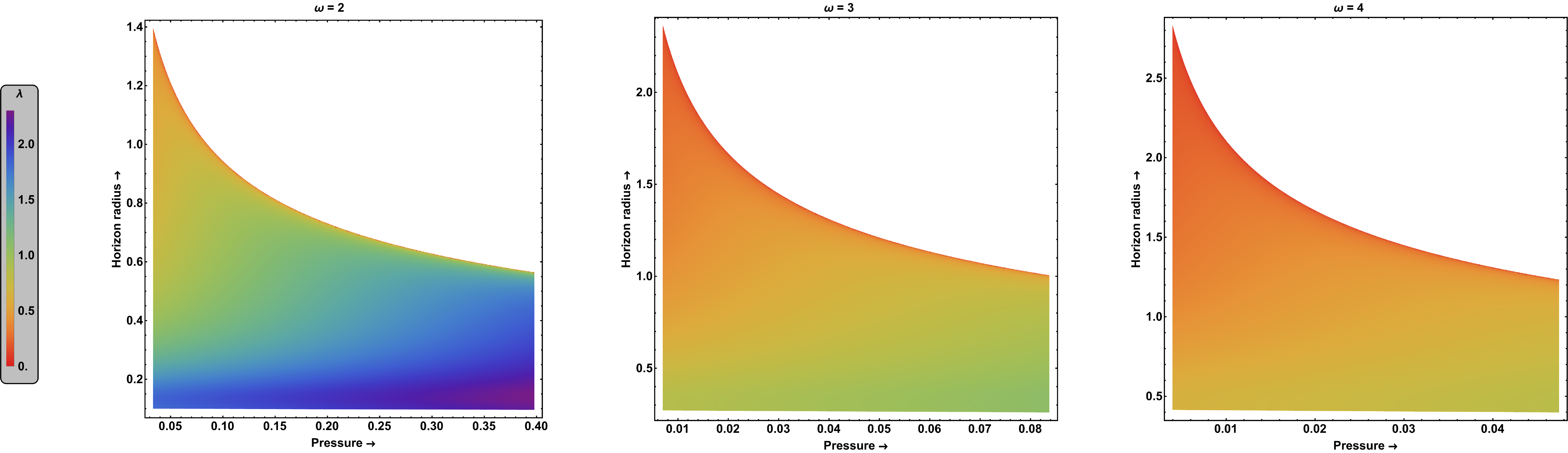}
	\caption{\label{fig:MG_Lmassive_contour}Density plot of $\lambda$ for the massive particle as a function of the pressure $P$ and horizon radius $r_h$ for the AdS black hole in massive gravity. Here $\omega = 2$ (left), $\omega = 3$ (centre) and $\omega = 4$ (right) with fixed $Q = 0.1$ and $\gamma = -1$ are used.}
\end{figure}

The plot of the Lyapunov exponent as a function of temperature for different $P$ values is shown in Fig.~\ref{fig:MG_Lmassive_vsT_collage}. Since we are working in the extended phase space, $\lambda$ approaches zero at different $T$ values for different $P$ values. To highlight this point, the plot $\lambda$ vs $r_h$ is also shown in Fig.~\ref{fig:MG_Lmassive_vsrh_collage}, where we can see that the $r_h$ value where $\lambda$ approaches zero decreases as we increase $P$. This is further elaborated and visualised for different $\omega$ values in the density plot of $\lambda$ shown in Fig.~\ref{fig:MG_Lmassive_contour}. For a fixed $P$, going from bottom to top, i.e., increasing $r_h$ decreases $\lambda$ slowly until it approaches zero at some $r_h$. A red line in the density plot shows this. The same is true for $\omega=3$ and $\omega=4$ as well. Just like in the massless case, here, too, the effect of increasing the value of $\omega$ is to decrease the available range of $\lambda$.

\section{Critical exponent for charged black holes based on Lyapunov exponents}
\label{sec:order-parameter}
In the mean-field theories, critical exponents play an important role and are only defined as limiting power laws as $T\rightarrow T_c$~\cite{Goldenfeld2019Jun}, which means that if any observable, say, $a(t=|T-T_c|)$ has critical exponent $\xi$ then,
\begin{equation}\label{eq:criticalexponentbasics1}
    a(t) \sim t^{\xi}\,,
\end{equation}
which implies that,
\begin{equation}\label{eq:criticalexponentbasics2}
    \xi = \lim_{t\to 0} \frac{\log a(t)}{\log t} \,.
\end{equation}
Depending upon the physical observable under consideration, one can associate different mean-field critical exponents at the phase transition point, such as those associated with the specific heat, susceptibility, compressibility, etc. The one we are most interested in and whose existence forms the basis of the famous Landau theory is the order parameter critical exponent~\cite{Wang2021May}. The order parameter is the observable whose expectation value is nonzero below the transition temperature and zero above the transition temperature~\cite{Aron/SciPostPhysLectNotes.11}. For the van der Waals liquid/gas phase transition, the density shows discontinuity between the two phases and is considered the order parameter. The order parameter critical exponent for it was found to be 1/2~\cite{van2004continuity}.

The small/large black hole phase transition we saw in previous sections for different charged black holes is similar to the van der Waals liquid/gas transition~\cite{Chamblin:1999hg}. The temperature at which this first-order phase transition occurs is denoted by $T_p$ in our work. Let us also denote the Lyapunov exponent of small and large black holes, calculated at $T_p$, by $\lambda_s$ and $\lambda_l$, respectively and the one calculated at the second-order critical point by $\lambda_c$. Therefore, $\lambda_s = \lambda_l = \lambda_c$ at the second-order critical point. Following \cite{guo2022probing, Wang2021May}, one can consider the discontinuity in the Lyapunov exponent $\Delta\lambda = \lambda_s - \lambda_l$ at the first-order phase transition as an order parameter. In particular, we can try to find the critical exponent $\beta$, satisfying
$\Delta\lambda$,
\begin{equation}\label{eq:delta-lambda-op}
    \Delta\lambda \sim a\lvert T - T_c \rvert^\beta
\end{equation}

\subsection{Theoretical verification}
We use the method illustrated in~\cite{Banerjee:2012zm,lyu2023probing} to calculate this critical exponent $\beta$ related to the order parameter $\Delta\lambda$. First, we do a Taylor expansion of $\lambda(r_h)$ up to the first order around the critical point as,
\begin{equation}\label{eq:taylor-expand-lambda}
    \lambda(r_h) = \lambda_c +  \left[ \frac{\partial\lambda}{\partial r_h}\right]_c  \left(r_h - r_c\right) + \mathcal{O}(r_h)
\end{equation}
The horizon radius $r_{h}$ can be written in terms of horizon radius at the critical point $r_{h c}$ as,
\begin{equation}\label{eq:rh-rc-relation}
    r_{h} = r_{c}\left(1 + \Delta\right)\,,
\end{equation}
with $\lvert\Delta\rvert \ll 1$. Now Eq.~(\ref{eq:taylor-expand-lambda}) for small and large black holes is
\begin{eqnarray}\label{eq:taylor-expand-small-large}
    \lambda_s(r_{hs}) = \lambda_c +  \left[ \frac{\partial\lambda}{\partial r_h}\right]_c \left(r_{hs} - r_c\right) + \mathcal{O}(r_{hs}^2) \,, \nonumber\\
    \lambda_l(r_{hl}) = \lambda_c +  \left[ \frac{\partial\lambda}{\partial r_h}\right]_c \left(r_{hl} - r_c\right) + \mathcal{O}(r_{hl}^2) \,,
\end{eqnarray}
where the subscript $l$ and $s$ stands for large and small.
Using Eqs.~(\ref{eq:rh-rc-relation}) and~(\ref{eq:taylor-expand-small-large}), we can get,
\begin{equation}\label{eq:delta-lambda}
    \Delta\lambda = \lambda_s - \lambda_l = r_{h c}\left[ \frac{\partial\lambda}{\partial r_h}\right]_c \left(\Delta_s - \Delta_l\right)\,,
\end{equation}
where $\Delta_s$ and $\Delta_l$ are the values of $\Delta$ for small and large black hole branches respectively. Similarly, the Hawking temperature $T(r_h)$ can be written as
\begin{equation}\label{eq:temprh-tcritical-relation}
    T(r_h) = T_c \left(1+\delta\right)\,,
\end{equation}
where $T_c$ is the critical temperature and $\lvert\delta\rvert \ll 1$. The Taylor expansion of $T(r_h)$ up to the second order around the critical point is,
\begin{equation}\label{eq:taylor-expand-temp}
    T(r_h) = T_c + \left[ \frac{\partial T}{\partial r_h}\right]_c \left(r_h - r_c\right) + \frac{1}{2}\left[ \frac{\partial^2 T}{\partial r_h^2}\right]_c \left(r_h - r_c\right)^2 + \mathcal{O}(r_{h}^3) \,.
\end{equation}
Now, at the critical point, the slope $\big[\frac{\partial T}{\partial r_h}\big]_c \rightarrow 0$. Using Eq.~(\ref{eq:rh-rc-relation}), we can rewrite Eq.~(\ref{eq:taylor-expand-temp}), after ignoring higher order terms, as
\begin{equation}\label{eq:final-T}
    T(r_h) = T_c + \frac{1}{2}\left[ \frac{\partial^2 T}{\partial r_h^2}\right]_c {r_c}^2\Delta^2\,.
\end{equation}
Comparing Eq.~(\ref{eq:temprh-tcritical-relation}) and Eq.~(\ref{eq:final-T}), we get, suppressing for now the subscripts $s$ or $l$,
\begin{equation}\label{eq:Delta}
    \Delta = \frac{1}{r_c}\sqrt{\frac{T_c \delta}{\frac{1}{2}\left[ \frac{\partial^2 T}{\partial {r_h}^2}\right]_c}}\,.
\end{equation}
Finally, substituting the values of $\Delta_{s,l}$ from Eq.~(\ref{eq:Delta}) into Eq.~(\ref{eq:delta-lambda}) gives us the desired relation,
\begin{equation}\label{eq:op-ce-relation}
    \frac{\Delta\lambda}{\lambda_c} = k \lvert{t-1}\rvert^{1/2}\,,
\end{equation}
where,
\begin{equation}\label{eq:k-relation}
    k = \frac{\sqrt{T_c}}{\lambda_c}\left[ \frac{\partial\lambda}{\partial r_h}\right]_c\left[\frac{1}{2}\frac{\partial^2 T}{\partial r_h^2}\right]_c^{-1/2}\,,
\end{equation}
and
\begin{equation}\label{eq:t-relation}
    t = \frac{T(r_h)}{T_c}\,.
\end{equation}
Thus, we conclude that the critical exponent $\beta$ related to the order parameter $\Delta\lambda$ is $1/2$, the same as the van der Waals fluid~\cite{van2004continuity}. The same critical exponent of $1/2$ was also found in \cite{kumara2024lyapunov} for another class of black holes, stressing its quite universal value.

Next, we will provide numerical evidence to solidify our theoretical result further.\par

\subsection{Numerical verification}
\begin{figure}[htb!]
	\centering
	\includegraphics[width=\linewidth]{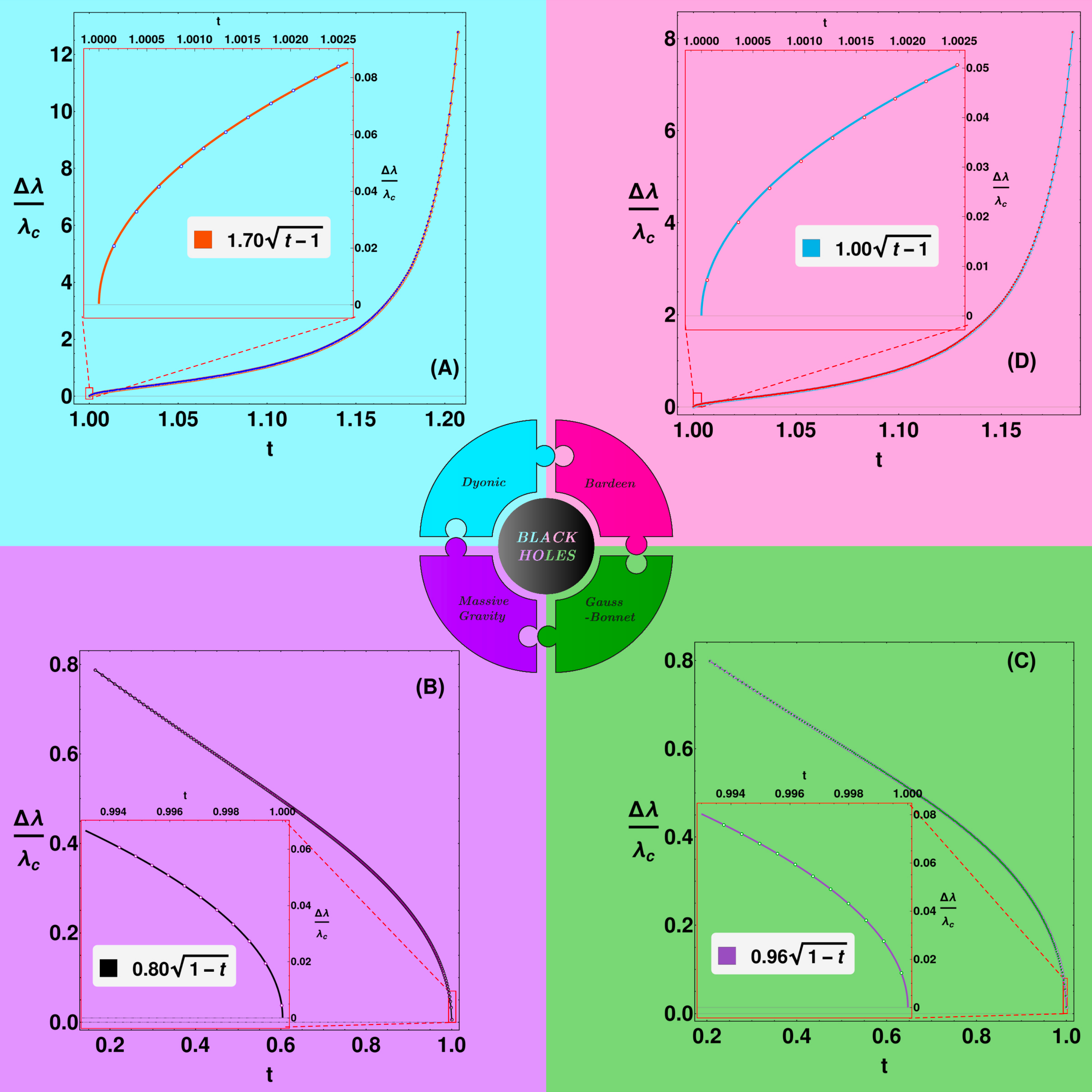}
	\caption{\label{fig:critical_exponent_massless}Plots of the rescaled discontinuity in the Lyapunov exponent $\Delta\lambda/\lambda_c$ vs rescaled phase transition temperature $t=T_p/T_c$ for the massless particle case.}
\end{figure}

The plot of the rescaled order parameter $\Delta\lambda/\lambda_c$ vs $t=T_p/T_c$ for the massless particle is shown in Fig.~\ref{fig:critical_exponent_massless}.
\begin{figure}[htb!]
	\centering
	\includegraphics[width=\linewidth]{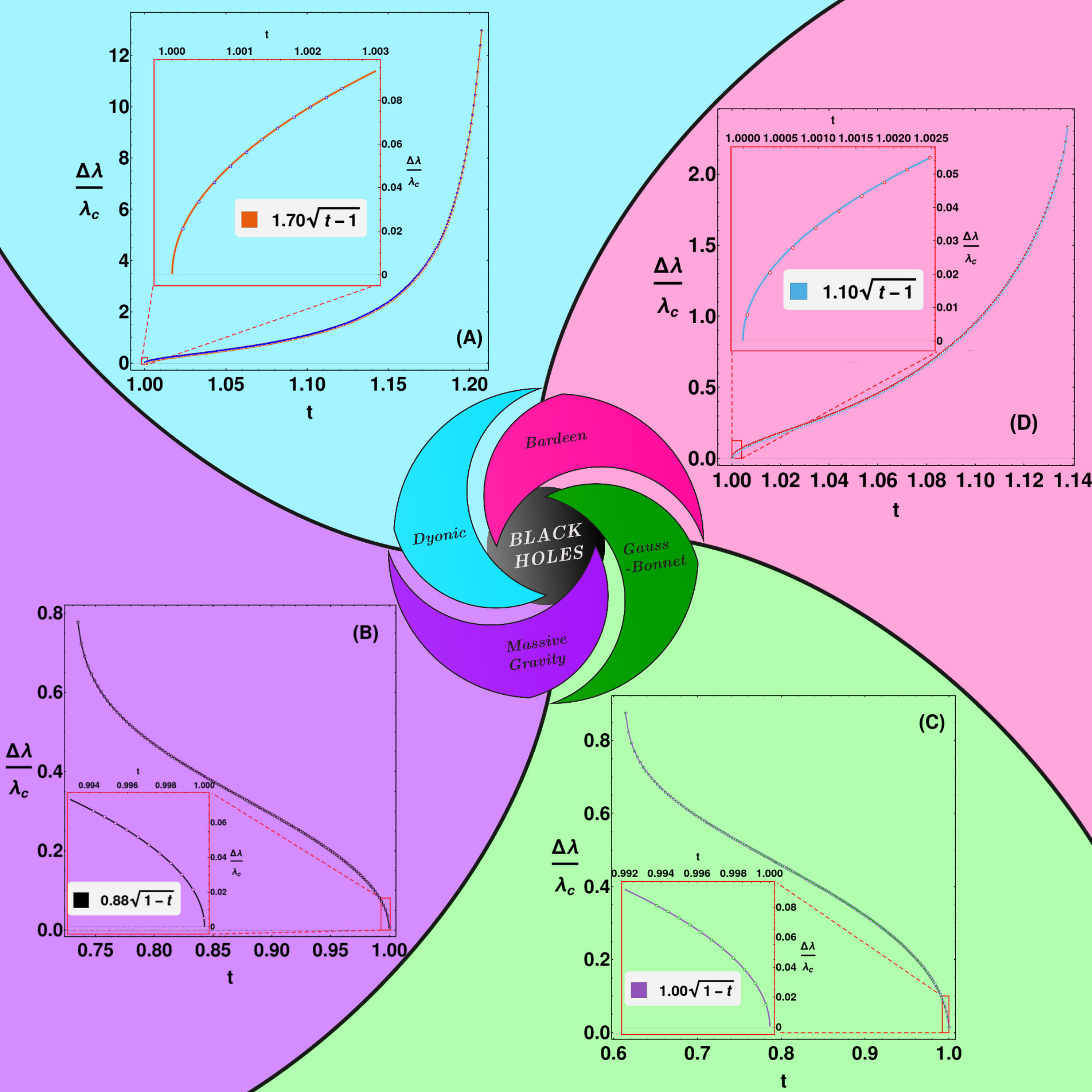}
	\caption{\label{fig:critical_exponent_massive}Plots of the rescaled discontinuity in the Lyapunov exponent $\Delta\lambda/\lambda_c$ vs rescaled phase transition temperature $t=T_p/T_c$ for the massive particle case. }
\end{figure}

Here, we have shown the diagram for all four charged black holes considered in the previous sections, i.e., dyonic, Bardeen, Gauss-Bonnet, and massive gravity. The plot near the critical value is shown in the overlay boxes, where open-circle markers show the data points. The solid lines represent the curve fitting. The expression $\Delta\lambda/\lambda_c=k \lvert{t-1}\rvert^{1/2}$ perfectly fits the numerical data points for all the considered charged black holes. Although the value of $k$ can differ, the critical exponent $\beta$ is always one-half for these black holes.
The same is also true for the massive particle. This is shown in Fig.~\ref{fig:critical_exponent_massive}. Thus, we can readily say that the discontinuity in the Lyapunov exponent $\Delta\lambda$ can be the order parameter with critical exponent $1/2$.


\section{Conclusions}\label{sec:conclusions}
In this work, we have further tested the suggested relationship between black hole phase transitions and the Lyapunov exponents for various physically motivated charged black holes in AdS spaces. In particular, we considered the dyonic, Bardeen, Gauss-Bonnet, and massive gravity black holes. These black holes were chosen not only because they have a rich phase structure by exhibiting interesting liquid/gas type van der Waals phase transition between the small and large black hole phases but also because they are of paramount importance in various (dual) gravitational studies. We tested this relationship for both extended and non-extended thermodynamics. In our analysis, we found that, like the thermal behaviour of the relevant free energy, when the control parameter is less than its critical value, the Lyapunov exponent exhibits multi-valuedness over some temperature range. In particular, the Lyapunov exponent displays distinct thermal behaviour in the small, large, and intermediate black hole phases. Similarly, the Lyapunov exponent becomes single-valued for all temperatures when the control parameter exceeds its critical value. This behaviour of the
Lyapunov exponent, specifically the transition from multi-to single-valuedness, correctly pinpoints the model-dependent second-order critical points, i.e., the critical point is reflected in the thermal behaviour of the Lyapunov exponent. This suggests that the information about the thermodynamic phase structure of considered AdS black holes, to some extent, is encoded in the Lyapunov exponent.

Our analysis further suggested that the Lyapunov exponent of the massive and massless particles behaves distinctly in the large black hole phase. Particularly, the Lyapunov exponent of the massless particle approaches a parameter-independent constant value asymptotically in the large radius limit. Interestingly, this asymptotic value scales with the AdS length $l$ and takes a unit value for $l=1$ for all gravity models considered here. The Lyapunov exponent of the massive particle goes to zero at a certain radius, which is also explained in the diagrams of the corresponding effective potential. We illustrated the complete dependence of the massive and massless particles' Lyapunov exponent on various parameters of the gravity models in density plots. We found that the Lyapunov exponents modify significantly in the small black hole phases compared to the large black hole phase when a certain parameter is varied. We then discussed the possibility of the difference between the Lyapunov exponents of the small and large black hole phases near the critical point as an order parameter. Interestingly, the corresponding critical exponent is $1/2$ for all the gravity models, the same as in the van der Waals phase transition.

Let us end this paper by pointing out a few limitations one faces by probing the black hole thermodynamic phase structure from Lyapunov exponents. An important point that we would like to mention, one which we think has not been discussed much, is that although the Lyapunov exponents of particles do seem to epitomise the presence of the black hole phase transition, they do not give any information about the transition temperature, especially if the transition is of the first order. The $\lambda~\text{vs}~T$ curves only emphasise the importance of $T_1$ around which $\lambda$ starts to show multi-valuedness. However, the first-order small/large black hole phase transition occurs at $T>T_1$, which cannot be determined from the thermal behaviour of the Lyapunov exponents. This should be contrasted with the information on the second-order critical points, which can be inferred from Lyapunov exponents. Another important point is that using the Lyapunov exponent presents a few issues, even within general relativity. In particular, since it amounts to the deviation of two nearby trajectories in time, the whole Lyapunov exponent analysis heavily depends on the considered time coordinate. This coordinate dependence can sometimes lead to wrong conclusions, for instance, by predicting zero Lyapunov exponents even for genuinely chaotic systems \cite{PhysRevLett.46.963, Barrow:1981sx, Cornish:1996yg}. However, this coordinate-dependent issue can be minimised when a preferred time direction exists, such as black hole geometries containing a timelike Killing vector, and meaningful results can be obtained  \cite{Cornish:2003ig}. In other words, if we are asymptotically far from the black hole, we can assign well-defined time coordinates in our observations. Our work also has a timelike Killing vector for all the black hole cases discussed. Moreover, we compare all timescales in the same coordinate system, making the whole Lyapunov exponent analysis credible.

\section*{Acknowledgements}
B.S.~would like to thank S.~Datta for the useful discussions on critical exponents. The work of S.M.~is supported by the core research grant from the Science and Engineering Research Board, a statutory body under the Department of Science and Technology, Government of India, under grant agreement number CRG/2023/007670.

\bibliography{biblio}

\begin{thebibliography}{141}%
\makeatletter
\providecommand \@ifxundefined [1]{%
 \@ifx{#1\undefined}
}%
\providecommand \@ifnum [1]{%
 \ifnum #1\expandafter \@firstoftwo
 \else \expandafter \@secondoftwo
 \fi
}%
\providecommand \@ifx [1]{%
 \ifx #1\expandafter \@firstoftwo
 \else \expandafter \@secondoftwo
 \fi
}%
\providecommand \natexlab [1]{#1}%
\providecommand \enquote  [1]{``#1''}%
\providecommand \bibnamefont  [1]{#1}%
\providecommand \bibfnamefont [1]{#1}%
\providecommand \citenamefont [1]{#1}%
\providecommand \href@noop [0]{\@secondoftwo}%
\providecommand \href [0]{\begingroup \@sanitize@url \@href}%
\providecommand \@href[1]{\@@startlink{#1}\@@href}%
\providecommand \@@href[1]{\endgroup#1\@@endlink}%
\providecommand \@sanitize@url [0]{\catcode `\\12\catcode `\$12\catcode
  `\&12\catcode `\#12\catcode `\^12\catcode `\_12\catcode `\%12\relax}%
\providecommand \@@startlink[1]{}%
\providecommand \@@endlink[0]{}%
\providecommand \url  [0]{\begingroup\@sanitize@url \@url }%
\providecommand \@url [1]{\endgroup\@href {#1}{\urlprefix }}%
\providecommand \urlprefix  [0]{URL }%
\providecommand \Eprint [0]{\href }%
\providecommand \doibase [0]{https://doi.org/}%
\providecommand \selectlanguage [0]{\@gobble}%
\providecommand \bibinfo  [0]{\@secondoftwo}%
\providecommand \bibfield  [0]{\@secondoftwo}%
\providecommand \translation [1]{[#1]}%
\providecommand \BibitemOpen [0]{}%
\providecommand \bibitemStop [0]{}%
\providecommand \bibitemNoStop [0]{.\EOS\space}%
\providecommand \EOS [0]{\spacefactor3000\relax}%
\providecommand \BibitemShut  [1]{\csname bibitem#1\endcsname}%
\let\auto@bib@innerbib\@empty
\bibitem [{\citenamefont {Lorenz}(1963)}]{Lorenz:1963yb}%
  \BibitemOpen
  \bibfield  {author} {\bibinfo {author} {\bibfnamefont {E.~N.}\ \bibnamefont
  {Lorenz}},\ }\bibfield  {title} {\bibinfo {title} {{Deterministic nonperiodic
  flow}},\ }\href
  {https://doi.org/10.1175/1520-0469(1963)020<0130:DNF>2.0.CO;2} {\bibfield
  {journal} {\bibinfo  {journal} {J. Atmos. Sci.}\ }\textbf {\bibinfo {volume}
  {20}},\ \bibinfo {pages} {130} (\bibinfo {year} {1963})}\BibitemShut
  {NoStop}%
\bibitem [{\citenamefont {Lorenzelli}(2014)}]{Lorenzelli2014Apr}%
  \BibitemOpen
  \bibfield  {author} {\bibinfo {author} {\bibfnamefont {F.}~\bibnamefont
  {Lorenzelli}},\ }\href {https://doi.org/10.1201/9781482288988} {\emph
  {\bibinfo {title} {{The Essence Of Chaos}}}}\ (\bibinfo  {publisher} {Taylor
  {\&} Francis},\ \bibinfo {address} {Andover, England, UK},\ \bibinfo {year}
  {2014})\BibitemShut {NoStop}%
\bibitem [{\citenamefont {Shenker}\ and\ \citenamefont
  {Stanford}(2014)}]{Shenker:2013pqa}%
  \BibitemOpen
  \bibfield  {author} {\bibinfo {author} {\bibfnamefont {S.~H.}\ \bibnamefont
  {Shenker}}\ and\ \bibinfo {author} {\bibfnamefont {D.}~\bibnamefont
  {Stanford}},\ }\bibfield  {title} {\bibinfo {title} {{Black holes and the
  butterfly effect}},\ }\href {https://doi.org/10.1007/JHEP03(2014)067}
  {\bibfield  {journal} {\bibinfo  {journal} {JHEP}\ }\textbf {\bibinfo
  {volume} {03}},\ \bibinfo {pages} {067}},\ \Eprint
  {https://arxiv.org/abs/1306.0622} {arXiv:1306.0622 [hep-th]} \BibitemShut
  {NoStop}%
\bibitem [{\citenamefont {Grozdanov}\ \emph {et~al.}(2018)\citenamefont
  {Grozdanov}, \citenamefont {Schalm},\ and\ \citenamefont
  {Scopelliti}}]{Grozdanov:2017ajz}%
  \BibitemOpen
  \bibfield  {author} {\bibinfo {author} {\bibfnamefont {S.}~\bibnamefont
  {Grozdanov}}, \bibinfo {author} {\bibfnamefont {K.}~\bibnamefont {Schalm}},\
  and\ \bibinfo {author} {\bibfnamefont {V.}~\bibnamefont {Scopelliti}},\
  }\bibfield  {title} {\bibinfo {title} {{Black hole scrambling from
  hydrodynamics}},\ }\href {https://doi.org/10.1103/PhysRevLett.120.231601}
  {\bibfield  {journal} {\bibinfo  {journal} {Phys. Rev. Lett.}\ }\textbf
  {\bibinfo {volume} {120}},\ \bibinfo {pages} {231601} (\bibinfo {year}
  {2018})},\ \Eprint {https://arxiv.org/abs/1710.00921} {arXiv:1710.00921
  [hep-th]} \BibitemShut {NoStop}%
\bibitem [{\citenamefont {Blake}(2016)}]{Blake:2016wvh}%
  \BibitemOpen
  \bibfield  {author} {\bibinfo {author} {\bibfnamefont {M.}~\bibnamefont
  {Blake}},\ }\bibfield  {title} {\bibinfo {title} {{Universal Charge Diffusion
  and the Butterfly Effect in Holographic Theories}},\ }\href
  {https://doi.org/10.1103/PhysRevLett.117.091601} {\bibfield  {journal}
  {\bibinfo  {journal} {Phys. Rev. Lett.}\ }\textbf {\bibinfo {volume} {117}},\
  \bibinfo {pages} {091601} (\bibinfo {year} {2016})},\ \Eprint
  {https://arxiv.org/abs/1603.08510} {arXiv:1603.08510 [hep-th]} \BibitemShut
  {NoStop}%
\bibitem [{\citenamefont {Lucas}\ and\ \citenamefont
  {Steinberg}(2016)}]{Lucas:2016yfl}%
  \BibitemOpen
  \bibfield  {author} {\bibinfo {author} {\bibfnamefont {A.}~\bibnamefont
  {Lucas}}\ and\ \bibinfo {author} {\bibfnamefont {J.}~\bibnamefont
  {Steinberg}},\ }\bibfield  {title} {\bibinfo {title} {{Charge diffusion and
  the butterfly effect in striped holographic matter}},\ }\href
  {https://doi.org/10.1007/JHEP10(2016)143} {\bibfield  {journal} {\bibinfo
  {journal} {JHEP}\ }\textbf {\bibinfo {volume} {10}},\ \bibinfo {pages}
  {143}},\ \Eprint {https://arxiv.org/abs/1608.03286} {arXiv:1608.03286
  [hep-th]} \BibitemShut {NoStop}%
\bibitem [{\citenamefont {Kudler-Flam}\ \emph {et~al.}(2021)\citenamefont
  {Kudler-Flam}, \citenamefont {Nozaki}, \citenamefont {Ryu},\ and\
  \citenamefont {Tan}}]{Kudler-Flam:2020yml}%
  \BibitemOpen
  \bibfield  {author} {\bibinfo {author} {\bibfnamefont {J.}~\bibnamefont
  {Kudler-Flam}}, \bibinfo {author} {\bibfnamefont {M.}~\bibnamefont {Nozaki}},
  \bibinfo {author} {\bibfnamefont {S.}~\bibnamefont {Ryu}},\ and\ \bibinfo
  {author} {\bibfnamefont {M.~T.}\ \bibnamefont {Tan}},\ }\bibfield  {title}
  {\bibinfo {title} {{Entanglement of local operators and the butterfly
  effect}},\ }\href {https://doi.org/10.1103/PhysRevResearch.3.033182}
  {\bibfield  {journal} {\bibinfo  {journal} {Phys. Rev. Res.}\ }\textbf
  {\bibinfo {volume} {3}},\ \bibinfo {pages} {033182} (\bibinfo {year}
  {2021})},\ \Eprint {https://arxiv.org/abs/2005.14243} {arXiv:2005.14243
  [hep-th]} \BibitemShut {NoStop}%
\bibitem [{\citenamefont {Jeong}\ \emph {et~al.}(2018)\citenamefont {Jeong},
  \citenamefont {Ahn}, \citenamefont {Ahn}, \citenamefont {Niu}, \citenamefont
  {Li},\ and\ \citenamefont {Kim}}]{Jeong:2017rxg}%
  \BibitemOpen
  \bibfield  {author} {\bibinfo {author} {\bibfnamefont {H.-S.}\ \bibnamefont
  {Jeong}}, \bibinfo {author} {\bibfnamefont {Y.}~\bibnamefont {Ahn}}, \bibinfo
  {author} {\bibfnamefont {D.}~\bibnamefont {Ahn}}, \bibinfo {author}
  {\bibfnamefont {C.}~\bibnamefont {Niu}}, \bibinfo {author} {\bibfnamefont
  {W.-J.}\ \bibnamefont {Li}},\ and\ \bibinfo {author} {\bibfnamefont {K.-Y.}\
  \bibnamefont {Kim}},\ }\bibfield  {title} {\bibinfo {title} {{Thermal
  diffusivity and butterfly velocity in anisotropic Q-Lattice models}},\ }\href
  {https://doi.org/10.1007/JHEP01(2018)140} {\bibfield  {journal} {\bibinfo
  {journal} {JHEP}\ }\textbf {\bibinfo {volume} {01}},\ \bibinfo {pages}
  {140}},\ \Eprint {https://arxiv.org/abs/1708.08822} {arXiv:1708.08822
  [hep-th]} \BibitemShut {NoStop}%
\bibitem [{\citenamefont {Ling}\ and\ \citenamefont
  {Xian}(2017)}]{Ling:2017jik}%
  \BibitemOpen
  \bibfield  {author} {\bibinfo {author} {\bibfnamefont {Y.}~\bibnamefont
  {Ling}}\ and\ \bibinfo {author} {\bibfnamefont {Z.-Y.}\ \bibnamefont
  {Xian}},\ }\bibfield  {title} {\bibinfo {title} {{Holographic Butterfly
  Effect and Diffusion in Quantum Critical Region}},\ }\href
  {https://doi.org/10.1007/JHEP09(2017)003} {\bibfield  {journal} {\bibinfo
  {journal} {JHEP}\ }\textbf {\bibinfo {volume} {09}},\ \bibinfo {pages}
  {003}},\ \Eprint {https://arxiv.org/abs/1707.02843} {arXiv:1707.02843
  [hep-th]} \BibitemShut {NoStop}%
\bibitem [{\citenamefont {Ling}\ \emph {et~al.}(2017)\citenamefont {Ling},
  \citenamefont {Liu},\ and\ \citenamefont {Wu}}]{Ling:2016ibq}%
  \BibitemOpen
  \bibfield  {author} {\bibinfo {author} {\bibfnamefont {Y.}~\bibnamefont
  {Ling}}, \bibinfo {author} {\bibfnamefont {P.}~\bibnamefont {Liu}},\ and\
  \bibinfo {author} {\bibfnamefont {J.-P.}\ \bibnamefont {Wu}},\ }\bibfield
  {title} {\bibinfo {title} {{Holographic Butterfly Effect at Quantum Critical
  Points}},\ }\href {https://doi.org/10.1007/JHEP10(2017)025} {\bibfield
  {journal} {\bibinfo  {journal} {JHEP}\ }\textbf {\bibinfo {volume} {10}},\
  \bibinfo {pages} {025}},\ \Eprint {https://arxiv.org/abs/1610.02669}
  {arXiv:1610.02669 [hep-th]} \BibitemShut {NoStop}%
\bibitem [{\citenamefont {Dong}\ \emph {et~al.}(2022)\citenamefont {Dong},
  \citenamefont {Wang}, \citenamefont {Weng},\ and\ \citenamefont
  {Wu}}]{Dong:2022ucb}%
  \BibitemOpen
  \bibfield  {author} {\bibinfo {author} {\bibfnamefont {X.}~\bibnamefont
  {Dong}}, \bibinfo {author} {\bibfnamefont {D.}~\bibnamefont {Wang}}, \bibinfo
  {author} {\bibfnamefont {W.~W.}\ \bibnamefont {Weng}},\ and\ \bibinfo
  {author} {\bibfnamefont {C.-H.}\ \bibnamefont {Wu}},\ }\bibfield  {title}
  {\bibinfo {title} {{A tale of two butterflies: an exact equivalence in
  higher-derivative gravity}},\ }\href
  {https://doi.org/10.1007/JHEP10(2022)009} {\bibfield  {journal} {\bibinfo
  {journal} {JHEP}\ }\textbf {\bibinfo {volume} {10}},\ \bibinfo {pages}
  {009}},\ \Eprint {https://arxiv.org/abs/2203.06189} {arXiv:2203.06189
  [hep-th]} \BibitemShut {NoStop}%
\bibitem [{\citenamefont {Alishahiha}\ \emph {et~al.}(2016)\citenamefont
  {Alishahiha}, \citenamefont {Davody}, \citenamefont {Naseh},\ and\
  \citenamefont {Taghavi}}]{Alishahiha:2016cjk}%
  \BibitemOpen
  \bibfield  {author} {\bibinfo {author} {\bibfnamefont {M.}~\bibnamefont
  {Alishahiha}}, \bibinfo {author} {\bibfnamefont {A.}~\bibnamefont {Davody}},
  \bibinfo {author} {\bibfnamefont {A.}~\bibnamefont {Naseh}},\ and\ \bibinfo
  {author} {\bibfnamefont {S.~F.}\ \bibnamefont {Taghavi}},\ }\bibfield
  {title} {\bibinfo {title} {{On Butterfly effect in Higher Derivative
  Gravities}},\ }\href {https://doi.org/10.1007/JHEP11(2016)032} {\bibfield
  {journal} {\bibinfo  {journal} {JHEP}\ }\textbf {\bibinfo {volume} {11}},\
  \bibinfo {pages} {032}},\ \Eprint {https://arxiv.org/abs/1610.02890}
  {arXiv:1610.02890 [hep-th]} \BibitemShut {NoStop}%
\bibitem [{\citenamefont {Maldacena}\ \emph {et~al.}(2016)\citenamefont
  {Maldacena}, \citenamefont {Shenker},\ and\ \citenamefont
  {Stanford}}]{Maldacena:2015waa}%
  \BibitemOpen
  \bibfield  {author} {\bibinfo {author} {\bibfnamefont {J.}~\bibnamefont
  {Maldacena}}, \bibinfo {author} {\bibfnamefont {S.~H.}\ \bibnamefont
  {Shenker}},\ and\ \bibinfo {author} {\bibfnamefont {D.}~\bibnamefont
  {Stanford}},\ }\bibfield  {title} {\bibinfo {title} {{A bound on chaos}},\
  }\href {https://doi.org/10.1007/JHEP08(2016)106} {\bibfield  {journal}
  {\bibinfo  {journal} {JHEP}\ }\textbf {\bibinfo {volume} {08}},\ \bibinfo
  {pages} {106}},\ \Eprint {https://arxiv.org/abs/1503.01409} {arXiv:1503.01409
  [hep-th]} \BibitemShut {NoStop}%
\bibitem [{\citenamefont {Kan}\ and\ \citenamefont {Gwak}(2022)}]{Kan:2021blg}%
  \BibitemOpen
  \bibfield  {author} {\bibinfo {author} {\bibfnamefont {N.}~\bibnamefont
  {Kan}}\ and\ \bibinfo {author} {\bibfnamefont {B.}~\bibnamefont {Gwak}},\
  }\bibfield  {title} {\bibinfo {title} {{Bound on the Lyapunov exponent in
  Kerr-Newman black holes via a charged particle}},\ }\href
  {https://doi.org/10.1103/PhysRevD.105.026006} {\bibfield  {journal} {\bibinfo
   {journal} {Phys. Rev. D}\ }\textbf {\bibinfo {volume} {105}},\ \bibinfo
  {pages} {026006} (\bibinfo {year} {2022})},\ \Eprint
  {https://arxiv.org/abs/2109.07341} {arXiv:2109.07341 [gr-qc]} \BibitemShut
  {NoStop}%
\bibitem [{\citenamefont {Gwak}\ \emph {et~al.}(2022)\citenamefont {Gwak},
  \citenamefont {Kan}, \citenamefont {Lee},\ and\ \citenamefont
  {Lee}}]{Gwak:2022xje}%
  \BibitemOpen
  \bibfield  {author} {\bibinfo {author} {\bibfnamefont {B.}~\bibnamefont
  {Gwak}}, \bibinfo {author} {\bibfnamefont {N.}~\bibnamefont {Kan}}, \bibinfo
  {author} {\bibfnamefont {B.-H.}\ \bibnamefont {Lee}},\ and\ \bibinfo {author}
  {\bibfnamefont {H.}~\bibnamefont {Lee}},\ }\bibfield  {title} {\bibinfo
  {title} {{Violation of bound on chaos for charged probe in Kerr-Newman-AdS
  black hole}},\ }\href {https://doi.org/10.1007/JHEP09(2022)026} {\bibfield
  {journal} {\bibinfo  {journal} {JHEP}\ }\textbf {\bibinfo {volume} {09}},\
  \bibinfo {pages} {026}},\ \Eprint {https://arxiv.org/abs/2203.07298}
  {arXiv:2203.07298 [gr-qc]} \BibitemShut {NoStop}%
\bibitem [{\citenamefont {Jahnke}(2019)}]{Jahnke:2018off}%
  \BibitemOpen
  \bibfield  {author} {\bibinfo {author} {\bibfnamefont {V.}~\bibnamefont
  {Jahnke}},\ }\bibfield  {title} {\bibinfo {title} {{Recent developments in
  the holographic description of quantum chaos}},\ }\href
  {https://doi.org/10.1155/2019/9632708} {\bibfield  {journal} {\bibinfo
  {journal} {Adv. High Energy Phys.}\ }\textbf {\bibinfo {volume} {2019}},\
  \bibinfo {pages} {9632708} (\bibinfo {year} {2019})},\ \Eprint
  {https://arxiv.org/abs/1811.06949} {arXiv:1811.06949 [hep-th]} \BibitemShut
  {NoStop}%
\bibitem [{\citenamefont {Lyapunov}(1992)}]{Lyapunov1992Mar}%
  \BibitemOpen
  \bibfield  {author} {\bibinfo {author} {\bibfnamefont {A.~M.}\ \bibnamefont
  {Lyapunov}},\ }\bibfield  {title} {\bibinfo {title} {{The general problem of
  the stability of motion}},\ }\bibfield  {journal} {\bibinfo  {journal} {Int.
  J. Control}\ }\href {https://doi.org/10.1080/00207179208934253}
  {10.1080/00207179208934253} (\bibinfo {year} {1992})\BibitemShut {NoStop}%
\bibitem [{\citenamefont {Sandri}(1996)}]{sandri1996}%
  \BibitemOpen
  \bibfield  {author} {\bibinfo {author} {\bibfnamefont {M.}~\bibnamefont
  {Sandri}},\ }\bibfield  {title} {\bibinfo {title} {Numerical calculation of
  lyapunov exponents},\ }\href@noop {} {\bibfield  {journal} {\bibinfo
  {journal} {Mathematica Journal}\ }\textbf {\bibinfo {volume} {6}},\ \bibinfo
  {pages} {78} (\bibinfo {year} {1996})}\BibitemShut {NoStop}%
\bibitem [{\citenamefont {Colangelo}\ \emph {et~al.}(2022)\citenamefont
  {Colangelo}, \citenamefont {Giannuzzi},\ and\ \citenamefont
  {Losacco}}]{Colangelo:2021kmn}%
  \BibitemOpen
  \bibfield  {author} {\bibinfo {author} {\bibfnamefont {P.}~\bibnamefont
  {Colangelo}}, \bibinfo {author} {\bibfnamefont {F.}~\bibnamefont
  {Giannuzzi}},\ and\ \bibinfo {author} {\bibfnamefont {N.}~\bibnamefont
  {Losacco}},\ }\bibfield  {title} {\bibinfo {title} {{Chaotic dynamics of a
  suspended string in a gravitational background with magnetic field}},\ }\href
  {https://doi.org/10.1016/j.physletb.2022.136949} {\bibfield  {journal}
  {\bibinfo  {journal} {Phys. Lett. B}\ }\textbf {\bibinfo {volume} {827}},\
  \bibinfo {pages} {136949} (\bibinfo {year} {2022})},\ \Eprint
  {https://arxiv.org/abs/2111.09441} {arXiv:2111.09441 [hep-th]} \BibitemShut
  {NoStop}%
\bibitem [{\citenamefont {Shukla}\ \emph {et~al.}(2023)\citenamefont {Shukla},
  \citenamefont {Dudal},\ and\ \citenamefont {Mahapatra}}]{Shukla:2023pbp}%
  \BibitemOpen
  \bibfield  {author} {\bibinfo {author} {\bibfnamefont {B.}~\bibnamefont
  {Shukla}}, \bibinfo {author} {\bibfnamefont {D.}~\bibnamefont {Dudal}},\ and\
  \bibinfo {author} {\bibfnamefont {S.}~\bibnamefont {Mahapatra}},\ }\bibfield
  {title} {\bibinfo {title} {{Anisotropic and frame dependent chaos of
  suspended strings from a dynamical holographic QCD model with magnetic
  field}},\ }\href {https://doi.org/10.1007/JHEP06(2023)178} {\bibfield
  {journal} {\bibinfo  {journal} {JHEP}\ }\textbf {\bibinfo {volume} {06}},\
  \bibinfo {pages} {178}},\ \Eprint {https://arxiv.org/abs/2303.15716}
  {arXiv:2303.15716 [hep-th]} \BibitemShut {NoStop}%
\bibitem [{\citenamefont {Ferriere}\ and\ \citenamefont
  {Gatto}(1995)}]{ferriere1995lyapunov}%
  \BibitemOpen
  \bibfield  {author} {\bibinfo {author} {\bibfnamefont {R.}~\bibnamefont
  {Ferriere}}\ and\ \bibinfo {author} {\bibfnamefont {M.}~\bibnamefont
  {Gatto}},\ }\bibfield  {title} {\bibinfo {title} {Lyapunov exponents and the
  mathematics of invasion in oscillatory or chaotic populations},\ }\href
  {https://doi.org/https://doi.org/10.1006/tpbi.1995.1024} {\bibfield
  {journal} {\bibinfo  {journal} {Theoretical Population Biology}\ }\textbf
  {\bibinfo {volume} {48}},\ \bibinfo {pages} {126} (\bibinfo {year}
  {1995})}\BibitemShut {NoStop}%
\bibitem [{\citenamefont {Dechert}\ and\ \citenamefont
  {Gen{\c{c}}ay}(1992)}]{dechert1992lyapunov}%
  \BibitemOpen
  \bibfield  {author} {\bibinfo {author} {\bibfnamefont {W.~D.}\ \bibnamefont
  {Dechert}}\ and\ \bibinfo {author} {\bibfnamefont {R.}~\bibnamefont
  {Gen{\c{c}}ay}},\ }\bibfield  {title} {\bibinfo {title} {Lyapunov exponents
  as a nonparametric diagnostic for stability analysis},\ }\href
  {https://doi.org/https://doi.org/10.1002/jae.3950070505} {\bibfield
  {journal} {\bibinfo  {journal} {Journal of Applied Econometrics}\ }\textbf
  {\bibinfo {volume} {7}},\ \bibinfo {pages} {S41} (\bibinfo {year}
  {1992})}\BibitemShut {NoStop}%
\bibitem [{\citenamefont {Vannitsem}(2017)}]{vannitsem2017predictability}%
  \BibitemOpen
  \bibfield  {author} {\bibinfo {author} {\bibfnamefont {S.}~\bibnamefont
  {Vannitsem}},\ }\bibfield  {title} {\bibinfo {title} {Predictability of
  large-scale atmospheric motions: Lyapunov exponents and error dynamics},\
  }\bibfield  {journal} {\bibinfo  {journal} {Chaos: An Interdisciplinary
  Journal of Nonlinear Science}\ }\textbf {\bibinfo {volume} {27}},\ \href
  {https://doi.org/https://doi.org/10.1063/1.4979042}
  {https://doi.org/10.1063/1.4979042} (\bibinfo {year} {2017})\BibitemShut
  {NoStop}%
\bibitem [{\citenamefont {Emary}\ and\ \citenamefont
  {Brandes}(2003{\natexlab{a}})}]{Emary:2003zza}%
  \BibitemOpen
  \bibfield  {author} {\bibinfo {author} {\bibfnamefont {C.}~\bibnamefont
  {Emary}}\ and\ \bibinfo {author} {\bibfnamefont {T.}~\bibnamefont
  {Brandes}},\ }\bibfield  {title} {\bibinfo {title} {{Chaos and the quantum
  phase transition in the Dicke model}},\ }\href
  {https://doi.org/10.1103/PhysRevE.67.066203} {\bibfield  {journal} {\bibinfo
  {journal} {Phys. Rev. E}\ }\textbf {\bibinfo {volume} {67}},\ \bibinfo
  {pages} {066203} (\bibinfo {year} {2003}{\natexlab{a}})},\ \Eprint
  {https://arxiv.org/abs/cond-mat/0301273} {arXiv:cond-mat/0301273}
  \BibitemShut {NoStop}%
\bibitem [{\citenamefont {Sorokhaibam}(2020)}]{Sorokhaibam:2019qho}%
  \BibitemOpen
  \bibfield  {author} {\bibinfo {author} {\bibfnamefont {N.}~\bibnamefont
  {Sorokhaibam}},\ }\bibfield  {title} {\bibinfo {title} {{Phase transition and
  chaos in charged SYK model}},\ }\href
  {https://doi.org/10.1007/JHEP07(2020)055} {\bibfield  {journal} {\bibinfo
  {journal} {JHEP}\ }\textbf {\bibinfo {volume} {07}},\ \bibinfo {pages}
  {055}},\ \Eprint {https://arxiv.org/abs/1912.04326} {arXiv:1912.04326
  [hep-th]} \BibitemShut {NoStop}%
\bibitem [{\citenamefont {Davis}\ and\ \citenamefont
  {Wang}(2023)}]{Davis:2022iqi}%
  \BibitemOpen
  \bibfield  {author} {\bibinfo {author} {\bibfnamefont {A.}~\bibnamefont
  {Davis}}\ and\ \bibinfo {author} {\bibfnamefont {Y.}~\bibnamefont {Wang}},\
  }\bibfield  {title} {\bibinfo {title} {{Quantum chaos and phase transition in
  the Yukawa\textendash{}Sachdev-Ye-Kitaev model}},\ }\href
  {https://doi.org/10.1103/PhysRevB.107.205122} {\bibfield  {journal} {\bibinfo
   {journal} {Phys. Rev. B}\ }\textbf {\bibinfo {volume} {107}},\ \bibinfo
  {pages} {205122} (\bibinfo {year} {2023})},\ \Eprint
  {https://arxiv.org/abs/2212.03265} {arXiv:2212.03265 [cond-mat.str-el]}
  \BibitemShut {NoStop}%
\bibitem [{\citenamefont {Miritello}\ \emph {et~al.}(2009)\citenamefont
  {Miritello}, \citenamefont {Pluchino},\ and\ \citenamefont
  {Rapisarda}}]{Miritello:2008zd}%
  \BibitemOpen
  \bibfield  {author} {\bibinfo {author} {\bibfnamefont {G.}~\bibnamefont
  {Miritello}}, \bibinfo {author} {\bibfnamefont {A.}~\bibnamefont
  {Pluchino}},\ and\ \bibinfo {author} {\bibfnamefont {A.}~\bibnamefont
  {Rapisarda}},\ }\bibfield  {title} {\bibinfo {title} {{Phase Transitions and
  Chaos in Long-Range Models of Coupled Oscillators}},\ }\href
  {https://doi.org/10.1209/0295-5075/85/10007} {\bibfield  {journal} {\bibinfo
  {journal} {EPL}\ }\textbf {\bibinfo {volume} {85}},\ \bibinfo {pages} {10007}
  (\bibinfo {year} {2009})},\ \Eprint {https://arxiv.org/abs/0807.1870}
  {arXiv:0807.1870 [cond-mat.stat-mech]} \BibitemShut {NoStop}%
\bibitem [{\citenamefont {Heiss}\ and\ \citenamefont
  {Sannino}(1991)}]{Heiss:1991zza}%
  \BibitemOpen
  \bibfield  {author} {\bibinfo {author} {\bibfnamefont {W.~D.}\ \bibnamefont
  {Heiss}}\ and\ \bibinfo {author} {\bibfnamefont {A.~L.}\ \bibnamefont
  {Sannino}},\ }\bibfield  {title} {\bibinfo {title} {{Transitional regions of
  finite Fermi systems and quantum chaos}},\ }\href
  {https://doi.org/10.1103/PhysRevA.43.4159} {\bibfield  {journal} {\bibinfo
  {journal} {Phys. Rev. A}\ }\textbf {\bibinfo {volume} {43}},\ \bibinfo
  {pages} {4159} (\bibinfo {year} {1991})}\BibitemShut {NoStop}%
\bibitem [{\citenamefont {Emary}\ and\ \citenamefont
  {Brandes}(2003{\natexlab{b}})}]{emary2003chaos}%
  \BibitemOpen
  \bibfield  {author} {\bibinfo {author} {\bibfnamefont {C.}~\bibnamefont
  {Emary}}\ and\ \bibinfo {author} {\bibfnamefont {T.}~\bibnamefont
  {Brandes}},\ }\bibfield  {title} {\bibinfo {title} {Chaos and phase
  transitions in quantum dots coupled to bosons},\ }\href
  {https://doi.org/https://www1.itp.tu-berlin.de/brandes/public_html/publications/EmaryICPS.pdf}
  {\bibfield  {journal} {\bibinfo  {journal} {Physical review E}\ }\textbf
  {\bibinfo {volume} {67}},\ \bibinfo {pages} {0662031} (\bibinfo {year}
  {2003}{\natexlab{b}})}\BibitemShut {NoStop}%
\bibitem [{\citenamefont {Da\u{g}}\ \emph {et~al.}(2019)\citenamefont
  {Da\u{g}}, \citenamefont {Sun},\ and\ \citenamefont
  {Duan}}]{daug2019detection}%
  \BibitemOpen
  \bibfield  {author} {\bibinfo {author} {\bibfnamefont {C.~B.}\ \bibnamefont
  {Da\u{g}}}, \bibinfo {author} {\bibfnamefont {K.}~\bibnamefont {Sun}},\ and\
  \bibinfo {author} {\bibfnamefont {L.~M.}\ \bibnamefont {Duan}},\ }\bibfield
  {title} {\bibinfo {title} {{Detection of Quantum Phases via Out-of-Time-Order
  Correlators}},\ }\href {https://doi.org/10.1103/PhysRevLett.123.140602}
  {\bibfield  {journal} {\bibinfo  {journal} {Phys. Rev. Lett.}\ }\textbf
  {\bibinfo {volume} {123}},\ \bibinfo {pages} {140602} (\bibinfo {year}
  {2019})},\ \Eprint {https://arxiv.org/abs/1902.05041} {arXiv:1902.05041
  [quant-ph]} \BibitemShut {NoStop}%
\bibitem [{\citenamefont {Sun}\ \emph {et~al.}(2020)\citenamefont {Sun},
  \citenamefont {Cai}, \citenamefont {Tang}, \citenamefont {Hu},\ and\
  \citenamefont {Fan}}]{sun2020out}%
  \BibitemOpen
  \bibfield  {author} {\bibinfo {author} {\bibfnamefont {Z.-H.}\ \bibnamefont
  {Sun}}, \bibinfo {author} {\bibfnamefont {J.-Q.}\ \bibnamefont {Cai}},
  \bibinfo {author} {\bibfnamefont {Q.-C.}\ \bibnamefont {Tang}}, \bibinfo
  {author} {\bibfnamefont {Y.}~\bibnamefont {Hu}},\ and\ \bibinfo {author}
  {\bibfnamefont {H.}~\bibnamefont {Fan}},\ }\bibfield  {title} {\bibinfo
  {title} {{Out-of-Time-Order Correlators and Quantum Phase Transitions in the
  Rabi and Dicke Models}},\ }\href {https://doi.org/10.1002/andp.201900270}
  {\bibfield  {journal} {\bibinfo  {journal} {Annalen Phys.}\ }\textbf
  {\bibinfo {volume} {532}},\ \bibinfo {pages} {1900270} (\bibinfo {year}
  {2020})}\BibitemShut {NoStop}%
\bibitem [{\citenamefont {Wang}\ and\ \citenamefont
  {P\'erez-Bernal}(2019)}]{wang2019probing}%
  \BibitemOpen
  \bibfield  {author} {\bibinfo {author} {\bibfnamefont {Q.}~\bibnamefont
  {Wang}}\ and\ \bibinfo {author} {\bibfnamefont {F.}~\bibnamefont
  {P\'erez-Bernal}},\ }\bibfield  {title} {\bibinfo {title} {{Probing an
  excited-state quantum phase transition in a quantum many-body system via an
  out-of-time-order correlator}},\ }\href
  {https://doi.org/10.1103/PhysRevA.100.062113} {\bibfield  {journal} {\bibinfo
   {journal} {Phys. Rev. A}\ }\textbf {\bibinfo {volume} {100}},\ \bibinfo
  {pages} {062113} (\bibinfo {year} {2019})},\ \Eprint
  {https://arxiv.org/abs/1812.01920} {arXiv:1812.01920 [quant-ph]} \BibitemShut
  {NoStop}%
\bibitem [{\citenamefont {Huh}\ \emph {et~al.}(2021)\citenamefont {Huh},
  \citenamefont {Ikeda}, \citenamefont {Jahnke},\ and\ \citenamefont
  {Kim}}]{huh2021diagnosing}%
  \BibitemOpen
  \bibfield  {author} {\bibinfo {author} {\bibfnamefont {K.-B.}\ \bibnamefont
  {Huh}}, \bibinfo {author} {\bibfnamefont {K.}~\bibnamefont {Ikeda}}, \bibinfo
  {author} {\bibfnamefont {V.}~\bibnamefont {Jahnke}},\ and\ \bibinfo {author}
  {\bibfnamefont {K.-Y.}\ \bibnamefont {Kim}},\ }\bibfield  {title} {\bibinfo
  {title} {{Diagnosing first- and second-order phase transitions with probes of
  quantum chaos}},\ }\href {https://doi.org/10.1103/PhysRevE.104.024136}
  {\bibfield  {journal} {\bibinfo  {journal} {Phys. Rev. E}\ }\textbf {\bibinfo
  {volume} {104}},\ \bibinfo {pages} {024136} (\bibinfo {year} {2021})},\
  \Eprint {https://arxiv.org/abs/2010.07478} {arXiv:2010.07478 [hep-th]}
  \BibitemShut {NoStop}%
\bibitem [{\citenamefont {Anegawa}\ \emph {et~al.}(2024)\citenamefont
  {Anegawa}, \citenamefont {Iizuka},\ and\ \citenamefont
  {Nishida}}]{anegawa2024krylov}%
  \BibitemOpen
  \bibfield  {author} {\bibinfo {author} {\bibfnamefont {T.}~\bibnamefont
  {Anegawa}}, \bibinfo {author} {\bibfnamefont {N.}~\bibnamefont {Iizuka}},\
  and\ \bibinfo {author} {\bibfnamefont {M.}~\bibnamefont {Nishida}},\
  }\bibfield  {title} {\bibinfo {title} {{Krylov complexity as an order
  parameter for deconfinement phase transitions at large $N$}},\ }\href@noop {}
  {\  (\bibinfo {year} {2024})},\ \Eprint {https://arxiv.org/abs/2401.04383}
  {arXiv:2401.04383 [hep-th]} \BibitemShut {NoStop}%
\bibitem [{\citenamefont {Hawking}\ and\ \citenamefont
  {Page}(1983)}]{hawking1983thermodynamics}%
  \BibitemOpen
  \bibfield  {author} {\bibinfo {author} {\bibfnamefont {S.~W.}\ \bibnamefont
  {Hawking}}\ and\ \bibinfo {author} {\bibfnamefont {D.~N.}\ \bibnamefont
  {Page}},\ }\bibfield  {title} {\bibinfo {title} {{Thermodynamics of Black
  Holes in anti-De Sitter Space}},\ }\href {https://doi.org/10.1007/BF01208266}
  {\bibfield  {journal} {\bibinfo  {journal} {Commun. Math. Phys.}\ }\textbf
  {\bibinfo {volume} {87}},\ \bibinfo {pages} {577} (\bibinfo {year}
  {1983})}\BibitemShut {NoStop}%
\bibitem [{\citenamefont {Chamblin}\ \emph
  {et~al.}(1999{\natexlab{a}})\citenamefont {Chamblin}, \citenamefont
  {Emparan}, \citenamefont {Johnson},\ and\ \citenamefont
  {Myers}}]{chamblin1999holography}%
  \BibitemOpen
  \bibfield  {author} {\bibinfo {author} {\bibfnamefont {A.}~\bibnamefont
  {Chamblin}}, \bibinfo {author} {\bibfnamefont {R.}~\bibnamefont {Emparan}},
  \bibinfo {author} {\bibfnamefont {C.~V.}\ \bibnamefont {Johnson}},\ and\
  \bibinfo {author} {\bibfnamefont {R.~C.}\ \bibnamefont {Myers}},\ }\bibfield
  {title} {\bibinfo {title} {{Charged AdS black holes and catastrophic
  holography}},\ }\href {https://doi.org/10.1103/PhysRevD.60.064018} {\bibfield
   {journal} {\bibinfo  {journal} {Phys. Rev. D}\ }\textbf {\bibinfo {volume}
  {60}},\ \bibinfo {pages} {064018} (\bibinfo {year} {1999}{\natexlab{a}})},\
  \Eprint {https://arxiv.org/abs/hep-th/9902170} {arXiv:hep-th/9902170}
  \BibitemShut {NoStop}%
\bibitem [{\citenamefont {Cai}\ \emph {et~al.}(2013)\citenamefont {Cai},
  \citenamefont {Cao}, \citenamefont {Li},\ and\ \citenamefont
  {Yang}}]{Cai:2013qga}%
  \BibitemOpen
  \bibfield  {author} {\bibinfo {author} {\bibfnamefont {R.-G.}\ \bibnamefont
  {Cai}}, \bibinfo {author} {\bibfnamefont {L.-M.}\ \bibnamefont {Cao}},
  \bibinfo {author} {\bibfnamefont {L.}~\bibnamefont {Li}},\ and\ \bibinfo
  {author} {\bibfnamefont {R.-Q.}\ \bibnamefont {Yang}},\ }\bibfield  {title}
  {\bibinfo {title} {{P-V criticality in the extended phase space of
  Gauss-Bonnet black holes in AdS space}},\ }\href
  {https://doi.org/10.1007/JHEP09(2013)005} {\bibfield  {journal} {\bibinfo
  {journal} {JHEP}\ }\textbf {\bibinfo {volume} {09}},\ \bibinfo {pages}
  {005}},\ \Eprint {https://arxiv.org/abs/1306.6233} {arXiv:1306.6233 [gr-qc]}
  \BibitemShut {NoStop}%
\bibitem [{\citenamefont {Dolan}\ \emph {et~al.}(2014)\citenamefont {Dolan},
  \citenamefont {Kostouki}, \citenamefont {Kubiznak},\ and\ \citenamefont
  {Mann}}]{Dolan:2014vba}%
  \BibitemOpen
  \bibfield  {author} {\bibinfo {author} {\bibfnamefont {B.~P.}\ \bibnamefont
  {Dolan}}, \bibinfo {author} {\bibfnamefont {A.}~\bibnamefont {Kostouki}},
  \bibinfo {author} {\bibfnamefont {D.}~\bibnamefont {Kubiznak}},\ and\
  \bibinfo {author} {\bibfnamefont {R.~B.}\ \bibnamefont {Mann}},\ }\bibfield
  {title} {\bibinfo {title} {{Isolated critical point from Lovelock gravity}},\
  }\href {https://doi.org/10.1088/0264-9381/31/24/242001} {\bibfield  {journal}
  {\bibinfo  {journal} {Class. Quant. Grav.}\ }\textbf {\bibinfo {volume}
  {31}},\ \bibinfo {pages} {242001} (\bibinfo {year} {2014})},\ \Eprint
  {https://arxiv.org/abs/1407.4783} {arXiv:1407.4783 [hep-th]} \BibitemShut
  {NoStop}%
\bibitem [{\citenamefont {L\"u}\ \emph {et~al.}(2013)\citenamefont {L\"u},
  \citenamefont {Pang},\ and\ \citenamefont {Pope}}]{lu2013ads}%
  \BibitemOpen
  \bibfield  {author} {\bibinfo {author} {\bibfnamefont {H.}~\bibnamefont
  {L\"u}}, \bibinfo {author} {\bibfnamefont {Y.}~\bibnamefont {Pang}},\ and\
  \bibinfo {author} {\bibfnamefont {C.~N.}\ \bibnamefont {Pope}},\ }\bibfield
  {title} {\bibinfo {title} {{AdS Dyonic Black Hole and its Thermodynamics}},\
  }\href {https://doi.org/10.1007/JHEP11(2013)033} {\bibfield  {journal}
  {\bibinfo  {journal} {JHEP}\ }\textbf {\bibinfo {volume} {11}},\ \bibinfo
  {pages} {033}},\ \Eprint {https://arxiv.org/abs/1307.6243} {arXiv:1307.6243
  [hep-th]} \BibitemShut {NoStop}%
\bibitem [{\citenamefont {Hartnoll}\ and\ \citenamefont
  {Kovtun}(2007)}]{Hartnoll:2007ai}%
  \BibitemOpen
  \bibfield  {author} {\bibinfo {author} {\bibfnamefont {S.~A.}\ \bibnamefont
  {Hartnoll}}\ and\ \bibinfo {author} {\bibfnamefont {P.}~\bibnamefont
  {Kovtun}},\ }\bibfield  {title} {\bibinfo {title} {{Hall conductivity from
  dyonic black holes}},\ }\href {https://doi.org/10.1103/PhysRevD.76.066001}
  {\bibfield  {journal} {\bibinfo  {journal} {Phys. Rev. D}\ }\textbf {\bibinfo
  {volume} {76}},\ \bibinfo {pages} {066001} (\bibinfo {year} {2007})},\
  \Eprint {https://arxiv.org/abs/0704.1160} {arXiv:0704.1160 [hep-th]}
  \BibitemShut {NoStop}%
\bibitem [{\citenamefont {Rasheed}(1995)}]{rasheed1995rotating}%
  \BibitemOpen
  \bibfield  {author} {\bibinfo {author} {\bibfnamefont {D.}~\bibnamefont
  {Rasheed}},\ }\bibfield  {title} {\bibinfo {title} {{The Rotating dyonic
  black holes of Kaluza-Klein theory}},\ }\href
  {https://doi.org/10.1016/0550-3213(95)00396-A} {\bibfield  {journal}
  {\bibinfo  {journal} {Nucl. Phys. B}\ }\textbf {\bibinfo {volume} {454}},\
  \bibinfo {pages} {379} (\bibinfo {year} {1995})},\ \Eprint
  {https://arxiv.org/abs/hep-th/9505038} {arXiv:hep-th/9505038} \BibitemShut
  {NoStop}%
\bibitem [{\citenamefont {Tseytlin}(1996)}]{tseytlin1996extreme}%
  \BibitemOpen
  \bibfield  {author} {\bibinfo {author} {\bibfnamefont {A.~A.}\ \bibnamefont
  {Tseytlin}},\ }\bibfield  {title} {\bibinfo {title} {{Extreme dyonic black
  holes in string theory}},\ }\href {https://doi.org/10.1142/S0217732396000709}
  {\bibfield  {journal} {\bibinfo  {journal} {Mod. Phys. Lett. A}\ }\textbf
  {\bibinfo {volume} {11}},\ \bibinfo {pages} {689} (\bibinfo {year} {1996})},\
  \Eprint {https://arxiv.org/abs/hep-th/9601177} {arXiv:hep-th/9601177}
  \BibitemShut {NoStop}%
\bibitem [{\citenamefont {Cheng}\ \emph {et~al.}(1994)\citenamefont {Cheng},
  \citenamefont {Hsu},\ and\ \citenamefont {Lin}}]{cheng1993dyonic}%
  \BibitemOpen
  \bibfield  {author} {\bibinfo {author} {\bibfnamefont {G.-J.}\ \bibnamefont
  {Cheng}}, \bibinfo {author} {\bibfnamefont {R.-R.}\ \bibnamefont {Hsu}},\
  and\ \bibinfo {author} {\bibfnamefont {W.-F.}\ \bibnamefont {Lin}},\
  }\bibfield  {title} {\bibinfo {title} {{Dyonic black holes in string
  theory}},\ }\href {https://doi.org/10.1063/1.530817} {\bibfield  {journal}
  {\bibinfo  {journal} {J. Math. Phys.}\ }\textbf {\bibinfo {volume} {35}},\
  \bibinfo {pages} {4839} (\bibinfo {year} {1994})},\ \Eprint
  {https://arxiv.org/abs/hep-th/9302065} {arXiv:hep-th/9302065} \BibitemShut
  {NoStop}%
\bibitem [{\citenamefont {Priyadarshinee}\ \emph {et~al.}(2021)\citenamefont
  {Priyadarshinee}, \citenamefont {Mahapatra},\ and\ \citenamefont
  {Banerjee}}]{priyadarshinee2021analytic}%
  \BibitemOpen
  \bibfield  {author} {\bibinfo {author} {\bibfnamefont {S.}~\bibnamefont
  {Priyadarshinee}}, \bibinfo {author} {\bibfnamefont {S.}~\bibnamefont
  {Mahapatra}},\ and\ \bibinfo {author} {\bibfnamefont {I.}~\bibnamefont
  {Banerjee}},\ }\bibfield  {title} {\bibinfo {title} {{Analytic topological
  hairy dyonic black holes and thermodynamics}},\ }\href
  {https://doi.org/10.1103/PhysRevD.104.084023} {\bibfield  {journal} {\bibinfo
   {journal} {Phys. Rev. D}\ }\textbf {\bibinfo {volume} {104}},\ \bibinfo
  {pages} {084023} (\bibinfo {year} {2021})},\ \Eprint
  {https://arxiv.org/abs/2108.02514} {arXiv:2108.02514 [hep-th]} \BibitemShut
  {NoStop}%
\bibitem [{\citenamefont {Chaturvedi}\ \emph {et~al.}(2017)\citenamefont
  {Chaturvedi}, \citenamefont {Das},\ and\ \citenamefont
  {Sengupta}}]{chaturvedi2017thermodynamic}%
  \BibitemOpen
  \bibfield  {author} {\bibinfo {author} {\bibfnamefont {P.}~\bibnamefont
  {Chaturvedi}}, \bibinfo {author} {\bibfnamefont {A.}~\bibnamefont {Das}},\
  and\ \bibinfo {author} {\bibfnamefont {G.}~\bibnamefont {Sengupta}},\
  }\bibfield  {title} {\bibinfo {title} {{Thermodynamic Geometry and Phase
  Transitions of Dyonic Charged AdS Black Holes}},\ }\href
  {https://doi.org/10.1140/epjc/s10052-017-4678-z} {\bibfield  {journal}
  {\bibinfo  {journal} {Eur. Phys. J. C}\ }\textbf {\bibinfo {volume} {77}},\
  \bibinfo {pages} {110} (\bibinfo {year} {2017})},\ \Eprint
  {https://arxiv.org/abs/1412.3880} {arXiv:1412.3880 [hep-th]} \BibitemShut
  {NoStop}%
\bibitem [{\citenamefont {Dutta}\ \emph {et~al.}(2013)\citenamefont {Dutta},
  \citenamefont {Jain},\ and\ \citenamefont {Soni}}]{dutta2013dyonic}%
  \BibitemOpen
  \bibfield  {author} {\bibinfo {author} {\bibfnamefont {S.}~\bibnamefont
  {Dutta}}, \bibinfo {author} {\bibfnamefont {A.}~\bibnamefont {Jain}},\ and\
  \bibinfo {author} {\bibfnamefont {R.}~\bibnamefont {Soni}},\ }\bibfield
  {title} {\bibinfo {title} {{Dyonic Black Hole and Holography}},\ }\href
  {https://doi.org/10.1007/JHEP12(2013)060} {\bibfield  {journal} {\bibinfo
  {journal} {JHEP}\ }\textbf {\bibinfo {volume} {12}},\ \bibinfo {pages}
  {060}},\ \Eprint {https://arxiv.org/abs/1310.1748} {arXiv:1310.1748 [hep-th]}
  \BibitemShut {NoStop}%
\bibitem [{\citenamefont {Balart}\ and\ \citenamefont
  {Vagenas}(2014)}]{Balart:2014cga}%
  \BibitemOpen
  \bibfield  {author} {\bibinfo {author} {\bibfnamefont {L.}~\bibnamefont
  {Balart}}\ and\ \bibinfo {author} {\bibfnamefont {E.~C.}\ \bibnamefont
  {Vagenas}},\ }\bibfield  {title} {\bibinfo {title} {{Regular black holes with
  a nonlinear electrodynamics source}},\ }\href
  {https://doi.org/10.1103/PhysRevD.90.124045} {\bibfield  {journal} {\bibinfo
  {journal} {Phys. Rev. D}\ }\textbf {\bibinfo {volume} {90}},\ \bibinfo
  {pages} {124045} (\bibinfo {year} {2014})},\ \Eprint
  {https://arxiv.org/abs/1408.0306} {arXiv:1408.0306 [gr-qc]} \BibitemShut
  {NoStop}%
\bibitem [{\citenamefont {Behrndt}\ \emph {et~al.}(1999)\citenamefont
  {Behrndt}, \citenamefont {Cvetic},\ and\ \citenamefont
  {Sabra}}]{Behrndt:1998jd}%
  \BibitemOpen
  \bibfield  {author} {\bibinfo {author} {\bibfnamefont {K.}~\bibnamefont
  {Behrndt}}, \bibinfo {author} {\bibfnamefont {M.}~\bibnamefont {Cvetic}},\
  and\ \bibinfo {author} {\bibfnamefont {W.~A.}\ \bibnamefont {Sabra}},\
  }\bibfield  {title} {\bibinfo {title} {{Nonextreme black holes of
  five-dimensional N=2 AdS supergravity}},\ }\href
  {https://doi.org/10.1016/S0550-3213(99)00243-6} {\bibfield  {journal}
  {\bibinfo  {journal} {Nucl. Phys. B}\ }\textbf {\bibinfo {volume} {553}},\
  \bibinfo {pages} {317} (\bibinfo {year} {1999})},\ \Eprint
  {https://arxiv.org/abs/hep-th/9810227} {arXiv:hep-th/9810227} \BibitemShut
  {NoStop}%
\bibitem [{\citenamefont {Duff}\ and\ \citenamefont {Liu}(1999)}]{Duff:1999gh}%
  \BibitemOpen
  \bibfield  {author} {\bibinfo {author} {\bibfnamefont {M.~J.}\ \bibnamefont
  {Duff}}\ and\ \bibinfo {author} {\bibfnamefont {J.~T.}\ \bibnamefont {Liu}},\
  }\bibfield  {title} {\bibinfo {title} {{Anti-de Sitter black holes in gauged
  N = 8 supergravity}},\ }\href {https://doi.org/10.1016/S0550-3213(99)00299-0}
  {\bibfield  {journal} {\bibinfo  {journal} {Nucl. Phys. B}\ }\textbf
  {\bibinfo {volume} {554}},\ \bibinfo {pages} {237} (\bibinfo {year}
  {1999})},\ \Eprint {https://arxiv.org/abs/hep-th/9901149}
  {arXiv:hep-th/9901149} \BibitemShut {NoStop}%
\bibitem [{\citenamefont {Cvetic}\ \emph {et~al.}(1999)\citenamefont {Cvetic},
  \citenamefont {Duff}, \citenamefont {Hoxha}, \citenamefont {Liu},
  \citenamefont {Lu}, \citenamefont {Lu}, \citenamefont {Martinez-Acosta},
  \citenamefont {Pope}, \citenamefont {Sati},\ and\ \citenamefont
  {Tran}}]{Cvetic:1999xp}%
  \BibitemOpen
  \bibfield  {author} {\bibinfo {author} {\bibfnamefont {M.}~\bibnamefont
  {Cvetic}}, \bibinfo {author} {\bibfnamefont {M.~J.}\ \bibnamefont {Duff}},
  \bibinfo {author} {\bibfnamefont {P.}~\bibnamefont {Hoxha}}, \bibinfo
  {author} {\bibfnamefont {J.~T.}\ \bibnamefont {Liu}}, \bibinfo {author}
  {\bibfnamefont {H.}~\bibnamefont {Lu}}, \bibinfo {author} {\bibfnamefont
  {J.~X.}\ \bibnamefont {Lu}}, \bibinfo {author} {\bibfnamefont
  {R.}~\bibnamefont {Martinez-Acosta}}, \bibinfo {author} {\bibfnamefont
  {C.~N.}\ \bibnamefont {Pope}}, \bibinfo {author} {\bibfnamefont
  {H.}~\bibnamefont {Sati}},\ and\ \bibinfo {author} {\bibfnamefont {T.~A.}\
  \bibnamefont {Tran}},\ }\bibfield  {title} {\bibinfo {title} {{Embedding AdS
  black holes in ten-dimensions and eleven-dimensions}},\ }\href
  {https://doi.org/10.1016/S0550-3213(99)00419-8} {\bibfield  {journal}
  {\bibinfo  {journal} {Nucl. Phys. B}\ }\textbf {\bibinfo {volume} {558}},\
  \bibinfo {pages} {96} (\bibinfo {year} {1999})},\ \Eprint
  {https://arxiv.org/abs/hep-th/9903214} {arXiv:hep-th/9903214} \BibitemShut
  {NoStop}%
\bibitem [{\citenamefont {Sahay}\ \emph {et~al.}(2010)\citenamefont {Sahay},
  \citenamefont {Sarkar},\ and\ \citenamefont {Sengupta}}]{Sahay:2010yq}%
  \BibitemOpen
  \bibfield  {author} {\bibinfo {author} {\bibfnamefont {A.}~\bibnamefont
  {Sahay}}, \bibinfo {author} {\bibfnamefont {T.}~\bibnamefont {Sarkar}},\ and\
  \bibinfo {author} {\bibfnamefont {G.}~\bibnamefont {Sengupta}},\ }\bibfield
  {title} {\bibinfo {title} {{On The Phase Structure and Thermodynamic Geometry
  of R-Charged Black Holes}},\ }\href {https://doi.org/10.1007/JHEP11(2010)125}
  {\bibfield  {journal} {\bibinfo  {journal} {JHEP}\ }\textbf {\bibinfo
  {volume} {11}},\ \bibinfo {pages} {125}},\ \Eprint
  {https://arxiv.org/abs/1009.2236} {arXiv:1009.2236 [hep-th]} \BibitemShut
  {NoStop}%
\bibitem [{\citenamefont {Priyadarshinee}\ and\ \citenamefont
  {Mahapatra}(2023)}]{Priyadarshinee:2023cmi}%
  \BibitemOpen
  \bibfield  {author} {\bibinfo {author} {\bibfnamefont {S.}~\bibnamefont
  {Priyadarshinee}}\ and\ \bibinfo {author} {\bibfnamefont {S.}~\bibnamefont
  {Mahapatra}},\ }\bibfield  {title} {\bibinfo {title} {{Analytic
  three-dimensional primary hair charged black holes and thermodynamics}},\
  }\href {https://doi.org/10.1103/PhysRevD.108.044017} {\bibfield  {journal}
  {\bibinfo  {journal} {Phys. Rev. D}\ }\textbf {\bibinfo {volume} {108}},\
  \bibinfo {pages} {044017} (\bibinfo {year} {2023})},\ \Eprint
  {https://arxiv.org/abs/2305.09172} {arXiv:2305.09172 [gr-qc]} \BibitemShut
  {NoStop}%
\bibitem [{\citenamefont {Hennigar}\ \emph {et~al.}(2015)\citenamefont
  {Hennigar}, \citenamefont {Brenna},\ and\ \citenamefont
  {Mann}}]{Hennigar:2015esa}%
  \BibitemOpen
  \bibfield  {author} {\bibinfo {author} {\bibfnamefont {R.~A.}\ \bibnamefont
  {Hennigar}}, \bibinfo {author} {\bibfnamefont {W.~G.}\ \bibnamefont
  {Brenna}},\ and\ \bibinfo {author} {\bibfnamefont {R.~B.}\ \bibnamefont
  {Mann}},\ }\bibfield  {title} {\bibinfo {title} {{$P - v$ criticality in
  quasitopological gravity}},\ }\href {https://doi.org/10.1007/JHEP07(2015)077}
  {\bibfield  {journal} {\bibinfo  {journal} {JHEP}\ }\textbf {\bibinfo
  {volume} {07}},\ \bibinfo {pages} {077}},\ \Eprint
  {https://arxiv.org/abs/1505.05517} {arXiv:1505.05517 [hep-th]} \BibitemShut
  {NoStop}%
\bibitem [{\citenamefont {Chen}\ \emph {et~al.}(2013)\citenamefont {Chen},
  \citenamefont {Liu}, \citenamefont {Liu},\ and\ \citenamefont
  {Jing}}]{Chen:2013ce}%
  \BibitemOpen
  \bibfield  {author} {\bibinfo {author} {\bibfnamefont {S.}~\bibnamefont
  {Chen}}, \bibinfo {author} {\bibfnamefont {X.}~\bibnamefont {Liu}}, \bibinfo
  {author} {\bibfnamefont {C.}~\bibnamefont {Liu}},\ and\ \bibinfo {author}
  {\bibfnamefont {J.}~\bibnamefont {Jing}},\ }\bibfield  {title} {\bibinfo
  {title} {{$P-V$ criticality of AdS black hole in $f(R)$ gravity}},\ }\href
  {https://doi.org/10.1088/0256-307X/30/6/060401} {\bibfield  {journal}
  {\bibinfo  {journal} {Chin. Phys. Lett.}\ }\textbf {\bibinfo {volume} {30}},\
  \bibinfo {pages} {060401} (\bibinfo {year} {2013})},\ \Eprint
  {https://arxiv.org/abs/1301.3234} {arXiv:1301.3234 [gr-qc]} \BibitemShut
  {NoStop}%
\bibitem [{\citenamefont {Kubiznak}\ and\ \citenamefont
  {Mann}(2012)}]{Kubiznak:2012wp}%
  \BibitemOpen
  \bibfield  {author} {\bibinfo {author} {\bibfnamefont {D.}~\bibnamefont
  {Kubiznak}}\ and\ \bibinfo {author} {\bibfnamefont {R.~B.}\ \bibnamefont
  {Mann}},\ }\bibfield  {title} {\bibinfo {title} {{P-V criticality of charged
  AdS black holes}},\ }\href {https://doi.org/10.1007/JHEP07(2012)033}
  {\bibfield  {journal} {\bibinfo  {journal} {JHEP}\ }\textbf {\bibinfo
  {volume} {07}},\ \bibinfo {pages} {033}},\ \Eprint
  {https://arxiv.org/abs/1205.0559} {arXiv:1205.0559 [hep-th]} \BibitemShut
  {NoStop}%
\bibitem [{\citenamefont {Dolan}(2015)}]{Dolan:2014jva}%
  \BibitemOpen
  \bibfield  {author} {\bibinfo {author} {\bibfnamefont {B.~P.}\ \bibnamefont
  {Dolan}},\ }\bibfield  {title} {\bibinfo {title} {{Black holes and Boyle's
  law \textemdash{} The thermodynamics of the cosmological constant}},\ }\href
  {https://doi.org/10.1142/S0217732315400027} {\bibfield  {journal} {\bibinfo
  {journal} {Mod. Phys. Lett. A}\ }\textbf {\bibinfo {volume} {30}},\ \bibinfo
  {pages} {1540002} (\bibinfo {year} {2015})},\ \Eprint
  {https://arxiv.org/abs/1408.4023} {arXiv:1408.4023 [gr-qc]} \BibitemShut
  {NoStop}%
\bibitem [{\citenamefont {Kastor}\ \emph {et~al.}(2009)\citenamefont {Kastor},
  \citenamefont {Ray},\ and\ \citenamefont {Traschen}}]{Kastor:2009wy}%
  \BibitemOpen
  \bibfield  {author} {\bibinfo {author} {\bibfnamefont {D.}~\bibnamefont
  {Kastor}}, \bibinfo {author} {\bibfnamefont {S.}~\bibnamefont {Ray}},\ and\
  \bibinfo {author} {\bibfnamefont {J.}~\bibnamefont {Traschen}},\ }\bibfield
  {title} {\bibinfo {title} {{Enthalpy and the Mechanics of AdS Black Holes}},\
  }\href {https://doi.org/10.1088/0264-9381/26/19/195011} {\bibfield  {journal}
  {\bibinfo  {journal} {Class. Quant. Grav.}\ }\textbf {\bibinfo {volume}
  {26}},\ \bibinfo {pages} {195011} (\bibinfo {year} {2009})},\ \Eprint
  {https://arxiv.org/abs/0904.2765} {arXiv:0904.2765 [hep-th]} \BibitemShut
  {NoStop}%
\bibitem [{\citenamefont {Dolan}(2011)}]{Dolan:2011xt}%
  \BibitemOpen
  \bibfield  {author} {\bibinfo {author} {\bibfnamefont {B.~P.}\ \bibnamefont
  {Dolan}},\ }\bibfield  {title} {\bibinfo {title} {{Pressure and volume in the
  first law of black hole thermodynamics}},\ }\href
  {https://doi.org/10.1088/0264-9381/28/23/235017} {\bibfield  {journal}
  {\bibinfo  {journal} {Class. Quant. Grav.}\ }\textbf {\bibinfo {volume}
  {28}},\ \bibinfo {pages} {235017} (\bibinfo {year} {2011})},\ \Eprint
  {https://arxiv.org/abs/1106.6260} {arXiv:1106.6260 [gr-qc]} \BibitemShut
  {NoStop}%
\bibitem [{\citenamefont {Dolan}(2012)}]{Dolan:2012jh}%
  \BibitemOpen
  \bibfield  {author} {\bibinfo {author} {\bibfnamefont {B.~P.}\ \bibnamefont
  {Dolan}},\ }\bibinfo {title} {{Where Is the PdV in the First Law of Black
  Hole Thermodynamics?}},\ in\ \href {https://doi.org/10.5772/52455} {\emph
  {\bibinfo {booktitle} {{Open Questions in Cosmology}}}}\ (\bibinfo
  {publisher} {INTECH},\ \bibinfo {year} {2012})\ \Eprint
  {https://arxiv.org/abs/1209.1272} {arXiv:1209.1272 [gr-qc]} \BibitemShut
  {NoStop}%
\bibitem [{\citenamefont {Kubiznak}\ \emph {et~al.}(2017)\citenamefont
  {Kubiznak}, \citenamefont {Mann},\ and\ \citenamefont
  {Teo}}]{Kubiznak:2016qmn}%
  \BibitemOpen
  \bibfield  {author} {\bibinfo {author} {\bibfnamefont {D.}~\bibnamefont
  {Kubiznak}}, \bibinfo {author} {\bibfnamefont {R.~B.}\ \bibnamefont {Mann}},\
  and\ \bibinfo {author} {\bibfnamefont {M.}~\bibnamefont {Teo}},\ }\bibfield
  {title} {\bibinfo {title} {{Black hole chemistry: thermodynamics with
  Lambda}},\ }\href {https://doi.org/10.1088/1361-6382/aa5c69} {\bibfield
  {journal} {\bibinfo  {journal} {Class. Quant. Grav.}\ }\textbf {\bibinfo
  {volume} {34}},\ \bibinfo {pages} {063001} (\bibinfo {year} {2017})},\
  \Eprint {https://arxiv.org/abs/1608.06147} {arXiv:1608.06147 [hep-th]}
  \BibitemShut {NoStop}%
\bibitem [{\citenamefont {Altamirano}\ \emph {et~al.}(2014)\citenamefont
  {Altamirano}, \citenamefont {Kubiznak}, \citenamefont {Mann},\ and\
  \citenamefont {Sherkatghanad}}]{Altamirano:2014tva}%
  \BibitemOpen
  \bibfield  {author} {\bibinfo {author} {\bibfnamefont {N.}~\bibnamefont
  {Altamirano}}, \bibinfo {author} {\bibfnamefont {D.}~\bibnamefont
  {Kubiznak}}, \bibinfo {author} {\bibfnamefont {R.~B.}\ \bibnamefont {Mann}},\
  and\ \bibinfo {author} {\bibfnamefont {Z.}~\bibnamefont {Sherkatghanad}},\
  }\bibfield  {title} {\bibinfo {title} {{Thermodynamics of rotating black
  holes and black rings: phase transitions and thermodynamic volume}},\ }\href
  {https://doi.org/10.3390/galaxies2010089} {\bibfield  {journal} {\bibinfo
  {journal} {Galaxies}\ }\textbf {\bibinfo {volume} {2}},\ \bibinfo {pages}
  {89} (\bibinfo {year} {2014})},\ \Eprint {https://arxiv.org/abs/1401.2586}
  {arXiv:1401.2586 [hep-th]} \BibitemShut {NoStop}%
\bibitem [{\citenamefont {Liu}\ \emph {et~al.}(2014)\citenamefont {Liu},
  \citenamefont {Zou},\ and\ \citenamefont {Wang}}]{Liu:2014gvf}%
  \BibitemOpen
  \bibfield  {author} {\bibinfo {author} {\bibfnamefont {Y.}~\bibnamefont
  {Liu}}, \bibinfo {author} {\bibfnamefont {D.-C.}\ \bibnamefont {Zou}},\ and\
  \bibinfo {author} {\bibfnamefont {B.}~\bibnamefont {Wang}},\ }\bibfield
  {title} {\bibinfo {title} {{Signature of the Van der Waals like small-large
  charged AdS black hole phase transition in quasinormal modes}},\ }\href
  {https://doi.org/10.1007/JHEP09(2014)179} {\bibfield  {journal} {\bibinfo
  {journal} {JHEP}\ }\textbf {\bibinfo {volume} {09}},\ \bibinfo {pages}
  {179}},\ \Eprint {https://arxiv.org/abs/1405.2644} {arXiv:1405.2644 [hep-th]}
  \BibitemShut {NoStop}%
\bibitem [{\citenamefont {Shen}\ \emph {et~al.}(2007)\citenamefont {Shen},
  \citenamefont {Wang}, \citenamefont {Lin}, \citenamefont {Cai},\ and\
  \citenamefont {Su}}]{Shen:2007xk}%
  \BibitemOpen
  \bibfield  {author} {\bibinfo {author} {\bibfnamefont {J.}~\bibnamefont
  {Shen}}, \bibinfo {author} {\bibfnamefont {B.}~\bibnamefont {Wang}}, \bibinfo
  {author} {\bibfnamefont {C.-Y.}\ \bibnamefont {Lin}}, \bibinfo {author}
  {\bibfnamefont {R.-G.}\ \bibnamefont {Cai}},\ and\ \bibinfo {author}
  {\bibfnamefont {R.-K.}\ \bibnamefont {Su}},\ }\bibfield  {title} {\bibinfo
  {title} {{The phase transition and the Quasi-Normal Modes of black Holes}},\
  }\href {https://doi.org/10.1088/1126-6708/2007/07/037} {\bibfield  {journal}
  {\bibinfo  {journal} {JHEP}\ }\textbf {\bibinfo {volume} {07}},\ \bibinfo
  {pages} {037}},\ \Eprint {https://arxiv.org/abs/hep-th/0703102}
  {arXiv:hep-th/0703102} \BibitemShut {NoStop}%
\bibitem [{\citenamefont {Rao}\ \emph {et~al.}(2007)\citenamefont {Rao},
  \citenamefont {Wang},\ and\ \citenamefont {Yang}}]{Rao:2007zzb}%
  \BibitemOpen
  \bibfield  {author} {\bibinfo {author} {\bibfnamefont {X.-P.}\ \bibnamefont
  {Rao}}, \bibinfo {author} {\bibfnamefont {B.}~\bibnamefont {Wang}},\ and\
  \bibinfo {author} {\bibfnamefont {G.-H.}\ \bibnamefont {Yang}},\ }\bibfield
  {title} {\bibinfo {title} {{Quasinormal modes and phase transition of black
  holes}},\ }\href {https://doi.org/10.1016/j.physletb.2007.04.049} {\bibfield
  {journal} {\bibinfo  {journal} {Phys. Lett. B}\ }\textbf {\bibinfo {volume}
  {649}},\ \bibinfo {pages} {472} (\bibinfo {year} {2007})},\ \Eprint
  {https://arxiv.org/abs/0712.0645} {arXiv:0712.0645 [gr-qc]} \BibitemShut
  {NoStop}%
\bibitem [{\citenamefont {He}\ \emph {et~al.}(2010)\citenamefont {He},
  \citenamefont {Wang}, \citenamefont {Cai},\ and\ \citenamefont
  {Lin}}]{He:2010zb}%
  \BibitemOpen
  \bibfield  {author} {\bibinfo {author} {\bibfnamefont {X.}~\bibnamefont
  {He}}, \bibinfo {author} {\bibfnamefont {B.}~\bibnamefont {Wang}}, \bibinfo
  {author} {\bibfnamefont {R.-G.}\ \bibnamefont {Cai}},\ and\ \bibinfo {author}
  {\bibfnamefont {C.-Y.}\ \bibnamefont {Lin}},\ }\bibfield  {title} {\bibinfo
  {title} {{Signature of the black hole phase transition in quasinormal
  modes}},\ }\href {https://doi.org/10.1016/j.physletb.2010.04.006} {\bibfield
  {journal} {\bibinfo  {journal} {Phys. Lett. B}\ }\textbf {\bibinfo {volume}
  {688}},\ \bibinfo {pages} {230} (\bibinfo {year} {2010})},\ \Eprint
  {https://arxiv.org/abs/1002.2679} {arXiv:1002.2679 [hep-th]} \BibitemShut
  {NoStop}%
\bibitem [{\citenamefont {Mahapatra}(2016)}]{Mahapatra:2016dae}%
  \BibitemOpen
  \bibfield  {author} {\bibinfo {author} {\bibfnamefont {S.}~\bibnamefont
  {Mahapatra}},\ }\bibfield  {title} {\bibinfo {title} {{Thermodynamics, Phase
  Transition and Quasinormal modes with Weyl corrections}},\ }\href
  {https://doi.org/10.1007/JHEP04(2016)142} {\bibfield  {journal} {\bibinfo
  {journal} {JHEP}\ }\textbf {\bibinfo {volume} {04}},\ \bibinfo {pages}
  {142}},\ \Eprint {https://arxiv.org/abs/1602.03007} {arXiv:1602.03007
  [hep-th]} \BibitemShut {NoStop}%
\bibitem [{\citenamefont {Priyadarshinee}(2023)}]{Priyadarshinee:2023exb}%
  \BibitemOpen
  \bibfield  {author} {\bibinfo {author} {\bibfnamefont {S.}~\bibnamefont
  {Priyadarshinee}},\ }\bibfield  {title} {\bibinfo {title} {{Quasi-normal mode
  of dyonic hairy black hole and its interplay with phase transitions}},\
  }\href@noop {} {\  (\bibinfo {year} {2023})},\ \Eprint
  {https://arxiv.org/abs/2308.05719} {arXiv:2308.05719 [gr-qc]} \BibitemShut
  {NoStop}%
\bibitem [{\citenamefont {Belhaj}\ \emph {et~al.}(2020)\citenamefont {Belhaj},
  \citenamefont {Chakhchi}, \citenamefont {El~Moumni}, \citenamefont
  {Khalloufi},\ and\ \citenamefont {Masmar}}]{Belhaj:2020nqy}%
  \BibitemOpen
  \bibfield  {author} {\bibinfo {author} {\bibfnamefont {A.}~\bibnamefont
  {Belhaj}}, \bibinfo {author} {\bibfnamefont {L.}~\bibnamefont {Chakhchi}},
  \bibinfo {author} {\bibfnamefont {H.}~\bibnamefont {El~Moumni}}, \bibinfo
  {author} {\bibfnamefont {J.}~\bibnamefont {Khalloufi}},\ and\ \bibinfo
  {author} {\bibfnamefont {K.}~\bibnamefont {Masmar}},\ }\bibfield  {title}
  {\bibinfo {title} {{Thermal Image and Phase Transitions of Charged AdS Black
  Holes using Shadow Analysis}},\ }\href
  {https://doi.org/10.1142/S0217751X20501705} {\bibfield  {journal} {\bibinfo
  {journal} {Int. J. Mod. Phys. A}\ }\textbf {\bibinfo {volume} {35}},\
  \bibinfo {pages} {2050170} (\bibinfo {year} {2020})},\ \Eprint
  {https://arxiv.org/abs/2005.05893} {arXiv:2005.05893 [gr-qc]} \BibitemShut
  {NoStop}%
\bibitem [{\citenamefont {Cai}\ and\ \citenamefont {Miao}(2021)}]{Cai:2021uov}%
  \BibitemOpen
  \bibfield  {author} {\bibinfo {author} {\bibfnamefont {X.-C.}\ \bibnamefont
  {Cai}}\ and\ \bibinfo {author} {\bibfnamefont {Y.-G.}\ \bibnamefont {Miao}},\
  }\bibfield  {title} {\bibinfo {title} {{Can we know about black hole
  thermodynamics through shadows?}},\ }\href@noop {} {\  (\bibinfo {year}
  {2021})},\ \Eprint {https://arxiv.org/abs/2107.08352} {arXiv:2107.08352
  [gr-qc]} \BibitemShut {NoStop}%
\bibitem [{\citenamefont {Zhang}\ and\ \citenamefont
  {Guo}(2020)}]{Zhang:2019glo}%
  \BibitemOpen
  \bibfield  {author} {\bibinfo {author} {\bibfnamefont {M.}~\bibnamefont
  {Zhang}}\ and\ \bibinfo {author} {\bibfnamefont {M.}~\bibnamefont {Guo}},\
  }\bibfield  {title} {\bibinfo {title} {{Can shadows reflect phase structures
  of black holes?}},\ }\href {https://doi.org/10.1140/epjc/s10052-020-8389-5}
  {\bibfield  {journal} {\bibinfo  {journal} {Eur. Phys. J. C}\ }\textbf
  {\bibinfo {volume} {80}},\ \bibinfo {pages} {790} (\bibinfo {year} {2020})},\
  \Eprint {https://arxiv.org/abs/1909.07033} {arXiv:1909.07033 [gr-qc]}
  \BibitemShut {NoStop}%
\bibitem [{\citenamefont {Wei}\ and\ \citenamefont {Liu}(2018)}]{Wei:2017mwc}%
  \BibitemOpen
  \bibfield  {author} {\bibinfo {author} {\bibfnamefont {S.-W.}\ \bibnamefont
  {Wei}}\ and\ \bibinfo {author} {\bibfnamefont {Y.-X.}\ \bibnamefont {Liu}},\
  }\bibfield  {title} {\bibinfo {title} {{Photon orbits and thermodynamic phase
  transition of $d$-dimensional charged AdS black holes}},\ }\href
  {https://doi.org/10.1103/PhysRevD.97.104027} {\bibfield  {journal} {\bibinfo
  {journal} {Phys. Rev. D}\ }\textbf {\bibinfo {volume} {97}},\ \bibinfo
  {pages} {104027} (\bibinfo {year} {2018})},\ \Eprint
  {https://arxiv.org/abs/1711.01522} {arXiv:1711.01522 [gr-qc]} \BibitemShut
  {NoStop}%
\bibitem [{\citenamefont {Chabab}\ \emph {et~al.}(2020)\citenamefont {Chabab},
  \citenamefont {El~Moumni}, \citenamefont {Iraoui},\ and\ \citenamefont
  {Masmar}}]{Chabab:2019kfs}%
  \BibitemOpen
  \bibfield  {author} {\bibinfo {author} {\bibfnamefont {M.}~\bibnamefont
  {Chabab}}, \bibinfo {author} {\bibfnamefont {H.}~\bibnamefont {El~Moumni}},
  \bibinfo {author} {\bibfnamefont {S.}~\bibnamefont {Iraoui}},\ and\ \bibinfo
  {author} {\bibfnamefont {K.}~\bibnamefont {Masmar}},\ }\bibfield  {title}
  {\bibinfo {title} {{Probing correlation between photon orbits and phase
  structure of charged AdS black hole in massive gravity background}},\ }\href
  {https://doi.org/10.1142/S0217751X19502312} {\bibfield  {journal} {\bibinfo
  {journal} {Int. J. Mod. Phys. A}\ }\textbf {\bibinfo {volume} {34}},\
  \bibinfo {pages} {1950231} (\bibinfo {year} {2020})},\ \Eprint
  {https://arxiv.org/abs/1902.00557} {arXiv:1902.00557 [hep-th]} \BibitemShut
  {NoStop}%
\bibitem [{\citenamefont {Han}\ \emph {et~al.}(2020)\citenamefont {Han},
  \citenamefont {Jiang}, \citenamefont {Zhang},\ and\ \citenamefont
  {Liu}}]{Han:2018ooi}%
  \BibitemOpen
  \bibfield  {author} {\bibinfo {author} {\bibfnamefont {S.-Z.}\ \bibnamefont
  {Han}}, \bibinfo {author} {\bibfnamefont {J.}~\bibnamefont {Jiang}}, \bibinfo
  {author} {\bibfnamefont {M.}~\bibnamefont {Zhang}},\ and\ \bibinfo {author}
  {\bibfnamefont {W.-B.}\ \bibnamefont {Liu}},\ }\bibfield  {title} {\bibinfo
  {title} {{Photon sphere and phase transition of d-dimensional (d
  \ensuremath{\geq} 5) charged Gauss\textendash{}Bonnet AdS black holes}},\
  }\href {https://doi.org/10.1088/1572-9494/aba259} {\bibfield  {journal}
  {\bibinfo  {journal} {Commun. Theor. Phys.}\ }\textbf {\bibinfo {volume}
  {72}},\ \bibinfo {pages} {105402} (\bibinfo {year} {2020})},\ \Eprint
  {https://arxiv.org/abs/1812.11862} {arXiv:1812.11862 [gr-qc]} \BibitemShut
  {NoStop}%
\bibitem [{\citenamefont {Johnson}(2014)}]{Johnson:2013dka}%
  \BibitemOpen
  \bibfield  {author} {\bibinfo {author} {\bibfnamefont {C.~V.}\ \bibnamefont
  {Johnson}},\ }\bibfield  {title} {\bibinfo {title} {{Large N Phase
  Transitions, Finite Volume, and Entanglement Entropy}},\ }\href
  {https://doi.org/10.1007/JHEP03(2014)047} {\bibfield  {journal} {\bibinfo
  {journal} {JHEP}\ }\textbf {\bibinfo {volume} {03}},\ \bibinfo {pages}
  {047}},\ \Eprint {https://arxiv.org/abs/1306.4955} {arXiv:1306.4955 [hep-th]}
  \BibitemShut {NoStop}%
\bibitem [{\citenamefont {Dey}\ \emph {et~al.}(2016)\citenamefont {Dey},
  \citenamefont {Mahapatra},\ and\ \citenamefont {Sarkar}}]{Dey:2015ytd}%
  \BibitemOpen
  \bibfield  {author} {\bibinfo {author} {\bibfnamefont {A.}~\bibnamefont
  {Dey}}, \bibinfo {author} {\bibfnamefont {S.}~\bibnamefont {Mahapatra}},\
  and\ \bibinfo {author} {\bibfnamefont {T.}~\bibnamefont {Sarkar}},\
  }\bibfield  {title} {\bibinfo {title} {{Thermodynamics and Entanglement
  Entropy with Weyl Corrections}},\ }\href
  {https://doi.org/10.1103/PhysRevD.94.026006} {\bibfield  {journal} {\bibinfo
  {journal} {Phys. Rev. D}\ }\textbf {\bibinfo {volume} {94}},\ \bibinfo
  {pages} {026006} (\bibinfo {year} {2016})},\ \Eprint
  {https://arxiv.org/abs/1512.07117} {arXiv:1512.07117 [hep-th]} \BibitemShut
  {NoStop}%
\bibitem [{\citenamefont {Dudal}\ and\ \citenamefont
  {Mahapatra}(2018)}]{Dudal:2018ztm}%
  \BibitemOpen
  \bibfield  {author} {\bibinfo {author} {\bibfnamefont {D.}~\bibnamefont
  {Dudal}}\ and\ \bibinfo {author} {\bibfnamefont {S.}~\bibnamefont
  {Mahapatra}},\ }\bibfield  {title} {\bibinfo {title} {{Interplay between the
  holographic QCD phase diagram and entanglement entropy}},\ }\href
  {https://doi.org/10.1007/JHEP07(2018)120} {\bibfield  {journal} {\bibinfo
  {journal} {JHEP}\ }\textbf {\bibinfo {volume} {07}},\ \bibinfo {pages}
  {120}},\ \Eprint {https://arxiv.org/abs/1805.02938} {arXiv:1805.02938
  [hep-th]} \BibitemShut {NoStop}%
\bibitem [{\citenamefont {Mahapatra}(2019)}]{Mahapatra:2019uql}%
  \BibitemOpen
  \bibfield  {author} {\bibinfo {author} {\bibfnamefont {S.}~\bibnamefont
  {Mahapatra}},\ }\bibfield  {title} {\bibinfo {title} {{Interplay between the
  holographic QCD phase diagram and mutual \textbackslash{}\& $n$-partite
  information}},\ }\href {https://doi.org/10.1007/JHEP04(2019)137} {\bibfield
  {journal} {\bibinfo  {journal} {JHEP}\ }\textbf {\bibinfo {volume} {04}},\
  \bibinfo {pages} {137}},\ \Eprint {https://arxiv.org/abs/1903.05927}
  {arXiv:1903.05927 [hep-th]} \BibitemShut {NoStop}%
\bibitem [{\citenamefont {Jain}\ \emph {et~al.}(2023)\citenamefont {Jain},
  \citenamefont {Jena},\ and\ \citenamefont {Mahapatra}}]{Jain:2022hxl}%
  \BibitemOpen
  \bibfield  {author} {\bibinfo {author} {\bibfnamefont {P.}~\bibnamefont
  {Jain}}, \bibinfo {author} {\bibfnamefont {S.~S.}\ \bibnamefont {Jena}},\
  and\ \bibinfo {author} {\bibfnamefont {S.}~\bibnamefont {Mahapatra}},\
  }\bibfield  {title} {\bibinfo {title} {{Holographic confining-deconfining
  gauge theories and entanglement measures with a magnetic field}},\ }\href
  {https://doi.org/10.1103/PhysRevD.107.086016} {\bibfield  {journal} {\bibinfo
   {journal} {Phys. Rev. D}\ }\textbf {\bibinfo {volume} {107}},\ \bibinfo
  {pages} {086016} (\bibinfo {year} {2023})},\ \Eprint
  {https://arxiv.org/abs/2209.15355} {arXiv:2209.15355 [hep-th]} \BibitemShut
  {NoStop}%
\bibitem [{\citenamefont {Jain}\ and\ \citenamefont
  {Mahapatra}(2020)}]{Jain:2020rbb}%
  \BibitemOpen
  \bibfield  {author} {\bibinfo {author} {\bibfnamefont {P.}~\bibnamefont
  {Jain}}\ and\ \bibinfo {author} {\bibfnamefont {S.}~\bibnamefont
  {Mahapatra}},\ }\bibfield  {title} {\bibinfo {title} {{Mixed state
  entanglement measures as probe for confinement}},\ }\href
  {https://doi.org/10.1103/PhysRevD.102.126022} {\bibfield  {journal} {\bibinfo
   {journal} {Phys. Rev. D}\ }\textbf {\bibinfo {volume} {102}},\ \bibinfo
  {pages} {126022} (\bibinfo {year} {2020})},\ \Eprint
  {https://arxiv.org/abs/2010.07702} {arXiv:2010.07702 [hep-th]} \BibitemShut
  {NoStop}%
\bibitem [{\citenamefont {Cardoso}\ \emph {et~al.}(2009)\citenamefont
  {Cardoso}, \citenamefont {Miranda}, \citenamefont {Berti}, \citenamefont
  {Witek},\ and\ \citenamefont {Zanchin}}]{Cardoso:2008bp}%
  \BibitemOpen
  \bibfield  {author} {\bibinfo {author} {\bibfnamefont {V.}~\bibnamefont
  {Cardoso}}, \bibinfo {author} {\bibfnamefont {A.~S.}\ \bibnamefont
  {Miranda}}, \bibinfo {author} {\bibfnamefont {E.}~\bibnamefont {Berti}},
  \bibinfo {author} {\bibfnamefont {H.}~\bibnamefont {Witek}},\ and\ \bibinfo
  {author} {\bibfnamefont {V.~T.}\ \bibnamefont {Zanchin}},\ }\bibfield
  {title} {\bibinfo {title} {{Geodesic stability, Lyapunov exponents and
  quasinormal modes}},\ }\href {https://doi.org/10.1103/PhysRevD.79.064016}
  {\bibfield  {journal} {\bibinfo  {journal} {Phys. Rev. D}\ }\textbf {\bibinfo
  {volume} {79}},\ \bibinfo {pages} {064016} (\bibinfo {year} {2009})},\
  \Eprint {https://arxiv.org/abs/0812.1806} {arXiv:0812.1806 [hep-th]}
  \BibitemShut {NoStop}%
\bibitem [{\citenamefont {Guo}\ \emph {et~al.}(2022{\natexlab{a}})\citenamefont
  {Guo}, \citenamefont {Wang}, \citenamefont {Wu},\ and\ \citenamefont
  {Yang}}]{Guo:2021enm}%
  \BibitemOpen
  \bibfield  {author} {\bibinfo {author} {\bibfnamefont {G.}~\bibnamefont
  {Guo}}, \bibinfo {author} {\bibfnamefont {P.}~\bibnamefont {Wang}}, \bibinfo
  {author} {\bibfnamefont {H.}~\bibnamefont {Wu}},\ and\ \bibinfo {author}
  {\bibfnamefont {H.}~\bibnamefont {Yang}},\ }\bibfield  {title} {\bibinfo
  {title} {{Quasinormal modes of black holes with multiple photon spheres}},\
  }\href {https://doi.org/10.1007/JHEP06(2022)060} {\bibfield  {journal}
  {\bibinfo  {journal} {JHEP}\ }\textbf {\bibinfo {volume} {06}},\ \bibinfo
  {pages} {060}},\ \Eprint {https://arxiv.org/abs/2112.14133} {arXiv:2112.14133
  [gr-qc]} \BibitemShut {NoStop}%
\bibitem [{\citenamefont {Guo}\ \emph {et~al.}(2022{\natexlab{b}})\citenamefont
  {Guo}, \citenamefont {Lu}, \citenamefont {Mu},\ and\ \citenamefont
  {Wang}}]{guo2022probing}%
  \BibitemOpen
  \bibfield  {author} {\bibinfo {author} {\bibfnamefont {X.}~\bibnamefont
  {Guo}}, \bibinfo {author} {\bibfnamefont {Y.}~\bibnamefont {Lu}}, \bibinfo
  {author} {\bibfnamefont {B.}~\bibnamefont {Mu}},\ and\ \bibinfo {author}
  {\bibfnamefont {P.}~\bibnamefont {Wang}},\ }\bibfield  {title} {\bibinfo
  {title} {{Probing phase structure of black holes with Lyapunov exponents}},\
  }\href {https://doi.org/10.1007/JHEP08(2022)153} {\bibfield  {journal}
  {\bibinfo  {journal} {JHEP}\ }\textbf {\bibinfo {volume} {08}},\ \bibinfo
  {pages} {153}},\ \Eprint {https://arxiv.org/abs/2205.02122} {arXiv:2205.02122
  [gr-qc]} \BibitemShut {NoStop}%
\bibitem [{\citenamefont {Yang}\ \emph {et~al.}(2023)\citenamefont {Yang},
  \citenamefont {Tao}, \citenamefont {Mu},\ and\ \citenamefont
  {He}}]{yang2023lyapunov}%
  \BibitemOpen
  \bibfield  {author} {\bibinfo {author} {\bibfnamefont {S.}~\bibnamefont
  {Yang}}, \bibinfo {author} {\bibfnamefont {J.}~\bibnamefont {Tao}}, \bibinfo
  {author} {\bibfnamefont {B.}~\bibnamefont {Mu}},\ and\ \bibinfo {author}
  {\bibfnamefont {A.}~\bibnamefont {He}},\ }\bibfield  {title} {\bibinfo
  {title} {{Lyapunov exponents and phase transitions of Born-Infeld AdS black
  holes}},\ }\href {https://doi.org/10.1088/1475-7516/2023/07/045} {\bibfield
  {journal} {\bibinfo  {journal} {JCAP}\ }\textbf {\bibinfo {volume} {07}},\
  \bibinfo {pages} {045}},\ \Eprint {https://arxiv.org/abs/2304.01877}
  {arXiv:2304.01877 [gr-qc]} \BibitemShut {NoStop}%
\bibitem [{\citenamefont {Lyu}\ \emph {et~al.}(2023)\citenamefont {Lyu},
  \citenamefont {Tao},\ and\ \citenamefont {Wang}}]{lyu2023probing}%
  \BibitemOpen
  \bibfield  {author} {\bibinfo {author} {\bibfnamefont {X.}~\bibnamefont
  {Lyu}}, \bibinfo {author} {\bibfnamefont {J.}~\bibnamefont {Tao}},\ and\
  \bibinfo {author} {\bibfnamefont {P.}~\bibnamefont {Wang}},\ }\bibfield
  {title} {\bibinfo {title} {{Probing the thermodynamics of charged Gauss
  Bonnet AdS black holes with the Lyapunov exponent}},\ }\href@noop {} {\
  (\bibinfo {year} {2023})},\ \Eprint {https://arxiv.org/abs/2312.11912}
  {arXiv:2312.11912 [gr-qc]} \BibitemShut {NoStop}%
\bibitem [{\citenamefont {Kumara}\ \emph {et~al.}(2024)\citenamefont {Kumara},
  \citenamefont {Punacha},\ and\ \citenamefont {Ali}}]{kumara2024lyapunov}%
  \BibitemOpen
  \bibfield  {author} {\bibinfo {author} {\bibfnamefont {A.~N.}\ \bibnamefont
  {Kumara}}, \bibinfo {author} {\bibfnamefont {S.}~\bibnamefont {Punacha}},\
  and\ \bibinfo {author} {\bibfnamefont {M.~S.}\ \bibnamefont {Ali}},\
  }\bibfield  {title} {\bibinfo {title} {{Lyapunov Exponents and Phase
  Structure of Lifshitz and Hyperscaling Violating Black Holes}},\ }\href@noop
  {} {\  (\bibinfo {year} {2024})},\ \Eprint {https://arxiv.org/abs/2401.05181}
  {arXiv:2401.05181 [gr-qc]} \BibitemShut {NoStop}%
\bibitem [{\citenamefont {Du}\ \emph {et~al.}(2024)\citenamefont {Du},
  \citenamefont {Li}, \citenamefont {Ma},\ and\ \citenamefont
  {Gu}}]{Du:2024uhd}%
  \BibitemOpen
  \bibfield  {author} {\bibinfo {author} {\bibfnamefont {Y.-Z.}\ \bibnamefont
  {Du}}, \bibinfo {author} {\bibfnamefont {H.-F.}\ \bibnamefont {Li}}, \bibinfo
  {author} {\bibfnamefont {Y.-B.}\ \bibnamefont {Ma}},\ and\ \bibinfo {author}
  {\bibfnamefont {Q.}~\bibnamefont {Gu}},\ }\bibfield  {title} {\bibinfo
  {title} {{Phase structure of the de Sitter Spacetime with KR field based on
  the Lyapunov exponent}},\ }\href@noop {} {\  (\bibinfo {year} {2024})},\
  \Eprint {https://arxiv.org/abs/2403.20083} {arXiv:2403.20083 [hep-th]}
  \BibitemShut {NoStop}%
\bibitem [{\citenamefont {Gogoi}\ \emph {et~al.}(2024)\citenamefont {Gogoi},
  \citenamefont {Acharjee},\ and\ \citenamefont {Phukon}}]{Gogoi:2024akv}%
  \BibitemOpen
  \bibfield  {author} {\bibinfo {author} {\bibfnamefont {N.~J.}\ \bibnamefont
  {Gogoi}}, \bibinfo {author} {\bibfnamefont {S.}~\bibnamefont {Acharjee}},\
  and\ \bibinfo {author} {\bibfnamefont {P.}~\bibnamefont {Phukon}},\
  }\bibfield  {title} {\bibinfo {title} {{Lyapunov Exponents and Phase
  Transition of Hayward AdS Black Hole}},\ }\href@noop {} {\  (\bibinfo {year}
  {2024})},\ \Eprint {https://arxiv.org/abs/2404.03947} {arXiv:2404.03947
  [hep-th]} \BibitemShut {NoStop}%
\bibitem [{\citenamefont {Dudal}\ and\ \citenamefont
  {Mahapatra}(2017)}]{Dudal:2017max}%
  \BibitemOpen
  \bibfield  {author} {\bibinfo {author} {\bibfnamefont {D.}~\bibnamefont
  {Dudal}}\ and\ \bibinfo {author} {\bibfnamefont {S.}~\bibnamefont
  {Mahapatra}},\ }\bibfield  {title} {\bibinfo {title} {{Thermal entropy of a
  quark-antiquark pair above and below deconfinement from a dynamical
  holographic QCD model}},\ }\href {https://doi.org/10.1103/PhysRevD.96.126010}
  {\bibfield  {journal} {\bibinfo  {journal} {Phys. Rev. D}\ }\textbf {\bibinfo
  {volume} {96}},\ \bibinfo {pages} {126010} (\bibinfo {year} {2017})},\
  \Eprint {https://arxiv.org/abs/1708.06995} {arXiv:1708.06995 [hep-th]}
  \BibitemShut {NoStop}%
\bibitem [{\citenamefont {Ayon-Beato}\ and\ \citenamefont
  {Garcia}(2000)}]{Ayon-Beato:2000mjt}%
  \BibitemOpen
  \bibfield  {author} {\bibinfo {author} {\bibfnamefont {E.}~\bibnamefont
  {Ayon-Beato}}\ and\ \bibinfo {author} {\bibfnamefont {A.}~\bibnamefont
  {Garcia}},\ }\bibfield  {title} {\bibinfo {title} {{The Bardeen model as a
  nonlinear magnetic monopole}},\ }\href
  {https://doi.org/10.1016/S0370-2693(00)01125-4} {\bibfield  {journal}
  {\bibinfo  {journal} {Phys. Lett. B}\ }\textbf {\bibinfo {volume} {493}},\
  \bibinfo {pages} {149} (\bibinfo {year} {2000})},\ \Eprint
  {https://arxiv.org/abs/gr-qc/0009077} {arXiv:gr-qc/0009077} \BibitemShut
  {NoStop}%
\bibitem [{\citenamefont {Boulware}\ and\ \citenamefont
  {Deser}(1985)}]{Boulware:1985wk}%
  \BibitemOpen
  \bibfield  {author} {\bibinfo {author} {\bibfnamefont {D.~G.}\ \bibnamefont
  {Boulware}}\ and\ \bibinfo {author} {\bibfnamefont {S.}~\bibnamefont
  {Deser}},\ }\bibfield  {title} {\bibinfo {title} {{String Generated Gravity
  Models}},\ }\href {https://doi.org/10.1103/PhysRevLett.55.2656} {\bibfield
  {journal} {\bibinfo  {journal} {Phys. Rev. Lett.}\ }\textbf {\bibinfo
  {volume} {55}},\ \bibinfo {pages} {2656} (\bibinfo {year}
  {1985})}\BibitemShut {NoStop}%
\bibitem [{\citenamefont {Deser}\ and\ \citenamefont
  {Tekin}(2003)}]{Deser:2002jk}%
  \BibitemOpen
  \bibfield  {author} {\bibinfo {author} {\bibfnamefont {S.}~\bibnamefont
  {Deser}}\ and\ \bibinfo {author} {\bibfnamefont {B.}~\bibnamefont {Tekin}},\
  }\bibfield  {title} {\bibinfo {title} {{Energy in generic higher curvature
  gravity theories}},\ }\href {https://doi.org/10.1103/PhysRevD.67.084009}
  {\bibfield  {journal} {\bibinfo  {journal} {Phys. Rev. D}\ }\textbf {\bibinfo
  {volume} {67}},\ \bibinfo {pages} {084009} (\bibinfo {year} {2003})},\
  \Eprint {https://arxiv.org/abs/hep-th/0212292} {arXiv:hep-th/0212292}
  \BibitemShut {NoStop}%
\bibitem [{\citenamefont {Gross}\ and\ \citenamefont
  {Sloan}(1987)}]{Gross:1986mw}%
  \BibitemOpen
  \bibfield  {author} {\bibinfo {author} {\bibfnamefont {D.~J.}\ \bibnamefont
  {Gross}}\ and\ \bibinfo {author} {\bibfnamefont {J.~H.}\ \bibnamefont
  {Sloan}},\ }\bibfield  {title} {\bibinfo {title} {{The Quartic Effective
  Action for the Heterotic String}},\ }\href
  {https://doi.org/10.1016/0550-3213(87)90465-2} {\bibfield  {journal}
  {\bibinfo  {journal} {Nucl. Phys. B}\ }\textbf {\bibinfo {volume} {291}},\
  \bibinfo {pages} {41} (\bibinfo {year} {1987})}\BibitemShut {NoStop}%
\bibitem [{\citenamefont {Fierz}\ and\ \citenamefont
  {Pauli}(1939)}]{Fierz:1939ix}%
  \BibitemOpen
  \bibfield  {author} {\bibinfo {author} {\bibfnamefont {M.}~\bibnamefont
  {Fierz}}\ and\ \bibinfo {author} {\bibfnamefont {W.}~\bibnamefont {Pauli}},\
  }\bibfield  {title} {\bibinfo {title} {{On relativistic wave equations for
  particles of arbitrary spin in an electromagnetic field}},\ }\href
  {https://doi.org/10.1098/rspa.1939.0140} {\bibfield  {journal} {\bibinfo
  {journal} {Proc. Roy. Soc. Lond. A}\ }\textbf {\bibinfo {volume} {173}},\
  \bibinfo {pages} {211} (\bibinfo {year} {1939})}\BibitemShut {NoStop}%
\bibitem [{\citenamefont {de~Rham}\ and\ \citenamefont
  {Gabadadze}(2010)}]{deRham:2010ik}%
  \BibitemOpen
  \bibfield  {author} {\bibinfo {author} {\bibfnamefont {C.}~\bibnamefont
  {de~Rham}}\ and\ \bibinfo {author} {\bibfnamefont {G.}~\bibnamefont
  {Gabadadze}},\ }\bibfield  {title} {\bibinfo {title} {{Generalization of the
  Fierz-Pauli Action}},\ }\href {https://doi.org/10.1103/PhysRevD.82.044020}
  {\bibfield  {journal} {\bibinfo  {journal} {Phys. Rev. D}\ }\textbf {\bibinfo
  {volume} {82}},\ \bibinfo {pages} {044020} (\bibinfo {year} {2010})},\
  \Eprint {https://arxiv.org/abs/1007.0443} {arXiv:1007.0443 [hep-th]}
  \BibitemShut {NoStop}%
\bibitem [{\citenamefont {de~Rham}\ \emph {et~al.}(2011)\citenamefont
  {de~Rham}, \citenamefont {Gabadadze},\ and\ \citenamefont
  {Tolley}}]{deRham:2010kj}%
  \BibitemOpen
  \bibfield  {author} {\bibinfo {author} {\bibfnamefont {C.}~\bibnamefont
  {de~Rham}}, \bibinfo {author} {\bibfnamefont {G.}~\bibnamefont {Gabadadze}},\
  and\ \bibinfo {author} {\bibfnamefont {A.~J.}\ \bibnamefont {Tolley}},\
  }\bibfield  {title} {\bibinfo {title} {{Resummation of Massive Gravity}},\
  }\href {https://doi.org/10.1103/PhysRevLett.106.231101} {\bibfield  {journal}
  {\bibinfo  {journal} {Phys. Rev. Lett.}\ }\textbf {\bibinfo {volume} {106}},\
  \bibinfo {pages} {231101} (\bibinfo {year} {2011})},\ \Eprint
  {https://arxiv.org/abs/1011.1232} {arXiv:1011.1232 [hep-th]} \BibitemShut
  {NoStop}%
\bibitem [{\citenamefont {Dvali}\ \emph {et~al.}(2000)\citenamefont {Dvali},
  \citenamefont {Gabadadze},\ and\ \citenamefont {Porrati}}]{Dvali:2000hr}%
  \BibitemOpen
  \bibfield  {author} {\bibinfo {author} {\bibfnamefont {G.~R.}\ \bibnamefont
  {Dvali}}, \bibinfo {author} {\bibfnamefont {G.}~\bibnamefont {Gabadadze}},\
  and\ \bibinfo {author} {\bibfnamefont {M.}~\bibnamefont {Porrati}},\
  }\bibfield  {title} {\bibinfo {title} {{4-D gravity on a brane in 5-D
  Minkowski space}},\ }\href {https://doi.org/10.1016/S0370-2693(00)00669-9}
  {\bibfield  {journal} {\bibinfo  {journal} {Phys. Lett. B}\ }\textbf
  {\bibinfo {volume} {485}},\ \bibinfo {pages} {208} (\bibinfo {year}
  {2000})},\ \Eprint {https://arxiv.org/abs/hep-th/0005016}
  {arXiv:hep-th/0005016} \BibitemShut {NoStop}%
\bibitem [{\citenamefont {Bergshoeff}\ \emph {et~al.}(2009)\citenamefont
  {Bergshoeff}, \citenamefont {Hohm},\ and\ \citenamefont
  {Townsend}}]{Bergshoeff:2009hq}%
  \BibitemOpen
  \bibfield  {author} {\bibinfo {author} {\bibfnamefont {E.~A.}\ \bibnamefont
  {Bergshoeff}}, \bibinfo {author} {\bibfnamefont {O.}~\bibnamefont {Hohm}},\
  and\ \bibinfo {author} {\bibfnamefont {P.~K.}\ \bibnamefont {Townsend}},\
  }\bibfield  {title} {\bibinfo {title} {{Massive Gravity in Three
  Dimensions}},\ }\href {https://doi.org/10.1103/PhysRevLett.102.201301}
  {\bibfield  {journal} {\bibinfo  {journal} {Phys. Rev. Lett.}\ }\textbf
  {\bibinfo {volume} {102}},\ \bibinfo {pages} {201301} (\bibinfo {year}
  {2009})},\ \Eprint {https://arxiv.org/abs/0901.1766} {arXiv:0901.1766
  [hep-th]} \BibitemShut {NoStop}%
\bibitem [{\citenamefont {de~Rham}(2014)}]{deRham:2014zqa}%
  \BibitemOpen
  \bibfield  {author} {\bibinfo {author} {\bibfnamefont {C.}~\bibnamefont
  {de~Rham}},\ }\bibfield  {title} {\bibinfo {title} {{Massive Gravity}},\
  }\href {https://doi.org/10.12942/lrr-2014-7} {\bibfield  {journal} {\bibinfo
  {journal} {Living Rev. Rel.}\ }\textbf {\bibinfo {volume} {17}},\ \bibinfo
  {pages} {7} (\bibinfo {year} {2014})},\ \Eprint
  {https://arxiv.org/abs/1401.4173} {arXiv:1401.4173 [hep-th]} \BibitemShut
  {NoStop}%
\bibitem [{\citenamefont {Hinterbichler}(2012)}]{Hinterbichler:2011tt}%
  \BibitemOpen
  \bibfield  {author} {\bibinfo {author} {\bibfnamefont {K.}~\bibnamefont
  {Hinterbichler}},\ }\bibfield  {title} {\bibinfo {title} {{Theoretical
  Aspects of Massive Gravity}},\ }\href
  {https://doi.org/10.1103/RevModPhys.84.671} {\bibfield  {journal} {\bibinfo
  {journal} {Rev. Mod. Phys.}\ }\textbf {\bibinfo {volume} {84}},\ \bibinfo
  {pages} {671} (\bibinfo {year} {2012})},\ \Eprint
  {https://arxiv.org/abs/1105.3735} {arXiv:1105.3735 [hep-th]} \BibitemShut
  {NoStop}%
\bibitem [{\citenamefont {Cai}\ \emph {et~al.}(2015)\citenamefont {Cai},
  \citenamefont {Hu}, \citenamefont {Pan},\ and\ \citenamefont
  {Zhang}}]{Cai:2014znn}%
  \BibitemOpen
  \bibfield  {author} {\bibinfo {author} {\bibfnamefont {R.-G.}\ \bibnamefont
  {Cai}}, \bibinfo {author} {\bibfnamefont {Y.-P.}\ \bibnamefont {Hu}},
  \bibinfo {author} {\bibfnamefont {Q.-Y.}\ \bibnamefont {Pan}},\ and\ \bibinfo
  {author} {\bibfnamefont {Y.-L.}\ \bibnamefont {Zhang}},\ }\bibfield  {title}
  {\bibinfo {title} {{Thermodynamics of Black Holes in Massive Gravity}},\
  }\href {https://doi.org/10.1103/PhysRevD.91.024032} {\bibfield  {journal}
  {\bibinfo  {journal} {Phys. Rev. D}\ }\textbf {\bibinfo {volume} {91}},\
  \bibinfo {pages} {024032} (\bibinfo {year} {2015})},\ \Eprint
  {https://arxiv.org/abs/1409.2369} {arXiv:1409.2369 [hep-th]} \BibitemShut
  {NoStop}%
\bibitem [{\citenamefont {Xu}\ \emph {et~al.}(2015)\citenamefont {Xu},
  \citenamefont {Cao},\ and\ \citenamefont {Hu}}]{Xu:2015rfa}%
  \BibitemOpen
  \bibfield  {author} {\bibinfo {author} {\bibfnamefont {J.}~\bibnamefont
  {Xu}}, \bibinfo {author} {\bibfnamefont {L.-M.}\ \bibnamefont {Cao}},\ and\
  \bibinfo {author} {\bibfnamefont {Y.-P.}\ \bibnamefont {Hu}},\ }\bibfield
  {title} {\bibinfo {title} {{P-V criticality in the extended phase space of
  black holes in massive gravity}},\ }\href
  {https://doi.org/10.1103/PhysRevD.91.124033} {\bibfield  {journal} {\bibinfo
  {journal} {Phys. Rev. D}\ }\textbf {\bibinfo {volume} {91}},\ \bibinfo
  {pages} {124033} (\bibinfo {year} {2015})},\ \Eprint
  {https://arxiv.org/abs/1506.03578} {arXiv:1506.03578 [gr-qc]} \BibitemShut
  {NoStop}%
\bibitem [{\citenamefont {Hendi}\ \emph
  {et~al.}(2017{\natexlab{a}})\citenamefont {Hendi}, \citenamefont {Mann},
  \citenamefont {Panahiyan},\ and\ \citenamefont
  {Eslam~Panah}}]{Hendi:2017fxp}%
  \BibitemOpen
  \bibfield  {author} {\bibinfo {author} {\bibfnamefont {S.~H.}\ \bibnamefont
  {Hendi}}, \bibinfo {author} {\bibfnamefont {R.~B.}\ \bibnamefont {Mann}},
  \bibinfo {author} {\bibfnamefont {S.}~\bibnamefont {Panahiyan}},\ and\
  \bibinfo {author} {\bibfnamefont {B.}~\bibnamefont {Eslam~Panah}},\
  }\bibfield  {title} {\bibinfo {title} {{Van der Waals like behavior of
  topological AdS black holes in massive gravity}},\ }\href
  {https://doi.org/10.1103/PhysRevD.95.021501} {\bibfield  {journal} {\bibinfo
  {journal} {Phys. Rev. D}\ }\textbf {\bibinfo {volume} {95}},\ \bibinfo
  {pages} {021501} (\bibinfo {year} {2017}{\natexlab{a}})},\ \Eprint
  {https://arxiv.org/abs/1702.00432} {arXiv:1702.00432 [gr-qc]} \BibitemShut
  {NoStop}%
\bibitem [{\citenamefont {Zou}\ \emph {et~al.}(2017)\citenamefont {Zou},
  \citenamefont {Yue},\ and\ \citenamefont {Zhang}}]{Zou:2016sab}%
  \BibitemOpen
  \bibfield  {author} {\bibinfo {author} {\bibfnamefont {D.-C.}\ \bibnamefont
  {Zou}}, \bibinfo {author} {\bibfnamefont {R.}~\bibnamefont {Yue}},\ and\
  \bibinfo {author} {\bibfnamefont {M.}~\bibnamefont {Zhang}},\ }\bibfield
  {title} {\bibinfo {title} {{Reentrant phase transitions of higher-dimensional
  AdS black holes in dRGT massive gravity}},\ }\href
  {https://doi.org/10.1140/epjc/s10052-017-4822-9} {\bibfield  {journal}
  {\bibinfo  {journal} {Eur. Phys. J. C}\ }\textbf {\bibinfo {volume} {77}},\
  \bibinfo {pages} {256} (\bibinfo {year} {2017})},\ \Eprint
  {https://arxiv.org/abs/1612.08056} {arXiv:1612.08056 [gr-qc]} \BibitemShut
  {NoStop}%
\bibitem [{\citenamefont {Hendi}\ \emph
  {et~al.}(2017{\natexlab{b}})\citenamefont {Hendi}, \citenamefont
  {Eslam~Panah},\ and\ \citenamefont {Panahiyan}}]{Hendi:2016vux}%
  \BibitemOpen
  \bibfield  {author} {\bibinfo {author} {\bibfnamefont {S.~H.}\ \bibnamefont
  {Hendi}}, \bibinfo {author} {\bibfnamefont {B.}~\bibnamefont {Eslam~Panah}},\
  and\ \bibinfo {author} {\bibfnamefont {S.}~\bibnamefont {Panahiyan}},\
  }\bibfield  {title} {\bibinfo {title} {{Topological charged black holes in
  massive gravity's rainbow and their thermodynamical analysis through various
  approaches}},\ }\href {https://doi.org/10.1016/j.physletb.2017.03.051}
  {\bibfield  {journal} {\bibinfo  {journal} {Phys. Lett. B}\ }\textbf
  {\bibinfo {volume} {769}},\ \bibinfo {pages} {191} (\bibinfo {year}
  {2017}{\natexlab{b}})},\ \Eprint {https://arxiv.org/abs/1602.01832}
  {arXiv:1602.01832 [gr-qc]} \BibitemShut {NoStop}%
\bibitem [{\citenamefont {Fernando}(2016)}]{Fernando:2016sps}%
  \BibitemOpen
  \bibfield  {author} {\bibinfo {author} {\bibfnamefont {S.}~\bibnamefont
  {Fernando}},\ }\bibfield  {title} {\bibinfo {title} {{P-V criticality in AdS
  black holes of massive gravity}},\ }\href
  {https://doi.org/10.1103/PhysRevD.94.124049} {\bibfield  {journal} {\bibinfo
  {journal} {Phys. Rev. D}\ }\textbf {\bibinfo {volume} {94}},\ \bibinfo
  {pages} {124049} (\bibinfo {year} {2016})},\ \Eprint
  {https://arxiv.org/abs/1611.05329} {arXiv:1611.05329 [gr-qc]} \BibitemShut
  {NoStop}%
\bibitem [{\citenamefont {Chandrasekhar}(1991)}]{chandrasekhar1991selected}%
  \BibitemOpen
  \bibfield  {author} {\bibinfo {author} {\bibfnamefont {S.}~\bibnamefont
  {Chandrasekhar}},\ }\bibfield  {title} {\bibinfo {title} {Selected papers,
  volume 6: The mathematical theory of black holes and of colliding plane
  waves},\ }\bibfield  {journal} {\bibinfo  {journal} {Selected papers}\ }\href
  {https://doi.org/https://books.google.co.in/books?id=GzrV2PHVs0oC}
  {https://books.google.co.in/books?id=GzrV2PHVs0oC} (\bibinfo {year}
  {1991})\BibitemShut {NoStop}%
\bibitem [{\citenamefont {Cornish}\ and\ \citenamefont
  {Levin}(2003)}]{Cornish:2003ig}%
  \BibitemOpen
  \bibfield  {author} {\bibinfo {author} {\bibfnamefont {N.~J.}\ \bibnamefont
  {Cornish}}\ and\ \bibinfo {author} {\bibfnamefont {J.~J.}\ \bibnamefont
  {Levin}},\ }\bibfield  {title} {\bibinfo {title} {{Lyapunov timescales and
  black hole binaries}},\ }\href {https://doi.org/10.1088/0264-9381/20/9/304}
  {\bibfield  {journal} {\bibinfo  {journal} {Class. Quant. Grav.}\ }\textbf
  {\bibinfo {volume} {20}},\ \bibinfo {pages} {1649} (\bibinfo {year}
  {2003})},\ \Eprint {https://arxiv.org/abs/gr-qc/0304056}
  {arXiv:gr-qc/0304056} \BibitemShut {NoStop}%
\bibitem [{\citenamefont {Bardeen}(1968)}]{bardeen1968non}%
  \BibitemOpen
  \bibfield  {author} {\bibinfo {author} {\bibfnamefont {J.}~\bibnamefont
  {Bardeen}},\ }\bibfield  {title} {\bibinfo {title} {Non-singular general
  relativistic gravitational collapse},\ }in\ \href@noop {} {\emph {\bibinfo
  {booktitle} {Proceedings of the 5th International Conference on Gravitation
  and the Theory of Relativity}}}\ (\bibinfo {year} {1968})\ p.~\bibinfo
  {pages} {87}\BibitemShut {NoStop}%
\bibitem [{\citenamefont {Bueno}\ \emph {et~al.}(2024)\citenamefont {Bueno},
  \citenamefont {Cano},\ and\ \citenamefont {Hennigar}}]{Bueno:2024dgm}%
  \BibitemOpen
  \bibfield  {author} {\bibinfo {author} {\bibfnamefont {P.}~\bibnamefont
  {Bueno}}, \bibinfo {author} {\bibfnamefont {P.~A.}\ \bibnamefont {Cano}},\
  and\ \bibinfo {author} {\bibfnamefont {R.~A.}\ \bibnamefont {Hennigar}},\
  }\bibfield  {title} {\bibinfo {title} {{Regular Black Holes From Pure
  Gravity}},\ }\href@noop {} {\  (\bibinfo {year} {2024})},\ \Eprint
  {https://arxiv.org/abs/2403.04827} {arXiv:2403.04827 [gr-qc]} \BibitemShut
  {NoStop}%
\bibitem [{\citenamefont {Hayward}(2006)}]{Hayward:2005gi}%
  \BibitemOpen
  \bibfield  {author} {\bibinfo {author} {\bibfnamefont {S.~A.}\ \bibnamefont
  {Hayward}},\ }\bibfield  {title} {\bibinfo {title} {{Formation and
  evaporation of regular black holes}},\ }\href
  {https://doi.org/10.1103/PhysRevLett.96.031103} {\bibfield  {journal}
  {\bibinfo  {journal} {Phys. Rev. Lett.}\ }\textbf {\bibinfo {volume} {96}},\
  \bibinfo {pages} {031103} (\bibinfo {year} {2006})},\ \Eprint
  {https://arxiv.org/abs/gr-qc/0506126} {arXiv:gr-qc/0506126} \BibitemShut
  {NoStop}%
\bibitem [{\citenamefont {Bronnikov}(2001)}]{Bronnikov:2000vy}%
  \BibitemOpen
  \bibfield  {author} {\bibinfo {author} {\bibfnamefont {K.~A.}\ \bibnamefont
  {Bronnikov}},\ }\bibfield  {title} {\bibinfo {title} {{Regular magnetic black
  holes and monopoles from nonlinear electrodynamics}},\ }\href
  {https://doi.org/10.1103/PhysRevD.63.044005} {\bibfield  {journal} {\bibinfo
  {journal} {Phys. Rev. D}\ }\textbf {\bibinfo {volume} {63}},\ \bibinfo
  {pages} {044005} (\bibinfo {year} {2001})},\ \Eprint
  {https://arxiv.org/abs/gr-qc/0006014} {arXiv:gr-qc/0006014} \BibitemShut
  {NoStop}%
\bibitem [{\citenamefont {Berej}\ \emph {et~al.}(2006)\citenamefont {Berej},
  \citenamefont {Matyjasek}, \citenamefont {Tryniecki},\ and\ \citenamefont
  {Woronowicz}}]{Berej:2006cc}%
  \BibitemOpen
  \bibfield  {author} {\bibinfo {author} {\bibfnamefont {W.}~\bibnamefont
  {Berej}}, \bibinfo {author} {\bibfnamefont {J.}~\bibnamefont {Matyjasek}},
  \bibinfo {author} {\bibfnamefont {D.}~\bibnamefont {Tryniecki}},\ and\
  \bibinfo {author} {\bibfnamefont {M.}~\bibnamefont {Woronowicz}},\ }\bibfield
   {title} {\bibinfo {title} {{Regular black holes in quadratic gravity}},\
  }\href {https://doi.org/10.1007/s10714-006-0270-9} {\bibfield  {journal}
  {\bibinfo  {journal} {Gen. Rel. Grav.}\ }\textbf {\bibinfo {volume} {38}},\
  \bibinfo {pages} {885} (\bibinfo {year} {2006})},\ \Eprint
  {https://arxiv.org/abs/hep-th/0606185} {arXiv:hep-th/0606185} \BibitemShut
  {NoStop}%
\bibitem [{\citenamefont {Dymnikova}(2004)}]{Dymnikova:2004zc}%
  \BibitemOpen
  \bibfield  {author} {\bibinfo {author} {\bibfnamefont {I.}~\bibnamefont
  {Dymnikova}},\ }\bibfield  {title} {\bibinfo {title} {{Regular electrically
  charged structures in nonlinear electrodynamics coupled to general
  relativity}},\ }\href {https://doi.org/10.1088/0264-9381/21/18/009}
  {\bibfield  {journal} {\bibinfo  {journal} {Class. Quant. Grav.}\ }\textbf
  {\bibinfo {volume} {21}},\ \bibinfo {pages} {4417} (\bibinfo {year}
  {2004})},\ \Eprint {https://arxiv.org/abs/gr-qc/0407072}
  {arXiv:gr-qc/0407072} \BibitemShut {NoStop}%
\bibitem [{\citenamefont {Ayon-Beato}\ and\ \citenamefont
  {Garcia}(1998)}]{Ayon-Beato:1998hmi}%
  \BibitemOpen
  \bibfield  {author} {\bibinfo {author} {\bibfnamefont {E.}~\bibnamefont
  {Ayon-Beato}}\ and\ \bibinfo {author} {\bibfnamefont {A.}~\bibnamefont
  {Garcia}},\ }\bibfield  {title} {\bibinfo {title} {{Regular black hole in
  general relativity coupled to nonlinear electrodynamics}},\ }\href
  {https://doi.org/10.1103/PhysRevLett.80.5056} {\bibfield  {journal} {\bibinfo
   {journal} {Phys. Rev. Lett.}\ }\textbf {\bibinfo {volume} {80}},\ \bibinfo
  {pages} {5056} (\bibinfo {year} {1998})},\ \Eprint
  {https://arxiv.org/abs/gr-qc/9911046} {arXiv:gr-qc/9911046} \BibitemShut
  {NoStop}%
\bibitem [{\citenamefont {Bronnikov}\ and\ \citenamefont
  {Fabris}(2006)}]{Bronnikov:2005gm}%
  \BibitemOpen
  \bibfield  {author} {\bibinfo {author} {\bibfnamefont {K.~A.}\ \bibnamefont
  {Bronnikov}}\ and\ \bibinfo {author} {\bibfnamefont {J.~C.}\ \bibnamefont
  {Fabris}},\ }\bibfield  {title} {\bibinfo {title} {{Regular phantom black
  holes}},\ }\href {https://doi.org/10.1103/PhysRevLett.96.251101} {\bibfield
  {journal} {\bibinfo  {journal} {Phys. Rev. Lett.}\ }\textbf {\bibinfo
  {volume} {96}},\ \bibinfo {pages} {251101} (\bibinfo {year} {2006})},\
  \Eprint {https://arxiv.org/abs/gr-qc/0511109} {arXiv:gr-qc/0511109}
  \BibitemShut {NoStop}%
\bibitem [{\citenamefont {Tzikas}(2019{\natexlab{a}})}]{tzikas2019bardeen}%
  \BibitemOpen
  \bibfield  {author} {\bibinfo {author} {\bibfnamefont {A.~G.}\ \bibnamefont
  {Tzikas}},\ }\bibfield  {title} {\bibinfo {title} {{Bardeen black hole
  chemistry}},\ }\href {https://doi.org/10.1016/j.physletb.2018.11.036}
  {\bibfield  {journal} {\bibinfo  {journal} {Phys. Lett. B}\ }\textbf
  {\bibinfo {volume} {788}},\ \bibinfo {pages} {219} (\bibinfo {year}
  {2019}{\natexlab{a}})},\ \Eprint {https://arxiv.org/abs/1811.01104}
  {arXiv:1811.01104 [gr-qc]} \BibitemShut {NoStop}%
\bibitem [{\citenamefont {Singh}\ and\ \citenamefont
  {Siwach}(2020)}]{singh2020thermodynamics}%
  \BibitemOpen
  \bibfield  {author} {\bibinfo {author} {\bibfnamefont {D.~V.}\ \bibnamefont
  {Singh}}\ and\ \bibinfo {author} {\bibfnamefont {S.}~\bibnamefont {Siwach}},\
  }\bibfield  {title} {\bibinfo {title} {{Thermodynamics and P-v criticality of
  Bardeen-AdS Black Hole in 4$D$ Einstein-Gauss-Bonnet Gravity}},\ }\href
  {https://doi.org/10.1016/j.physletb.2020.135658} {\bibfield  {journal}
  {\bibinfo  {journal} {Phys. Lett. B}\ }\textbf {\bibinfo {volume} {808}},\
  \bibinfo {pages} {135658} (\bibinfo {year} {2020})},\ \Eprint
  {https://arxiv.org/abs/2003.11754} {arXiv:2003.11754 [gr-qc]} \BibitemShut
  {NoStop}%
\bibitem [{\citenamefont {Pu}\ \emph {et~al.}(2020)\citenamefont {Pu},
  \citenamefont {Guo}, \citenamefont {Jiang},\ and\ \citenamefont
  {Zu}}]{guo2020joule}%
  \BibitemOpen
  \bibfield  {author} {\bibinfo {author} {\bibfnamefont {J.}~\bibnamefont
  {Pu}}, \bibinfo {author} {\bibfnamefont {S.}~\bibnamefont {Guo}}, \bibinfo
  {author} {\bibfnamefont {Q.-Q.}\ \bibnamefont {Jiang}},\ and\ \bibinfo
  {author} {\bibfnamefont {X.-T.}\ \bibnamefont {Zu}},\ }\bibfield  {title}
  {\bibinfo {title} {{Joule-Thomson expansion of the regular(Bardeen)-AdS black
  hole}},\ }\href {https://doi.org/10.1088/1674-1137/44/3/035102} {\bibfield
  {journal} {\bibinfo  {journal} {Chin. Phys. C}\ }\textbf {\bibinfo {volume}
  {44}},\ \bibinfo {pages} {035102} (\bibinfo {year} {2020})},\ \Eprint
  {https://arxiv.org/abs/1905.02318} {arXiv:1905.02318 [gr-qc]} \BibitemShut
  {NoStop}%
\bibitem [{\citenamefont {Zhang}\ and\ \citenamefont
  {Gao}(2018)}]{Zhang:2016ilt}%
  \BibitemOpen
  \bibfield  {author} {\bibinfo {author} {\bibfnamefont {Y.}~\bibnamefont
  {Zhang}}\ and\ \bibinfo {author} {\bibfnamefont {S.}~\bibnamefont {Gao}},\
  }\bibfield  {title} {\bibinfo {title} {{First law and Smarr formula of black
  hole mechanics in nonlinear gauge theories}},\ }\href
  {https://doi.org/10.1088/1361-6382/aac9d4} {\bibfield  {journal} {\bibinfo
  {journal} {Class. Quant. Grav.}\ }\textbf {\bibinfo {volume} {35}},\ \bibinfo
  {pages} {145007} (\bibinfo {year} {2018})},\ \Eprint
  {https://arxiv.org/abs/1610.01237} {arXiv:1610.01237 [gr-qc]} \BibitemShut
  {NoStop}%
\bibitem [{\citenamefont {Guo}\ and\ \citenamefont {Miao}(2022)}]{Guo:2021wcf}%
  \BibitemOpen
  \bibfield  {author} {\bibinfo {author} {\bibfnamefont {Y.}~\bibnamefont
  {Guo}}\ and\ \bibinfo {author} {\bibfnamefont {Y.-G.}\ \bibnamefont {Miao}},\
  }\bibfield  {title} {\bibinfo {title} {{Weinhold geometry and thermodynamics
  of Bardeen AdS black holes}},\ }\href
  {https://doi.org/10.1016/j.nuclphysb.2022.115839} {\bibfield  {journal}
  {\bibinfo  {journal} {Nucl. Phys. B}\ }\textbf {\bibinfo {volume} {980}},\
  \bibinfo {pages} {115839} (\bibinfo {year} {2022})},\ \Eprint
  {https://arxiv.org/abs/2107.01866} {arXiv:2107.01866 [gr-qc]} \BibitemShut
  {NoStop}%
\bibitem [{\citenamefont {Tzikas}(2019{\natexlab{b}})}]{Tzikas:2018cvs}%
  \BibitemOpen
  \bibfield  {author} {\bibinfo {author} {\bibfnamefont {A.~G.}\ \bibnamefont
  {Tzikas}},\ }\bibfield  {title} {\bibinfo {title} {{Bardeen black hole
  chemistry}},\ }\href {https://doi.org/10.1016/j.physletb.2018.11.036}
  {\bibfield  {journal} {\bibinfo  {journal} {Phys. Lett. B}\ }\textbf
  {\bibinfo {volume} {788}},\ \bibinfo {pages} {219} (\bibinfo {year}
  {2019}{\natexlab{b}})},\ \Eprint {https://arxiv.org/abs/1811.01104}
  {arXiv:1811.01104 [gr-qc]} \BibitemShut {NoStop}%
\bibitem [{\citenamefont {Li}\ \emph {et~al.}(2019)\citenamefont {Li},
  \citenamefont {Fang}, \citenamefont {He}, \citenamefont {Ding},\ and\
  \citenamefont {Deng}}]{li2019thermodynamics}%
  \BibitemOpen
  \bibfield  {author} {\bibinfo {author} {\bibfnamefont {C.}~\bibnamefont
  {Li}}, \bibinfo {author} {\bibfnamefont {C.}~\bibnamefont {Fang}}, \bibinfo
  {author} {\bibfnamefont {M.}~\bibnamefont {He}}, \bibinfo {author}
  {\bibfnamefont {J.}~\bibnamefont {Ding}},\ and\ \bibinfo {author}
  {\bibfnamefont {J.}~\bibnamefont {Deng}},\ }\bibfield  {title} {\bibinfo
  {title} {{Thermodynamics of the Bardeen Black Hole in Anti-de Sitter
  Space}},\ }\href {https://doi.org/10.1142/S021773231950336X} {\bibfield
  {journal} {\bibinfo  {journal} {Mod. Phys. Lett. A}\ }\textbf {\bibinfo
  {volume} {34}},\ \bibinfo {pages} {1950336} (\bibinfo {year} {2019})},\
  \Eprint {https://arxiv.org/abs/1812.02567} {arXiv:1812.02567 [hep-th]}
  \BibitemShut {NoStop}%
\bibitem [{\citenamefont {Cai}(2002)}]{Cai:2001dz}%
  \BibitemOpen
  \bibfield  {author} {\bibinfo {author} {\bibfnamefont {R.-G.}\ \bibnamefont
  {Cai}},\ }\bibfield  {title} {\bibinfo {title} {{Gauss-Bonnet black holes in
  AdS spaces}},\ }\href {https://doi.org/10.1103/PhysRevD.65.084014} {\bibfield
   {journal} {\bibinfo  {journal} {Phys. Rev. D}\ }\textbf {\bibinfo {volume}
  {65}},\ \bibinfo {pages} {084014} (\bibinfo {year} {2002})},\ \Eprint
  {https://arxiv.org/abs/hep-th/0109133} {arXiv:hep-th/0109133} \BibitemShut
  {NoStop}%
\bibitem [{\citenamefont {Pani}\ and\ \citenamefont
  {Cardoso}(2009)}]{Pani:2009wy}%
  \BibitemOpen
  \bibfield  {author} {\bibinfo {author} {\bibfnamefont {P.}~\bibnamefont
  {Pani}}\ and\ \bibinfo {author} {\bibfnamefont {V.}~\bibnamefont {Cardoso}},\
  }\bibfield  {title} {\bibinfo {title} {{Are black holes in alternative
  theories serious astrophysical candidates? The Case for
  Einstein-Dilaton-Gauss-Bonnet black holes}},\ }\href
  {https://doi.org/10.1103/PhysRevD.79.084031} {\bibfield  {journal} {\bibinfo
  {journal} {Phys. Rev. D}\ }\textbf {\bibinfo {volume} {79}},\ \bibinfo
  {pages} {084031} (\bibinfo {year} {2009})},\ \Eprint
  {https://arxiv.org/abs/0902.1569} {arXiv:0902.1569 [gr-qc]} \BibitemShut
  {NoStop}%
\bibitem [{\citenamefont {Clunan}\ \emph {et~al.}(2004)\citenamefont {Clunan},
  \citenamefont {Ross},\ and\ \citenamefont {Smith}}]{Clunan:2004tb}%
  \BibitemOpen
  \bibfield  {author} {\bibinfo {author} {\bibfnamefont {T.}~\bibnamefont
  {Clunan}}, \bibinfo {author} {\bibfnamefont {S.~F.}\ \bibnamefont {Ross}},\
  and\ \bibinfo {author} {\bibfnamefont {D.~J.}\ \bibnamefont {Smith}},\
  }\bibfield  {title} {\bibinfo {title} {{On Gauss-Bonnet black hole
  entropy}},\ }\href {https://doi.org/10.1088/0264-9381/21/14/009} {\bibfield
  {journal} {\bibinfo  {journal} {Class. Quant. Grav.}\ }\textbf {\bibinfo
  {volume} {21}},\ \bibinfo {pages} {3447} (\bibinfo {year} {2004})},\ \Eprint
  {https://arxiv.org/abs/gr-qc/0402044} {arXiv:gr-qc/0402044} \BibitemShut
  {NoStop}%
\bibitem [{\citenamefont {Wei}\ and\ \citenamefont {Liu}(2014)}]{Wei:2014hba}%
  \BibitemOpen
  \bibfield  {author} {\bibinfo {author} {\bibfnamefont {S.-W.}\ \bibnamefont
  {Wei}}\ and\ \bibinfo {author} {\bibfnamefont {Y.-X.}\ \bibnamefont {Liu}},\
  }\bibfield  {title} {\bibinfo {title} {{Triple points and phase diagrams in
  the extended phase space of charged Gauss-Bonnet black holes in AdS space}},\
  }\href {https://doi.org/10.1103/PhysRevD.90.044057} {\bibfield  {journal}
  {\bibinfo  {journal} {Phys. Rev. D}\ }\textbf {\bibinfo {volume} {90}},\
  \bibinfo {pages} {044057} (\bibinfo {year} {2014})},\ \Eprint
  {https://arxiv.org/abs/1402.2837} {arXiv:1402.2837 [hep-th]} \BibitemShut
  {NoStop}%
\bibitem [{\citenamefont {Zhou}\ \emph {et~al.}(2020)\citenamefont {Zhou},
  \citenamefont {Liu},\ and\ \citenamefont {Wei}}]{Zhou:2020vzf}%
  \BibitemOpen
  \bibfield  {author} {\bibinfo {author} {\bibfnamefont {R.}~\bibnamefont
  {Zhou}}, \bibinfo {author} {\bibfnamefont {Y.-X.}\ \bibnamefont {Liu}},\ and\
  \bibinfo {author} {\bibfnamefont {S.-W.}\ \bibnamefont {Wei}},\ }\bibfield
  {title} {\bibinfo {title} {{Phase transition and microstructures of
  five-dimensional charged Gauss-Bonnet-AdS black holes in the grand canonical
  ensemble}},\ }\href {https://doi.org/10.1103/PhysRevD.102.124015} {\bibfield
  {journal} {\bibinfo  {journal} {Phys. Rev. D}\ }\textbf {\bibinfo {volume}
  {102}},\ \bibinfo {pages} {124015} (\bibinfo {year} {2020})},\ \Eprint
  {https://arxiv.org/abs/2008.08301} {arXiv:2008.08301 [gr-qc]} \BibitemShut
  {NoStop}%
\bibitem [{\citenamefont {de~Rham}\ \emph {et~al.}(2017)\citenamefont
  {de~Rham}, \citenamefont {Deskins}, \citenamefont {Tolley},\ and\
  \citenamefont {Zhou}}]{deRham:2016nuf}%
  \BibitemOpen
  \bibfield  {author} {\bibinfo {author} {\bibfnamefont {C.}~\bibnamefont
  {de~Rham}}, \bibinfo {author} {\bibfnamefont {J.~T.}\ \bibnamefont
  {Deskins}}, \bibinfo {author} {\bibfnamefont {A.~J.}\ \bibnamefont
  {Tolley}},\ and\ \bibinfo {author} {\bibfnamefont {S.-Y.}\ \bibnamefont
  {Zhou}},\ }\bibfield  {title} {\bibinfo {title} {{Graviton Mass Bounds}},\
  }\href {https://doi.org/10.1103/RevModPhys.89.025004} {\bibfield  {journal}
  {\bibinfo  {journal} {Rev. Mod. Phys.}\ }\textbf {\bibinfo {volume} {89}},\
  \bibinfo {pages} {025004} (\bibinfo {year} {2017})},\ \Eprint
  {https://arxiv.org/abs/1606.08462} {arXiv:1606.08462 [astro-ph.CO]}
  \BibitemShut {NoStop}%
\bibitem [{\citenamefont {Dubovsky}(2004)}]{Dubovsky:2004sg}%
  \BibitemOpen
  \bibfield  {author} {\bibinfo {author} {\bibfnamefont {S.~L.}\ \bibnamefont
  {Dubovsky}},\ }\bibfield  {title} {\bibinfo {title} {{Phases of massive
  gravity}},\ }\href {https://doi.org/10.1088/1126-6708/2004/10/076} {\bibfield
   {journal} {\bibinfo  {journal} {JHEP}\ }\textbf {\bibinfo {volume} {10}},\
  \bibinfo {pages} {076}},\ \Eprint {https://arxiv.org/abs/hep-th/0409124}
  {arXiv:hep-th/0409124} \BibitemShut {NoStop}%
\bibitem [{\citenamefont {Rubakov}\ and\ \citenamefont
  {Tinyakov}(2008)}]{Rubakov:2008nh}%
  \BibitemOpen
  \bibfield  {author} {\bibinfo {author} {\bibfnamefont {V.~A.}\ \bibnamefont
  {Rubakov}}\ and\ \bibinfo {author} {\bibfnamefont {P.~G.}\ \bibnamefont
  {Tinyakov}},\ }\bibfield  {title} {\bibinfo {title} {{Infrared-modified
  gravities and massive gravitons}},\ }\href
  {https://doi.org/10.1070/PU2008v051n08ABEH006600} {\bibfield  {journal}
  {\bibinfo  {journal} {Phys. Usp.}\ }\textbf {\bibinfo {volume} {51}},\
  \bibinfo {pages} {759} (\bibinfo {year} {2008})},\ \Eprint
  {https://arxiv.org/abs/0802.4379} {arXiv:0802.4379 [hep-th]} \BibitemShut
  {NoStop}%
\bibitem [{\citenamefont {Bebronne}\ and\ \citenamefont
  {Tinyakov}(2009)}]{Bebronne:2009mz}%
  \BibitemOpen
  \bibfield  {author} {\bibinfo {author} {\bibfnamefont {M.~V.}\ \bibnamefont
  {Bebronne}}\ and\ \bibinfo {author} {\bibfnamefont {P.~G.}\ \bibnamefont
  {Tinyakov}},\ }\bibfield  {title} {\bibinfo {title} {{Black hole solutions in
  massive gravity}},\ }\href {https://doi.org/10.1007/JHEP06(2011)018}
  {\bibfield  {journal} {\bibinfo  {journal} {JHEP}\ }\textbf {\bibinfo
  {volume} {04}},\ \bibinfo {pages} {100}},\ \bibinfo {note} {[Erratum: JHEP
  06, 018 (2011)]},\ \Eprint {https://arxiv.org/abs/0902.3899} {arXiv:0902.3899
  [gr-qc]} \BibitemShut {NoStop}%
\bibitem [{\citenamefont {Comelli}\ \emph {et~al.}(2011)\citenamefont
  {Comelli}, \citenamefont {Nesti},\ and\ \citenamefont
  {Pilo}}]{Comelli:2010bj}%
  \BibitemOpen
  \bibfield  {author} {\bibinfo {author} {\bibfnamefont {D.}~\bibnamefont
  {Comelli}}, \bibinfo {author} {\bibfnamefont {F.}~\bibnamefont {Nesti}},\
  and\ \bibinfo {author} {\bibfnamefont {L.}~\bibnamefont {Pilo}},\ }\bibfield
  {title} {\bibinfo {title} {{Stars and (Furry) Black Holes in Lorentz Breaking
  Massive Gravity}},\ }\href {https://doi.org/10.1103/PhysRevD.83.084042}
  {\bibfield  {journal} {\bibinfo  {journal} {Phys. Rev. D}\ }\textbf {\bibinfo
  {volume} {83}},\ \bibinfo {pages} {084042} (\bibinfo {year} {2011})},\
  \Eprint {https://arxiv.org/abs/1010.4773} {arXiv:1010.4773 [hep-th]}
  \BibitemShut {NoStop}%
\bibitem [{\citenamefont {Goldenfeld}(2019)}]{Goldenfeld2019Jun}%
  \BibitemOpen
  \bibfield  {author} {\bibinfo {author} {\bibfnamefont {N.}~\bibnamefont
  {Goldenfeld}},\ }\href {https://doi.org/10.1201/9780429493492} {\emph
  {\bibinfo {title} {{Lectures On Phase Transitions And The Renormalization
  Group}}}}\ (\bibinfo  {publisher} {Taylor {\&} Francis},\ \bibinfo {address}
  {Andover, England, UK},\ \bibinfo {year} {2019})\BibitemShut {NoStop}%
\bibitem [{\citenamefont {Wang}(2021)}]{Wang2021May}%
  \BibitemOpen
  \bibfield  {author} {\bibinfo {author} {\bibfnamefont {J.-S.}\ \bibnamefont
  {Wang}},\ }\href {https://doi.org/10.1142/12414} {\emph {\bibinfo {title}
  {{Advanced Statistical Mechanics}}}}\ (\bibinfo  {publisher} {World
  Scientific Publishing Company},\ \bibinfo {address} {Singapore},\ \bibinfo
  {year} {2021})\BibitemShut {NoStop}%
\bibitem [{\citenamefont {Beekman}\ \emph {et~al.}(2019)\citenamefont
  {Beekman}, \citenamefont {Rademaker},\ and\ \citenamefont {van
  Wezel}}]{Aron/SciPostPhysLectNotes.11}%
  \BibitemOpen
  \bibfield  {author} {\bibinfo {author} {\bibfnamefont {A.~J.}\ \bibnamefont
  {Beekman}}, \bibinfo {author} {\bibfnamefont {L.}~\bibnamefont {Rademaker}},\
  and\ \bibinfo {author} {\bibfnamefont {J.}~\bibnamefont {van Wezel}},\
  }\bibfield  {title} {\bibinfo {title} {{An introduction to spontaneous
  symmetry breaking}},\ }\href
  {https://doi.org/10.21468/SciPostPhysLectNotes.11} {\bibfield  {journal}
  {\bibinfo  {journal} {SciPost Phys. Lect. Notes}\ ,\ \bibinfo {pages} {11}}
  (\bibinfo {year} {2019})}\BibitemShut {NoStop}%
\bibitem [{\citenamefont {Van Der~Waals}\ and\ \citenamefont
  {Rowlinson}(2004)}]{van2004continuity}%
  \BibitemOpen
  \bibfield  {author} {\bibinfo {author} {\bibfnamefont {J.}~\bibnamefont {Van
  Der~Waals}}\ and\ \bibinfo {author} {\bibfnamefont {J.}~\bibnamefont
  {Rowlinson}},\ }\href {https://books.google.co.in/books?id=5Iuxbxvg7l0C}
  {\emph {\bibinfo {title} {On the Continuity of the Gaseous and Liquid
  States}}},\ Dover Books on Physics Series\ (\bibinfo  {publisher} {Dover
  Publications},\ \bibinfo {year} {2004})\BibitemShut {NoStop}%
\bibitem [{\citenamefont {Chamblin}\ \emph
  {et~al.}(1999{\natexlab{b}})\citenamefont {Chamblin}, \citenamefont
  {Emparan}, \citenamefont {Johnson},\ and\ \citenamefont
  {Myers}}]{Chamblin:1999hg}%
  \BibitemOpen
  \bibfield  {author} {\bibinfo {author} {\bibfnamefont {A.}~\bibnamefont
  {Chamblin}}, \bibinfo {author} {\bibfnamefont {R.}~\bibnamefont {Emparan}},
  \bibinfo {author} {\bibfnamefont {C.~V.}\ \bibnamefont {Johnson}},\ and\
  \bibinfo {author} {\bibfnamefont {R.~C.}\ \bibnamefont {Myers}},\ }\bibfield
  {title} {\bibinfo {title} {{Holography, thermodynamics and fluctuations of
  charged AdS black holes}},\ }\href
  {https://doi.org/10.1103/PhysRevD.60.104026} {\bibfield  {journal} {\bibinfo
  {journal} {Phys. Rev. D}\ }\textbf {\bibinfo {volume} {60}},\ \bibinfo
  {pages} {104026} (\bibinfo {year} {1999}{\natexlab{b}})},\ \Eprint
  {https://arxiv.org/abs/hep-th/9904197} {arXiv:hep-th/9904197} \BibitemShut
  {NoStop}%
\bibitem [{\citenamefont {Banerjee}\ and\ \citenamefont
  {Roychowdhury}(2012)}]{Banerjee:2012zm}%
  \BibitemOpen
  \bibfield  {author} {\bibinfo {author} {\bibfnamefont {R.}~\bibnamefont
  {Banerjee}}\ and\ \bibinfo {author} {\bibfnamefont {D.}~\bibnamefont
  {Roychowdhury}},\ }\bibfield  {title} {\bibinfo {title} {{Critical behavior
  of Born Infeld AdS black holes in higher dimensions}},\ }\href
  {https://doi.org/10.1103/PhysRevD.85.104043} {\bibfield  {journal} {\bibinfo
  {journal} {Phys. Rev. D}\ }\textbf {\bibinfo {volume} {85}},\ \bibinfo
  {pages} {104043} (\bibinfo {year} {2012})},\ \Eprint
  {https://arxiv.org/abs/1203.0118} {arXiv:1203.0118 [gr-qc]} \BibitemShut
  {NoStop}%
\bibitem [{\citenamefont {Barrow}(1981)}]{PhysRevLett.46.963}%
  \BibitemOpen
  \bibfield  {author} {\bibinfo {author} {\bibfnamefont {J.~D.}\ \bibnamefont
  {Barrow}},\ }\bibfield  {title} {\bibinfo {title} {Chaos in the einstein
  equations},\ }\href {https://doi.org/10.1103/PhysRevLett.46.963} {\bibfield
  {journal} {\bibinfo  {journal} {Phys. Rev. Lett.}\ }\textbf {\bibinfo
  {volume} {46}},\ \bibinfo {pages} {963} (\bibinfo {year} {1981})}\BibitemShut
  {NoStop}%
\bibitem [{\citenamefont {Barrow}(1982)}]{Barrow:1981sx}%
  \BibitemOpen
  \bibfield  {author} {\bibinfo {author} {\bibfnamefont {J.~D.}\ \bibnamefont
  {Barrow}},\ }\bibfield  {title} {\bibinfo {title} {{Chaotic behavior in
  general relativity}},\ }\href {https://doi.org/10.1016/0370-1573(82)90171-5}
  {\bibfield  {journal} {\bibinfo  {journal} {Phys. Rept.}\ }\textbf {\bibinfo
  {volume} {85}},\ \bibinfo {pages} {1} (\bibinfo {year} {1982})}\BibitemShut
  {NoStop}%
\bibitem [{\citenamefont {Cornish}\ and\ \citenamefont
  {Levin}(1997)}]{Cornish:1996yg}%
  \BibitemOpen
  \bibfield  {author} {\bibinfo {author} {\bibfnamefont {N.~J.}\ \bibnamefont
  {Cornish}}\ and\ \bibinfo {author} {\bibfnamefont {J.~J.}\ \bibnamefont
  {Levin}},\ }\bibfield  {title} {\bibinfo {title} {{The Mixmaster universe is
  chaotic}},\ }\href {https://doi.org/10.1103/PhysRevLett.78.998} {\bibfield
  {journal} {\bibinfo  {journal} {Phys. Rev. Lett.}\ }\textbf {\bibinfo
  {volume} {78}},\ \bibinfo {pages} {998} (\bibinfo {year} {1997})},\ \Eprint
  {https://arxiv.org/abs/gr-qc/9605029} {arXiv:gr-qc/9605029} \BibitemShut
  {NoStop}%
\end{thebibliography}%

\end{document}